\documentclass[12pt]{ullthesis2017}
\usepackage[vlined,linesnumbered,ruled,boxed]{algorithm2e}
 \usepackage{amsmath}
 \usepackage{amsfonts}
 \usepackage{amssymb}
 \usepackage{enumitem}
 \usepackage{fancyvrb}
 \usepackage[usenames]{color}
 \usepackage{layout}
 \usepackage{mydef}
 \usepackage{url}
\usepackage{bibentry}
\usepackage{tabularx} %table
\usepackage{algorithmic}

\nobibliography*

\usepackage[bookmarks=false]{hyperref}
 \usepackage{wrapfig,color,caption,subcaption,balance,multirow}
 \usepackage[vlined,linesnumbered,ruled,boxed]{algorithm2e}
 \usepackage{indentfirst}
\usepackage{etoolbox}% http://ctan.org/pkg/etoolbox
\makeatletter
\patchcmd{\@chapter}{\addtocontents{lof}{\protect\addvspace{10\p@}}}{}{}{}% LoF
\patchcmd{\@chapter}{\addtocontents{lot}{\protect\addvspace{10\p@}}}{}{}{}% LoT
\makeatother

\makeatletter
\AtBeginEnvironment{thebibliography}{%
   \clubpenalty10000
   \@clubpenalty \clubpenalty
   \widowpenalty10000}
\makeatother

\newcommand{\etal}{{\it et~al. }}
\newcommand{\comments}[1]{}

\newlength{\boxfigwidth}
\newcommand{\boxfig}[1]{
\begin{figure}[htbp]
\begin{center}
\begin{small}
\setlength{\boxfigwidth}{\textwidth}
\addtolength{\boxfigwidth}{0in}
\noindent\framebox{\quad\begin{minipage}{\boxfigwidth}
#1
\end{minipage}\quad}
\end{small}
\end{center}
\end{figure}
}

\usepackage{titlesec}
 \titleformat{\chapter}[block]
{\singlespacing\bfseries\filcenter}{\chaptertitlename\ \thechapter: \ }{0pt}{}
\titlespacing*{\chapter}
{0pt}{-0.5\baselineskip}{0pt}  

\titleformat{\section}
{\bfseries}{\thesection\ }{1em}{}
\titlespacing*{\section}{0pt}{*0}{0pt}

\titleformat{\subsection}[runin]
{\bfseries}{\thesubsection}{1em}{} [ ]
\titlespacing{\subsection}
{\parindent}{*0}{0pt}

\titleformat{\subsubsection}[runin]
{\itshape\bfseries}{\thesubsubsection}{1em}{}[ ] 
\titlespacing{\subsubsection}
{\parindent}{*0}{0pt}
\raggedright
\usepackage{tocloft}  

 \def\BibTeX{{\rm B\kern-.05em{\sc i\kern-.025em b}\kern-.08em
    T\kern-.1667em\lower.7ex\hbox{E}\kern-.125emX}}

\begin{document}

    \title{Cost- and QoS-Efficient Serverless Cloud Computing}
    \author{Chavit Denninnart}
    \convocationdate{Fall}
    \gradyear{2020}
    \degree{Doctor of Science}
    \major{Computer Science}
    \supervisor{Mohsen Amini Salehi}
   \ranksupervisor{Assistant Professor of Computer Science \\ The Center for Advanced Computer Studies}
%   \cosupervisor{Co--supervisor}      % remove "%" if applicable
%   \rankcosupervisor{Professor of Statistics}
    \deanofgraduateschool{Mary Farmer-Kaiser}
    \firstcommitteemember{Nian-Feng Tzeng}
  \rankfirstcommitteemember{Professor of Computer Science \\ The Center for Advanced Computer Studies}
    \secondcommitteemember{Sheng Chen}
  \ranksecondcommitteemember{Assistant Professor of Computer Science \\ The Center for Advanced Computer Studies}
    \thirdcommitteemember{Miao Jin}       
  \rankthirdcommitteemember{Associate Professor of Computer Science \\ The Center for Advanced Computer Studies}
 %\fourthcommitteemember{Riel E. Hestra}     % remove "%" if applicable
  %\rankfourthcommitteemember{Assistant Professor of  Mathematics}
 %\fifthcommitteemember{Axel R. Ator}       % remove "%" if applicable
  %\rankfifthcommitteemember{Professor of Physics}

  \filefordedication{B-dedicatory}
    \fileforacknowledgement{B-acknowledgement}

%  \nolistoffigures    % remove "%" to switch off list of figures, just removed list of figure
  \nolistoftables     % remove "%" to switch off list of tables

% \fileforpreface{B-preface}

\prefatorypages

% All includes for chapters go here.
    \chapter{Introduction}
%%% another sentence to highlight server -> IaaS -> PaaS -> Serverless before this paragraph?

\begin{figure}[hb]
    \centering
    \includegraphics[width=\textwidth]{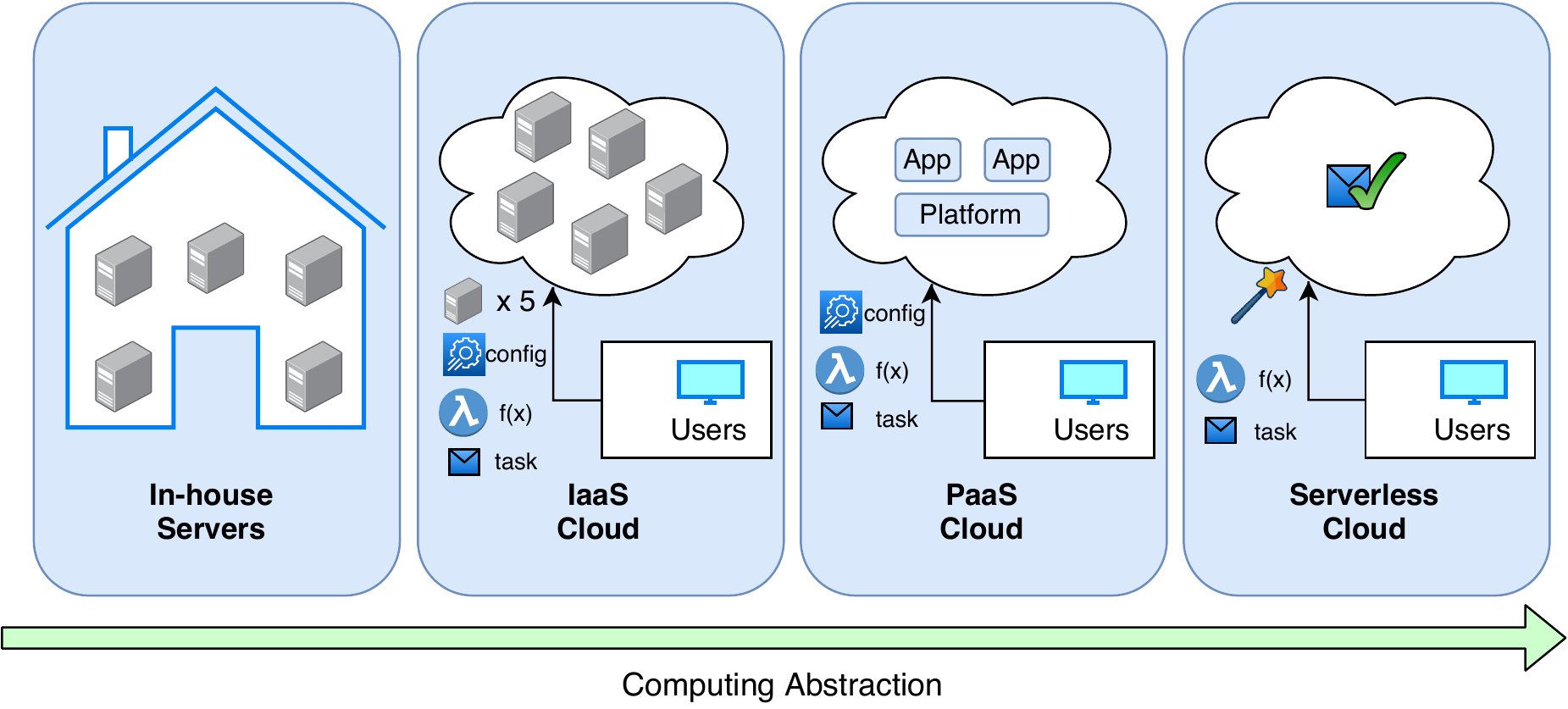}
    
    \caption{Serverless computing abstract the server management and configurations from the users. It allow the users to focus on their function codes and task requests without worrying about system details to get their tasks completed in a timely manner. 
    The system provider makes the illusion of tasks seamlessly executed without the user's involvement. 
    % The system provider handles the job of optimize task executions on their computing resources. 
    }
    \label{fig:C1Serverless}
\end{figure}

Serverless Computing or Function as a Service gaining more popularity as the practical low overhead on-demand computing system. The common practice to utilize serverless computing effectively is to break the monolithic application into multiple micro-service~\cite{lloyd2018serverless} functions. Each user provides his/her executable functions and the conditions to trigger them (\eg based on a timer or upon arrival of a web request). Once triggered, the task requests are formed, and it has to be completed in a timely manner. The serverless cloud providers aim to provide the illusion of such requests being executed on infinite resources and abstract users from details of allocation and management decisions. 
Figure~\ref{fig:C1Serverless} highlights the abstraction of serverless cloud against other more conventional types of cloud service offering paradigms. 
%While there are multiple forms of deploying serverless computing systems, 

From a provider's perspective, a scheme to efficiently utilize cloud resources is based on a central queue of arriving micro-service (henceforth, termed task) requests with a scheduler that allocates these task requests to a elastic pool of computing resources behind the scene. This shared computing resource approach reduces resource start-up overhead and amortizes the spare reserve of the computing resources. The service requests often have individual deadlines that failure to meet them compromises the Quality of Service (QoS) expected by the end-users. 
The increased number of service requests in the system and the overhead of moving data around implies that the platform is getting busy. The need for efficient use of these platforms is of paramount importance both for users and the cloud providers. 

The fact that serverless computing users are not involved in the execution of service requests on the shared resource pools allows the serverless cloud provider to utilize techniques to gain extra efficiency. Specifically, \emph{approximating} and \emph{reusing} techniques across various users' requests in the serverless systems have a significant and untapped potential that can avoid \emph{redundant computation} in these systems. The abstraction from user-managed resource allocation also allows the scheduler to utilize heterogeneous computing resources. Different task types of various users have different affinities (\ie matching) with heterogeneous machines that are available in cloud datacenters. Furthermore, each of these task types can be consistently heterogeneous within itself. For instance, it takes a longer time to change the resolution of 10-second video, compared to the time of a 5-second long segment. %This uncertainty for a given task's execution time can cause inefficiency of resource allocation~\cite{ali2000} and oversubscription of the whole system or certain computing resources.
These uncertainties pose a great demand on an efficient task scheduler to predict the task execution time and plan an efficient task mapping scheme. 

%%% skip the story about QoS? Just optimize?
Although large public cloud providers can supply virtually unlimited resources, users often have budget constraints, thus, they are not allowed to lavishly use cloud resources.
Similarly, a fog computing system, deployed in a rural area, or a private serverless computing system can fall short on elasticity and resource scaling.
Such budget or resource limitation from end users and the providers raise the \emph{oversubscription} problem on the computing resource, particularly, when there is a surge in the requests arriving from multiple users. An \emph{oversubscribed system} is formally defined as a system that is overwhelmed with arriving requests to the extent that there is no way to meet the deadlines of all the requests, thus, violating end-users' QoS. 

%remove this?
% A large body of research has been dedicated to alleviate the oversubscription problem in computing system. The approaches undertaken in these research works follow two main lines of thinking; First, allocation-based approaches (\eg \cite{liu2016dynamic,alfayly2015qoe,hou2015qoe}) that try to minimize the impact of oversubscription through efficient mapping (scheduling) of the requests to the resources. Second, approaches based on computational reuse (\eg \cite{zhang2012distributed,casas2017balanced}) that avoid or alleviates the oversubscription through efficient caching of the computational results. 
In this dissertation, we investigate an efficient serverless computing platform that takes advantage of serverless cloud characteristics to gain resource efficiency as well as mitigating the side-effects of oversubscription. Such that the overall incurred cost are minimize and the user QoS is enhanced.

\vspace{14pt}
\section{Motivational Context}
 \begin{figure}[ht]
    \centering
    \includegraphics[width=\textwidth]{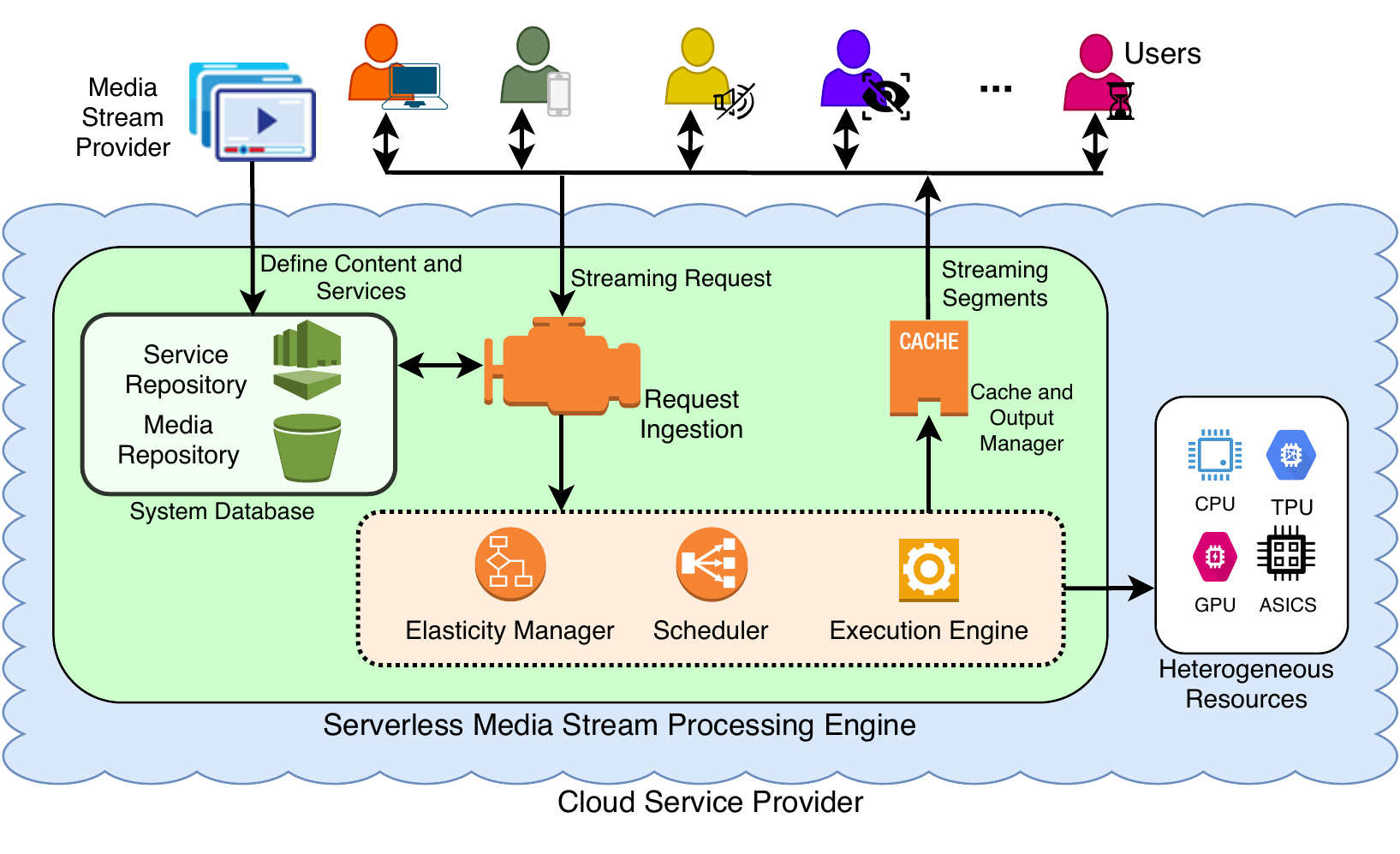}
    
    \caption{ Bird-eye view of the Serverless Multi-Media Stream Procressing Engine we aims to develop. }
    \label{fig:C1Serverless2}
\end{figure}

Our motivational application in this dissertation is an \emph{interactive} media (video, virtual reality, augmented reality, and/or audio) streaming engine that processes media to the viewers' personal requirements on the cloud before streaming them to viewers~\cite{ahmad2005video}. While the conventional media streaming platforms offer a few media quality levels to fit the bandwidth or device's specification, media processing in interactive streaming addresses the user's personal requirement. Each viewer can add one or more processing services (\eg audio translation, censoring contents, color correction, context summarization) they want to apply to the media they are using. Then streaming media will process and stream them to the viewer interactively.  
 %\textcolor{red}{THIS IS NOT A MOTIVATING EXAMPLE-->(\eg downsizing resolution or bit-rate). TRY TO EXPLAIN INTERACTIVE VIDEO STREAMING HERE. MAKE ONE OR MORE SENTENCES TO EXPLAIN ATTRACTIVE MOTIVATING SCENARIO LIKE ONLINE EDUCATION, SUMMARIZATION, VISUALLY IMPAIRED, ETC.} 

As shown in figure~\ref{fig:C1Serverless2}, we aims to utilize a private serverless computing platform with heterogeneous resources. The platform allows both on-demand media stream and live media stream processing. Both of which creates different opportunity for optimization. On the on-demand media streaming side, it is possible that multiple viewers request the same media to be streamed, hence, creating similar or identical requests in the system. In particular, when the system is oversubscribed, the likelihood of having similar requests increases. For example, two viewers with no special requirement using the same type of display device may request to stream the same media with the same specifications. Alternatively, two viewers with dissimilar display devices (\eg different resolution and compression standard) or personal requirements (\eg audio translation or graphic censorship) may stream the same video but with different specifications. 
The former case creates \emph{identical} requests in the system whereas the latter one creates \emph{similar} requests. Both of which can be combined into a compound request to save computing resources. 

\begin{figure}[ht]
    \centering
    \includegraphics[width=0.8\textwidth]{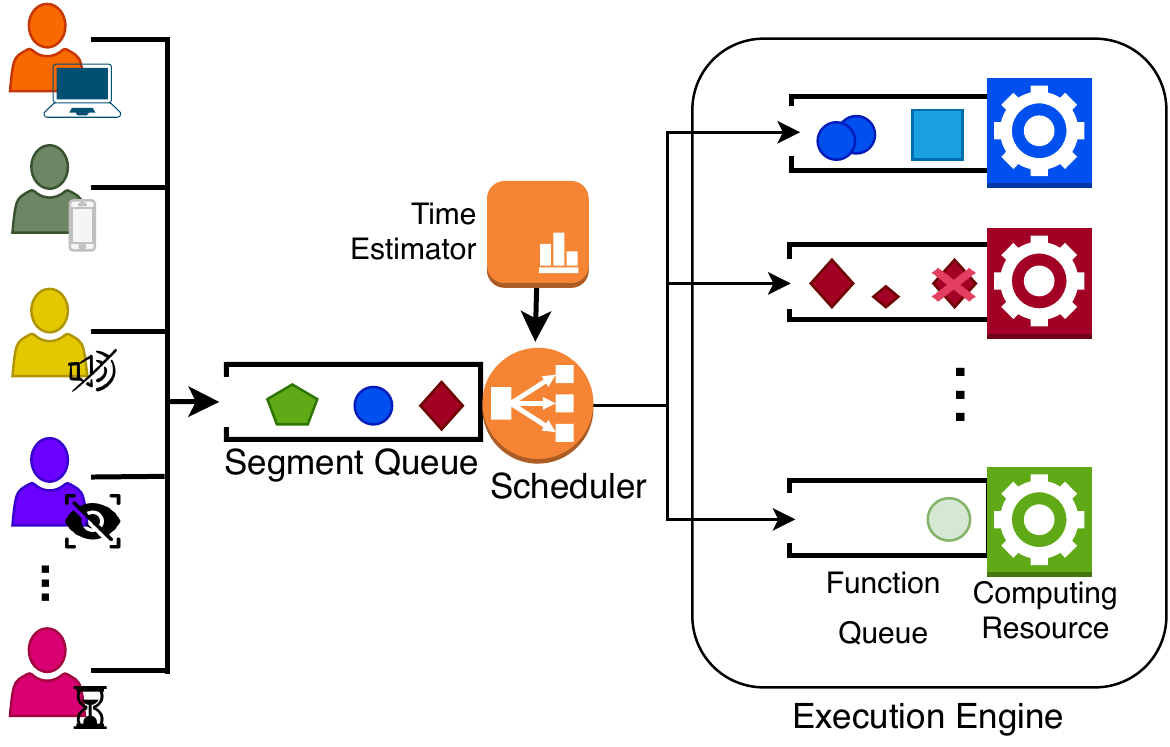}
    
    \caption{Tasks from multiple users are sent to a shared scheduling queue to be executed on computing resources. The time estimator and scheduler allows efficient use of computing machines. Geometries of different shapes, color, and size represent different processing tasks. }
    \label{fig:C1intro_multiuser}
\end{figure}

On the live media stream processing side, the computational reuse may not be possible. However, the liveness nature of the live streaming means any media segment that miss its deadline is no longer useful and can be disgarded. To minimize impact to the user QoS if the system gets highly oversubscribed, %when the system cannot complete all service requests on time by the requests' deadlines, 
certain media processing requests must be dropped to free up the resource for other task requests to meet their deadlines. This technique of dropping certain task requests, in a certain cases, can also apply for on-demand stream processing. On-demand stream processing, although naturally should not skip a media segment streaming to the user, also allows the providers to change specification of the media during the playback. In a case that a high-quality processed media segment cannot be obtained in a timely manner, a backup low-quality version of the media segment is then transmitted to the user instead. While dropping some media processing requests and supplying the viewer with lower quality media segments can impact the user's QoS, this is arguably better than having a stuttering media playback. Therefore this dissertation investigates the possibility of augmenting scheduling systems with request merging and pruning mechanisms. To maintain overall QoS of multiple users, the mechanisms should also maintain fairness across the user base. 
Figure~\ref{fig:C1intro_multiuser} shows a simplified scheduling system for an interactive multimedia streaming engine. In this system, task requests from multiple users are scheduled to be processed on heterogeneous computing resources. Similar requests can be combined and the stream can be approximated by proactively dropping (\ie discard processing) infeasible requests (segments).

%%%%%%as shown in Figure~\ref{fig:C1intro_multiuser}
%% draft of figure

\section{Research Problem and Objectives}
With the aim of constructing a cost- and QoS-efficient serverless cloud platform for media streaming, in this dissertation, we address the following research problems:
\begin{enumerate}
    \item How to predict the gain of task aggregation for a given configuration? 
    \item How to perform request aggregation with minimal scheduling overhead to avoid redundant processing in a serverless platform and reduce deadline violations in the system? 
    \item How to perform a lightweight task pruning for unlikely-to-succeed tasks in a serverless platform to avoid wasting cloud resources?
    \item How to design and implement a serverless media streaming system that is both modular (hence, expandable) and efficient?
\end{enumerate}

\section{Contributions}

Considering the research questions described in the previous section, the major \textbf{contributions} of this dissertation are:
\begin{itemize}
%%%c3
 \item Identifying potential benefits of task request merging and then developing a resource saving prediction module.
%%%c4
 \item Proposing an efficient way of identifying potentially mergeable requests.
 %\item Proposing methods for proper positioning of merged requests in the scheduling queue.
 \item Determining appropriateness and potential side-effects of aggregating requests.
 %\item Evaluating request merging decisions based on the request and system-wide factors.
 
 %\item Analyzing the performance of the request aggregation mechanism on the viewer's QoS and the total time of utilizing the processing units.
 
%%%c5
	\item Mathematically modeling the impact of request dropping on the success probability of other tasks scheduled to execute after the dropped one.
	%combination of below
	\item Proposing the request pruning mechanism as a generic mechanism that can be applied to existing resource allocation systems.
%	\item Developing a pruning-aware and a fairness-aware mapping heuristic for an HC system.
%	\item Reducing incurred overhead of deploying pruning mechanism via approximate computing and computational reuse.
	%\item Analyzing the impact of request pruning mechanisms on the robustness of various types of computing systems and under varying workload characteristics.
%%%c6	
	\item Implementing a prototype of on-demand media processing platform operating in a serverless computing manner. %The architecture is modular and expandable to cope with different needs.
 
\end{itemize}

%\section{Methodology}

\section{Dissertation Organisation}
\begin{itemize}

\item Chapter~\ref{chap:bg} explores the related studies and provide background for serverless computing, reusing and approximating techniques for serverless computing platform, and commonly used task mapping heuristics. 
%, and task requests combining methods. %%% included in the reusing part
\begin{itemize}
     \item \textbf{Chavit Denninnart}, Mohsen Amini Salehi, \textit{ A survey on Approximate Computing in Serverless platforms }, preparing to submit  
\end{itemize}

\item Chapter~\ref{section:MergeSaving} studies the potential benefits and feasibility of computational reuse by schedule similar task requests to be executed together. A range of video processing task are benchmarked in various configurations to study the resource usage saving of task executed as a group against individually. Result of the benchmark is used to train a machine-learning based resource saving predictor.
\begin{itemize}
     \item Shangrui Wu, \textbf{Chavit Denninnart}, Xiangbo Li, Yang Wang, Mohsen Amini Salehi, \textit{ Descriptive and Predictive Analysis of Aggregating Functions in Serverless Clouds: the Case of Video Streaming }, accepted in 22nd IEEE International Conference on High Performance Computing and Communications (HPCC '20)
     \item A dataset of video merging benchmark which is available in the following GitHub repository: \url{https://github.com/hpcclab/VideoStreaming_Workload} 
     \item GitHub repository for the developed Merge saving predictor module: \url{https://github.com/hpcclab/Merge-Saving-Prediction}

\end{itemize}

\item Chapter~\ref{section:Reusing} explores the first mechanism to remedy the redundant processing issue in serverless cloud computing systems by creating request aggregation mechanism inside Admission Control component of the system. For an arriving request, Admission Control can recognize if it is mergeable with the ones exist in the batch queue.
Assuming that a request in the batch queue can be modified (or cancelled and resubmitted), the Admission Control decides if the arriving request can and should be merged with one of the existing request. 
\begin{itemize}
     \item \textbf{Chavit Denninnart}, Mohsen Amini Salehi, \textit{Improving Cost-Efficiency and QoS of Serverless Computing via Computational Reuse }, Submitted to IEEE Transactions on Parallel and Distributed Systems
     %\item \bibentry{denninnart2018leveraging}
     \item \textbf{Chavit Denninnart}, Mohsen Amini Salehi, Adel Nadjaran Toosi, and Xiangbo Li, \textit{Leveraging computational reuse for cost- and qos-efficient task scheduling in clouds}, Proceedings of the 16th International Conference on Service-Oriented Computing (ICSOC '18), Nov. 2018,pp. 828–836
    \item GitHub repository for the prototype of Media streaming system with task merging components: \url{https://github.com/hpcclab/adaptivemerging}
\end{itemize}

\item Chapter~\ref{section:DropDefer}, first, provides a mathematical model to estimate a task's probability of meeting its deadline in the presence of task dropping. Then, investigate methods for engaging probabilistic dropping. %We propose methods to dynamically determine task dropping and deferring threshold probabilities.
Based on the proposed model, a pruning system and a pruning-aware mapping heuristic is implemented with an extension to engender fairness across various task types. The pruning mechanism is presented as an independent component that can be applied to any mapping heuristic to improve the system robustness. To reduce overhead of the pruning mechanism, approximation methods that remarkably reduce the number of mathematical calculations and improve the practicality of deploying the mechanism in heterogeneous or homogeneous computing systems are developed. %Cost and energy gains of the pruning mechanism on both system with commonly used mapping heuristics and system with specifically created pruning-aware mapping heuristics is shown.
\begin{itemize}
    \item \textbf{Chavit Denninnart}, James Gentry, Ali Mokhtari, and Mohsen Amini Salehi, \textit{Efficient Task Pruning Mechanism to Improve Robustness of Heterogeneous Computing Systems}, Accepted in Journal of Parallel and Distributed Computing, 2020
    
     %\item \bibentry{denninnart2019improving}
    \item \textbf{Chavit Denninnart}, James Gentry, and Mohsen Amini Salehi,\textit{Improving robustness of heterogeneous serverless computing systems via probabilistic task pruning}, 28$^{th}$ Heterogeneity in Computing Workshop (HCW 2019), in the proceedings of the IPDPS 2019 Workshops \& PhD Forum (IPDPSW), May 2019
    %\item \bibentry{ipdps19}
    \item James Gentry, \textbf{Chavit Denninnart}, and Mohsen Amini Salehi,\textit{Robust dynamic resource allocation via probabilistic task pruning in heterogeneous computing systems}, Proceedings of the 33$^{rd}$ IEEE International Parallel \& Distributed Processing Symposium, IPDPS ’19, May 2019
\end{itemize}

\item Chapter~\ref{section:platform} explains the implementation of the interactive media processing platform (SMSE) that operates in a serverless computing manner. This platform is to explore the feasibility of improvements developed in other chapters.
\begin{itemize}
     \item \textbf{Chavit Denninnart}, Mohsen Amini Salehi, \textit{SMSE: A Serverless-Based Platform for Interactive Multimedia Streaming Engine }, preparing for submission 
      \item GitHub repository for the prototype of interactive media streaming system: \url{https://github.com/hpcclab/CVSS_impl} 

\end{itemize}

\item Chapter~\ref{section:thesiscon} concludes the dissertation with a discussion of our main findings and future research directions in the area of efficient serverless cloud computing platform.
\end{itemize}
    %\chapter{Survey of Related Literature} \label{chap:survey}
%\vspace{14pt}
\chapter{Background and Literature Study} \label{chap:bg}
\vspace{14pt}

%\section{Survey of Related Literature}\label{sec:survey}
This section provides background and survey of other research works undertaken in the fields most related to this work. %and position the contribution of our works against them.

\vspace{15px}
\section{Serverless Cloud Computing Paradigm}
\label{sec:serverlessIntro}
\indent Cloud computing offload the application deployment details to be in the hand of cloud providers. Refers to Figure~\ref{fig:C1Serverless}, Infrastructure as a Service (IaaS) cloud offers the infrastructure that the user requires while abstracting the burden steps of obtaining the infrastructure. Platform as a Service (PaaS) abstracts the deployment further by hiding details of their infrastructure. Users only have a view of the software platform they want to use. Finally, Function as a Service (FaaS)~\cite{scheuner2020function} or serverless cloud takes the abstraction to another level by abstracting the whole server and platform management details; the user can simply view their application as the collection of functional codes which will be handled by the cloud providers.

In a serverless computing platform, system providers manage all execution environments, such as resource allocating, scaling, scheduling, and ensuring availability. To use serverless computing, the end-user provides their application as one or more %(stateless standalone ?) 
 application functions with a predefined function trigger that can be timer- or event-based~\cite{peekserverless} (such as a web request).

Since the users do not explicitly obtain their computing resources, Service requests of multiple users can be scheduled on a shared resource pool. This approach improves cost efficiency~\cite{eismann2020review} over the traditional IaaS model by reducing idle resources wastage (idle resource on each active machine and also the number of warm spare machines.) Furthermore, the approach allows for data and computations to be shared across multiple users, increase the cost efficiency even further. In this section, we explore the concepts, feasibility and related studies for computational reuse and approximate computing to enable efficient serverless computing platform.

\subsection{Function as a Service (FaaS).}~
Function as a Service is a type of cloud service model where cloud provider provide a shared system that handle the execution of user's code invocation on the system's shared computing resource. 
%relation between FaaS to PaaS (vs SaaS)
Function as a Service often gets included to be a sub-type of Platform as a Service (PaaS) or Software as a Service (SaaS). This is because FaaS cloud can be viewed as a platform/software that serve users' functional code execution. 
Unlike IaaS and PaaS where user still have view and management tasks of the platform, FaaS hide away such details to the cloud provider's responsibility. 
Unlike Software as a Service cloud where providers provide and abstract the complete solution including the functional codes to serve end user who just want to use the software. In Function as a Service, users still provide the functional code and one or more ways to trigger the execution (\eg trigger by timer, by web request or by other events such as completion of another function)~\cite{shahrad2020serverless}. Once the function and trigger is defined, the serverless computing platform handle all other steps to realize task execution in a timely manner. Therefore, FaaS has more flexibility to cope with vast application requirements than the software pre-defined in the SaaS platform.

%what is micro services/ DevOPS

\subsection{Stateless versus stateful functions.}~
In a classic definition, serverless functions are stateless. That is, the `function' part of the execution does not maintain any data between consecutive runs. Each function invocation receives input data and performs the user-defined function, then output the results before termination. This scheme makes sure the platform only has to handle data dependency on the input data before function execution rather than maintaining data consistency during multiple concurrent runs. 
And thus, simplify the function scheduler and management. However, various use cases require the state data to be maintained between each execution. And therefore, such functions have to rely on external data storage or simply store the state data with function input and output. Both choices add a significant overhead to the execution. Pu~\etal found that using stateless serverless computing with external storage to perform data analytic works can be up to 500 times slower than running the same task on clustered VMs~\cite{pu2019shuffling}. Therefore, they proposed a specialized serverless computing platform with a two-tier (fast and slow) storage system to remedy the over-head.

To remedy the needs of functions that need to maintain state data, certain serverless platforms provide the built-in capability to maintain state in an efficient manner~\cite{klimovic2018pocket}.
There are two main categories of state storage: key-value storage and system storage. %%%%% actually, three, object storage, Key-value Storage, API for both?

Sreekanti~\etal explores a stateful serverless cloud computing platform in Cloudburst~\cite{sreekanti2020cloudburst}. Cloudburst utilizes Anna~\cite{wu2020autoscaling}, an autoscaling key-value store, for its state maintaining. 
%Pu\etal, found the inefficiency of using stateless FaaS for data analytic works, and then propose Shuffling~\cite{pu2019shuffling}, a scaleable specialized serverless computing platform for data-analytics. Shuffling creates two layers of data storage, a fast one cached by memory, and a slower main storage.
Schleier-Smith~\etal developed a dedicated file storage system for stateful serverless computing named FAASFS~\cite{schleier2020faas}. Their work tackles multiple challenges of providing a shared storage file system across serverless functions, including cache consistency~\cite{yu2018sundial} and transactional consistency~\cite{sreekanti2020fault}. Their storage system can be accessed with POSIX-like commands. 
Shillaker and Pietzuch experiment stateful serverless execution platform in FASSM~\cite{shillaker2020faasm}. This platform stores and shares state data as both files and memory segments in their software-based serverless computing system.
 
 %It also have built-in caching system.

In both ways of realizing stateful serverless computing by centralizing the state data management, it also provides the side benefit of enabling state data sharing across different functions. That is, centrally store state data by serverless execution platform enables more data to be shared and reused.

\subsection{Containerization in FaaS.}~ %and virtualization? / uniKernel
%how to/ why
In a pure form of FaaS, containerization is not necessary. A user can call a function by a command that looks like client.invoke(FunctionName=`F',Payload=Data) \cite{mcgrath2017serverless,perez2018serverless}. The serverless cloud platform can interpret the function code embedded in the execution request and, therefore, do not require a predefined service container.  
However, if the function is not easily interpreted in a timely manner (\ie function takes a long time to compile) or require any specific environment configuration (such as software packages), this scheme can be inefficient as the overhead of making the service available before the execution can be excessively high. Therefore, a predefined and preloaded function can reduce the start up overhead. Such a prepared software package is usually provided as a container image for several reasons~\cite{perez2018serverless}. 
First, the container image is very portable. There are predefined standards of containerization that are publicly available and widely supported by multiple software stacks. Second, containerization improves system security by isolating each functions to their own space rather than running them directly on the system platform. Third, unlike Virtual machine images, containers are light-weight and can be deployed with minimal overhead.

\subsection{Bare-metal FaaS.}~
As an alternative to container-based serverless, there are research projects and initiatives to create software-based serverless computing platforms with the benefits of a container-based serverless platform. Shillaker and Pietzuch proposed FAASM~\cite{shillaker2020faasm}, a software-based serverless computing platform that provides isolation through the usage of WebAssembly's software-fault isolation. Their implementation also allows for data sharing across serverless functions and make the functions stateful rather than stateless. Their performance evaluation shows promising results. However, their platform still requires functions to compile to WebAssembly, and therefore, support only a subset of all possible application implementations supported in a container form. We believe WebAssembly is a potent technology that can allow software-based serverless computing to compete with container-based serverless computing in terms of security and portability while also sporting performance advantage.

%a figure...

\subsection{Reusing of containers and container images to reduce overhead.}~
\label{subsec:reusingcontainer}
The most naïve way to execute a function in a container is to create a single-use container image based on each function invocation request. %which also provides full isolation. 
However, the overhead of creating one container image for every single request can be excessive. The more practical approach is to predefine a function as a container image ahead of time. Each of the stateless container images is pre-loaded ready to run in the memory. This way, the isolation is still maintained, and the start-up overhead of repeated function invocation can be reduced. Due to memory limitations, however, not all images can be kept in the memory for the rapid start of the function. Infrequently used functions must be offload from memory to the storage system (\eg hard-drive). A container that can start rapidly from memory is generally referred to as a warm start container. In contrast, the function that must be load from the storage system is referred to as cold start container~\cite{lloyd2018serverless}. Cold start containers face additional container image loading overhead in comparison to the warm start counterpart. Such memory limitation spawns research problems. Specifically, (1) how to reduce start-up the cold start overhead; (2) how to keep more containers in the same amount of the memory; (3) how to minimize the number of cold starts strategically through memory allocation.

%first and 2nd question, 
An approach to tackle both start-up overhead of the cold start container and allows more container images to be stored in the memory is to minimize the amount of data required for each container image.  %(also available to VM images~\cite{yan2019z}). 
The technique commonly deployed by container platforms includes union mounting. In such a platform with union mounting, each container image is not stored in one big image but rather as multiple separate components that can mount together to form an image. This allows the components to be build in layers. The top layer overwrite and expand certain parts of the file stored in lower layers. The same layer can be used by multiple container images that require the same components, and therefore, less amount of data must be obtained to load a container image. In addition, more container images can be maintained in the same amount of memory. 
Multiple researches improves the container layer loading system over the years. Zhao~\etal propose Duphunter~\cite{zhao2020duphunter}, a replacement layer loading module on docker platform. The architecture is more effective in deduplicating similar layers across multiple docker images with less deduplication overhead than prior designs.

\subsection{Resource allocation and scheduling.}~
The third approach to reduce cold start overhead is to reduce the frequency of the cold start by strategically allocate which container to remain in the memory, either for a certain time after an execution, or preload the container in expectation of a request in an immediate future.
Shahrad~\etal~\cite{shahrad2020serverless} collected and studied a 14 days period of function invocation patterns on serverless cloud usage in Microsoft Cloud. From their analysis, they propose a strategic plan to reduce amount of cold starts by categorize each functions into categories. Each category have its pattern of \emph{pre-warm} and \emph{keep-alive} period. After an invocation, function is removed from the memory for the pre-warm period as the system don't expect to get another invocation request until the pre-warm period expires. Then the container is loaded back into the memory ready for the next warm-start function invocation. If the function stay in the memory for more than keep-alive period without an invocation, then the function is again removed from memory. Such strategy reduce the number of cold start for majority of the functions. There are still some functions that their invocation pattern is too unpredictable and does not benefit from strategic predictive-base allocation. Nevertheless, benefit from correctly predicting the trigger period of most functions certainly worth the complication of maintaining historical profiling and prediction.

\subsection{Commercially available and open-source serverless cloud.}~
Serverless computing become widely commercially available for the first time on Amazon AWS lambda service~\cite{awslambda}. AWS lambda execute each serverless computing code in a container upon user-defined trigger. Amazon charges for actual resource usage by the functions. Their implementation arguably pioneer and shape the view of commercially viable serverless computing.

Throughout the years. Multiple competing serverless computing services emerges. Such as Azure Functions~\cite{msAzurefunctions}, (Google) Cloud Functions~\cite{gcloudfunctions}, Iron.io~\cite{ironio}, Webtask.io~\cite{webtask} and many more.

%\subsection{Open Source Products}
On the open-source side, Apache OpenWhisk~\cite{openwhisk} is the most popular open-source serverless cloud platform. OpenWhisk can manage and dynamically scale infrastructure on multiple scales. In fact, the OpenWhisk itself can be deployed as containers that can be managed by popular container management systems. The platform accepts functions in a wide range of languages and has an active software development community.

\section{Computational Reuse in Serverless Cloud}
\label{sec:reuse}

\begin{figure}
    \includegraphics[width=\textwidth]{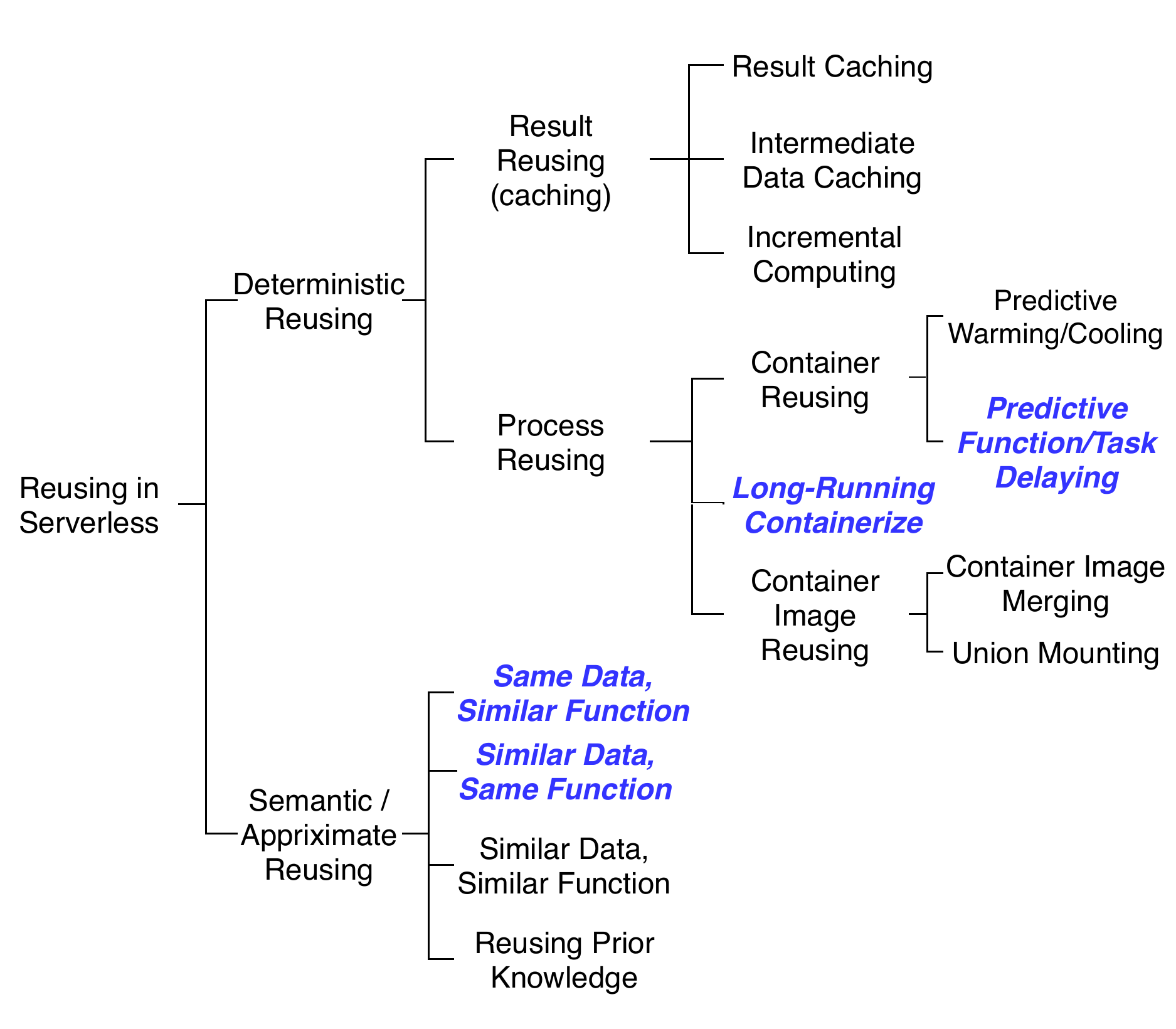}
        \caption{Taxonomy of approximate reusing techniques in serverless computing. Emphasized texts highlight the techniques we explore in this dissertation. }
                \label{fig:ReuseTaxo}
\end{figure}

\indent Computational reuse is conventionally deployed to gain speed up by deduplicate data or computation. From the earliest day of programming, programmers save computing results in some variable to be reused later. Various kinds of data caching are developed and deployed to deduplicate the computation. In this section, We first define the goals of performing computational reuse in serverless platform. Then we explore various approaches of computational reuse shown in Figure~\ref{fig:ReuseTaxo}. 
%cite the taxonomy figure

%%%%%%%%% Not here, this is motivational???
\subsection{Goals and benefits of reusing in serverless.}~ Reusing provides several benefits to help the efficiency of serverless systems. Specifically: cost efficiency, quality of services, and energy efficiency. 

\paragraph*{\textbf{Cost efficiency:}}
Serverless computing is a cost-efficient computing paradigm where users only pay for the resources they really use. The providers can offer high availability and low latency computing services at a low cost by sharing multiple users' tasks on the same sharing resource pools. In some ways, many existing serverless computing platforms already adopt resource reusing and process reusing (such as reusing container images) in their systems. To leverage the economy of scale further, We envision serverless computing platform to enable computational reuse across multiple users. 

\paragraph*{\textbf{Quality of services (QoS):}}
Computational reuse deduplicate data and computation redundancy. Other than the apparent benefit to the cost-saving by saving computing resources. Reusing also has both a direct and indirect effect on the perceived QoS of the user. Computational reuse can improve the QoS directly by allowing common tasks to finish earlier with cached results. For uncached results, Reusing free up computing resources to be less busy; thus, more resources can be assigned to each task.

%%% write about memoization
%%%reusing FuncX (somewhere else, not this section)

\paragraph*{\textbf{Energy efficiency:}}
Similar to cost, Reusing can reduce the amount of energy consumption. As the requirement of the computing is still on the rising trend and accelerated by the digital shift necessitate by the recent global events. The energy consumption of the computing system and green computing has become more in focus. Centralized Cloud allows the computing to be done in places where energy management and cooling system are favorable (such as under water~\cite{cutler2017dunking}). Serverless computing eases the migration of functions across locations to reduce data transfer wastage. 
More importantly, Serverless platforms enable a further reduction in energy consumption by reducing the wastage of standby resources through the sharing of readily available resource pools. Reusing in serverless computing improves energy efficiency by sharing the computations and storage for tasks on the same platform. 
% \label{subsec:reuse_characteristics}

\subsection{Taxonomy of reusing in serverless computing.}~ In this part, we explore the different approaches to achieve reusing in serverless computing.

\paragraph*{\textbf{Semantic reusing versus deterministic reusing:}}
Deterministic reusing refers to the type of computation or data reusing that result is deduplicates redundant tasks or data. The majority of the reusing techniques focus on detecting %(or design the architecture to ease the detection of) 
the computation or data that is used frequently and then cache it to be reused later. Deterministic reusing does not impact the computation results of tasks in the system.

Semantic reusing takes a step further on the reusing aggressiveness; rather than reusing the obvious reusable, some research works try to find the semantic relationship between the not-so-obvious similar data and reuse them together. In a sense, the semantic relationship can also be considered as approximate reusing (a part of approximate computing). An example of approximate reusing approach is proposed by Krishnan~\etal~\cite{krishnan2016incapprox}. Their work can reuse partial results from similar tasks for incremental computing. 
By the nature of implying the semantic, misinterpretation can happen, and it leads to incorrect results. Most of the common semantic relation finding systems imply the semantic similarity between data or functions. 

An example of semantic reusing on functions is to imply function $A$ and function $B$ to be similar, and therefore, the result of the function $A$ on data $x$ (denoted $A(x)$) can be reused by function $B$ on data $x$ ($B(x)$). Additionally, one of the functions can be replaced by another rather than keeping both functions available. Serverless computing systems accept functions from multiple users. Usually, the functions are defined in a short and for single purpose following the Micro-service architecture to ease the  continuous development process. And therefore, the chance of multiple users define their own functions that practically do the same thing is not unlikely. A similar approach can also be applied to imply a similarity between input data of the same function~\cite{guo2018potluck}. However, when the reusing approach not only tries to reuse the results of similar tasks. But also predict and make a correction to the results, then the approach is generally considered as an approximate computing technique rather than computational reuse.

%The remaining of this section, we explore some examples of common approaches of reusing in both deterministic and semantic reusing categorized in Figure~\ref{fig:ReuseTaxo}.

\paragraph*{\textbf{Result reusing:}}
\label{subsec:reuse_result}
%traditional result caching
Data reusing (especially caching) is an integral part of efficient programming. The distinct characteristic of caching (against other types of data storage) is that caching data is optional. Caching is maintained to speed up the execution. However, the program can still function if it misses the caching data and has to retrieve it in other ways.

Since the storage is not infinite or free, a well-designed system must balance the proper cache size and complexity. A sophisticated and extensive caching system imposes its own cost. However, a caching too small or too simple can miss a large portion of reusable data, which result in recomputing cost. This trade-off appears on all levels, from hardware to request transaction level. In the context of a serverless cloud platform, the serverless platform can offer the service to cache and reuse computation results from each of the task request execution.

There are numerous data and computational reusing researches that allow the application to reuse the result to reduce resource usage~\cite{jiang2017secure} and to speed-up the response time~\cite{arteaga2016cloudcache}. We found most of the researches to fall into three categories based on how they store and reuse the data: 1) result caching, 2) intermediate data caching, and 3) incremental computing solution.

\paragraph*{\textbf{Result Caching:}}
Result caching can be recognized when the platform captures the end result of the task execution to be reused. Cache lines are usually indexed by the task request signature (function and certain or all parts of the input data). To address security concerns of the cache poisoning, some works separate cache bucket with tasks same user or same function only~\cite{SecureDedup}.

\paragraph*{\textbf{Caching Intermediate Data:}}
Intermediate result caching are those that, as the name suggests, cache partial results. The technique is usually suited to a large function that has multiple computing steps, in which caching the intermediate result is more effective than caching the final result. Intermediate result caching can be offered as key-value storage for the function to access. Rather than transparent cache, the current implementations~\cite{arteaga2016cloudcache,wu2020autoscaling,sreekanti2020cloudburst,memcached,carlson2013redis} usually need the function code to actively store and look for the stored partial result for reuse from a centralized caching system. Other variations of the intermediate result caching include scheduling systems that maximizes the reusability of workflow task processing~\cite{popa2009dryadinc,casas2017balanced}, and those that fuse multiple functions together~\cite{wang2020accelerating} to form a larger function that eliminate the data movement between functions (reusing data already in the execution platform).

\paragraph*{\textbf{Incremental Computing:}}
The third category is incremental computing. This technique usually caches the task result similar to result caching. However, it is not only reusable by the task that have the same input data. Rather, it utilizes the system-defined or user-defined correction function to change the result of one input data to the data of another input data. A common use case of incremental processing is in the big data query~\cite{tiwari2014mapreuse}. Consider a database that updates daily with minor changes from one day to the other. Supposed there is a repeated query that is executed daily. Rather than starting the query from the beginning each day, the incremental computing can collect the result from yesterday, subtract partial results of the input that no longer valid today and add the partial results from the new input to form the new result.

%Secure Deduplication,

%intermediate result caching
%Secure Deduplication,14, MapReuse,Accelerating

%incremental computing
%DryadInc, 

\paragraph*{\textbf{Process reusing:}}
\label{subsec:reuse_process}
In serverless computing, the process (function) loading and maintaining are also where reusing can significantly improve system efficiency. Earlier in Section~\ref{subsec:reusingcontainer}\, , we discussed that Union mounting help reusing part of the container image. Also, predictive warming/cooling of the containers can maximize the reuse of the container image in the memory. In this part, we discuss three more techniques to increase the process reusing, namely, container image merging, predictive function delaying and converting frequently used functions to a long-running container.

\paragraph*{\textbf{Container Image Merging:}} 
Consider two or more function container images that are very similar to each other. Union mounting can capture and share a high percentage of the image when starting them from a cold start. However, the scheduler still sees them as separate functions. And therefore, treat them separately. If each of those functions is not frequently used, each function can get evicted from the memory. Those infrequently used functions that do not have a library requirement (and hence don't require a specific container environment) can share the same general propose container image.

However, for functions that require function libraries and specific container images, we envision some form of container image merging that is recognizable in the scheduler. Such that two functions can use the same container image (but with a different command to start the function execution). And therefore, multiple infrequently used functions can share the same container image in the memory for a warm start. 

\paragraph*{\textbf{Long-Running Container:}}
Based on studies~\cite{eismann2020review,shahrad2020serverless}, certain percentages of functions are invoked very frequently. And if those functions incur a significant loading overhead even as a warm start (\ie functions that require big libraries such as machine learning frameworks), the repeating container warm start for each request alone can compound to a significant resource usage. We envision an efficient serverless computing system to recognize those tasks and transform the container to become a stateful long running daemon with its own task queue to reduce the start-up overhead further. %We deploy multiple function containers in this manner in Section~\ref{section:platform}.

\paragraph*{\textbf{Predictive Task Delaying:}}
Commonly, a serverless cloud tries to serve every request as soon as possible. In fact, turnaround time is one of the main criteria when considering the performance of FaaS cloud system. However, in reality, not every application needs the task to be completed as soon as possible. Some task types can tolerate some delay as long as it still meet its deadline, and some task types may not even have a hard deadline. 
To maximize reusing in serverless computing, the user or the system should have a way to declare task urgency. An urgent task request is scheduled to be completed as soon as possible, with warm start containers ready to execute the task. While the less urgent task can potentially wait %for a certain period until the resource become more freely available or 
to tag along with other similar arriving requests to maximize container reuse and hence, reduce incurred cost. The scheduler should have the ability to predict the latest time to schedule the task such that each of them waits as long as possible to share the container with other tasks yet still meet its deadline.

\section{Approximate Computing in Serverless Cloud}
\label{sec:approxcloud}
\subsection{Goals and benefits of approximate computing in serverless cloud.}~
Approximate computing allows tasks that has undesirable demands in response time, energy, or cost to be completed in a timely, energy~\cite{jayakumar2016energy}, and cost-effective manner. However, by the nature of approximate computing, the precision and accuracy of the result are compromised. In an efficient serverless computing platform design, we envision the system to have integrated capabilities to approximately compute tasks that are not feasible otherwise.

We envision approximate computing to be offered to the users in three options: 
\begin{itemize}
    \item[1)] Provide an approximate result early, then following with an accurate result later.
    \item[2)] Provide only approximate results to save expenses.
    \item[3)] Provide an approximate result only if the system cannot complete the accurate task on time.
\end{itemize}

% Approximate computing can speeds up the response time and improve QoS by getting an inaccurate result earlier than accurate one.

% Applying approximate computing to a certain parts of the workflow can improve the response time (hence QoS) of the user. 
%interactive websites. Many websites have predictive caching system in place to predict what resource might be request in an immediate moment. 

% \subsubsection{Cost Efficiency}
% Approximate computing generally utilized to allow tasks that naturally too computing prohibitive to compute accurately in an acceptable time to be completed in a timely and cost-effective manner. In serverless computing platforms, we envision that the platform should provide an option to perform approximate reusing and approximate computing as an option to help saving user's expense.

% \subsubsection{Energy Efficiency}
% From the system provider perspective, energy consumption can be a major concern to the sustainability of data centers. Approximate in serverless computing can reduce the energy consumption in a similar ways that it improve cost efficiency and QoS, by enable more reusing and reduce computational requirement to finish a task. 

\subsection{Difference between approximation and reusing.}~
The main difference in approximation against computational or data reusing is the impact on the result's accuracy and precision. Computational reuse accelerates the turnaround time or saves the computing resource without impacting the result. Approximate computing decrease the result's accuracy and precision for more time- and resource-saving benefits. While approximate computing can be applied to the task individually, many of the approximating techniques reuse information gathered from other tasks. Such information can either be collected in run time or predefined ahead of time (such as models trained for prediction). Approximate computing techniques that directly reuse the result of tasks with the similarly close-enough specification is also called approximate reusing.

\subsection{Requirement for approximate computing.}~
%%%%% this figure is too simple?
% \begin{figure}
%     \includegraphics[width=0.5\textwidth]{Figures/ReqApprox.pdf}
%         \label{fig:reqApprox}
% \end{figure}
Approximate computing model exploit \textit{resilient} property of the system by getting the inexact but acceptable result at a lower cost. %The error rate of the result must be in a range that the system can handle. 
A resilient system~\cite{jayakumar2016energy} or application must be able to \textit{tolerate} a certain amount of errors~\cite{chippa2013analysis}. A successful approximate computing model must not cause error that exceed the resiliency of the system in both error \textit{magnitude} and error \textit{chance}.

\paragraph*{\textbf{Error magnitude:}}
Error magnitude can be identified by the variation of the result value against the actual result. Each application can tolerate a different amount of the error magnitude. %Approximate computing techniques consistently increase the error magnitude are those that sampling some data rather than calculating the whole (skip data points) or reduced calculating precision.

\paragraph*{\textbf{Error chance:}}
On the other hand, some approximation techniques have a trait that it has a high percentage chance (but not guaranteed) to give an accurate result. However, there is also a certain chance that the approximation can be totally wrong. The chance of getting an overly inaccurate approximation is called the error chance. Correction function can be applied to fix the error result once the error is detected. If the chance of getting an error is too high, then the overhead of correcting results frequently may degrade or exceeds the benefits of the approximate computing. %The most common technique that causes this kind of error is those that have a prediction as part of the approximation scheme. 

%  The inaccuracy can be in term of the precision (\eg lower precision decimals or suboptimal solution that give near optimal result) or the accuracy (\eg approximate result have some small chance to be totally incorrect.) The later case may require follow up correction after the incorrect result is detected. 
 
%The inaccuracy can be in form of most result are in lower precision but close to the actual result. Or most result are correctly represented, while some certain percentage of the result are completely incorrect. This case require some follow up correction later.

%which level? hardware, platform, software

 %A more developed form of approximate computing intentionally forego some computation accuracy for speed up after with consideration of the situation. Most approximate computing include the usage of computational reuse of similar but non-identical data with some correction function and/or some form of calculating precision that determined on the run time. 

\subsection{Approximation approaches.}
\begin{figure}[htbp]
    \includegraphics[width=0.8\textwidth]{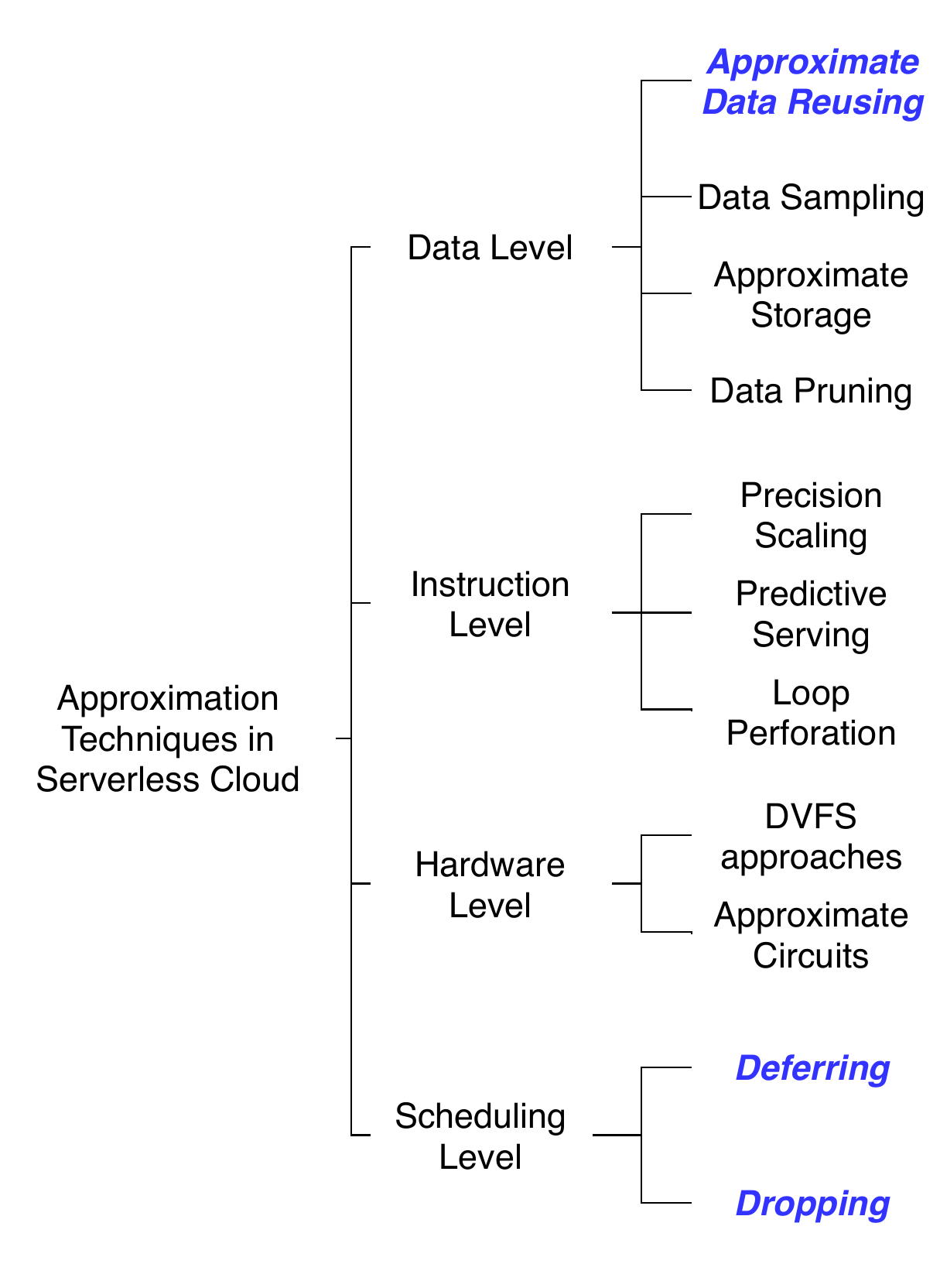}
        \caption{Taxonomy of approximation techniques in serverless computing. Emphasized texts highlight the techniques we explore in this dissertation.}
                \label{fig:ApproxTaxo}
\end{figure}

~
There are various approximating techniques that can benefit the serverless computing systems. We categorize them in four categories as shown in Figure~\ref{fig:ApproxTaxo}.

\paragraph*{\textbf{Approximate data reusing:}}
Samadi~\etal~\cite{samadi2014paraprox} proposed a system that identifies potentially computationally reuseable tasks and uses them for software-only $approximate$ computation (\ie reusing the computations that would give close enough results rather than re-processing to get accurate results) to improve performance, especially in Video processing context where accurate computational results are not required.

\paragraph*{\textbf{Data sampling:}}
For tasks that work on a batch of data, such as data analysis works, it is possible to reduce the input data size by sampling the dataset. There are various techniques to take data samples, perform the computations, and then analyze the variability and the error rate in comparison to the complete data analysis over the whole data. Wen~\etal proposed ApproxIoT~\cite{wen2018approxiot}, in their work, they propose a sampling algorithm and quantifying the error bound to pick data from the stream of unknown data size. Sampled data are stored in the `reservoir', which is limited in size. New data can randomly replace the existing data in the reservoir.

\paragraph*{\textbf{Approximate data storage and data pruning:}}
Approximate data storage can save a portion of the data footprint by either storing only part of the data or scramble multiple data points together (such as various hash table techniques without hash collision correction mechanisms). Alternative to approximate data storage inside the users' application, the cloud platform can also offer a service to store and deduplicate `similar' data together~\cite{paulo2014distributed}. We believe the deduplication-capable data storage system should be offered as a standard service in the efficient serverless platform.

The process of retrieving data can also be approximated, usually in the form of returning the result before ensuring the data correctness. In a storage system where data can be modified by multiple parties concurrently, the system can either check for validity before returning the data (strict consistency) or return the data first then have a following correction if the data is found to be invalidated (eventual consistency).

%\subsubsection{Precision scaling:}
\paragraph*{\textbf{Precision scaling and stochastic computing:}}
The earliest form of approximate computing are created by necessity. Computer storage cannot store an infinite decimal points, hence the numbers are approximately stored and represented by a `close enough' value. Those early precision limitation are applied either system wide or by programmer's decision (\ie use smaller floating point variable type). The concept developed further to a more deliberate dynamic precision scaling~\cite{zamani2017edge} where the calculating precision are scaled by the framework based on multiple factors, including the trade off between computing precision and energy requirement or turnaround time requirement.

Stochastic computing is a famous collection of techniques to achieve precision scaling. Stochastic computing represents continuous values by streams of bits. Calculation precision can be scaled by altering the number of bits in the bitstreams. This technique generally requires a specialized processing unit specifically designed for stochastic computing.
% \begin{itemize}
%     \item Hardware level;
%     \item Step skipping;
    
%     \item Data skipping;
%     \item Memoization;
% \end{itemize}
 
 %Ali Reza Zamani~\etal
%rather than a careful consideration of where and how to perform approximate computing.

%We envision that modern serverless computing runtime (framework) should have a built in precision scaling...

\paragraph*{\textbf{Predictive serving:}}
%need citation
The Serverless computing platform can offer an optional service to approximate the user request ahead of time. This applies at the cloud level by predicting which function or container the user will request next and also at the application level by predicting the user interaction. In a case that predictive serving miss-predict, the optional correction function can trigger to correct the misprediction. 

\paragraph*{\textbf{Loop perforation and instruction replacement:}}
In serverless computing platform, Functions can either be provided as  containers, or as the functional codes. It is possible for the platform provider to analyze the code and apply approximate computing techniques to save the computing resources.
Approxilyzer~\cite{venkatagiri2016approxilyzer} framework proposed by Venkatagiri~\etal analyzes the machine code and dynamically replaces parts of the instruction with the approximated version. The aggressiveness of approximation is tuned with consideration of appropriate quality, resiliency, and overhead.

%\subsection{Hardware Level}

\paragraph*{\textbf{Approximation in the hardware level:}}
%\subsubsection{DVFS approaches}
On the hardware side, a specialized serverless platform can offer a user opt-in special computing platform that reduced computing costs by using approximate computing. The offering can be especially attractive for big data and machine learning tasks that require a large batch of data~\cite{du2014leveraging}. Two of the main hardware-level approximating approaches are DVFS~\cite{rahimi2015approximate} and approximate computing hardware.

By designing specialized hardware to compute in an approximated way, both hardware design complexity and energy consumption can be significantly reduced~\cite{du2014leveraging}. Alternatively, rather than using specialized hardware to perform approximate computing, DVFS is a technique that strategically undervolt certain parts of the common hardware. Such undervoltage can induce errors in computation. However, by controlling the error rate against the resiliency of the software, the result can be acceptable in certain applications. An example of such an approach is shown in the study of Rahimi\etal~\cite{rahimi2015approximate}. Rahimi strategically undervolt the GPU to gain energy efficiency. To allow more error tolerance, they employ humming distance to improve application resiliency.

%\subsubsection{Approximate Circuits}

%\subsection{Task Level}
\paragraph*{\textbf{Approximation in the task level:}}
In serverless computing cloud, each of the processing request can be a part of the bigger workflow. In some cases, such workflow can tolerate the loss of certain sub tasks and therefore, we can exploit the workflow's resilient property in the scheduling level. We explore task deferring and dropping extensively in Chapter~\ref{section:DropDefer}.

\section{Heterogeneous Computing in Serverless Cloud}
\indent Serverless computing hides away the execution details from the end user. Therefore, it open up the opportunity for the providers to utilize heterogeneous computing system to improve efficiency. 
A Heterogeneous Computing (HC) system can be described by two types of heterogeneity: inconsistent and consistent~\cite{ali2000,li2018cost}. Inconsistent machine heterogeneity refers to differences in machine architecture (\eg CPU versus GPU versus FPGA~\cite{zahaf2017het, zhao2017fpga,hong2017gpu}). Consistent machine heterogeneity describes the differences among machines of a certain architecture (\eg different clock speeds). Compute services offered by cloud providers are a good example of an HC system. Amazon cloud~\cite{aws} offers inconsistent heterogeneity in form of various Virtual Machine (VM) types, such as CPU-Optimized, Memory-Optimized, Disk-Optimized, and Accelerated Computing (GPU and FPGA). Within each type, various VMs are offered with consistent performance scaling with price~\cite{aws}. Moreover, both consistent and inconsistent heterogeneity can exist in arriving tasks. For example, an HC system dedicated to processing live video streams is responsible for many categorically different types of tasks: changing video stream resolution, changing the compression standard, changing video bit-rate~\cite{li2018cost}. Each of these task types can be consistently heterogeneous within itself (\eg it takes longer to change resolution of 10 seconds of video, compared to 5).

Many HC systems (\eg~\cite{zong2017marcher,LONI}) present both consistent and inconsistent heterogeneity in machines used and task types processed~\cite{smith09}. These systems present cases where each task type can execute differently on each machine type, where machine type $A$ performs task type 1 faster than machine type $B$ does, but is slower than other machine types for task type 2. Specifically, compute intensive tasks run faster on (\ie matches better with) a GPU machine whereas tasks with memory and disk accesses bottlenecks (\eg in-memory databases~\cite{wang2014using,dos2015smart,malensek2016minerva}) runs faster on a CPU-based machine. 

All of this heterogeneity results in uncertainty for a given task's execution time, thus, inefficiency of resource allocation~\cite{ali2000}. Accordingly, a major challenge in HC systems is to assign tasks to machines to optimize performance goal of the system~\cite{ali2000}.

%%%%%%BG style

\section{Task Request Mapping Heuristics}
\label{section:bgTaskMap}
\begin{figure}[htbp]
  \centering
  \includegraphics[width=0.8\textwidth]{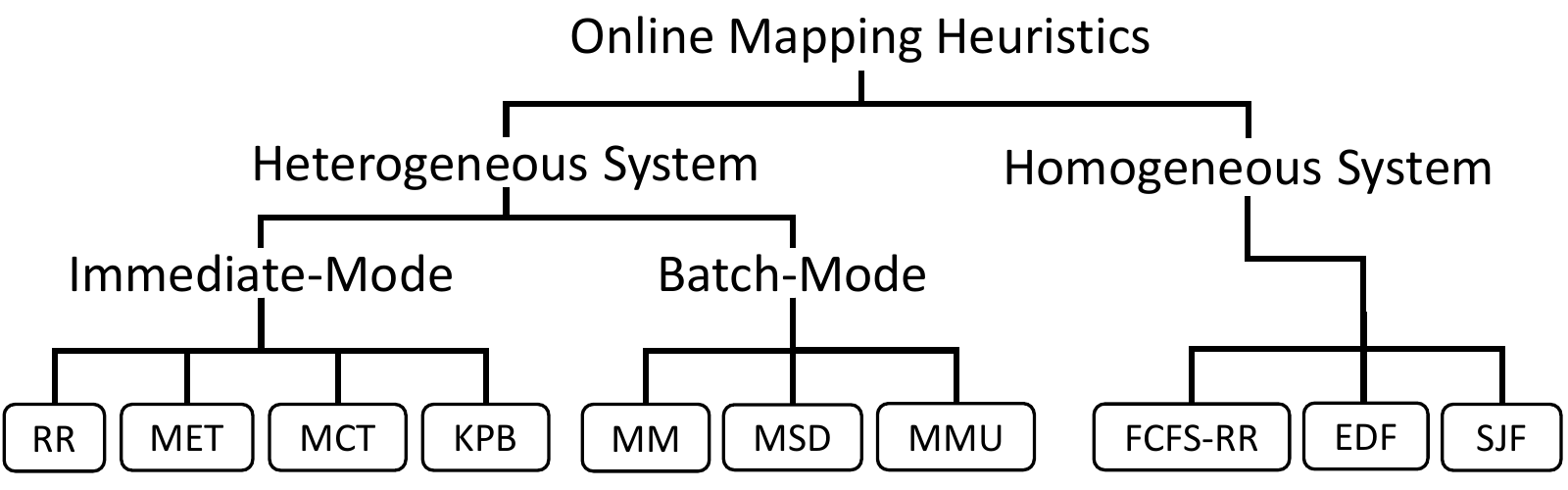}
  \caption{Overview of mapping heuristics widely-used in various types of distributed systems. \label{fig:convtree} }
\end{figure}

\indent Serverless (and cloud computing system in general) have such a wide variety of performance characteristics and QoS requirement. 
In both homogenerous computing system and heterogeneous computing system, optimal task-machine mapping is proven to be an NP-complete problem \cite{coffman76,Ibarra77}. Therefore, a large body of mapping heuristics (\eg~\cite{Braun01,Ibarra77,madni2017performance}) have been developed for these systems. Figure~\ref{fig:convtree} provides an overview of mapping heuristics commonly used in heterogeneous and homogeneous systems. In particular, mapping heuristics of HC systems can be further categorized based on those operate in immediate-mode and in batch-mode resource allocation systems. 

Immediate-mode mapping heuristics do not hold tasks in an arrival queue and they are simpler to develop. In batch-mode heuristics, however, mapping occurs both upon task arrival (when machine queues are not full) and upon task completion. 
Batch-mode heuristics generally use an auxiliary structure, known as \emph{virtual queue} (and also task queue or request queue), where arriving tasks are examined on different machine queues. These heuristics commonly use a two-phase process for decision making. In the first phase, the heuristic finds the best machine for each task, by virtue of a per-heuristic objective. In the second phase, from task-machine pairs obtained in the first phase, the heuristic attempts to choose the best machine-task pairs for each available machine queue slot. After all slots are filled, or when the unmapped queue is emptied, the virtual mappings are pushed (assigned) to the machine queues, and the mapping method is complete.

Although mapping heuristics used in homogeneous computing systems are of batch nature, their logic is simpler than those used in batch-mode of HC systems. Here we review widely-used heuristics in both heterogeneous and homogeneous computing systems.

\subsection{Immediate-mode mapping heuristics for heterogeneous computing systems.}~

\begin{figure}[htbp]
    \centering 
    \includegraphics[width=0.6\textwidth]{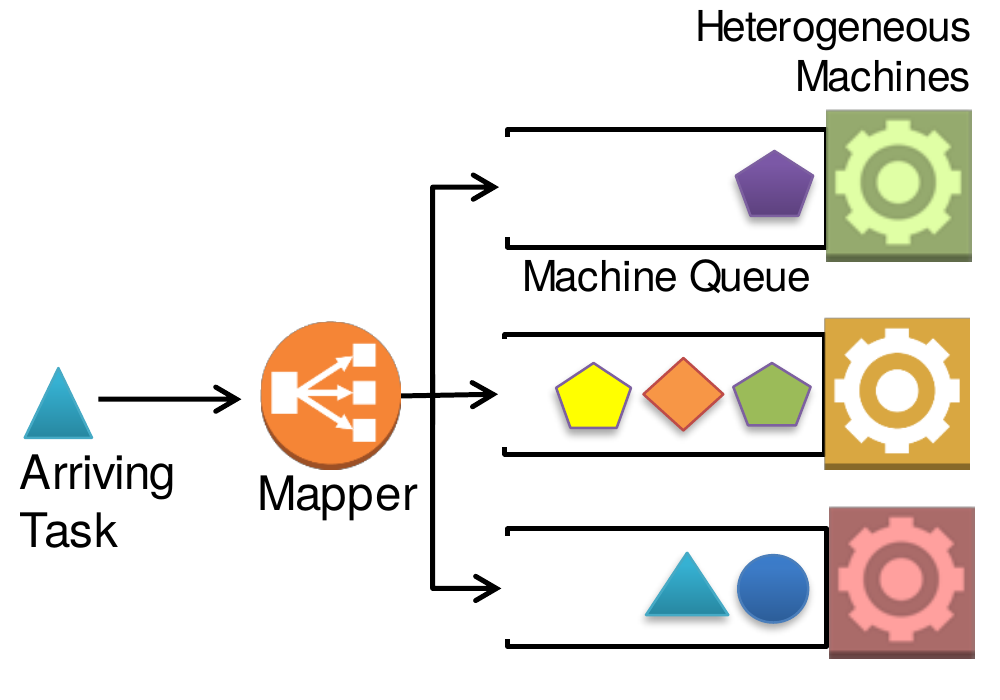} 
    \label{fig:ImmSchedule}
     \caption{Immediate-mode resource allocation operates immediately, upon arrival of a task to the system.}
\end{figure}

\paragraph*{\textbf{Round Robin (RR):}}  %changed from FCFS
In RR~\cite{li2016high}, incoming tasks are assigned in a round robin manner to an available machine, from Machine 0 to Machine $n$.

\paragraph*{\textbf{Minimum expected Execution Time (MET):}} 
In MET, the incoming task $i$ is assigned to the machine that offers the minimum expected execution time (\ie the average of the $PET(i,j)$ for task $i$ on machine $j$). 

%recheck def, if convolve or sum of avg
\paragraph*{\textbf{Minimum expected Completion Time (MCT):}} In MCT, the incoming task is assigned to the machine that offers the minimum expected completion time. The completion time is obtained based on the accumulated expected execution time of tasks queued in a given machine.

\paragraph*{\textbf{K-Percent Best (KPB):}} 
KPB is a combination of MCT and MET. It only considers MCT amongst the $K$ percent of machines with the lowest expected execution times for an incoming task.

\subsection{Batch-mode mapping heuristics for heterogeneous systems.}~

\label{subsec:baseline}
\begin{figure}[htbp]%
\centering
\includegraphics[width=0.8\textwidth]{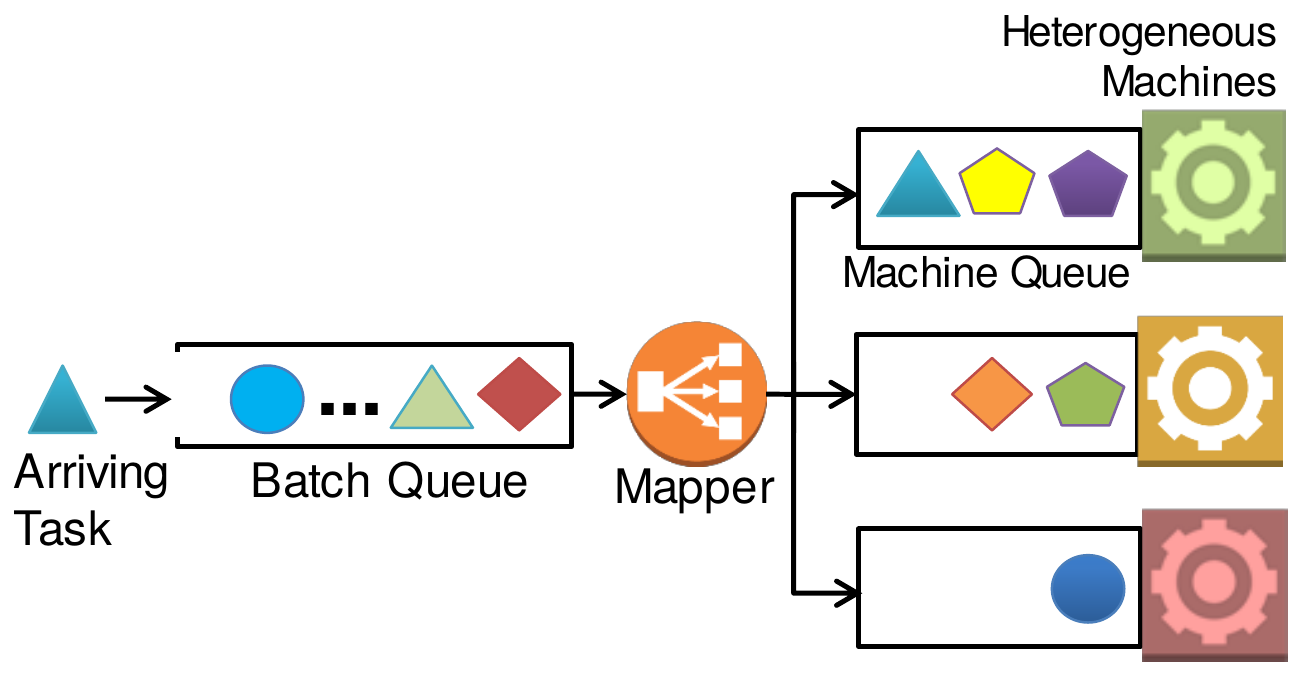}
    \label{fig:BatchSchedule}
    \caption{Batch-mode resource allocation operates on batches of tasks upon task completion (and task arrival when machines queues are not full).}    
\end{figure}

\paragraph*{\textbf{MinCompletion-MinCompletion (MM):}}
This heuristic has been extensively used in the literature~\cite{he2003qos,pedemonte2016accelerating,ezzatti2013efficient, salehi2016stochastic}. In the first phase of the heuristic, the virtual queue is traversed, and for each task in that queue, the machine with the minimum expected completion time is found, and a pair is made. In the second phase, for each machine with a free slot, the provisional mapping pairs are examined to find the machine-task pair with the minimum completion time, and the assignment is made to the machine queues. The process repeats itself until all machine queues are full, or until the batch queue is exhausted.

\paragraph*{\textbf{MinCompletion-Soonest Deadline (MSD):}}
Phase one is as in MM. Phase two selects the tasks for each machine with the soonest deadline. In the event of a tie, the task with the minimum expected completion time is selected. As with MM, after each machine with an available queue slot receives a task from the provisional mapping in phase two, the process is repeated until either the virtual machine queues are full, or the unmapped task queue is empty.

\paragraph*{\textbf{MinCompletion-MaxUrgency (MMU):}}
Urgency of task $i$ on machine $j$ is defined as $U_{ij}={1}/{(\delta_i-\mathbb{E}[C_{ij}])}$, where $\delta_i$ is the deadline of task $i$, $\mathbb{E}[C_{ij}]$ is the expected completion time of task $i$ on machine $j$. 

Phase one of MMU is the same as MM. Using the urgency equation, phase two selects the task-machine pair that has the greatest urgency, and adds that mapping to the virtual queue. The process is repeated until either the batch queue is empty, or until the virtual machine queues are full.

\paragraph*{\textbf{Max Ontime Completions (MOC):}}
The MOC heuristic was developed in~\cite{salehi2016stochastic}. It uses the PET matrix to calculate robustness of task-machine mappings. The first mapping phase finds, for each task, the machine offering the highest robustness value. The culling phase clears the virtual queue of any tasks that fail to meet a pre-defined ($30\%$) robustness threshold. The last phase finds the three virtual mappings with the highest robustness and permutes them to find the task-machine pair that maximizes the overall robustness and maps it to that machine's virtual queue. %PAM, in contrast, chooses the pair with the minimum completion time. 
The process repeats until either all tasks in the batch queue are mapped or dropped, or until the virtual machine queues are full.

\subsection{Mapping heuristics for homogeneous systems.}~

\paragraph*{\textbf{First Come First Served - Round Robin (FCFS-RR):}}
In FCFS-RR, a task is selected in first come first serve manner and is assigned to the first available machine in a round robin manner, from machine 0 to the last machine.

\paragraph*{\textbf{Earliest Deadline First (EDF):}}
EDF is functionally similar to MSD heuristic for HC systems. The first phase finds the machine with the least expected completion time. Then, the second phase sorts the arrival queue in an ascending order based on tasks' deadlines. Next, the task in the head of the arrival queue is assigned to the machine with minimum expected completion time. This process is repeated until all task are mapped or the machine queues are full.

\paragraph*{\textbf{Shortest Job First (SJF):}}
SJF is functionally similar to MM heuristic for HC systems. The first phase finds the machine with the least expected completion time. Then, the second phase sorts the arrival queue in an ascending order based on tasks' expected execution time. Next, the task in the head of the arrival queue is assigned to the machine with minimum expected completion time. This process is repeated until all task are mapped or the machine queues are full.

\section{Probabilistic-Based Mapping Heuristics}%\label{sec:related}
\indent Mapping tasks in HC systems have been shown to be an NP-complete problem~\cite{coffman76,Ibarra77}. As such, there are multiple prior efforts that achieve sub-optimal solutions. Here are some notable mentions where they are either being similar or have some influence on our work.

To model task execution times, Shestak \etal, instead of using a scalar value, lay the groundwork for the use of probability mass functions (aka PMF)~\cite{Shestak08}. The method for convolution of execution times to form completion times for a queue of tasks is established. Our work in Chapter~\ref{section:DropDefer} builds upon their use of PMFs and robustness measurement, while also adding the conditions of probabilistically drop executing tasks and pending tasks.
Khemka \etal~\cite{khemka2015utility} investigate resource allocation in oversubscribed heterogeneous systems. They test task utility functions based on priority, utility class, and urgency. They use a matrix with deterministic execution times, whereas we model the times probabilistically. Also, unlike our approach of probabilistically determining if a task should be dropped, their task dropping occurs only after a task's utility goes below a static threshold. In~\cite{salehi2016stochastic}, Salehi \etal model the stochastic nature of the heterogeneous task types on heterogeneous machine types using a matrix of probability mass functions (PMFs) to improve robustness of dynamic resource allocation.  A mathematical model for calculating the completion time of stochastically modeled tasks in the presence of task dropping is provided. However, Salehi \etal only consider dropping tasks after their deadlines have passed. %Our work builds upon the matrix of PMFs, and the completion time calculations to enable probabilistic task dropping. 

Delimitrou and Kozyrakis~\cite{paragon13} propose Paragon which is an immediate (\ie not batch) dynamic scheduling system for heterogeneous data centers. They use singular value decomposition of historical data to classify incoming tasks based on their heterogeneity. The classifications are used in a greedy algorithm to select a list of candidate resources based on interference, and then from that, the best fit based on heterogeneity~\cite{delimitrou2014quality}. 
Unlike our work in Chapter~\ref{section:DropDefer} that considers probabilistic execution times for decision making, their mapping heuristics operates based on scalar execution times. The performance metrics are also different, as their tasks do not have deadline to consider, Paragon is only concerned about system throughput.

In~\cite{liperformanceanalysis}, Li \etal introduce the affinity (\ie match) of heterogeneous cloud VMs to change coding of video streams. They observed that depending on their content types, video files have different performances on heterogeneous VM types. Particularly, they notice that slow-motion video contents gain from compute intensive VMs, such as GPUs, whereas fast-motion videos do not gain much from such VMs.
They concluded that categorizing videos based on their content types and deploying an inconsistently heterogeneous set of cloud VMs can reduce the incurred cost of using cloud without compromising quality. In another work~\cite{li2018cost}, Li \etal dynamically composes an inconsistently HC system to process a heterogeneous set of video streaming tasks. However, they do not consider the case of task dropping.% because they focus on Video On Demand where tasks continue until completion. 

Malawski \etal~\cite{malawski2015algorithms} evaluate dynamic mapping of deadline- and cost-constrained tasks in cloud. They support dropping workflows that would result in a loss of high priority tasks completion, however, their metrics to quantify and evaluate each task's worthiness are different. Unlike our work, they focus on homogeneous cloud VMs. %, and do not have a priority consideration. 
Tetrisched~\cite{tumanov2016tetrisched} is a mapping method for consistent HC systems used for YARN and MapReduce. It operates based on mixed integer linear programming and considers task execution time on different machines types. Our system uses a similar set of information to for mapping, however, it also leverages task deferring to find a better match for tasks and considers task dropping to alleviate oversubscription and improve robustness. %Mage~\cite{romero2018mage} is a two-tier task mapper based on online machine learning techniques. It maps a task to a group of machines and then to a specific machine within the group. Our work differs from Mage because we consider inconsistent heterogeneity along with the pruning mechanism.

\section{Media Stream Processing}

\indent Our motivational application is a serverless-based on-demand multimedia processing system. Media streams (including video streams and audio streams), either in form of on-demand streaming or live streaming, usually have to be processed before sending to the viewer. The conversion processing to fit characteristics of clients' devices~\cite{liccgrid16,li2016vlsc} can be called transcoding. Transcoding can encompass operations such as \emph{bit rate adjustment, spatial resolution reduction, temporal resolution (aka frame rate) reduction, and video/audio compression standard (codec) conversion}. Transcoding can be either performed offline or in an online (\ie on-demand) manner for media that are rarely accessed~\cite{li2018cost}. However, in the case of live media streaming, it is compulsory to transcode media in an online manner because they are not available for offline processing~\cite{li2016vlsc}. Our motivational application can perform media transcoding operations, but it is also open to other more general processing on the media stream as well. We believe a more personalized media processing (such as on-demand customizable video and audio censorship) can become more common in the future. 

For video media stream, A video stream %, as shown in \ref{fig:2}
 is composed of several sequences. Each sequence itself is composed of multiple \emph{Group Of Pictures} (GOPs) with sequence header at the beginning. Each GOP contains series of frames that begin with $I$ (intra) frame, followed by a number of $P$ (predicted) frames or $B$ (bi-directional predicted) frames. In practice, each GOP is considered as a video streaming processing request with an individual deadline \cite{jokhio2011analysis,matin_paper}. Deadline violation of any request reduces QoS (Quality of Experience) of the viewer. %In this study, serverless functions are modelled after some kind of video processing task.

For workload arrival patterns,
Baccour \etal~\cite{facebookvideolive18} collects and analyzes the number of lives, viewer, location, and other metadata of more than 1,500,000 live video streaming on Facebook during June and July 2018. They found the number of video streams to follow two peaks distribution daily. The difference in workload between each day of the week is small. Therefore, to accurately replicate the arrival pattern, our workload that emulates video streaming requests are generated as the series of valleys, which include a high load period and low load periods.

% \subsection{Cloud-Based Video Transcoding for Video On Demand (VOD)}
% \vspace{1px}
% TO BE WRITTEN (from paper of Chapter~\ref{section:platform})

% \subsection{Cloud-Based Video Transcoding for Live Streaming}
% \vspace{1px}
% TO BE WRITTEN (from paper of Chapter~\ref{section:platform})

\section{Cloud-based Video Streaming Engine (CVSE) and Serverless-based Media Streaming Engine (SMSE)}
\label{sec:CVSEbg}
\indent While storing multiple versions of the same video to serve different needs is a conventional practice, Cloud-based Video Streaming Engine (CVSE) enables on-demand (\ie lazy) transcoding of video streams~\cite{li2018cost}. This is particularly useful for video versions that are rarely accessed. In fact, it has been proven that video streams have long-tail access pattern~\cite{darwich16} where most of video streams are rarely accessed and only a few percentage of videos are popular (hot)~\cite{miranda}. For instance, in the case of YouTube, it has been reported that only 5\% of videos are frequently accessed and the rest are rarely accessed~\cite{gill2007youtube}. This lazy on-demand processing approach enables a wider variety of video customization than what currently offered by a pre-transcoding system.

%On top of various devices dependent specification, we also expect the user requirements to be more vary in the future. For example, a user might need the audio translation or subtitle or auto censorship to screen out unwanted graphics.

In this dissertation, we take the conceptual design clues of CVSE to create Serverless-based Media Stream Processing Engine (SMSE) to efficiently perform media processing on serverless computing system cloud.

\begin{figure}[htbp]
\centering
    \includegraphics[width=0.8\textwidth]{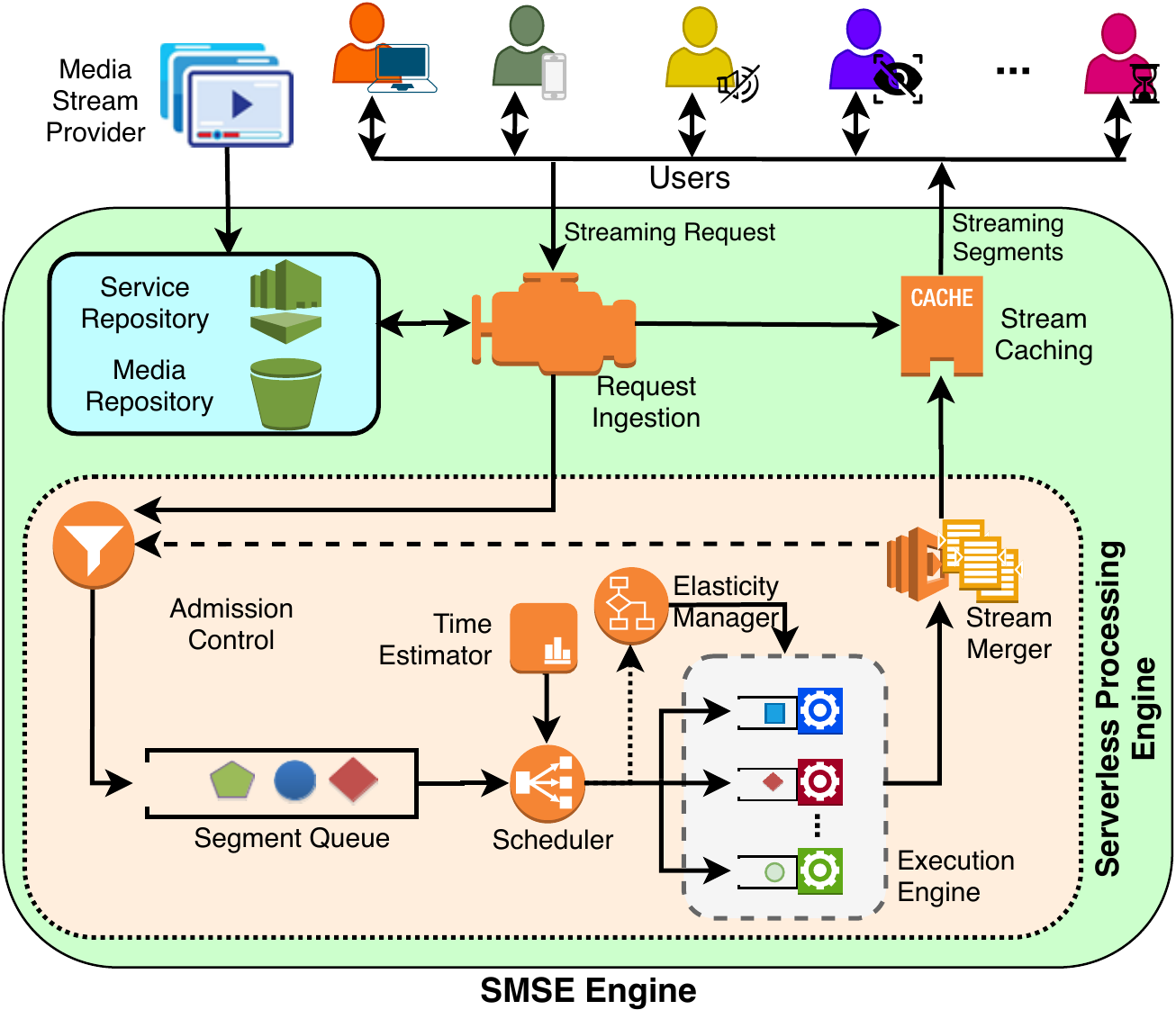}
    \caption{Overview of Serverless-based Media Streaming Engine (SMSE) that is used for on-demand processing of media streams.}
    \label{fig:CVSE}
  \end{figure}
In the SMSE architecture, as shown in Figure~\ref{fig:CVSE}, the Request Ingestion dispatches segments of the requested media stream. A media processing request includes the operation required along with the corresponding parameters bound to that request. Each arriving request is assigned an individual deadline and a priority by the Admission Control component. Then, the Admission Control sends the request to the shared batch queue where the request waits to be assigned by the Scheduler to one of multiple processing unit's queue. Scheduler's batch queue is managed based on a certain scheduling policy, such as FCFS (First-Come-First-Serve), Earliest-Deadline-First (EDF), Max-Urgency-First and Highest-Priority-First~\cite{matin_paper}. Most of scheduling policies are reliant on the Time Estimator component that is aware of the expected execution time of each request type (\ie different transcoding operation) on each Transcoding VMs. Such estimation can be obtained based on historic execution time information of different transcoding operations~\cite{liccgrid16}. 

Once a request is on a processing unit's segment queue, its operational function and required data (such as the media segment itself) is fetched to that processing unit and finally the request gets executed. Stream Merger component is in charge of merging processed segments and transfer them to the viewer. Stream Merger considers an output window for each media stream to keep track of processed segments. The segments that are missing (\eg because of a failure) are requested to be resubmitted by the Admission Control component. The Segments that still require additional processing re-entrance to the Admission Contol. Finally, the segments that are getting popular are recognized by the Stream Cachine of SMSE and are stored to enable caching-based computational reuse.

\section{Summary and Positioning of this Dissertation}
%%% ref to taxonomy for highlighted 
%%look at marcos's work about positioning
%\section{Conclusion}
\indent In this section, we introduced serverless cloud computing paradigm and it goals. We discussed that there are significant redundancy of computation. In particular, we explored the scope for reusing functions and approximately processing them.

Although there are multiple prior studies in task scheduling, serverless cloud computing, and task deduplication, we noticed that there is a huge scope to  improve the efficiency in the serverless cloud computing systems. Accordingly, the positioning of this dissertation in the serverless cloud computing research are two-folds: (a) we devise approximate processing approaches that operate based on dropping and deferring requests proactively; (b) we develop computational reuse mechanism via combining similar task requests together based on their similarity.
    \chapter{Evaluating the Benefit and Feasibility of Computational Reuse}
\label{section:MergeSaving}

\section{Overview}
%need revise
In the previous chapter, our preliminary research found that computational reuse is possible in serverless computing system. Aggregating (a.k.a. merging) of multiple tasks brings about multiple performance benefits, in terms of reducing the makespan time, and incurred cost requirement. However, task merging can have a side-effect of degrading the users' Quality of Service (QoS). In particular, rearranging and aggregating multiple small tasks create large tasks whose execution can potentially lead to deadline violation of either the merged task or other pending tasks scheduled behind it.

\label{sec:intro}
\begin{figure}%[hb]
    \centering
    \includegraphics[width=0.8\textwidth]{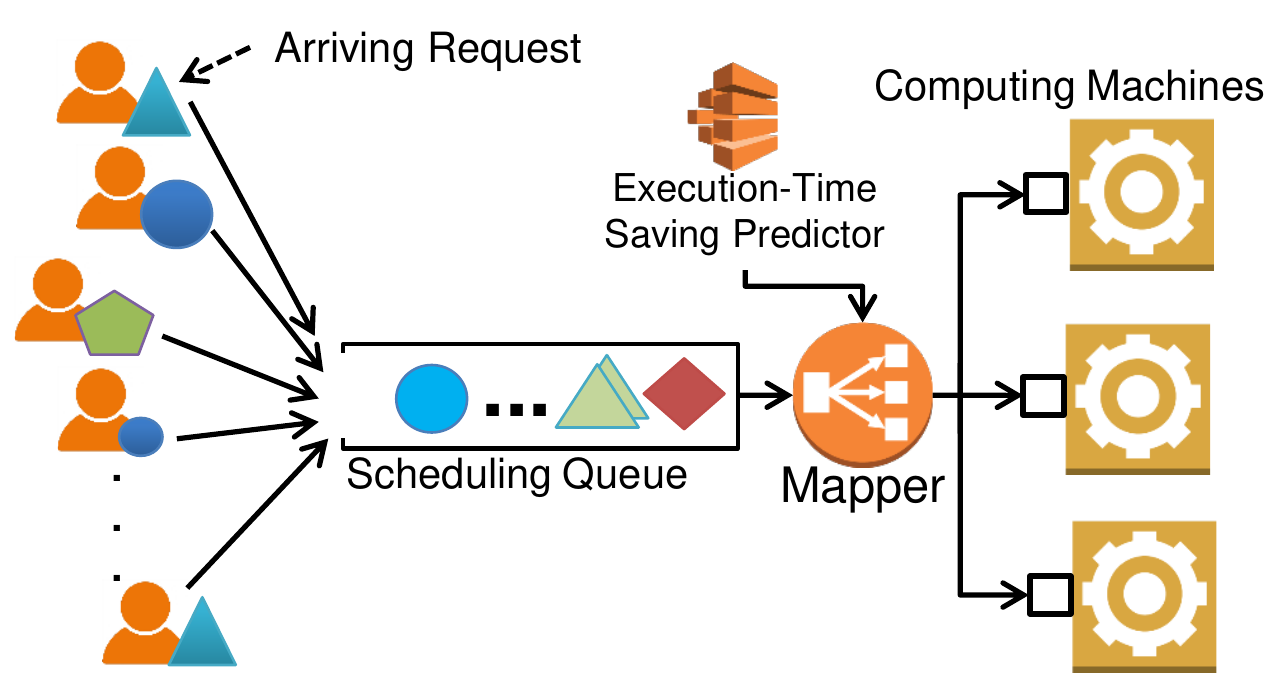}
    \caption{Tasks from multiple users are sent to a shared scheduling queue to be executed on computing resources. The execution-time saving predictor allows efficient use of computing machines. Geometries of different shapes, color, and size represent different (but can be similar) processing tasks.}
    \label{fig:C3intro_multiuser}
    \vspace{-10pt}
\end{figure}

To avoid the side-effect of task merging and deadline violation, informed merging decisions should be made. Specifically, the mapper needs to know how much saving can be accomplished by merging two or more tasks and then, the merging is carried out, only if it is worthwhile. However, to date, a little attention has been paid in the literature to profile the execution-time of the merged tasks and understand their behavior. 

The challenge in profiling the task merging is that the number of possible combinations (\ie merging cases) is interactable and it is not feasible to examine and understand the behavior of all possible cases. Therefore, a method that can predict the execution-time of the merged task is required.

Accordingly, in this chapter, we \emph{first} strategically benchmark a variety of merging cases to understand the influential factors on merging effectiveness. 
Then, in the \emph{second} part, we develop a method (shown as Execution-Time Saving Predictor in Fig.~\ref{fig:C3intro_multiuser}) to estimate the execution-time saving resulted from merging any two or more given tasks. The proposed method operates based on a machine learning model that is trained using our observations in the first part.

\section{Analysis of Video Task Merging Operation}
%While eventually we want to develop a system that can process all the multimedia data type and processing types, we first start from only transcoding video in this part.

\label{sec:exp}
\subsection{Video benchmark dataset.}~
We used 3,159 video segments to construct the benchmark dataset. The video segments are gathered from a set of 100 open-license videos in YouTube \cite{youtube}. To build a representative dataset, we assured that the chosen videos cover diverse content types with distinct motion patterns (\ie fast or slow pace) and various object categories. 

\begin{table}[htb]
\centering
\normalsize
\setlength\tabcolsep{1pt}

\begin{tabularx}{0.9\textwidth} { 
   >{\centering\arraybackslash}X 
  | >{\centering\arraybackslash}X 
  | >{\centering\arraybackslash}X 
  | >{\centering\arraybackslash}X 
  | >{\centering\arraybackslash}X
   >{\centering\arraybackslash}X}
% \hline
  & \textbf{Codec} & \textbf{Frame-rate} & \textbf{Resolution} & \textbf{Container}  \\
 \hline
 \hline
 \textbf{\small{Standardized format}} & H.264 (High) & 30 fps & 1280$\times$ 720   & MPEG transport stream (TS)\\
\end{tabularx}
\\
\vspace{5pt}
\caption{Standardized specifications for videos in the collected video benchmark dataset.}
\label{table:1}
\vspace{-20pt}
\end{table}%
%%%%%%%%%%%%%%%%%%%%%%%%%%%%%%%%%%%%%%%

To systematically analyze the evaluation results and eliminate the impact of different video formats that affect the execution-time, we split all the videos to two-second video segments with the standardized format detailed in Table \ref{table:1}. It is noteworthy that segmenting videos is a common practice in stream providers and the two-second is to comply with the MPEG transport streaming standard~\cite{fairhurst2005unidirectional,alzahrani2018impact}. We choose H.264 as the unified codec, because it is still the most common and widely compatible format for video streaming. It is worth noting that we selected libx264 \cite{x264} as the encoders to change all the proposed video formats. The benchmark dataset contains 3,159 video segments that are publicly available\footnote{\url{https://bit.ly/3gKNijT}} 
%\url{https://github.com/hpcclab/videostreamingBenchmark}} 
for reproducibility purposes, 
with detailed description of the each video\footnote{\url{https://bit.ly/2YMIwwb}}.

\subsection{Benchmarking execution-time of video transcoding tasks.}~
\label{subsec:collection}
Based on the video segments of the collected dataset, we perform a set of benchmark services that consists of four primary video transcoding operations (tasks), namely changing bit-rate, frame-rate, resolution, and codec. Early evaluation of the collected execution-time revealed a remarkable variation in the execution-time of some task types. Specifically, we noticed that codec execution-time is far beyond the other three task types.
Accordingly, we categorize the tasks types into two groups: \emph{First} group is called Video Information Conversion (\emph{VIC}) that includes changing bit-rate, frame-rate, or resolution task types. Tasks of this group have a low variation in their execution-times, when processing different video segments on the same machine type. \emph{Second} group is Video Compression Conversion that only includes the codec task type (hence, we call it the Codec group). In contrast to the first group, the codec execution-time  (and subsequently its merge-saving) for different video segments varies remarkably even on the same machine.

\begin{table}[htb]
\normalsize
\centering
\setlength\tabcolsep{1pt}
\scalebox{1}{\begin{tabular}{c|c|c|c|l|l}
                                                                               \multicolumn{3}{c|}{\small{\textbf{\small{Video Information Conversion (VIC)}}}}                                                                    & \multicolumn{3}{c}{\multirow{2}{*}{\small{\textbf{Codec}}}} \\ \cline{1-3}
                                                                               \multicolumn{1}{c|}{\small{\textbf{Bit-rate}}} & \multicolumn{1}{c|}{\small{\textbf{Frame-rate}}} & \multicolumn{1}{c|}{\small{\textbf{Resolution}}} & \multicolumn{3}{c}{}                       \\ \hline \hline 
                                                                    
                                                                             384K                          & 10 fps                             & 352$\times$288                         & \multicolumn{3}{c}{\small{MPEG-4}}                 \\ %\hline
                                                                              512K                          & 15 fps                             & 680$\times$320                         & \multicolumn{3}{c}{\small{H.265/HEVC}}             \\ %\hline
                                                                              768K                          & 20 fps                             & 720$\times$480                         & \multicolumn{3}{c}{\small{VP9}}                    \\ %\hline
                                                                              1024K                         & 30 fps                             & 1280$\times$800                        & \multicolumn{3}{c}{-}                       \\ %\hline
                                                                              1536K                         & 40 fps                             & 1920$\times$1080                       & \multicolumn{3}{c}{-}                       %\\ \hline
\end{tabular}}
\vspace{5pt}
\caption{The list of parameters employed to form various transcoding tasks. Each transcoding task changes only one specification of the videos in the standardized benchmark dataset. Accordingly, there are collectively 18 transcoding tasks: 5 for bit-rate changing, 5 for frame-rate changing, 5 for resolution changing, and 3 for codec changing.} 
\label{table:2}
\vspace{-10pt}
\end{table}

To limit the degree of freedom in execution-time, we configured each transcoding task to change only one specification of the videos in the benchmark dataset. The characteristics (parameters) of the evaluated transcoding tasks are listed in Table~\ref{table:2}. According to the table, there are 4 task types and collectively 18 transcoding tasks, including 5 different parameters in tasks changing bit-rate, 5 parameter for tasks changing frame-rate, 5 parameters in tasks that change resolution, and 3 parameters in tasks changing codec. 

To evaluate a variety of task merging cases, we compare the time difference between executing the 18 video transcoding tasks individually against executing them in various merged forms. Our preliminary evaluations  %~\cite{denninnart2018leveraging,Chavit2020Leveraging} %cite paper of the next chapter???
showed that there is little gain in merging more than five tasks. In addition, we observed that it is unlikely to find more than five (similar, but not identical) 
mergeable tasks at any given moment in the system. As such, in the benchmarking, the maximum number of merged tasks (a.k.a. degree of merging) is limited to five. Even with this limitation, exhaustively examining all possible permutations of merging 18 tasks (in batches of 2, 3, 4, 5 tasks) collectively leads to $C(18,2)+C(18,3)+C(18,4)+C(18,5)$ cases, where $C(x,y)$ 
refers to $y$-combinations from a set of $x$ tasks. That entails 12,597 experiments per video segment.
As performing this many experiments is time prohibitive, we reduce the number of possible test cases to some highly representative merging cases for each video segment. Details of the conducted benchmarking is as follows:

\begin{enumerate}[label=(\Alph*)]
    \item We measured the execution-time of the 18 tasks on each one of the 3,159 video segments in the dataset individually. This means that, in this step, we collected 56,862 execution-times for individual tasks. 
    \item We measured the execution-time of merged tasks with the same operation and 2---5 various parameters. That is, each merged transcoding task is composed of one operation (\eg changing resolution) with two to five different parameters (\eg based on the possible values of resolution, mentioned in Table~\ref{table:2}). Then, to measure the magnitude of saving resulted by the task merging (henceforth, referred to as \emph{merge-saving}), the resulting execution-times are compared against execution-time of individual tasks, generated in Step (A).
   \item In our initial evaluations, we observed more consistent behavior in merge-saving of the VIC group, as opposed those mergings included codec. As such, our evaluations were focused on the merging cases with various operations within the VIC group. Each operation can have various parameters. For instance, consider video $A$ with bit-rate $b_1$, frame-rate $f_1$, and resolution $r_1$. We merge multiple transcoding tasks on $A$ to change: its resolution to $r_2$, its bit-rate to $b_2$ and its frame-rate to $f_2$ and $f_3$. Then to measure the magnitude of merge-saving, the resulting execution-times are compared against execution-time of individual transcoding time from (A).
    \item We benchmark and analyze execution-time of merged tasks with codec operation and operations from the VIC group. The process is similar to (C). However, each merged task is composed of one codec changing operation with one or more VIC class operations.
 \end{enumerate}

\subsection{Analyzing the impact of task merging on execution-time.}
\begin{figure}[htbp]
\centering
\subfloat[Bit-rate]{
    %\begin{minipage}[t]{7cm}
    \centering
    \includegraphics[width=0.314\textwidth]{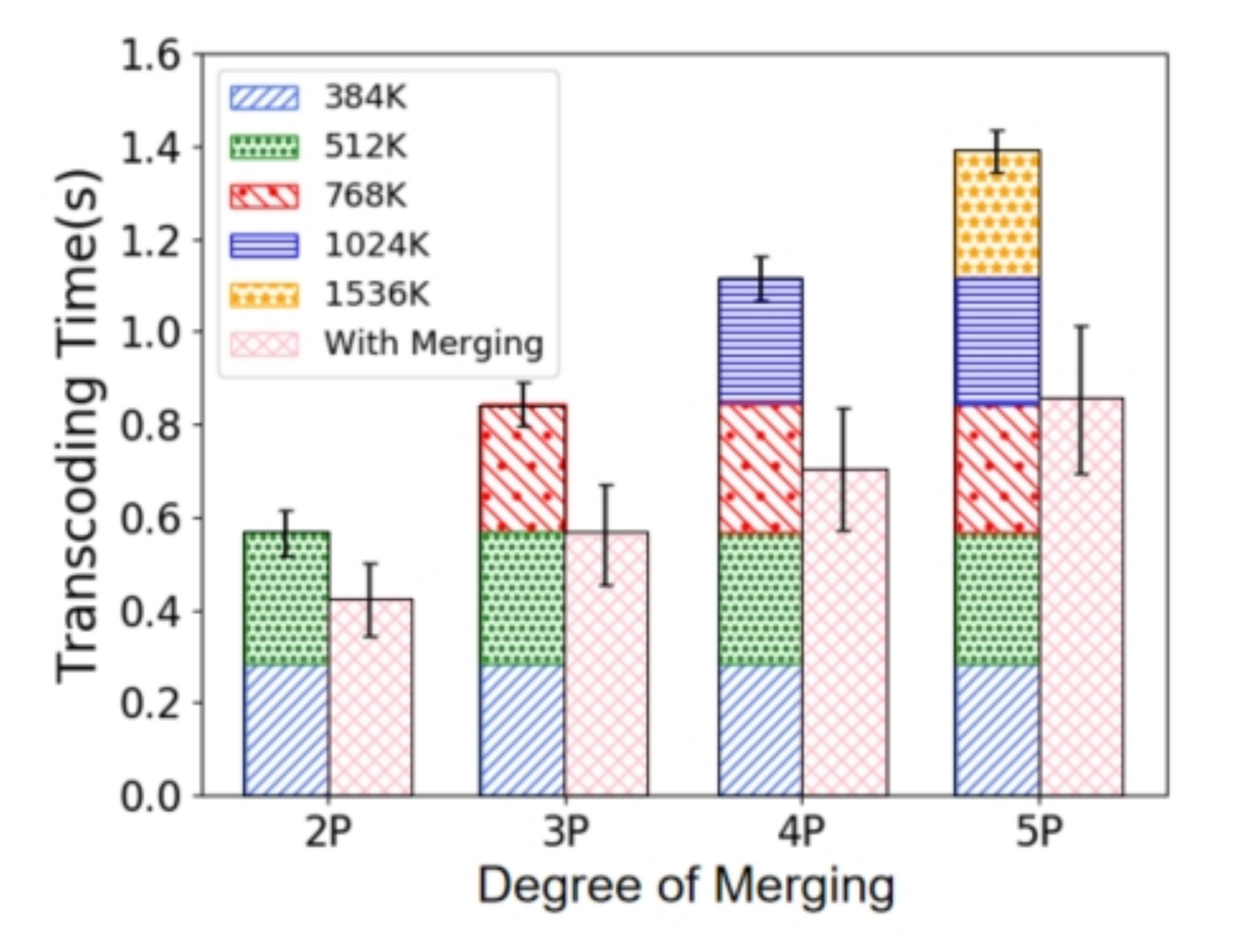}
    %\hspace{1in}
    \label{fig:result-singlea}
    %\end{minipage}
}
%\hspace{0.1}
\subfloat[Frame-rate]{
    %\begin{minipage}[t]{7cm}
    \centering
    \includegraphics[width=0.307\textwidth]{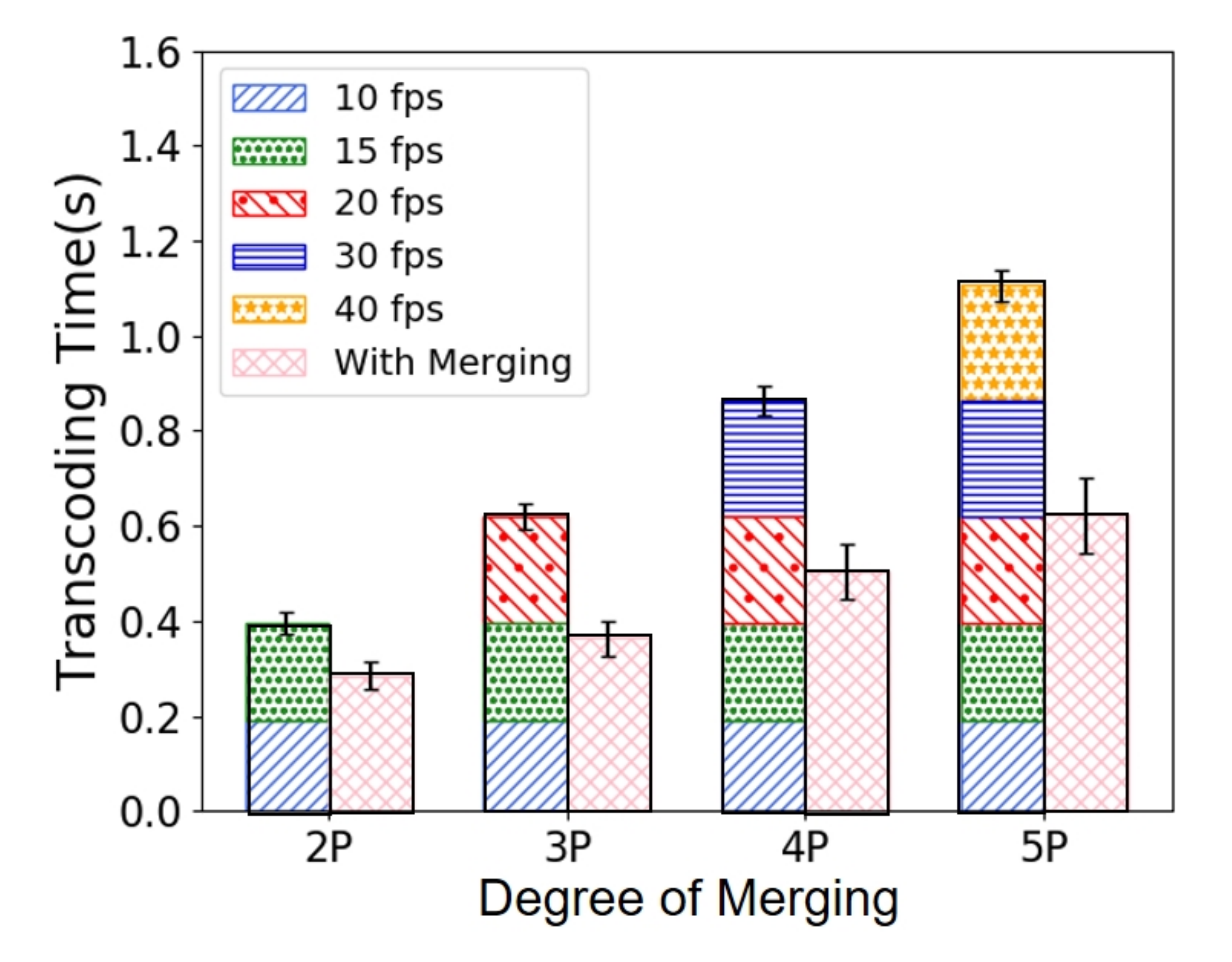}
    \label{fig:result-singleb}
    %\end{minipage}
}
%\hspace{0.1}
\subfloat[Resolution]{
    %\begin{minipage}[t]{7cm}
    \centering
    \includegraphics[width=0.295\textwidth]{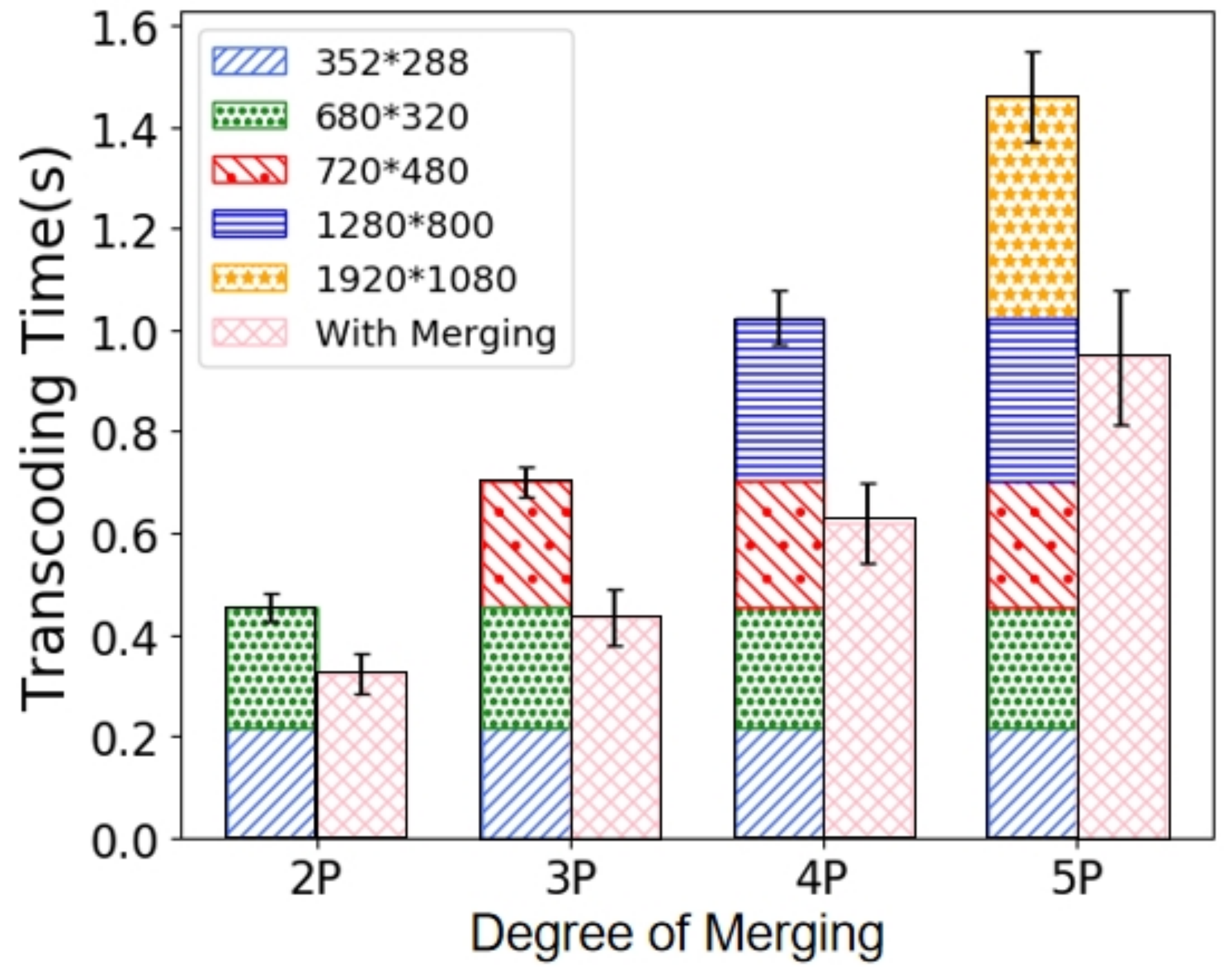}
    \label{fig:result-singlec}
    %\end{minipage}
}

\caption{Comparison of the total transcoding time (\ie makespan) (in seconds) to execute multiple tasks with two to five parameters (2P---5P in the horizontal axes) within the VIC group in two scenarios: executing individual tasks sequentially (without task merging) versus executing them as a merged task. Sub-figures (a), (b), and (c) represent transcoding time of bit-rate changing operation, frame-rate changing operation, and resolution changing operation, respectively.}
\label{fig:result-single}
\end{figure}

~
\paragraph*{Evaluating the impact on the makespan time:}
To understand the task merging performance behavior, we evaluate the total transcoding time (a.k.a. makespan) of the tasks in the VIC group under two scenarios: transcoding with and without merging. We consider merging of two to five parameters for bit-rate, frame-rate, and resolution separately shown as $2P$ to $5P$ in the horizontal axes of Fig.~\ref{fig:result-single}. 

%It is needless to say that %%%%Chavit: remove this phase reduce a line
The difference between transcoding time when executing each task individually versus when the tasks are merged represents the merge-saving. 

We observe that, in all cases, there is an increasing trend in the merge-saving when the degree of merging is increased.
Interestingly, we observe that the ratio of merge-saving generally increases for the higher degrees of merging. The only exception is in Fig.~\ref{fig:result-singlec} (changing resolution) that by increasing the degree of merging from 4P to 5P, the merge-saving ratio is not increased. 
In general, we can conclude that all task merging with operations within the VIC group consistently and substantially save the execution-time. 

\begin{figure}[htbp]
    \centering
    \includegraphics[width=0.8\textwidth]{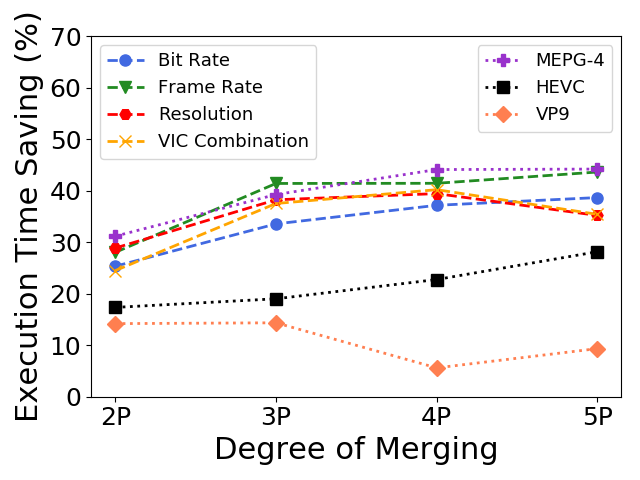}
    \caption{The result of merge-saving across varying numbers of the videos transcoding tasks. Figure (a) and (b) show the makespan saving when tasks merged within the VIC group and the makespan saving when codec transcoding tasks merged with VIC group, respectively.}
    \label{fig:result-sum}
    \vspace{-12pt}
\end{figure}

\paragraph*{Evaluating the impact on execution-time saving:}
\label{subsec:C3Eval}
Changing the view to focus on execution-time saving percentage, Fig.~\ref{fig:result-sum} shows that, on average, when two tasks in the VIC group are merged ($2P$), the execution-time is saved by 26\%. 
The saving increases to 37\% when three tasks merged together. From there, the saving taper off to around 40\% for four and five tasks merging (4P and 5P).  We do not observe significant extra merge-savings after 5P. In addition, forming a large merged task complicates the scheduling and increase the potential side-effects (in the form of delaying) the completion of the large task itself or other pending tasks \cite{Chavit2020Leveraging}. This observation holds for the merged tasks compose of multiple different operations within VIC group (denoted as \textit{VIC Combination}).

For merged tasks that include codec changing operations, the results are far from consistent. Merge-saving of tasks that include MPEG-4 codec changing behave similarly to pure VIC group operations. Merge-savings of tasks with HEVC codec changing operation are consistently lower than any aforementioned cases for every degree of merging. 
The minimum saving is observed when the merged task includes VP9 codec changing operation. In which case, the saving is even reduced when the degree of merging increased from 3P to 4P.

The results suggest that the significant gain in merging takes place in the first three tasks merging.
We can conclude that, to strike a balance between efficiency gain and potential side-effects of task merging, the system should target to form groups of about three tasks, rather than forming the biggest possible group of task merging. 
It is also worth mentioning that codec changing operations have a significantly (up to eight times) longer execution-time than VIC group operations. Merging a codec changing task to VIC group tasks does not necessarily offer a significant merge-saving, yet can jeopardizes the users' QoS. That is, merging a short task from the VIC group to a large task from the codec group can significantly delay the completion time of the short task and degrades its QoS (\eg in terms of missing the task's deadline).

%%%%%%%%%%%%%% section learning
\section{Predicting the Execution-Time Saving of Task Merging}
\label{sec:learning}

\subsection{A model to predict execution-time saving.}~
In the benchmarking process, %we realized that the merge saving has a correlation with factors such as the video segment's characteristics, the task type, and its parameters. In addition,
we noticed that the number of cases that tasks can be merged in a system is interactable (see Section~\ref{subsec:collection} ). That is, it is not feasible to pre-generate the knowledge of the merge-saving of all task types with all possible parameter values and for all video files. However, such a knowledge is crucial to decide about performing a task merging case \cite{Chavit2020Leveraging}.  
As such, our goal in this part is to leverage our findings in the benchmarking section and develop a machine learning model that can predict the merge-saving of any given set of mergeable tasks based on the task types and characteristics of the video segments.

In total, 81,327 data points, obtained from the benchmarking, were used to train the proposed model. For training and validating the model, we extracted metadata of the benchmark videos and transcoding configurations. A short sample of these metadata is shown in Table~\ref{table:3}.
%and the whole training dataset and the trained model are also made publicly available\footnote{\url{https://github.com/hpcclab/Merge-Saving-Prediction}}. 
As we can see in the table, for each video, we collected its essential static features, including duration, segment size, frame-rate (FR), width, and height (for the sake of better presentation, only few columns are shown in the table). 
Then, we concatenate the static features to the specification of merged task's transcoding configuration. 
The transcoding configuration includes the number of bit-rate changing (B), spatial resolution/frame-rate changing (S), resolution changing (R), and the type of codec changing included in the merged task. 
The output of the machine learning model is the merge-saving, \ie the percentage of improvement in execution-time upon merging several tasks versus not merging them. %The duration property represents the length of the video segments in seconds, which is two seconds exception for the last segment of each video. The segment size is the volume of the segment in bytes. 

Since the three codec transcoding parameters behave significantly different, the codec operation parameters are marked separately in Table~\ref{table:3}, as MPEG4, VP9, and HEVC columns. 
In contrast, for the ones in the VIC group, we observed that their configurations (\ie parameter values) have little influence on the merge-saving, in compare with their degree of merging.  As such, for elements of the VIC group, we consider the number of operations (sub-tasks) in the merged task as opposed to the value of their parameters. Accordingly, the integer values in the B, S, and R columns represents the number of those operations included in the merged task. The main benefit of marking the table in this manner is to create a robust model that can infer the merge-saving even for unforeseen parameters. Arguably, if we bind the elements of VIC group to their parameter values in the training, then the model cannot efficiently predict the merge-saving of a merge request whose parameter values are out of the scope of the training dataset.

\begin{table}[htb]
\centering
\setlength\tabcolsep{0.5pt}
\begin{tabular}{c|c|c|c|c|c|c|c|c|c|c|c}
\textbf{\small{Dura-}} & \textbf{\small{Size}} &  \textbf{\small{FR}} & \textbf{\small{Width}} & \textbf{\small{Height}}  & \textbf{\small{B}}   & \textbf{\small{S}}   & \textbf{\small{R}}   & \textbf{\small{MP-}} & \textbf{\small{VP9}} & \textbf{\small{HEVC}} & \textbf{\small{Saving}} 
\\
\textbf{\small{tion (s)}} & \textbf{\small{(KB)}} &  &  &   &   &    & &\textbf{\small{EG-4}}  &  & &
\\ \hline
\hline

2.0 & 876  & 30  & 1280  & 720    & 1   & 0   & 0   & 1     & 0   & 0    & 33.60\% \\

2.0 & 1085 & 30  & 1280  & 720    & 1   & 2   & 1   & 0     & 0   & 0    & 39.17\% \\

2.0 & 1231 & 30  & 1280  & 720    & 1   & 1   & 1   & 0     & 1   & 0    & 20.22\%  \\

1.2 & 969  & 30  & 1280  & 720    & 0   & 0   & 1   & 0     & 1   & 0    & 27.89\% \\

2.0 & 864  & 30  & 1280  & 720    & 1   & 3   & 1   & 0     & 0   & 0    & 23.33\% \\

2.0 & 1091 & 30  & 1280  & 720    & 1   & 1   & 1   & 0     & 0   & 1    & 21.95\%  \\

0.9 & 347  & 30  & 1280  & 720    & 1   & 0   & 1   & 0     & 0   & 0    & 31.32\% \\

...      & ...  & ...       & ...   & ...    & ... & ... & ... & ...   & ... & ...  & ...\\
\end{tabular}
\vspace{4pt}
 \caption{A sample of the training dataset. Left side columns show static features of videos, such as duration, size, frame-rate (FR), and dimensions. B, S, and R columns
 represent bit-rates, frame-rate, and resolution changing operation sub-tasks in the particular merged task. Codec changing operation parameters are marked separately with one possible parameter per column (as MPEG-4, VP9, and HEVC.) The Saving column indicates the merge-saving caused by a particular task merging.}
\label{table:3}
\vspace{-10pt}
\end{table}

\subsection{Gradient boosting decision tree (GBDT) to predict the execution-time saving.}~
Decision tree~\cite{vadim2018overview} is a known form of prediction model that functions based on a tree-based structure. Starting from the head node, the model performs a test on a feature at each one of its internal nodes. Ultimately, the traversal leads to a leaf node that includes the prediction~\cite{magerman1995statistical}. In particular, decision trees are proven to be appropriate for predicting numerical of unknown data~\cite{kotsiantis2013decision}. Because merge-saving prediction can be considered as a kind of numerical prediction problem, we choose decision trees to predict the saving. However, solutions based on a single decision tree are generally prone to the over-fitting problem~\cite{kotsiantis2013decision}. That means, the model is excessively attached to the training dataset such that, at the inference time, its prediction cannot cover slight variations in the input. %citation removal du2002building (replaced with kotsiantis2013decision)

Accordingly, to devise a prediction model that is robust against over-fitting, we utilize a optimal method of decision trees, known as Gradient Boosted Decision Trees (GBDT)~\cite{ friedman2002stochastic}. This is an iterative construct based on boosted ensemble of weak-learner decision trees. In fact, GBDT combine the multiple boosted weak-learners into a high accuracy and robust model. The boosting technique uses a process in which subsequent predictors learn from errors of the previous predictors. The objective of each iteration is to reduce the prediction error, which is calculated by a loss function~\cite{friedman2002stochastic}, to the minimum possible.

The pseudo-code, shown in Algorithm~\ref{fig:boosting}, elaborates on how the merge-saving prediction model is trained based on GBDT. On Step 2 of the pseudo-code, a subset of the benchmark dataset, explained in Section \ref{sec:exp}\,~ , is generated and is used as the training dataset, denoted as $t$. We considered 80\% of the benchmarked dataset in $t$. The initial decision tree, denoted as $B_0(x)$, is created with random number and trained based on $t$ on Step 3. On Step 4, the main loop of the training model aims at creating one weak model based (decision tree) per iteration. 
%Within the loop, the initial decision tree, denoted as $B_0(x)$, is created with random number and trained based on $t$ in Step 4.1. 
Note that $x$ represents the input features of the merged task, as expressed in Table~\ref{table:3}.
In this step, there are various hyper-parameters that affect form of the decision tree being created. Notable hyper-parameters (among many others~\cite{kotsiantis2013decision}) that impact the accuracy of the prediction model are the learning rate (denoted as $L$), maximum depth of the individual regression estimators (denoted as $D$), the minimum number of samples required to split an internal node (denoted as $S$), and the minimum number of samples needed to be at a leaf node (denoted as $J$). In Sections~\ref{subsec:evalL}\, ~---\ref{subsec:evalSJ}\,~~, we elaborate on the appropriate values of these hyper-parameters such that the prediction accuracy of the merge-saving prediction model is maximize.

\begin{algorithm}[h]
    \caption{Pseudo-code of the method to build the prediction model of the execution-time saving of a merged task.}
    \label{fig:boosting}
    \begin{algorithmic}[1]
        \REQUIRE
            The merge-saving benchmark dataset $T$, obtained from Section~\ref{sec:exp};
        \ENSURE
            Execution-time saving predictor $B_M(x)$;
        \STATE Let $M$ be the number of decision trees (and iterations)
        \STATE Create training dataset t, where $t \subset T$;
        \STATE Initialize decision tree $B_0(x)$ from $t$;
        \FOR{$m \leftarrow 1$ to $M$} 
            \STATE $r_{mi} \leftarrow$ Compute the prediction error of the $B_{m-1}(x)$;
            \STATE Utilize ($x_{i}$, $r_{mi}$) to fit a regression tree, calculating the fitted values for each terminal region;%, giving terminal regions $R_{mj}$, j = $1,2,...,J_m$;
            %\FOR{$j \leftarrow 1$ to $J$}
                %\STATE Calculate the fitted values $c_{mj}$ for each terminal regions:
                %\STATE $c_{mj} \leftarrow arg\underset{c}{min}\sum_{x_i\in R_{mj}}^{}L(y_i, B_{m-1}(x_i)+c)$;
            %\ENDFOR
            \STATE Update $B_m(x)$ based on the $B_{m-1}(x)$;
            %\STATE $B_m(x) \leftarrow$ Update $B_m(x)$ based on the $B_{m-1}(x)$ and $c_{mj}$;
        \ENDFOR 
        \RETURN The merge-saving prediction model $B_M(x)$;
    \end{algorithmic}
\end{algorithm}
%%%%%%%%%%%%%%%% Figure from evaluation, 

%%%%%%%%%%%%%%%%%
%Once we create $B_m(x)$, in Step 4.2, the prediction error for records of the training is calculated. 

Let $r_{mi}$ denote the prediction error of record $i \in t$. Recall that the core idea of GBDT is to learn from and improve upon the mistakes of the previous iteration. Accordingly, on Step 5, we calculate $r_{mi}$ of the model created in the previous iteration (\ie $B_{m-1}(x)$). %Note that for the first iteration, we define $B_{0}(x)=B_{0}(x)$. %%the way we now define it, this statement is not needed (no B' anymore)
The value of $r_{mi}$ is calculated based on Equation~\ref{eq:gbdgrad}. In this equation, $y_i$ is the ground truth (\ie actual saving in Table~\ref{table:3}) for the prediction made by $B_{m-1}(x_i)$. Also, $L(y_{i}, B_{m-1}(x_{i}))$ denotes the loss function and it is calculated as explained in~\cite{friedman2002stochastic}. 

\begin{equation}
    r_{mi} = -{\begin{bmatrix}
          \frac{\partial L(y_{i} , B_{m-1}(x_{i}))}{\partial B_{m-1}(x_{i})}
          \end{bmatrix}} %_{B(x) = B_{m-1}(x)}
          \label{eq:gbdgrad}
\end{equation}

On Step 7, the decision tree is updated (called $B_m(x)$) based on the value of $r_{mi}$. 
On Step 9, the ensemble of created decision trees form the merge-saving prediction model. Details of forming the ensemble can be found in \cite{friedman2002stochastic}.
% In Step 4.3, the decision tree is updated (called $B_m(x)$) based on the value of $r_{mi}$. 
% At the end of each iteration (Step 4.4) the updated decision tree is saved. In Step 5, the ensemble of created decision trees form the merge-saving prediction model. Details of forming the ensemble can be found in \cite{friedman2002stochastic}.

%The equation to form the ensemble is presented in~\ref{eq:gbdtfinal}.  In this equation, $j = 1,2,3, ...,J$, is the corresponding leaf node area, $J$ is the number of regression tree leaf node used in step 4.2. $T_j$ is the optimal fitting value to the area of corresponding leaf node $j$,and $I$ is an identity matrix. 

%%%%%%%%%%%%%% section Eval

%%% the figure code have to move to previous page, if it fall to another page.

\section{Performance Evaluation of the Execution-Time Saving Predictor}
\label{sec:C3eval}
To maximize the prediction accuracy and efficiency, it is critical to determine the optimal combination of parameter values used in the GBDT model. As such, in this section, first, we examine various parameters that influence the accuracy of the prediction model.  
The best performance is achieved by deliberately selecting the fittest combination of these parameters. 
The predicted time-saving is primarily used for scheduling purposes where prediction errors can perturb the scheduler. As such, we consider Root Mean Square Error (RMSE) as the primary performance evaluation metric.

Once we optimally configure the proposed GBDT model, in the second part, we measure and analyze its prediction accuracy with respect to other methods that can alternatively employed to predict the merge-saving.

\subsection{Tuning the learning rate of the predictor method.}~
\label{subsec:evalL}
Gradient boosting predictors become robust when the model is sufficiently learned. However, over-fitting can occur, if they learn too fast with too little variation in the input. The learning rate ($L$) of the predictor indicates how fast it can learn at each iteration. This parameter is generally considered along with the number of trees (denoted as $M$) that is used to train the model. Parameter $M$ is also known as the iterations parameter, because each iteration generates one tree.  

In this part, our goal is to tune the predictor with the appropriate learning rate. For that purpose, we examine the RMSE metric when the learning rate $L$ changes in the range of [0.5 , 0.005]. Each learning rate is examined when number of trees varies in the range of [350 , 6,000].

\begin{figure}[htbp]
\centering

\subfloat[Number of Trees (M)]{
    \includegraphics[width=0.31\textwidth]{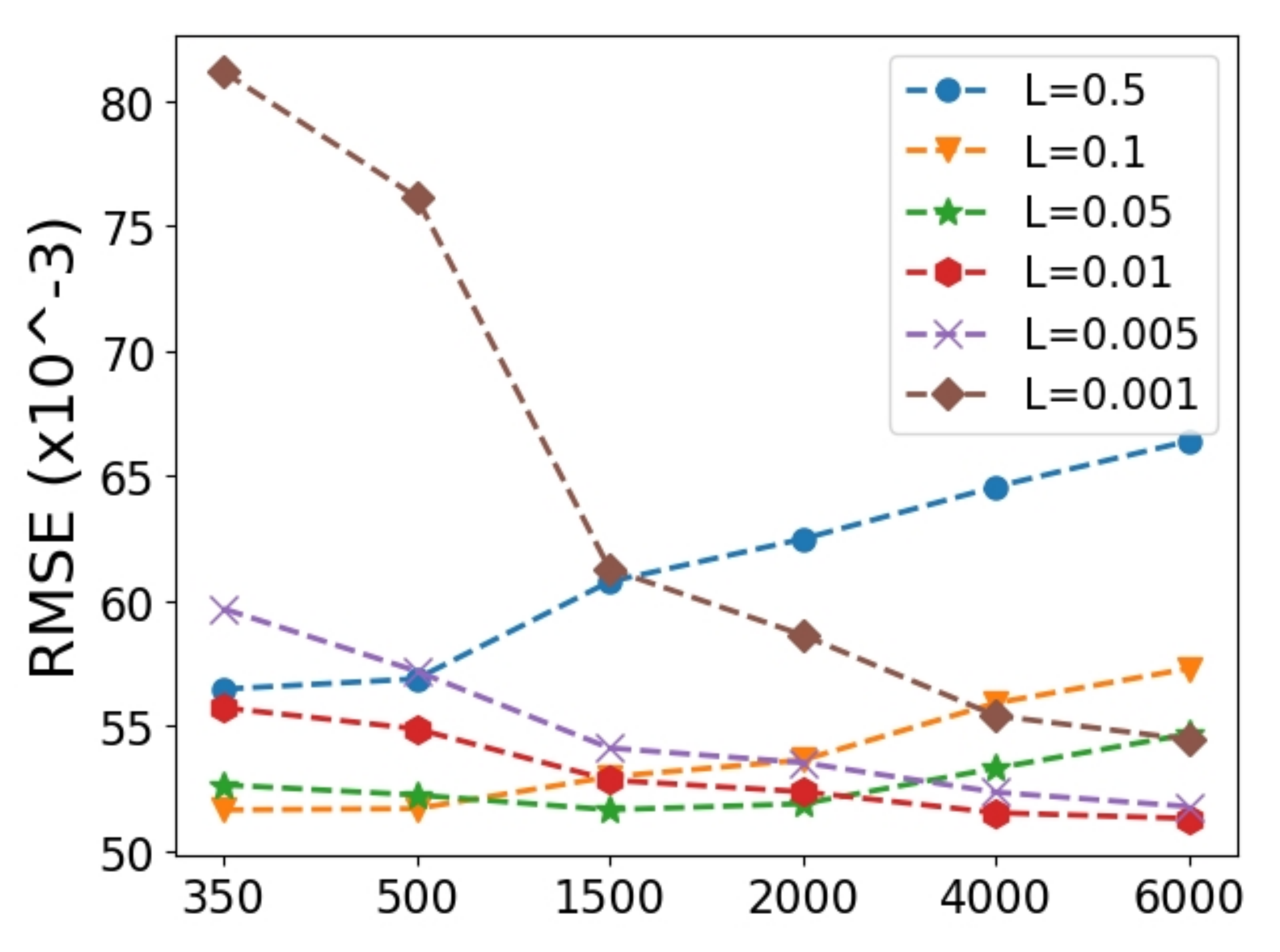}
    \label{fig:estimators}
}
\subfloat[Maximum Depth (D)]{
    \includegraphics[width=0.31\textwidth]{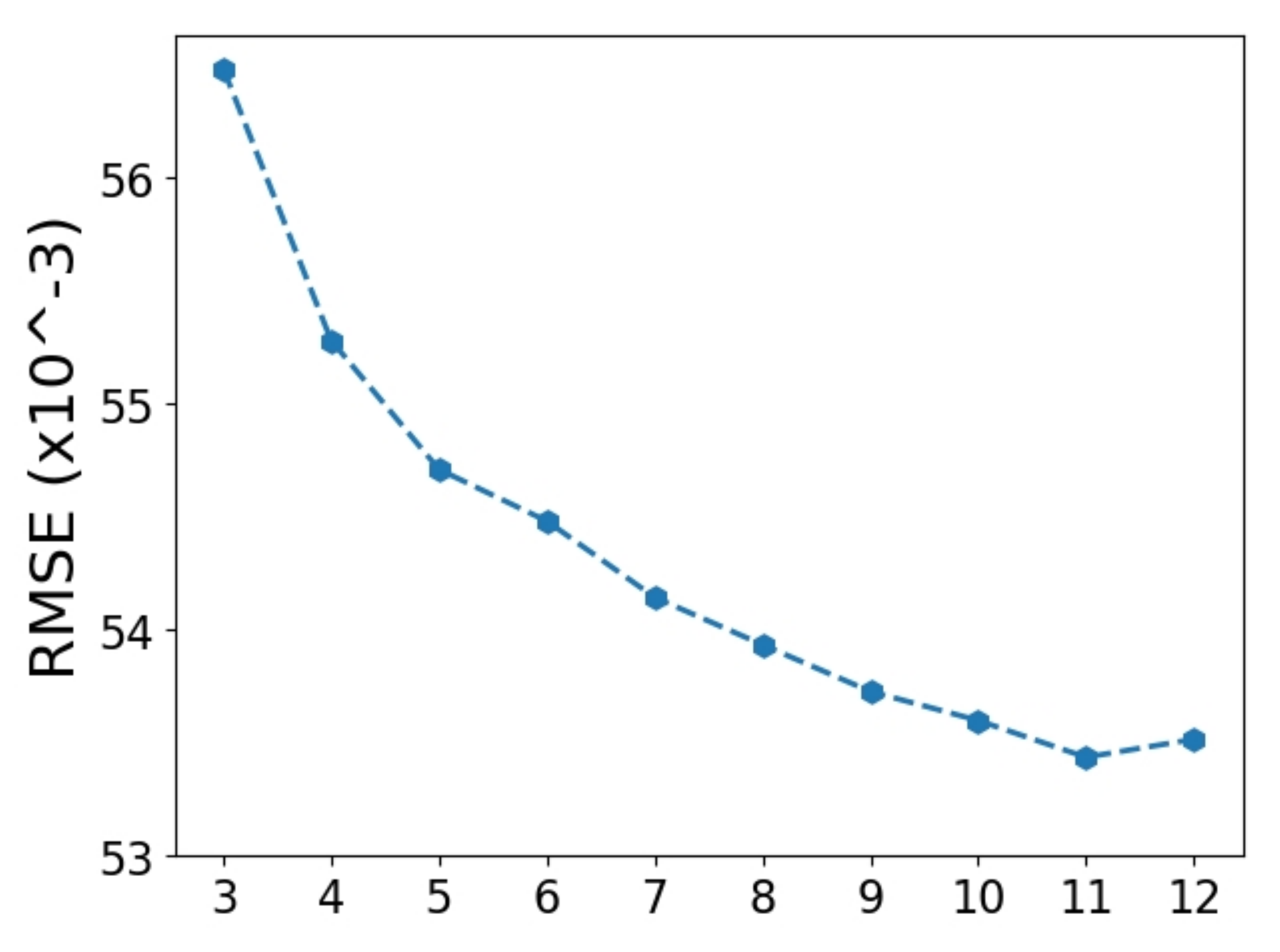}
    \label{fig:depth}
}\subfloat[Minimum number of samples to split an internal node (S)]{
    \includegraphics[width=0.31\textwidth]{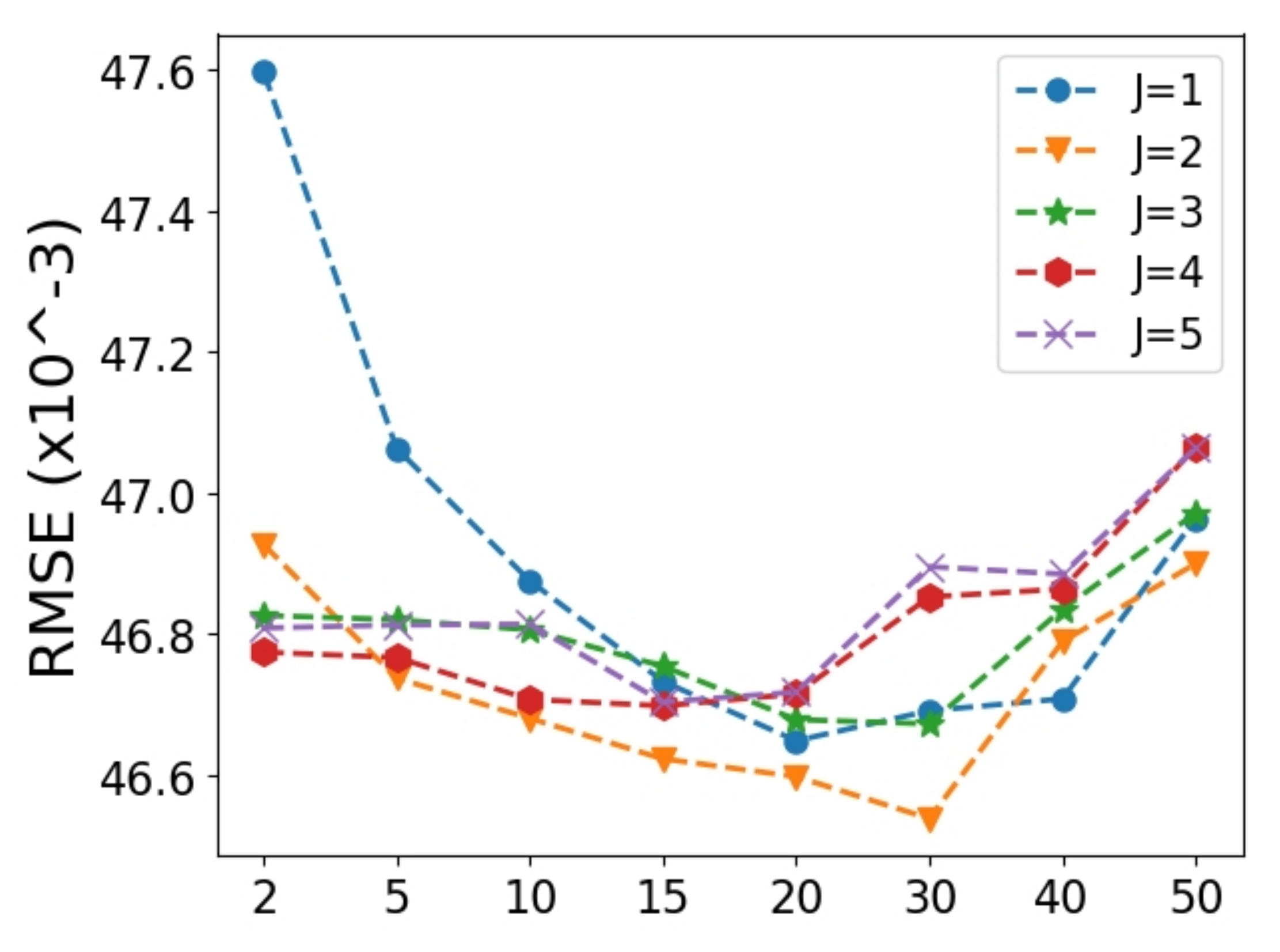}
    \label{fig:evalleaf-split}
}
\caption{Effect of various learning parameters on the accuracy of the prediction. Y-axis represents the error rate. X-axis of (a), (b), and (c) represent the number of trees in the GBDT algorithm, maximum depth of the decision tree, and the minimum number of samples to split a node (parameter S). Each line of (a) and (c) represent learning rate L and J values respectively. }
\end{figure}

Fig.~\ref{fig:estimators} demonstrates the relationship between RMSE and $M$ for different values of $L$.
We observe that when the number of trees is low (\ie short training), higher learning rates lead to a faster converge of the model. Therefore, the model achieves high accuracy in a lower number of iterations. However, the high learning rate can be susceptible to noise on the gradient that impacts the accuracy when leaned with a relative high number of tree.

We observe the maximum prediction accuracy for low learning rates and high number of trees. 
Increasing $M$ and decreasing $L$ make the model less susceptible to the noise, however, it make the model more complex and time consuming. 
Accordingly, to strike a balance between accuracy and the model complexity, we configure $M=350$ and $L=0.1$. 

\subsection{Tuning the value of regression estimator maximum depth.}~ Maximum Depth ($D$) is a parameter that controls the number of decision trees allowed in the model. The optimal value of $D$ varies from one model to another, depending on the interaction of features within the training dataset and other training parameters. This parameter can be ignored when there are only few features. However, in our model, the optimal depth value should be limited based on the interplay of the input parameters.

Fig.~\ref{fig:depth} shows the correlation between maximum depth of the tree in the range of [3, 12] in the horizontal axis and its corresponding error rate (RMSE). We notice that, as the value of $D$ increases, the prediction accuracy continues to increase until $D$ reaches 12 where we have an inflection point and we observe over-fitting. Therefore, we set $D=11$ as the appropriate value for the task merging prediction method.

\subsection{Tuning the value of minimum samples to create internal- and leaf-node.}~
\label{subsec:evalSJ}
In this part, we evaluate the parameters that control the minimum sample to create a new internal node and the minimum sample to create a new leaf node ($S$ and $J$ parameters, respectively) and measure their impact on the accuracy of the prediction model.

The value of $J$ parameter correlates with the  value of $S$ parameter. Accordingly,
in Fig.~\ref{fig:evalleaf-split}, we explore the prediction accuracy (by means of the RMSE value in the vertical axis) obtained when the values of $S$ varies in the range of [2 , 50]. The experiment is conducted for different values of $J$ (in the range of [1 , 5]). 

We observe that regardless of the $J$ value, by increasing the value of $S$ a reverse bell curve shape is emerged. The lowest error rate, however, varies depending on the value of $J$ parameter. The rebound of error rate indicates overfitting and should be avoided. From this experiment, we configure $J=2$ and $S=30$ that offer the lowest error rate.

\subsection{Evaluating improvement in the prediction accuracy.}~
In this part, we evaluate accuracy of the proposed prediction model (when configured as: \{ $M=350$, $L=0.1$, $D=11$, $S=30$, $J=2$ \}) against two alternative prediction methods. The first baseline approach, called \emph{Na\"ive} predictor, carries out the prediction based on a lookup table of mean execution-time saving for each operation. Another baseline approach is based on machine learning and uses a multi-layer perceptron (MLP)~\cite{plonis2020prediction} for prediction.

\begin{figure*}%[h]
 \centering
 \subfloat[Compare predictions with $\tau$ = 0.12]{
     \includegraphics[width=0.314\textwidth]{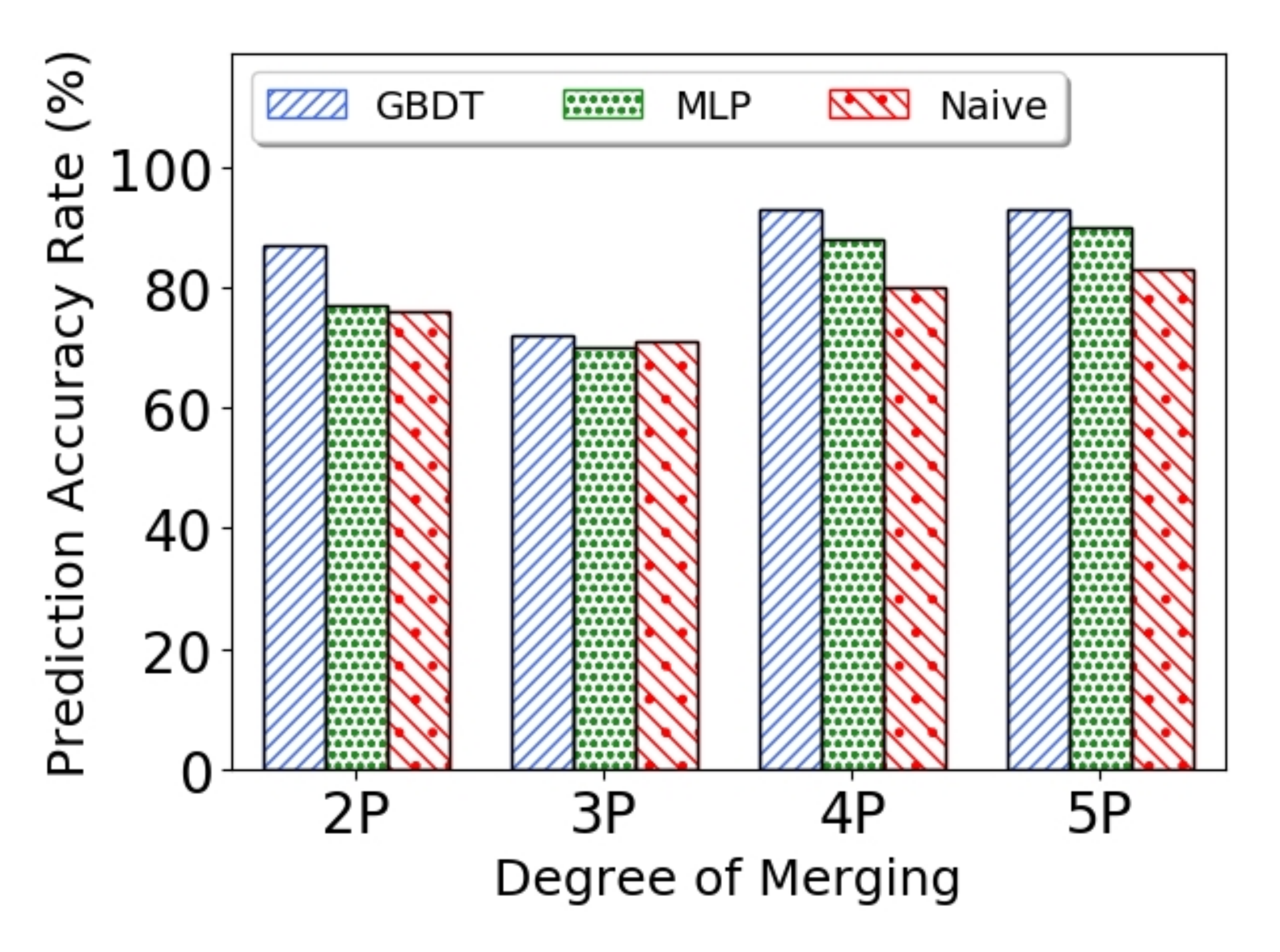}
     \label{fig:acct12}
 }
 \subfloat[Compare predictions with $\tau$ = 0.1]{
     \includegraphics[width=0.314\textwidth]{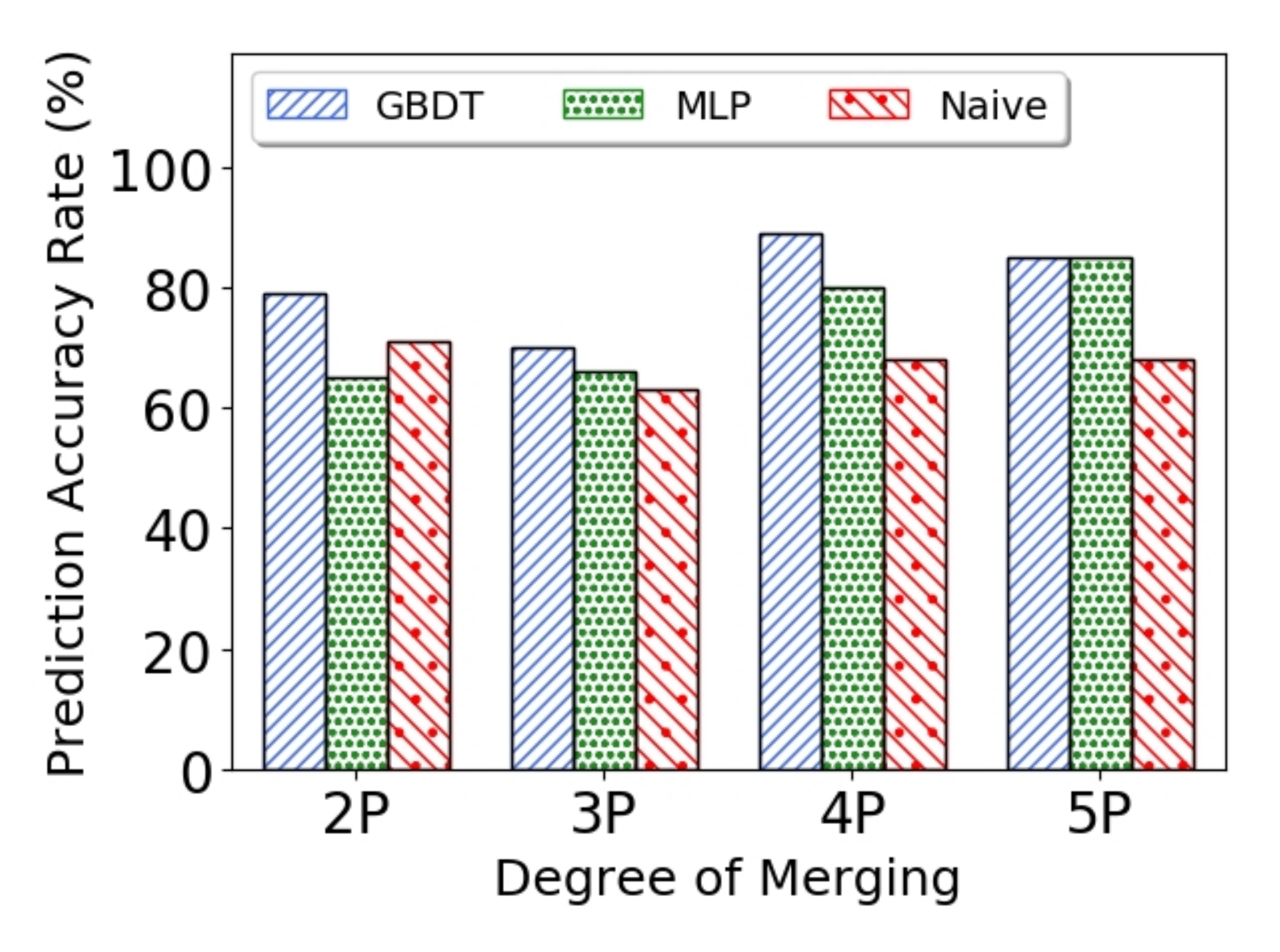}
     \label{fig:acct10}
 }\subfloat[Compare predictions with $\tau$ = 0.08]{
     \includegraphics[width=0.314\textwidth]{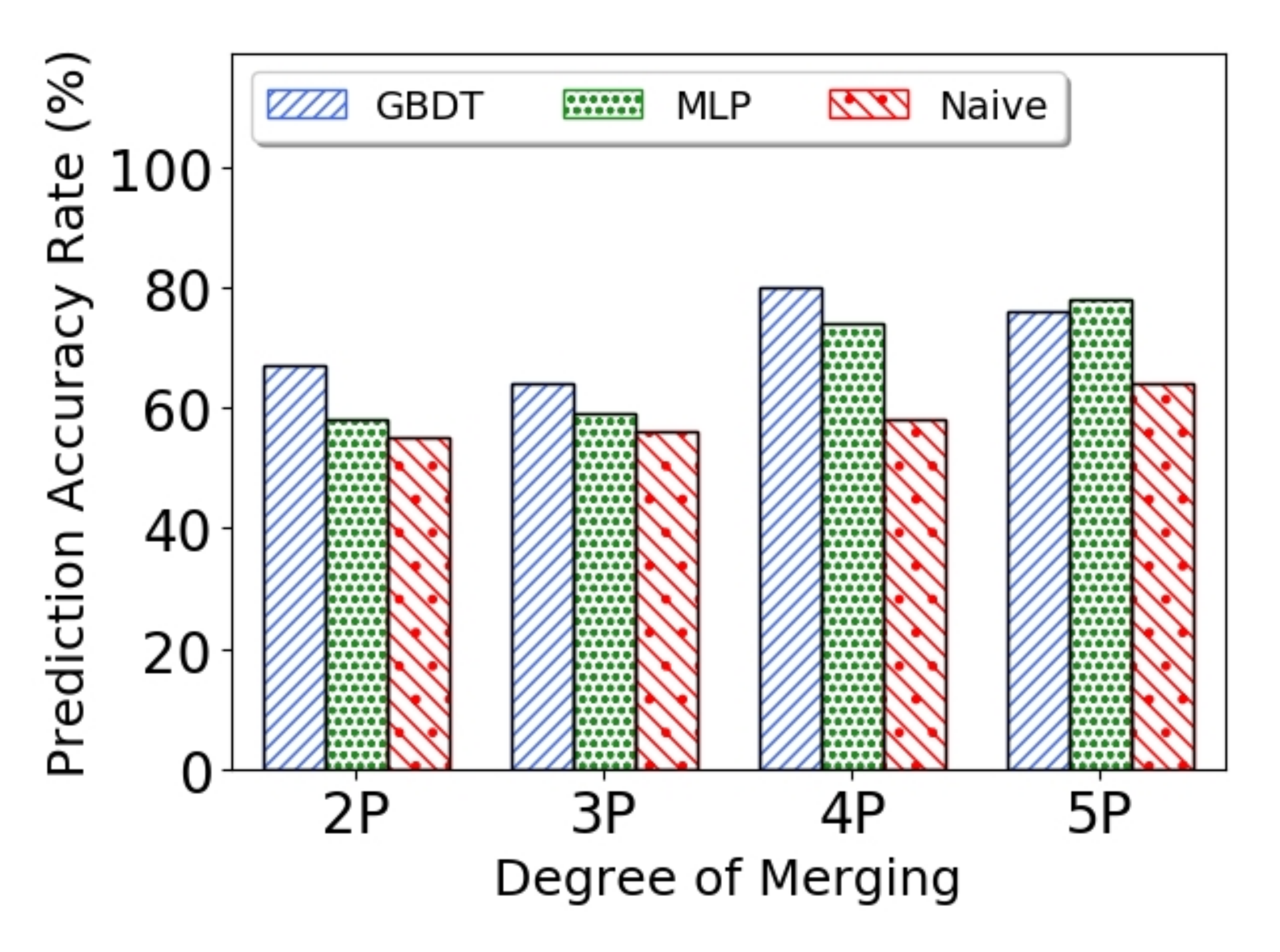}
     \label{fig:acct08}
 }
 \caption{Comparing the prediction accuracy of proposed execution-time saving prediction model (GBDT) against MLP and Na\"ive approaches. The horizontal axis represents the number of tasks merged to create a merged task and vertical axis represents the percentage of cases accurately predicted. }
 \label{fig:comparison}
 \end{figure*}

The prediction accuracy is reported as the percentage of correct predictions, denoted as $C$ and is defined based on Equation \ref{eq:accuracy}. In this equation, $A$ represents the total number of test cases, $P$ is the predicted execution-time saving ratio, $E$ is the observed execution-time saving ratio, and $\tau$ is the acceptable error rate, which is set to  0.12 in Fig.~\ref{fig:comparison}.

\begin{equation}
    C = 100\% \times \frac{1}{A}\sum_{i=1}^{A}
    \begin{cases}
    0, & |P_i- E_i| > \tau \\
    1, & |P_i- E_i| \leq \tau
    \end{cases}
    \label{eq:accuracy}
\end{equation}

We observed that %, regardless of the acceptable accuracy threshold, 
the GBDT model significantly outperforms the prediction accuracy of MLP and Na\"ive approaches, regardless of   merging degree. % cases.
%, except for two of the 5P cases in Figure~\ref{fig:acct10} and Figure~\ref{fig:acct08} where MLP approach performs similar to GBDT. 
Both MLP and GBDT significantly perform more accurate for higher degrees of merging (4P and 5P) than the lower ones (2P and 3P). The primary reason is that, the lower degree of merging saves relatively low amount of execution-time, which is difficult to accurately predict. The maximum prediction accuracy is 93\% when GBDT is employed in 4P.

\section{Summary}
\label{C3Summary}
In this chapter, we studied the potential of reusing computation via merging similar tasks to reduce their overall execution-time in the clouds.
Considering video processing context, we built a video benchmarking dataset and evaluated the parameters that influence the merge-saving. We observed that merging similar video processing tasks can save up to 31\% (for merging two tasks) of the execution-time that implies a significant cost saving in the cloud. We also learned that the merge-saving gain becomes negligible, when degree of merging is greater than three.
Then, we leveraged the collected observations to train a machine learning method based on Gradient Boosting Decision Trees (GBDT) to predict the merge-saving of unforeseen task merging cases. The fine-tuned prediction model can provide up to 93\% %was 94 for some reason??
accurate resource saving prediction.

We found the resource saving as a result of reusing in the media processing context to be sizable and worth pursuing. The next chapter, we explore the approach to efficiently find and merge similar tasks together.

%develop a serverless-based video processing system to efficiently find and merge similar tasks together.

    \chapter{Reusing Computation in Serverless Clouds}
\label{section:Reusing}
%\section{Overview} %will have one once 2nd part of the chapter (heterogeneous) is finished.
\vspace{14pt}
%\section{Request Merging Mechanism for Homogeneous Serverless Cloud Computing System}
%, we propose a mechanism based on the computational reuse approach that aims at alleviating oversubscription by aggregating similar tasks in the task scheduling queue of the serverless platforms. As shown in Figure~\ref{fig:arch}, the mechanism can aggregate (\ie merge) not only identical tasks, but also those that partially share their computation. We note that our mechanism complements existing allocation- and caching-based approaches and is not a replacement for them. In fact, the merging mechanism makes the scheduling queue less busy and potentially lighten up the scheduling process. Caching-based approaches are also complemented by capturing the in-progress tasks and those that are partially similar.

\begin{figure}[h!]
\centering
\includegraphics[width=0.8\textwidth ]{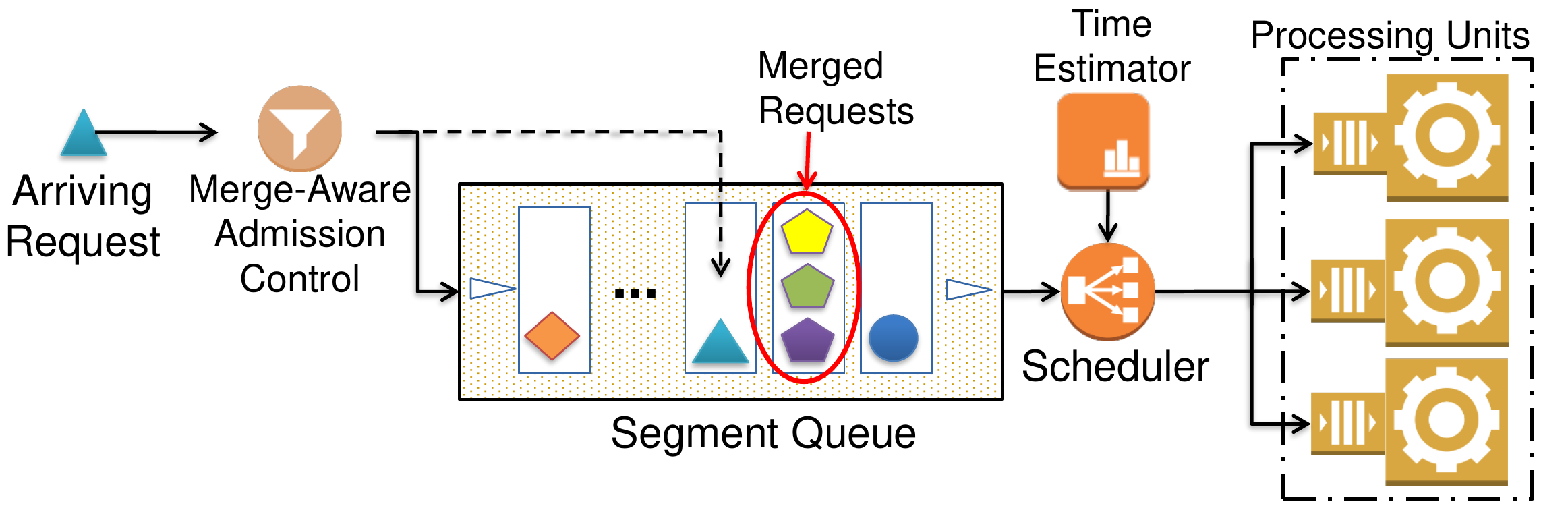}
    \caption{The overview of task aggregation procedure. New task arrives to Admission Control can be merged to an existing task in the Batch Queue. Task shapes represents different task types in the system and shape color represents different configurations of a task.}
\label{fig:arch} %,height=60mm after width
\end{figure}

In this chapter, we propose a mechanism based on the computational reuse approach that aims at alleviating oversubscription by aggregating similar tasks in the task scheduling queue of the serverless platforms. As shown in Figure~\ref{fig:arch}, the mechanism can aggregate (\ie merge) not only identical tasks, but also those that partially share their computation. We note that our mechanism complements existing allocation- and caching-based approaches and is not a replacement for them. In fact, the merging mechanism makes the scheduling queue less busy and potentially lighten up the scheduling process. Caching-based approaches are also complemented by capturing the in-progress tasks and those that are partially similar.

To reuse part of the computation, a question that needs to be addressed is how to identify \emph{mergeable} tasks?  An arriving task can potentially have multiple mergeable pairs with varying levels of similarity. Also, the solution for task similarity detection should not impose an extensive overhead to the system. The other concern in merging tasks is to form large compound tasks that potentially causes missing the deadline of either the merged tasks or other pending tasks waiting behind the merged task. As such, merging tasks raises the following two problems: (A) \emph{What are different types of mergeable tasks and how to detect them?} (B) \emph{How to perform merging without endangering other tasks in the system? 
}

Recalling that our motivational application in this dissertation is a media streaming engine that needs to process media segments in the cloud before streaming them to viewers~\cite{li2018cost}. Multiple viewers can stream medias in various configurations, hence, creating similar or identical tasks in the system. In particular, when the system is oversubscribed, the likelihood of having mergeable tasks increases. 
In this context, our proposed mechanism can detect identical and similar tasks and reuse the whole or part of the computation by merging them. Intelligently achieving task merging can benefit both the viewers, by enabling more tasks to meet their deadlines, and the stream providers, by improving resource utilization and reducing their incurred cost of using services.

% %In this chapter, 
% We develop an Admission Control module (see Figure \ref{fig:arch})
% that detects different levels of similarity between tasks and performs merging by considering the tasks' deadlines. 
% In summary, the \textbf{key contributions} of this research are as follows:
% \begin{itemize}
%  \item Proposing an efficient method to identify mergeable tasks.
%  \item Proposing methods for proper positioning of merged tasks in the scheduling queue.
%  \item Determining appropriateness and potential side-effects of merging tasks considering the oversubscription level of the system.
%  \item Analyzing the performance of merging on the viewer's QoS and the cost of utilizing the processing units.
% \end{itemize}

Although we develop this mechanism in the context of media processing system on serverless platform, the idea of task aggregation and research findings of this work are valid for other domains. However,
we note that identifying mergeable tasks is domain-specific and requires task profiling for each particular system.
%%%%%%%%%%%%%%%%%%%%%%%%

\section{Overview of the Admission Control Mechanism to Reuse Computation via Task Merging}
\label{sec:arch}
\vspace{1px}
%%%%%%%%%%%%%%very different to ICSOC version, check both

%\subsection{Overview of Task Merging Mechanism} 
Refers to Chapter~\ref{sec:CVSEbg}\, , Admission Control is the front gate of the batch queue and it is in charge of performing merging arriving tasks with the ones already in the batch queue. The reason we do not perform the merging in the batch queue (\ie after the task admission) is that, in that case, to find mergeable tasks, we need to scan the entire queue and perform a pair-wise matching between the queued tasks, which is inefficient and implies a significant number of redundant comparisons.  

\begin{figure}[htbp]
\centering
    \includegraphics[width=\textwidth]{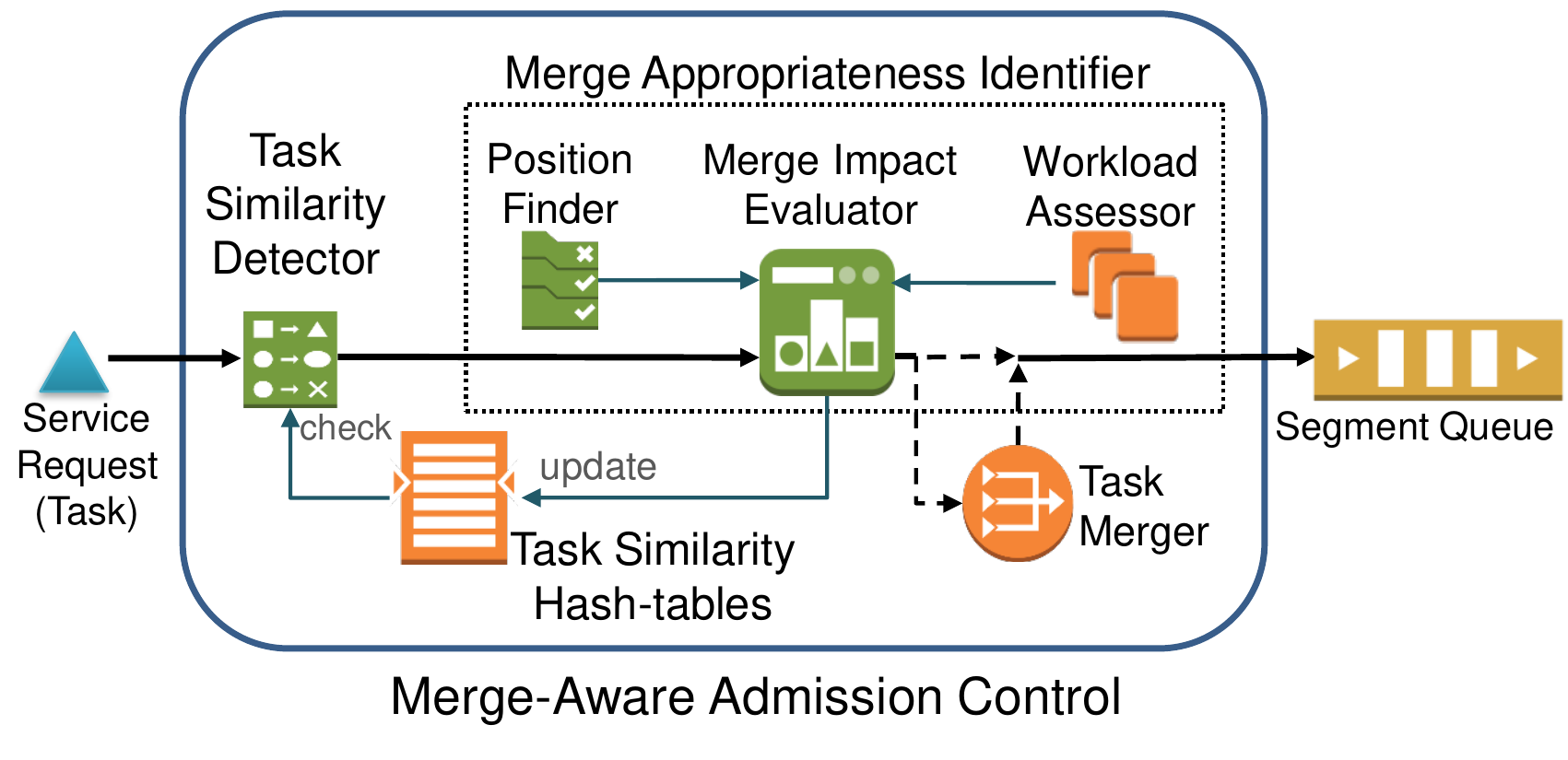}
    \caption{\small{Task aggregation mechanism inside Admission Control of SMSE. Before adding a task to the batch queue, it is checked if it is mergeable with any other queued tasks and whether or not the merging operation is appropriate to be achieved.} }
    \label{fig:adc}
  \end{figure}
  
The proposed task merging mechanism, shown in Figure \ref{fig:adc}, consists of three main components as follow: (A) Task similarity detector; (B) Merging appropriateness identifier; and (C) Task merger. 
\textbf{Task Similarity Detector} is a lightweight method based on hashing techniques to identify mergeable tasks. As detailed in Section~\ref{sec:detection}\, , it maintains multiple hash tables to cover multiple levels of tasks' mergeability. If the arriving task is identified mergeable with an existing task, then the system employs the \textbf{Merge appropriateness identifier} to assess if performing the merge on the identified tasks can impact other tasks in the system or not. Merge appropriateness identifier has three cooperating modules. \textit{Position Finder} locates the suitable position for merged tasks in the scheduling queue, such that the other tasks are not affected. To examine each position, Position finder consults with \textit{Merge Impact Evaluator} to estimate which and how many tasks can potentially miss their deadlines as the result of merging. The task merging decision is made based on system oversubscription level obtained from \textit{Workload Assessor} (see Section~\ref{sec:appropriateness}\,~~ and~\ref{sec:adapt}\,~~ for further details). Once the merging is confirmed as appropriate in a certain position of the batch queue, \textbf{Task merger} component carries out the merge operation on the two tasks.

% Although the idea of request aggregation is general and can be developed in many Serverless computing and HPC systems, we note that the implementation details of the merge operation and the system support for merging are application-specific. In this work, we discuss task aggregation in the specific context of Video On Demand (VOD) streaming and SMSE. In particular, integrating Request aggregation within Admission Control of SMSE requires support from Time Estimator and Media Merger components. It also implies modification to request data structure and the GOP Scheduler to understand the compounded requests. 

\section{Categories of Mergeable Requests} 
\label{sec:detection}
Mergeability of two given requests can be explained based on the amount of computation the two requests share. In particular, mergeability of two or more requests can be achieved in the following levels: 
\begin{enumerate}[label=(\Alph*)]

 \item \emph{Task level}: This is when more than one task which create the same processing task present to the scheduling queue. Therefore, this level is also known as 
 \emph{Identical tasks} and can achieve maximum computational reusability. For instance, consider two viewers without personalized requirements stream the same media and need it to be transcoded with the same resolution to be displayed on compatible devices. As these tasks lead to identical media processing, merging them consumes the same resources required for only one task, hence, reducing both cost and processing delay.

 %swap Data and Operation?

 \item \emph{Data-and-Operation level}:   %increases in compare to the execution time of a single task.
This is when two or more tasks perform the same operation on the same data but with different configurations. This level of merging results in a combined processing task for output equivalent to processing each task individually.
Computational reusability can be achieved through the sharing of function loading overhead and common processing steps. For instance, consider two viewers who stream the same video with different resolutions. Without merging, the two tasks need to load the video, decode it, and encode it separately. However, by merging the two tasks, the loading and decoding operations can be shared, then the encoding operation is carried out separately.

\item \emph{Data-only level}: 
This is when the only common sharing specification between two tasks is only on the data. Tasks that share the same data can reduce the data retrieval overhead. 
This third tier task similarity level saves the least amount of processing time in comparison to other cases.

 \end{enumerate}
 %MAKE CHANGE
 %%or Operation-only level ....

It is noteworthy that, although merging increases the execution time of the merged task (except in the Task level), our study in Chapter~\ref{section:MergeSaving} shows that the execution time of the merged task is remarkably (up to 40\%) shorter than the combined execution time of the unmerged tasks.

In SMSE, the Admission Control component can achieve task level reusability for the same segments that need to be processed with the same function, but for different viewers. Data-and-Operation level reusability is achieved for segments that perform the same processing function, but with different configurations. Finally, Data-only reusability is achieved for the same segments that are served by different functions.

\section{Detecting Similar Tasks}
%\subsection{Detecting Similar Requests}
\label{sec:mergable-detection}

Assuming there are $n$ tasks in the queue, for each arriving task, a na\"{i}ve mergeable task detection method has the overhead of performing $n$ comparisons to find the mergeable tasks. To reduce the overhead, we propose a method that functions based on hashing techniques. The general idea of the proposed method is to generate a hash key from the arriving task request signature (\eg media segment id, processing type, and their parameters). Then, the Admission Control finds mergeable tasks by searching for a matching key in the hash table of existing tasks in the scheduling queue.

The explained method can detect Task level mergeability. We need to expand it to detect other levels of task mergeabilities. To maximize the computational reusability, an arriving task is first verified against Task level mergeability. If there is no match in the Task level, then the method proceeds with checking the next levels of mergeability, namely Data-and-operation level and Data-only level, respectively. To achieve the multiple levels of mergeability, we create three hash-tables---each covers one level of mergeability. The hash-keys in each level are constructed from the tasks' characteristics that are relevant in deciding mergeability at such level. For instance, in the media streaming case study, keys in the hash-table that verifies task level mergeability are constructed from media segment id, processing type, and their parameters. While, keys in the hash-table that verifies Data-and-operation level mergeability are constructed from media segment id and processing type. Similarly, keys in the hash-table of Data-only level mergeability are constructed from segment id.

\boxfig{
Upon arrival of task $j$: 
\begin{itemize}

\item[(1)] if $j$ merges with existing task $i$ on Task level similarity:
\begin{itemize}
\item[--] No update on hash-table is required
\end{itemize}
\item[(2)] if $j$ merges with existing task $i$ on operation-and-data or Data-only level similarity:
\begin{itemize}
\item[--] Add an entry to each hash-table with hash-keys of task $j$ and point them to merged task $i+j$ 
\end{itemize}
\item[(3)] if $j$ matches with existing task $i$ but the system chooses not to merge them:
\begin{itemize}
\item[--] Add an entry to each hash-table with hash-keys of task $j$ and point them to task $j$
\end{itemize}
\item[(4)] if $j$ does not match with any of the existing tasks:
\begin{itemize}
\item[--] Hash-keys of task $j$ are added to the respective hash-tables 
\end{itemize}
\end{itemize}

Upon task $j$ completing execution (\ie dequeuing task $j$):
\begin{itemize}
\item[--] Remove all entries pointing to task $j$ from hash-tables
\end{itemize}
\caption{\small{The procedure to update hash-tables upon arrival or completion of tasks.}} 
\label{fig:htupdate}

}

Each entry of the hash-tables includes a hash-key and a pointer to the corresponding task. Entries of the three hash-tables must be updated upon a task arrival and execution. The only exception is Task level merging, which does not require updating the hash-tables. Figure \ref{fig:htupdate} shows the procedure for updating the hash-tables for a given task $j$. 

When the system merges task $j$ with existing task $i$, the merged task, denoted as $i+j$, is essentially the object of task $i$ that is augmented with task information (\eg processing parameters) of task $j$. In this case, as shown in Step (2) of this procedure, the system only adds an entry to each hash-table with hash-key of task $j$ pointing to merged task $i+j$ as existing key for task $i$ already pointed to task $i+j$. 
When task $j$ is mergeable with existing task $i$, but the system decides to add task $j$ to the batch queue without merging. In this case, task $j$ has a higher likelihood of merging with other arriving tasks. The reason is that %it is likely that task $j$ stays in the queue for a longer time than task $i$. In addition, because 
task $i$ has not merged with task $j$, and it does not merge with other arriving tasks. Hence, as shown in Step (3) of the procedure, the matching entry pointing to task $i$ is redirected and points to task $j$. 
It is worth noting that if the arriving task does not match with any of the existing tasks, as shown in Step (4), its hash-keys must be generated and added to the respective hash-tables. Also, when a task is served (processed), its corresponding entries are removed from the hash-tables.

%%%%%%%%%%
\section{Identifying Merging Appropriateness}
\label{sec:appropriateness}
\subsection{Overview.}~
Assume that arriving task $j$ has Data-and-operation or Data-only similarity with existing task $i$. Also, assume that task $i$ is scheduled ahead of at least one other task, denoted task $k$, in the scheduling queue. Merging task $j$ with $i$ either delays the execution of task $k$ or task $i$. Such an imposed delay can potentially cause task $k$ or $i$ to miss their deadlines. Therefore, it is critical to assess the impact of merging tasks before performing the merge. 
%The merge is carried out only if it does not cause deadline violation for other tasks.
The merge should be carried out only if it does not cause more QoS violations than it improves.
It is noteworthy that Task level merging does not delay the execution of other tasks; thus, it always can be performed.

Accordingly, in this section, we first introduce Merge Impact Evaluator component whose job is to assess the impact of the merging arriving task on existing tasks. Later, we introduce Position Finder, whose job is to position the arriving task in the scheduling queue, either through merging with other tasks or as a new entry in the scheduling queue.

%alg table here?
\subsection{Evaluating impact of merging.}~
\label{sec:impact}
Ideally, task aggregation should be performed without causing deadline violations for other tasks. Accordingly, the impact of merging two or more tasks is evaluated based on the number of tasks missing their deadlines due to the merging. The evaluation requires the Time Estimator component (see Figure~\ref{fig:arch}) to estimate the mean and standard deviation of execution time of the tasks. To evaluate the impact of merging, a temporary structure, called \textit{virtual queue}, is constructed that contains a copy of machine queues. Then, we assume the merging has taken place on the tasks in the batch queue and schedule them to the virtual queue according to the scheduling policy. This enables us to estimate the number of tasks missing their deadlines in the presence of merging.

To assure the minimal impact of the merging, by default, a worst-case analysis is performed on the completion time of the affected tasks to estimate the number of tasks missing their deadlines. For a given task $i$, we assume its execution time follows a Normal distribution \cite{li2018cost,hussain2019federated} and $\mu_i$ and $\sigma_i$ represent the mean and standard deviation of its execution time. Let $E_i$ be the estimated execution time of task $i$. In the worst-case analysis, we consider $E_i$ to be large enough that with a high probability (97.7\%), the real execution time is less than $E_i$. As such, $E_i$ is formally defined based on Equation~\ref{eq:e_i}.

\begin{equation}\label{eq:e_i}
E_i=\mu_i + \alpha\cdotp \sigma_i    
\end{equation}
 
In this equation, $\alpha$ is the standard deviation coefficient and its value equals 2, such that with 97.7\% chance task $i$ is not affected by the merging. Note that to encourage more aggressive merging under oversubscription, we can relax the severity of the worst-case analysis by diminishing the value of  $\alpha$ (see Section \ref{sec:adapt} for further details).

Once we know $E_i$, we can leverage it to estimate the completion of task $i$ on a given machine $m$, denoted as $C_i^m$. We know that calculating $C_i^m$  involves the summation of the following four factors: (A) current time, denoted $\tau$; (B) estimated remaining time to complete the task currently executing on machine $m$, denoted $e_r^m$; (C) sum of the estimated execution times of $N$ tasks that are pending in machine queue $m$, ahead of task $i$. This is calculated as $\sum_{p=1}^{N}(\mu_p + \alpha\cdotp \sigma_p)$; (D) estimated execution time of task $i$. 
The formal definition of $C_i^m$ is shown in Equation~\ref{eq:compl},

 \begin{equation}\label{eq:compl}
  C_i^m = \tau + e_r^m + \sum_{p=1}^{N} (\mu_p +\alpha\cdotp \sigma_p)  + (\mu_i + \alpha\cdotp \sigma_i)
 \end{equation}

In the tie situation that the number of tasks missing their deadlines with and without merging is the same, we choose to perform merging to reduce the overall time of using cloud resources. However, one may argue an alternative approach to not perform the merging, because merging can marginally increase the chance of missing deadline for other tasks. 

\subsection{Positioning aggregated tasks in the scheduling queue.}~
%\label{sec:positioning}

Once two tasks are detected as mergeable, the next question is: where should the merged task be placed in the batch queue? The number of possible answers depends on the scheduling policy of the underlying serverless computing platform. 
If manipulating the order of tasks in the batch queue is allowed, then the Position Finder examines possible locations for the merged tasks in the queue. For each location, it consults with the Merge Impact Evaluator component (see Figure \ref{fig:adc}) to identify if the merge has potential side-effects on the involved tasks or not. Once Position Finder locates an appropriate position, it notifies Task Merger to construct the merged task.

Scheduling policies usually sort tasks in the batch queue based on a certain metric (known as the queuing policy). For instance, Earliest Deadline First~\cite{liccgrid16} sorts the queued tasks based on their deadlines. This assumption restricts the number of positions can be identified for the merged tasks that in turn limits the performance gain of task merging. 
To conduct a comprehensive study, in this section, we investigate two main scenarios: (A) when the queuing policy is mandated, (elaborated in Sub-section~\ref{sec:posrespect}\,~ ); (B) when the queuing policy is relaxed (elaborated in Sub-section~\ref{sec:posignore}\,~).

\subsection{Task positioning while queuing policy is maintained.}~
\label{sec:posrespect}
In this part, we study three commonly used queuing policies: \textbf{(a)} First Come First Served (FCFS); \textbf{(b)} Earliest Deadline First (EDF); and \textbf{(c)} Max Urgency. While FCFS and EDF are known queuing policies, Max Urgency sorts the tasks in the queue based on tasks' deadline and execution time. More specifically, for task $i$, urgency is calculated as $U_i=1/(\delta_i - E_i)$, where $U_i$ is urgency score of task $i$, $\delta_i$ is its deadline, and $E_i$ is its estimated execution time. 

\paragraph*{\textbf{FCFS:}} Let $j$ be the arriving task and $i$ a matching task already exists in the queue. We can merge tasks by either augment task $i$ with $j$'s specification or cancel task $i$ and reinsert $i+j$ into the queue. Therefore, the arrival time of the merged task ($i+j$) can be either the arrival time of task $i$ or task $j$. In the former case, $i+j$ delays completion time of all tasks located behind $i$. In the latter case, $i+j$ only delays completion time of $i$. In either case, the delayed task(s) can potentially miss their deadline(s) due to the merge operation. A compromise between these two extreme positions is possible and is described in Sub-section \ref{sec:posignore}.

\paragraph*{\textbf{EDF:}} In this policy, tasks with an earlier deadline are positioned earlier in the queue. When two or more tasks are merged, each of them still keeps its individual deadline. However, only the earliest deadline is considered for the queuing policy. Assuming that task $i$ has an earlier deadline than $j$, task $i+j$ can be only  positioned in task $i$'s spot.

\paragraph*{\textbf{Max Urgency:}}
Recall that except in Task level merging, other levels of merging increase the execution time of the merged task. In this case, the urgency of $i+j$ is: $U_{i+j}=1/(min(\delta_{i},\delta_{j})-E_{i+j})$. This means the urgency of the merged task is increased. Thus, the merged task can potentially move forward in the queue and get executed earlier. As such, tasks merging in max urgency queue can potentially cause missing the deadline of tasks located ahead of $i$ in the scheduling queue as well.

\subsection{Task positioning while queuing policy is relaxed.}~
\label{sec:posignore}
Queuing policies mentioned in the previous part are not aware of task merging. Except for Max Urgency that moves the merged task forward in the queue due to the increase in the merged task urgency, other policies do not relocate the merged task. However, a more suitable position for the merged task can be found by relaxing the queuing policy. In this case, assuming there are $n$ tasks in the batch queue, the merged task, $i+j$, has to be examined against $n+1$ possible locations to find the position that maximizes the chance of all tasks meeting their deadlines. Examining each possible location implies evaluating the impact of merging, hence, calling the scheduling method. Assuming there are $m$ machines in the system, each impacts evaluation costs $n\cdotp m$ and performing such evaluation for all $n+1$ possible locations implies $(n^2+n)\cdotp m$ complexity. This makes the time complexity of finding an optimal solution as approximately $O(n^3)$.

Such overhead itself is a burden to the system that is already oversubscribed. As such, in the rest of this section, we propose two Position Finding heuristics and analyze them. The objectives of these heuristics are: first, not to allow the merged task to miss its deadline; and second, do not cause other tasks to miss their deadlines.% In this part, we let $j$ to be arriving task; $i$ to be the existing task matches $j$; and $i+j$ to be the merged task. 

\paragraph*{\textbf{Logarithmic probing heuristic:}} This heuristic evaluates the impact of merging when $i+j$ is in the middle of the queue. The evaluation result dictates how to proceed with the probe as follows:

(i) The position neither causes deadline violation for other tasks nor  $i+j$. Therefore, the appropriate position is found.

(ii) Task $i+j$ misses its deadline, but the number of other tasks missing their deadlines does not increase as a result of merging. This implies that $i+j$ should be executed earlier. Thus, the procedure continues to probe in the first half of the queue. 

(iii) Task $i+j$ meets its deadline, but the number of other tasks missing their deadlines increases as a result of merging. This implies that $i+j$ should be executed later to reduce merging impact on other tasks. Thus, the procedure continues to probe in the latter half of the queue.

(iv) Task $i+j$ misses its deadline, and the number of other tasks missing their deadlines increases as a result of merging. Then, stop the procedure and cancel merging because the procedure cannot find an appropriate position for merging.

The aforementioned steps are repeated until it terminates or there is no position left to be examined in the batch queue. In the latter case, we stop the procedure and cancel merging.

\paragraph*{\textbf{Linear probing heuristic:}} 
%For the sake of efficiency, this procedure is designed to operate in two phases. 
In the FCFS policy, we know that the order of tasks in the batch queue implies the order of their execution. That is, placing a task in position $p$ of the queue only delays tasks located behind $p$. That said, the \emph{first} phase of this heuristic aims at finding the latest position for task $i+j$ in the batch queue so that it does not miss its deadline. The latest position for $i+j$ in the queue implies the minimum number of tasks are affected ---those located behind the merged task. 

To carry out the first phase, the procedure constructs virtual queues to find the latest position for $i+j$. For that purpose, it alternates the position of $i+j$ in the batch queue, starting from the head of the queue. In each position, the completion time of $i+j$ is calculated based on the tasks located ahead of it and is examined if $i+j$ misses its deadline. Once task $i+j$ misses its deadline, the previous position is the latest possible location for it not to miss its deadline.

Once we found the latest position for $i+j$, we need to verify if the insertion of $i+j$ causes any deadline violation for the tasks behind it or not.
For that purpose, in the \emph{second} phase, we only need to evaluate the merging (via Merging Impact Evaluator) once. If there is no impact, then the found position is confirmed. Otherwise, the merging is canceled.

 It is noteworthy that this procedure is efficient because the virtual queue is created only once. Also, after each task assignment to the virtual queue, it simply adds one more checking to calculate $i+j$ completion time. 

\paragraph*{\textbf{Analysis of the heuristics:}}
In this part, we analyze Logarithmic Probing and Linear Probing heuristics in terms of their complexity and optimality of the position they find. 

\paragraph*{\textbf{Complexity Analysis:}} 
Phase one of Linear Probing Heuristic examines $n$ tasks to be scheduled on $m$ machines with an additional check if $i+j$ can be scheduled on time directly after each of the $n$ tasks. That results in $n\cdotp m$ complexity to provide a single position for Phase two to verify. Phase two is essentially evaluating the impact of merging, which again needs $n$ tasks to be scheduled on $m$ machines. The combined complexity of the two phases is $2\cdotp n\cdotp m$. Alternatively, Logarithmic Probing Heuristic spends trivial computation of $O(1)$ to pick a position in the batch queue to verify the appropriateness. If the position identified as inappropriate, the search continues for up to $\log n$ positions. Since the complexity of evaluating each position is $n\cdotp m$, the total complexity is $n \cdotp m\cdotp\log n$. As the complexity of evaluating impact of merging dominates the total complexity, the Linear Probing Heuristic which spends less time evaluating the position is more efficient.

\paragraph*{\textbf{Optimality Analysis:}} Assume that there are multiple appropriate positions for task $i+j$. Logarithmic Probing Heuristic returns the first position it finds and meets the criteria, thus, is not biased to any certain appropriate position for the merged task. Alternatively, Linear Probing Heuristic always finds the latest appropriate position in the batch queue for task $i+j$. This ensures that task $i+j$ has the least impact on other tasks' completion times. Being the last possible position, however, increases the likelihood of $i+j$ to miss its deadline. In addition, this makes it unlikely for other tasks to be scheduled in front of $i+j$, hence, limiting the chance of future merging operations.

\section{Adapting Task Merging based on the oversubscription level}
\label{sec:adapt}
\subsection{Overview.}~
In Section \ref{sec:appropriateness}\, , we discussed the merge appropriateness of each task by considering a worst-case analysis to assure no task is affected by the merging. However, when the system is oversubscribed, we can compromise the worst-case analysis and make the system more permissive to task merging in order to mitigate the oversubscription. In fact, sacrificing a few tasks in favor of more merging can lighten the system oversubscription and ultimately cause fewer tasks missing their deadlines. For that purpose, in this section, we develop the Workload Assessor component (see Figure~\ref{fig:adc}) that is in charge of assessing the oversubscription level of the system and accordingly adjusting the aggression level of applying the task merging.

%from experiment of previous section???, we found that it is beneficial to be conservative to task aggregation when the system is not oversubscribed, in the other hand more aggressive task aggregation is beneficial when system is heavilyy oversubscribed. Therefore it is beneficial that task aggregating decision making procedure should  adapt to degree or level of oversubscription.

\subsection{Quantifying oversubscription of a computing system.}~
Level of oversubscription in the system can be quantified based various factors, such as the rate of missing deadline and the task arrival rate. The quantification can be achieved in a reactive manner (\ie from known metadata) or in a proactive manner (\ie based on the factors that suggest the system is about to get oversubscribed in the near future). In this part, we provide a method for Workload Assessor that uses decisive indicators of oversubscription to quantify the oversubscription level of a serverless computing system.

The first intuitive idea to quantify oversubscription is based on the (measured or expected) ratio of the task arrival rate to the processing rate \cite{mahato2018reliability}. In this case, a system is oversubscribed, only if it cannot process tasks as fast as it receives them. This idea has two main limitations: (A) It requires the knowledge of processing rate, which is difficult to accurately measure; (B) It is prone to report false negative in the oversubscription evaluation. In particular, it cannot discriminate between different circumstances that the ratio tends to one. Such a circumstance can occur when the tasks' arrival and processing rates are similar, however, the batch queue may be congested (\ie the system is oversubscribed) or may not be (\ie the system is not oversubscribed).

Another idea is to use the ratio of number of tasks missing their deadlines to the total number of tasks executed \cite{mahato2018reliability}. This is based on the fact that an oversubscribed system cannot complete all its tasks on time, thus, missing a high number of task deadlines suggests an oversubscribed situation. Although this idea has a good potential, yet it falls short in quantifying the degree of oversubscription. That is, it cannot discriminate between a system that completes tasks a short time after their deadlines versus the one that completes tasks a long time after their deadlines.

Improving on the shortcomings of the aforementioned methods, we propose to quantify the oversubscription level of the system in a given time window based on the \emph{deadline miss severity ratio}. We define \emph{waitable time} of task $i$, denoted $W_i$, as the maximum time it can wait in the queue without missing its deadline. Let $A_i$ denote the arrival time of task $i$, then its waitable time is calculated as: $W_i = \delta_i-A_i- E_i$ . To quantify the oversubscription level, denoted $OSL$, in the first place, we discard the contribution of infeasible tasks (\ie those with $W_i < 0$) and those that can complete on time (\ie the ones with $C_i^m\leq\delta_i$). Next, the tasks that complete after their deadlines contribute to the oversubscription level based on the severity of their deadline miss. For a given task $i$, this is calculated based on the proximity between its completion time and its deadline (\ie $C_i^m-\delta_i$) and with respect to its waitable time (\ie $W_i$). Equation~\ref{eq:weightedmiss} formally shows how $OSL$ is calculated. Recall that $C_i^m$ is estimated based on Equation~\ref{eq:compl} to quantify the oversubscription in the current time window and $N_a$ represents the total number of tasks across all the machine queues. To adapt Equation \ref{eq:weightedmiss} for quantifying the oversubscription of a past time window, we need to replace the estimated completion time with the observed completion time of the tasks. %and $N_a$ with task completed in a recent time window.

 \begin{equation}\label{eq:weightedmiss}
  OSL = \frac{1}{N_a}\sum_{i=1}^{N_a} 
  \begin{cases}
  0,                     & W_i \le 0 \\
  0,                     & C_i^m \leq \delta_i \\
  \frac{C_i^m - \delta_i}{W_i} ,& C_i^m > \delta_i 
  \end{cases}
 \end{equation}

\subsection{Adaptive task merging aggressiveness.}~
The method explained in Section~\ref{sec:appropriateness}\,~~ estimates the side-effect of merging on other tasks in a conservative manner to assure that the merging does not cause their deadlines violated. 
In the face of oversubscription, estimation of the side-effect can be relaxed from the worst-case analysis to allow more aggressive task merging, hence, mitigating the oversubscription and increasing the overall QoS.

To make the aggressiveness of task merging adaptive, based on the measured oversubscription level of the system, we modify the acceptable probability that a merge operation does not cause deadline violation on other tasks of the system. More specifically, for higher values of the oversubscription level, the acceptable probability that other tasks meet their deadlines should be diminished and vice versa. For that purpose, we set the acceptable probability of meeting deadline for the tasks affected by merging to vary in the range of [2.3\% , 97.7\%], depending on the oversubscription intensity. To this end, the coefficient of standard deviation ($\alpha$) has to be in [-2 , 2] range. Specifically, to adapt the value of $\alpha$ based on the oversubscription level, we determine $\alpha$ as: $\alpha = 2-4 \cdotp OSL$. 
    \section{Performance Evaluation}

 \label{sec:evltn4}

 \subsection{Experimental setup.}~
We use a prototype of the SMSE platform (developed in Chapter~\ref{section:platform} with task merging mechanism in place. %We made SMSE publicly available\footnote{https://github.com/hpcclab/adaptivemerging} for the research community and reproducibility purposes. 
In this chapter, to comprehensively examine various workloads with different configurations, we used SMSE in the emulation mode. The task merging mechanism, proposed in this paper, is implemented as the Admission Control component of SMSE. We evaluated the proposed mechanism using eight homogeneous processing units modeled after Chameleon Cloud \cite{keahey2019chameleon} small VMs. 
% as the system's only processing units.  

The video repository we used for evaluation includes multiple replicas of a set of benchmark videos. Videos in the benchmarking set are diverse both in terms of the content types and length. The length of the videos in the benchmark varies in the range of [10, 220] seconds splitting into 5-110 video segments. The benchmark videos are publicly available for reproducibility purposes\footnote{\url{https://github.com/hpcclab/videostreamingBenchmark}}. More details about the characteristics of the benchmark videos can be found in Chapter~\ref{section:MergeSaving}. For each segment of the benchmark videos, we obtained their execution times by executing each micro-service 30 times. The benchmarked micro-services are: reducing resolution, adjusting bit rate, adjusting frame rate, and changing codec. In each case, two conversion parameters are examined. For example, frame rate is changed from 60 fps down to either 30 fps or 24 fps. Note that a codec changing micro-service can take up to 8x longer to execute than other more trivial processing operations (see Section~\ref{subsec:C3Eval}).

To evaluate the system under various workload intensities, we generate [1,000, 2,500] video segment processing tasks within a fixed time interval. All transcoding micro-services are available in the processing units (\ie warm starting micro-services). Transcoding tasks arrive to the system in a group of 5 consecutive segments at a time. To accurately emulate common workload behavior observed in the real video steaming systems, each workload repeatedly toggle their arrival rate between base period and high load period where the arrival rate is increased by two folds. Each base period is approximately three times longer than the high load period. Each simulation case spans up to 15 of high and base period cycles. 
In each simulation case, if all tasks arrive simultaneously, there is approximately 30\% chance for some tasks to find a mergeable pair. However, as the tasks are dynamically arriving to the system throughout the simulation time, the chance of task merging reduces to be less than 20\%.

We collect the deadline miss-rate (DMR) and makespan (\ie execution time to finish all tasks) of completing all tasks. For the sake of better visualization of the miss rate reduction, DMR of each configuration with merging policy is normalized against a nearly identical configuration without the task merging in place. We conducted each experiment 30 times, each time with different randomized task arrival time and order. Mean and 95\% confidence interval of the results are reported. In every experiment, all tasks must be completed, even if they miss their deadlines.
\begin{figure}[ht]
	\centering
	\includegraphics[width=0.8\textwidth]{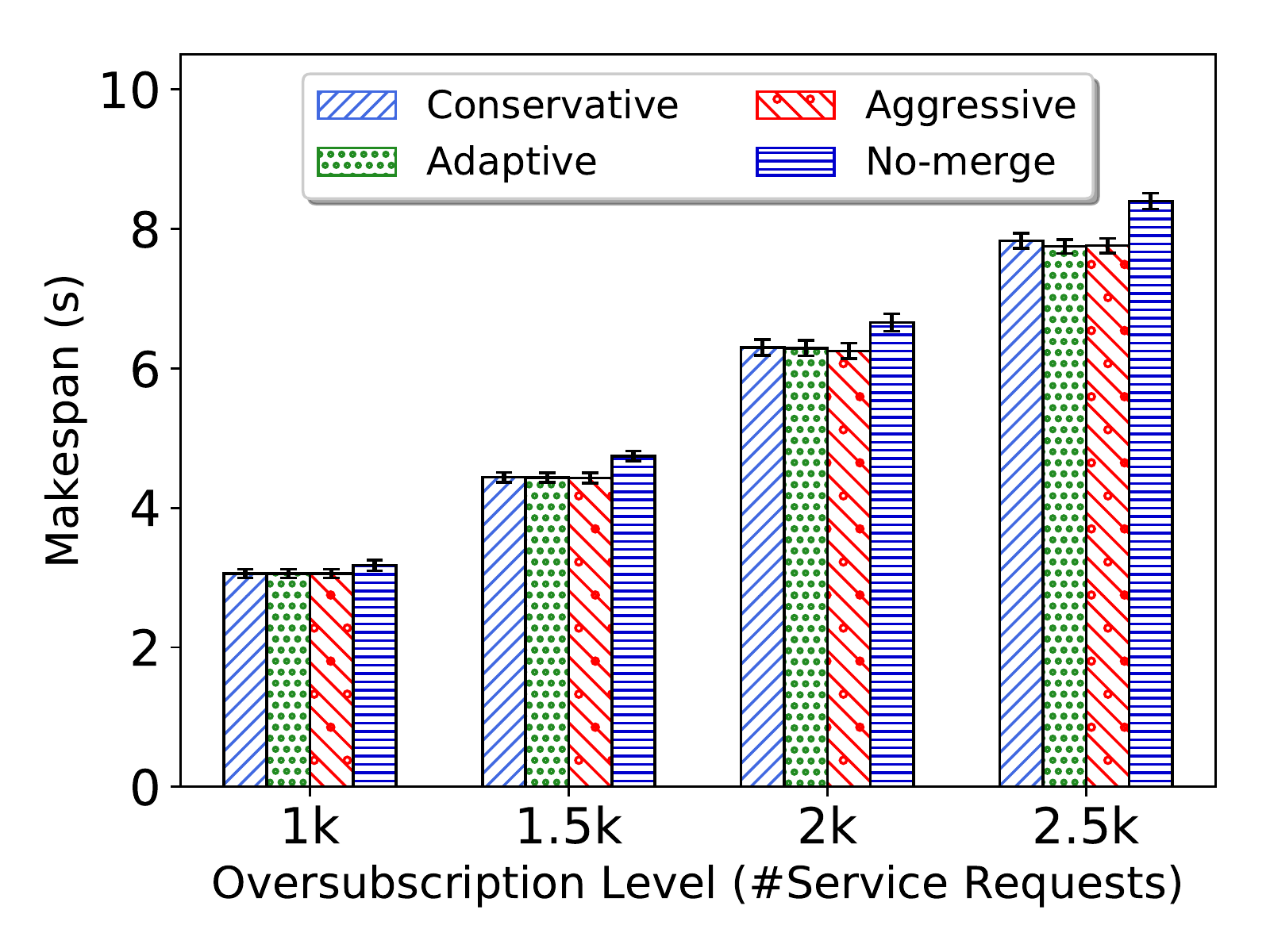} 
	\caption{Comparing the total time to complete all tasks (\ie makespan) under a varying number of arriving processing tasks (horizontal axis) in four scenarios: without task merging, with Adaptive, Conservative, and Aggressive merging in place.}%
	\label{fig:exetime}%
\end{figure}

   \begin{figure}[b] %[!b]%
     \centering 
    
	\subfloat[ FCFS queue]{{\includegraphics[width=0.33\textwidth]{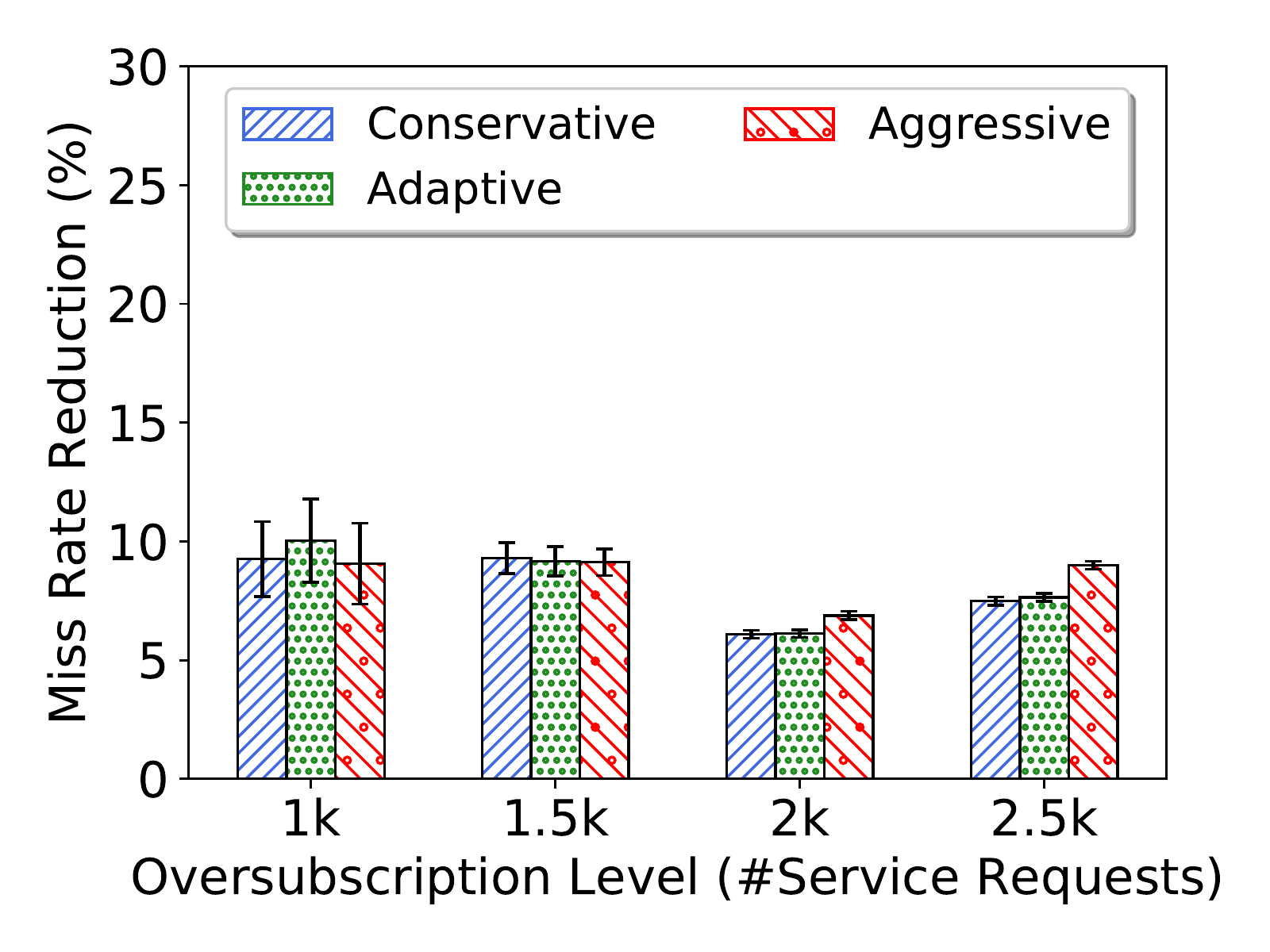} }} %\hspace{0.1cm}%
	\subfloat[ EDF  queue]{{\includegraphics[width=0.33\textwidth]{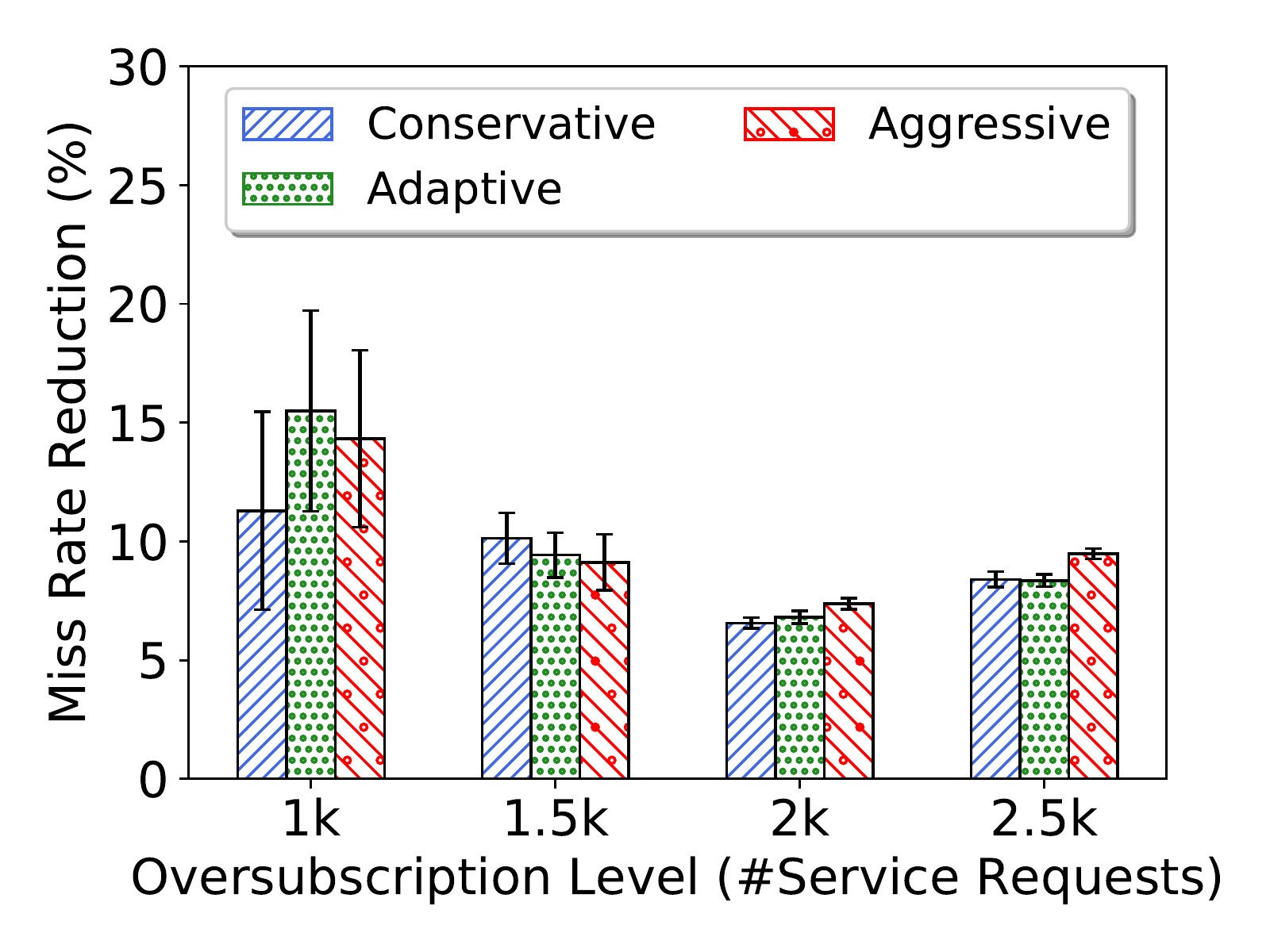} }} %\hspace{0.1cm}%
\subfloat[ MU queue]{{\includegraphics[width=0.33\textwidth]{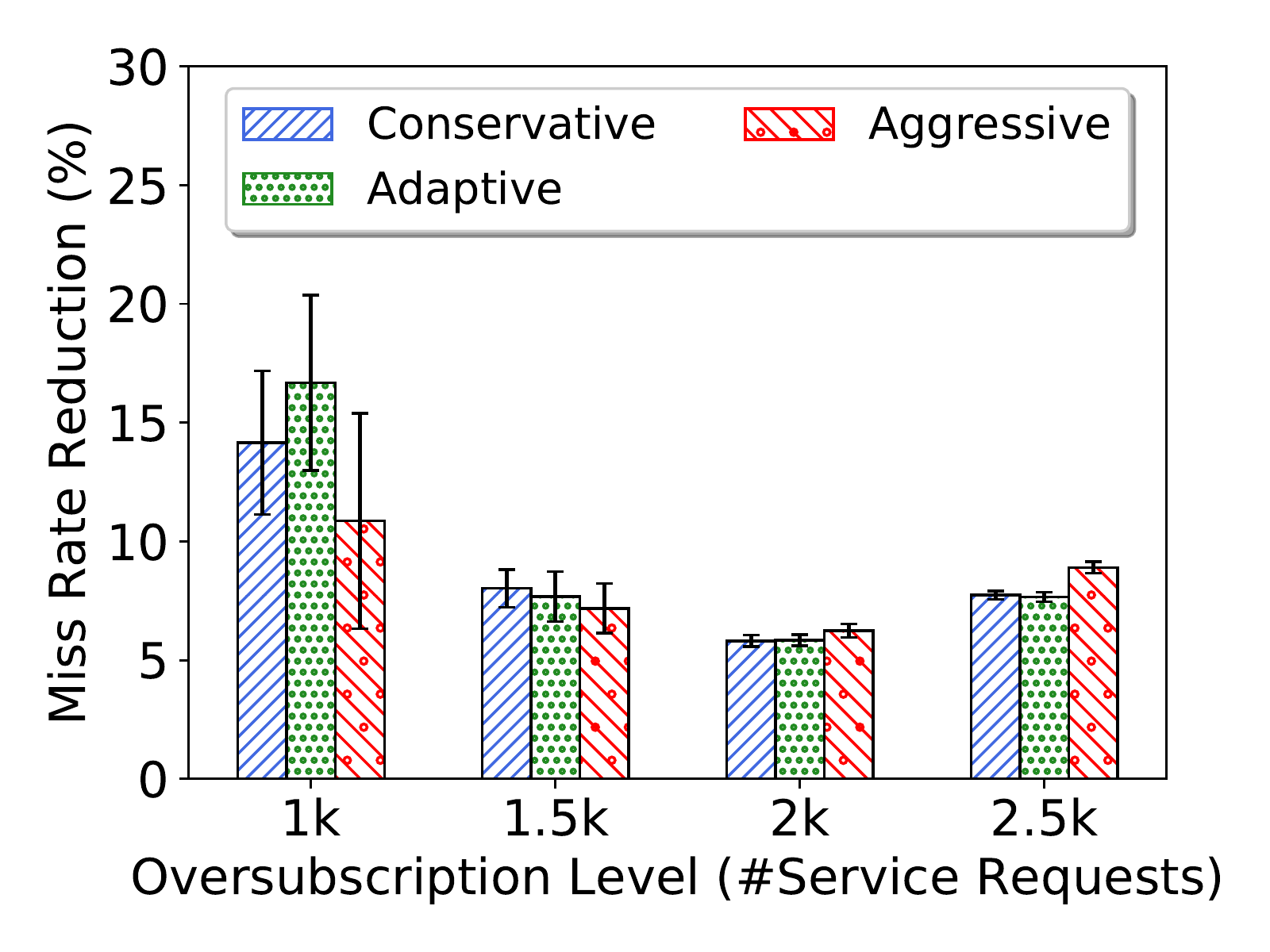} }}%
     \caption{\small{Comparing the deadline miss rate reduction under a varying number of tasks (horizontal axes) using Conservative, Aggressive, and Adaptive merging policies. Subfigures (a), (b), and (c) show the reduction under FCFS, EDF, and Max Urgency (MU) queuing policies.}}   
     \label{fig:deadline}%
 \end{figure}

For each experiment, we examine the system in four scenarios: (A) Without task merging; (B) Conservative task merging policy (\ie by considering merge appropriateness to strictly not cause additional deadline miss); (C) Aggressive task merging policy (\ie without considering merge appropriateness); and (D) Adaptive task merging policy (\ie an adaptive system that work either similar to considerate or aggressive depending on the situation). However, for the sake of better presentation, only some parts of the results are shown in each experiment.

% %floating result figure of execution time moved to previous section, to make it appear before Evaluation section.

 \subsection{Impact of the task merging on the makespan time.}~
In the first experiment, our goal is to see the impact of the task merging on makespan. This metric implies the time cloud resources are deployed, and subsequently, the cost incurred to execute all the tasks. We examine the system under various subscription levels (from 1,000 to 2,500 requests) arriving within the time interval. In this experiment, we examine systems with three queuing policies, namely FCFS, EDF, and MU (Max Urgency). Also, the Position Finder component is disabled for this experiment. That means all the merged tasks are placed in the position of the existing task in the batch queue.
In this system, because tasks are not dropped and computing resources are homogeneous, the scheduling policies do not make a significant change in the makespan. Therefore, only the results of the FCFS queuing policy are presented. 

As we can see in Figure~\ref{fig:exetime}, our proposed merging mechanism saves the makespan between 4\% to 9.1\%. Saving in the makespan time is more pronounced when the system is highly oversubscribed. This is because, there is more backlog of tasks in the scheduling queue at any moment, hence, there is a higher chance of a new arriving task to find its mergeable pair. %It is worth noting that makespan does not vary under different scheduling policies.

Comparing different task merging policies in the figure reveals that their difference in the total makespan is mostly marginal. The conservative merging policy is more reluctant to perform task merging that can result in tasks missing their deadlines. However, that has an unintended positive effect on the total makespan by piling up more tasks in the early stage of its execution, which subsequently increases the chance of a new task to find a suitable mergeable pair in the later stage.
Nonetheless, at a higher level of oversubscription, such effect is diminished, as there is a sufficient number of merge candidates in the batch queue regardless of the merging policy being employed. Thus, in a highly oversubscribed system, the makespan saving of the Conservative merging slightly lags behind other more Aggressive merging policies.

 \subsection{Impact of the task merging on QoS.}~
 In this experiment, our goal is to evaluate the viewers' QoS. For that purpose, we measure the deadline miss rate reduction resulted from merging tasks and compare it with a system that has no task merging under various oversubscribed levels.% and with different scheduling policies. 
 As shown in Figure~\ref{fig:deadline}, we observe that task merging significantly reduces the deadline miss rate for all the scheduling policies. We observe that the improvement in deadline miss rate of FCFS is less than the EDF and MU scheduling policies. It is also more consistent in compare with the other policies. This is because FCFS, by nature, causes a larger average waiting time and does not schedule tasks based on their deadlines. That is why the performance of task merging mechanism, when combined with FCFS, is lower than other scheduling policies.

The comparison across different merging policies reveals that, for low oversubscription levels, Conservative and Adaptive merging result in a higher deadline miss rate reduction than Aggressive merging. The reason is that the Aggressive merging makes inappropriate merging decisions that lead to deadline violation. However, as the oversubscription level increases, aggressively merging tasks seems to be the best approach.

Comparing the results shown in Figure~\ref{fig:exetime} with those in Figure~\ref{fig:deadline} reveals that the difference in deadline miss rate can be larger than the difference in makespan (\ie up to 18\% miss rate reduction compare to up to 9\% makespan reduction). This is because a small reduction in completion time can cause the merged tasks to meeting their deadlines, instead of missing it. We can conclude that the impact of task aggregation mechanism on viewers' QoS becomes more remarkable when it is combined with efficient scheduling policies.

 \subsection{Evaluating the impact of the position finder.}~
 In this part, we examine the effect of the merge position finder module from Section~\ref{sec:posrespect}\,~~ on the deadline miss rate reduction. We assume the system to schedule tasks in the FSCS manner while each of the merged tasks has a chance to be placed outside of their original order in the queue (using Linear Probing heuristic). We apply different merging policies without and with the position finder module (represented as \texttt{+Pfind} in %Figure \ref{fig:Pfind})
 the result figure) 
 in place.
 
   \begin{figure}[htbp]
	\centering
	%\vspace{-10pt}
	\includegraphics[width=0.8\textwidth]{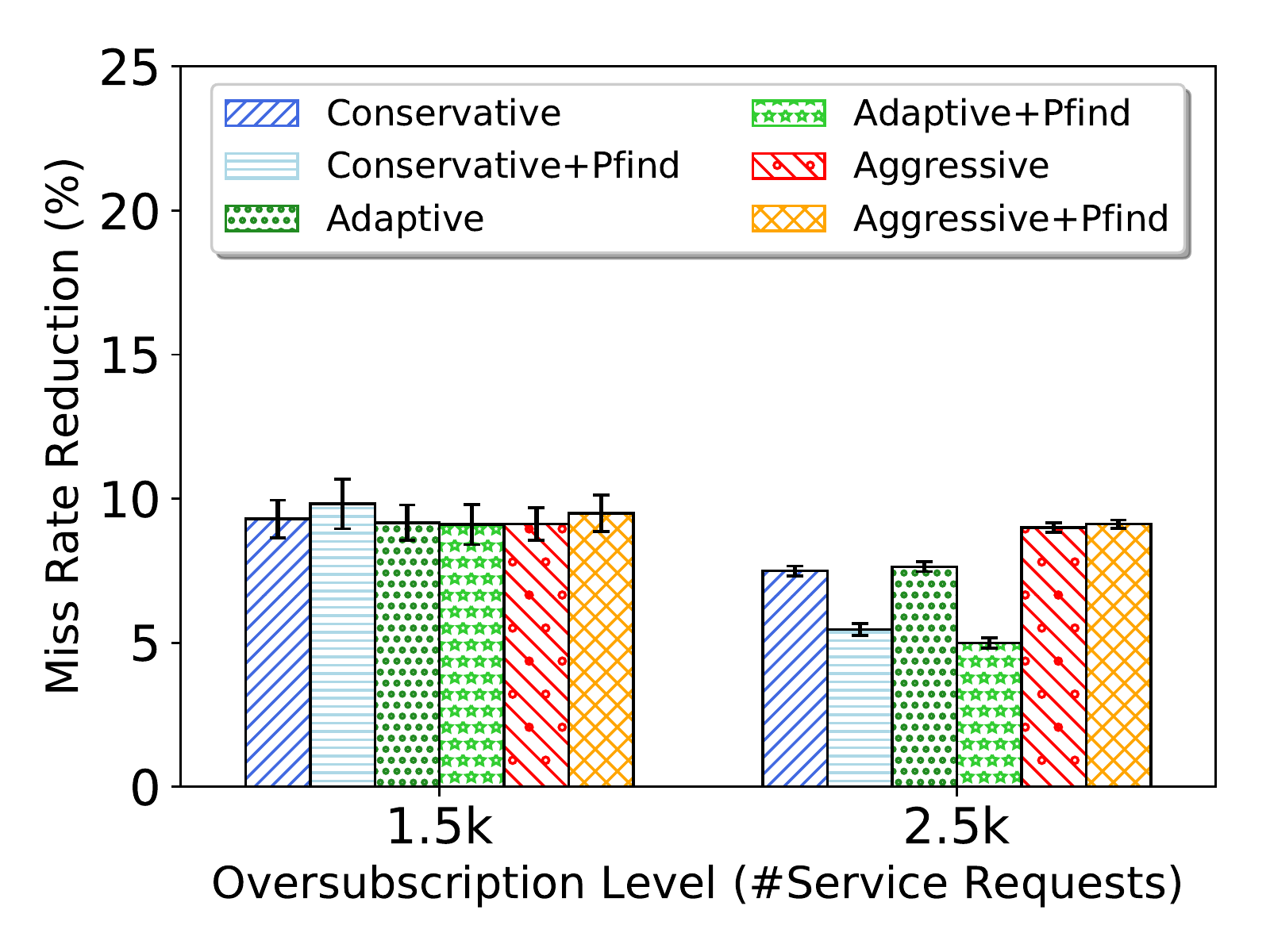} 
	\caption{\small{Comparing the effect of position finder module in term of deadline miss rate reduction under varying number of arriving tasks (horizontal axis) with three merge aggressiveness levels that are applied without and with the position finder, shown as \texttt{Pfind} in the chart. }}%
	\label{fig:Pfind}%
\end{figure}

 Figure~\ref{fig:Pfind} shows an interesting result where position finding module not only improves the deadline miss rate reduction in most cases, but also degrades the performance of Conservative and Adaptive merging policies under highly oversubscribed condition. This is because the position finder module places each merged task in a position that introduces the least amount of impact to other tasks. However, such a position that puts the least amount of impact on other tasks is also a position that is the closest to miss its own deadline. This, at the edge position, limits future mergeability should the other tasks want to merge in front of it. This is not the case for a system with low task arrival rate, because it is likely that the merged task completes its execution before another task merging in front of it. Also, Aggressive merging does not concern with the merge appropriateness, thus, future mergeability is not reduced by the merged task placement. Accordingly, we recommend against using Position Finder module with the Conservative and Adaptive merging policies in the face of high oversubscription levels.

 \subsection{Impact of the execution time uncertainty on task merging.}~
As we noticed in Section~\ref{sec:impact}\,~~, merging decisions are made based on their impacts on the completion time distribution of other tasks. However, the magnitude of uncertainty in the execution time distribution of the tasks can be a decisive factor on the accuracy of estimating the merging side-effects, and subsequently, the deadline miss rate resulted by them. Accordingly, in this experiment, our aim is to evaluate how the three task merging policies function in the face of different levels of uncertainty in the execution time. For that purpose, we increase the randomness of execution time when sampling from the mean execution times. The base level of uncertainty in execution time distribution, observed from the video transcoding services, is relatively low, as the standard deviation equals to approximately 4\% of the mean execution time. In this part, we examine the deadline miss reduction when the standard deviation of execution time distribution is increased by 5 and 10 times, expressed as \texttt{5SD} and \texttt{10SD} in the results, under different oversubscription levels in the system. 

 \begin{figure}[htbp]
	\centering
	\includegraphics[width=0.8\textwidth]{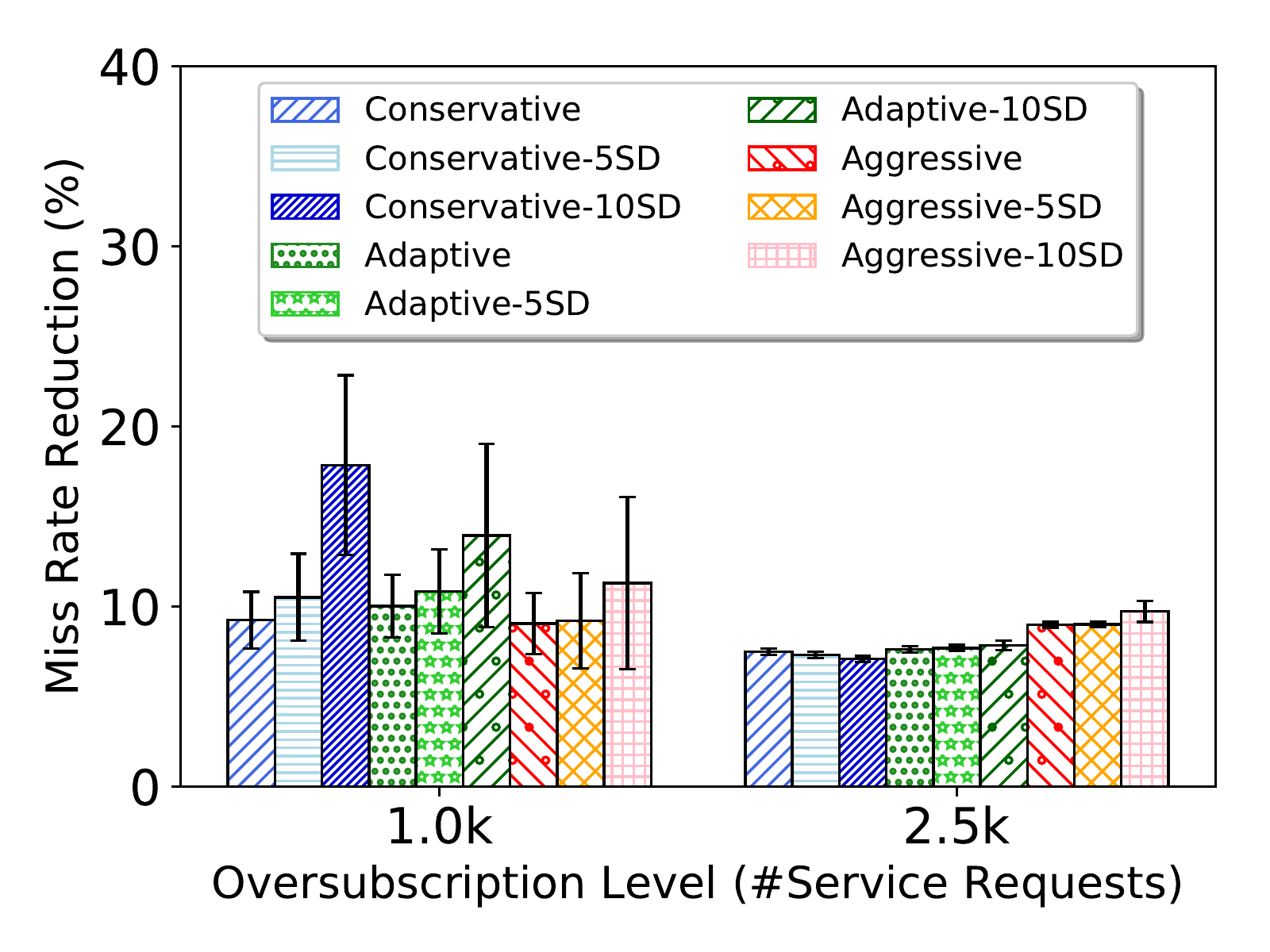} 
	\caption{\small{Comparing the deadline miss rate reduction for different number of arriving tasks (horizontal axis) using the three merging policies applied on tasks with different uncertainty in their execution time distribution. \texttt{5SD} and \texttt{10SD} designate five times and ten times the randomness than the regular dataset. }}%
	\label{fig:varySD}%
\end{figure}

The result of this experiment shown in Figure~\ref{fig:varySD} and includes some interesting observations. Specifically, we observe that as the level of uncertainty rises, there is more performance gain in performing merging. At the low oversubscription level, Conservative and Adaptive merging policies, both of which consider the standard deviation coefficient ($\alpha$) and the merge impact evaluation, gain more deadline miss reduction than the Aggressive policy. %by avoiding risky task merging
However, at a high oversubscription level (2.5k) with a high level of uncertainty, unlike other merging policies, Conservative merging often evaluates merging options as too risky to impacting other tasks. Therefore, the Conservative merging does not allow as many task merging as other policies, and thus performance gain is reduced as the uncertainty level rises. Adaptive merging does not exhibit such behavior and perform well in both situations.

\begin{figure}[htbp]
	\centering
	\includegraphics[width=0.8\textwidth]{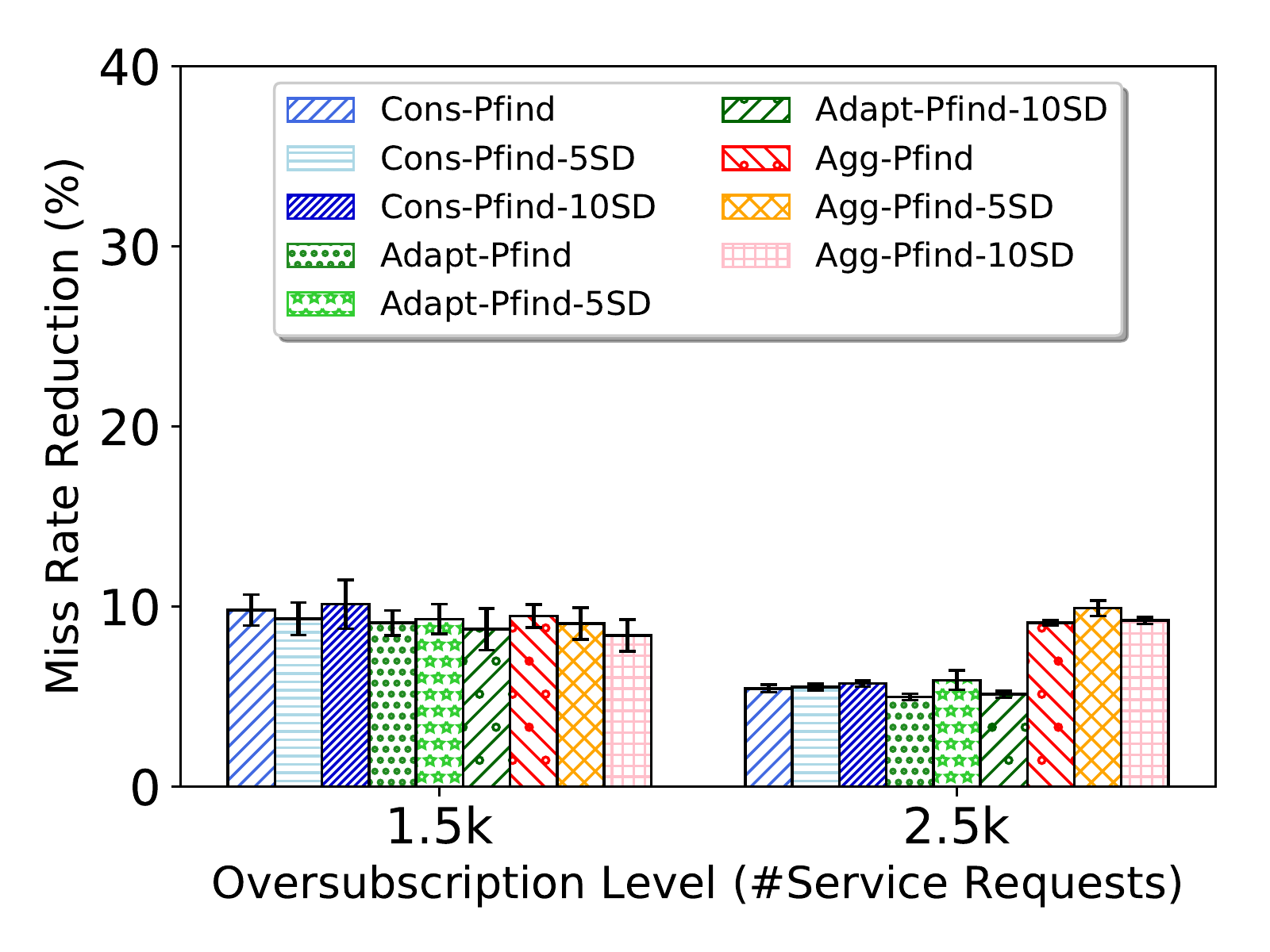} 
	\caption{\small{Comparing the deadline miss rate reduction under varying number of arriving tasks (horizontal axis) using the three merging policies and three levels of execution time uncertainty. In every case, the position finder module (\texttt{Pfind} in the chart) is activated. \texttt{5SD} and \texttt{10SD} designate five and ten times more uncertainty in execution time distribution than the regular workload trace. }}%
	\label{fig:varySDPOS}%
\end{figure}

Figure~\ref{fig:varySDPOS} shows the impact of increasing the uncertainty level on the performance of the position finder module. Comparing the result to those from Figure~\ref{fig:varySD}, we learn that when the position finder module is engaged, the increasing level of uncertainty only has a minimal impact on the deadline miss rate reduction. At 2.5k oversubscription level, the Aggressive policy with the help of position finder module still performs significantly better than other policies.

    % \section{Request Merging Mechanism in Heterogeneous Serverless Cloud Computing System}
%%%%%% Het Merge does not exist.

\section{Summary}
In this chapter, we investigate the computational reuse through task merging. Our goal is to gain resource efficiency and alleviate the oversubscription via merging arriving service requests (task) with other (exact or similar) tasks in the system. In that regard, we dealt with two challenges: \emph{First}, how to identify identical and similar tasks in an efficient manner? \emph{Second}, how to perform (or not perform) merging to achieve the best QoS in the system? 
To address the first challenge, we identified three main levels of similarity that tasks can be merged. Then, we developed a method to detect different levels of task similarity within a constant time complexity. To address the second challenge, we developed a method that determines, based on system oversubscription condition, how to perform the merge operation so that the deadlines of other tasks in the system are likely least affected. Experimental results demonstrate that task merging can reduce the overall execution time of tasks by more than 9\%. Hence, cloud resources can be deployed for a shorter time. Interestingly, this benefit comes with improving QoS of the users by up to 18\%. We concluded that when the level of oversubscription in the system is high, merging tasks aggressively (\ie without being considerate of the impact on other tasks) helps in improving the QoS. Conversely, with lower levels of oversubscription, merging should be carried out with consideration of the impact on other tasks, not to cause unnecessary impact on the QoS.

Although we implemented this system in the context of a video streaming platform, the concept can be applied to the computing platform of other domains as long as we can define similarity levels in those domains. In this chapter we investigated the impact of reusing on the efficiency of serverless cloud platforms. In the next section, we explore an alternative direction where the serverless computing proactively avoid processing task requests that are unlikely to succeed.

%investigate the impact of approximate computing on achieving the goals of serverless computing.

%In the future, we plan to extend this work by considering the impact of using heterogeneous computing resources in the system. Another interesting future research direction can be exploring the impact of marginally compromising tasks' specification accuracy (substitute some task parameters with similar value) to enable more computational sharing with other existing tasks.

%, in favor of a remarkable cost-saving on the cloud resources.

    \chapter{Approximate Computating in Serverless Clouds}
%Approximate Computating in Serverless Clouds
\label{section:DropDefer}
%\section{Overview}
\vspace{14pt}
In this chapter, we introduce another mechanism to alleviate the impact of oversubscription. 
%In the previous chapter, we try to gain resource usage efficiency through task merging to complete all tasks. 
Unlike previous chapter, here, we assume the serverless computing platform already tries to optimize the computing schemes to no avail. That is, even with the use of Heterogeneous Computing (HC) resources (making the computing platform HC system) and all other optimization techniques, it still unable to finish all tasks on time. Therefore, we minimize the impact of not able to finish all the tasks instead. We revise the scheduling system to postpone the mapping of tasks that are unlikely to succeed on time from utilizing the resource (deferring) and drop tasks that are hopeless to finish before the deadline (dropping). Such actions can reduce the impact of oversubscription to the overall perceived QoS of the end-users by allowing more on-time task completion. And hence, more robust against oversubscription of the resources. In our motivational application (SMSE), such task dropping triggers the media streaming to either skip some frames or use low-quality back-up segments in place of the requested media segment. Such an approximate processing approach allows the system to keep up with the demand while resulting in a slightly inaccurate (media streaming) result.

% The deferring and dropping together improves robustness (on average by $\simeq$22\%) and cost in an oversubscribed HC system by up to $\simeq$33\%.

%%%%%%%%%%%%%%%%
%\section{Introduction}

We define \emph{robustness} as the degree to which a system can maintain performance in the face of uncertainty~\cite{Shestak08}. The overall \emph{goal} of this study is to maximize the robustness of an HC system. Each task request is considered to have a hard individual deadline, past which, no value remains in executing the task. Hence, tasks are dropped (\ie removed) from the system when their deadline passes~\cite{khemka2015utility,khemka14utility}. %When the HC system is under load, such that it is impossible for all tasks to complete before their deadlines, the system is considered \emph{oversubscribed}. %%mentioned in prior chapter
The performance metric based on which we measure robustness of an HC system is the number of tasks that meet their deadlines in the system. Therefore, the specific goal of this study is to maximize the number of tasks meeting their deadlines in the HC system (referred to as \emph{task success}) in the face of uncertain execution times in an oversubscribed system. A model of machine and task heterogeneity~\cite{alebrahim2017het} must be available to the resource allocation system, and the system must harness this awareness to overcome with the uncertainty of the HC system.

When tasks have hard deadlines, time spent executing tasks that are ultimately dropped is wasted time. This wasted time cascades down the queue of tasks, delaying the execution of other tasks, and increasing the number of missed tasks in the future---decreasing system robustness. To mitigate this, tasks with a low probability of success should not be mapped, and if they are, they should be dropped before execution~\cite{hussain2019federated}. If probabilistically pruning these unlikely-to-succeed tasks yields more tasks completing on-time in oversubscribed HC systems, how do we maximize the robustness gained thereby? %The more specific research questions can be stated as: \emph{(1) Below what level of probability should we defer tasks? (2) When should the resource allocation transition to a more aggressive mode and drop tasks?}

\begin{figure}[ht]
  \centering
  \includegraphics[width=0.8\textwidth]{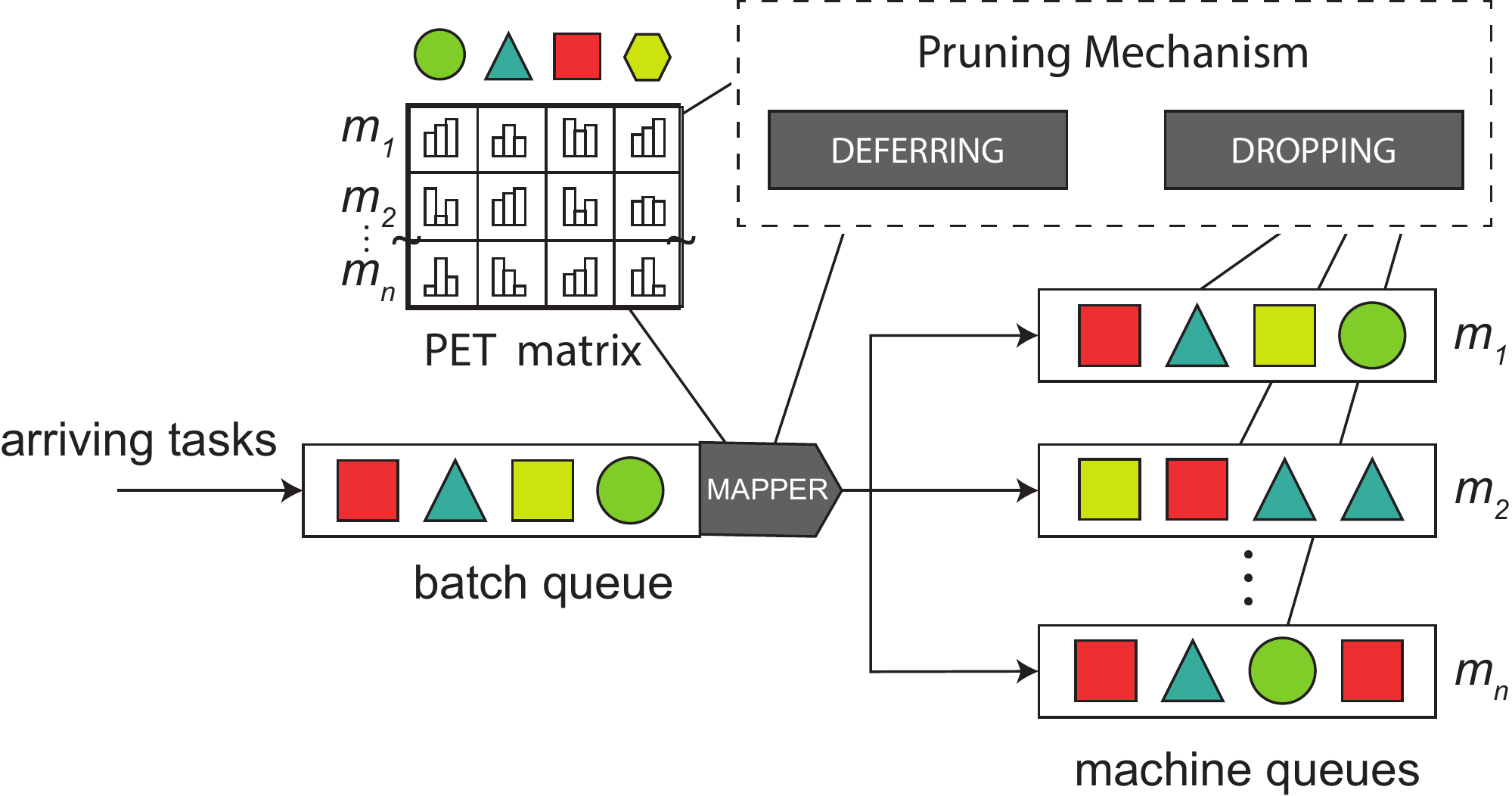}
  \caption{Pruning mechanism. Heterogeneous tasks are mapped to heterogeneous machines in batches. In each mapping, the pruner drops or defers tasks based on their probability of success.} \label{fig:introQueue}
\end{figure}

To address this question, in this research, we propose a pruning mechanism~\cite{denninnart2019improving} (as depicted in Figure~\ref{fig:introQueue}) that is composed of two methods, namely \emph{deferring} and \emph{dropping}. Task deferring deals with postponing assignment of unlikely-to-succeed tasks to a next mapping event with the hope that the tasks can be mapped to a machine that provides a higher chance of success for them. Alternatively, when the system is oversubscribed, the pruning mechanism transitions to a more aggressive mode and drops the tasks that are unlikely to succeed. Before determining deferring and dropping details, we need to model the impact of task dropping on the probability of success for the tasks scheduled to execute after the dropped task. Then, we determine the appropriate probability for dropping and deferring. We propose a method to dynamically determine when the resource allocation system should transition to a more aggressive mode and engage in task dropping. We compare and analyze robustness obtained from deploying our proposed pruning mechanism against an HC system that either does not perform pruning or has a basic pruning implemented. 

Maximizing robustness of HC systems in terms number of tasks meeting their deadlines can potentially cause bias towards executing certain task types and affects fairness of the system. As such, we develop a mapping method to maintain fairness while maximizing robustness.% and compare it against other fairness-oblivious methods. 

Our hypothesis is that the proposed pruning mechanism not only improves robustness of an HC system, but can impact the incurred cost and energy consumption of using resources. %The former is particularly important for users who deploy heterogeneous cloud VMs~\cite{salehi10}, whereas the latter can be appealing to the administrators of High Performance Computing (HPC) systems.
%Such cost saving can be passed from the serverless provider to the user.
As such, we investigate the impact of the proposed probabilistic pruning mechanism on the incurred cost and energy consumption of using heterogeneous cloud VMs and compare it against common mapping methods. Due to generality of the pruning idea, we implement it as an independent mechanism that can be applied to mapping heuristics of any type of (homogeneous or heterogeneous) computing system to improve its robustness. 

Naively implementing the theory behind the pruning decisions imposes a significant overhead to decide the fate of a given task. As such, to make the pruning mechanism practical, we develop methods based on approximation and caching that effectively mitigate the mechanism's overhead, without major impact on the effectiveness of pruning.

%%%%%%%%%%%%%%%%%%%%%%%%%%%%%%%%%%%%%%%%
\begin{figure}%
    \centering 
\includegraphics[width=0.7\textwidth]{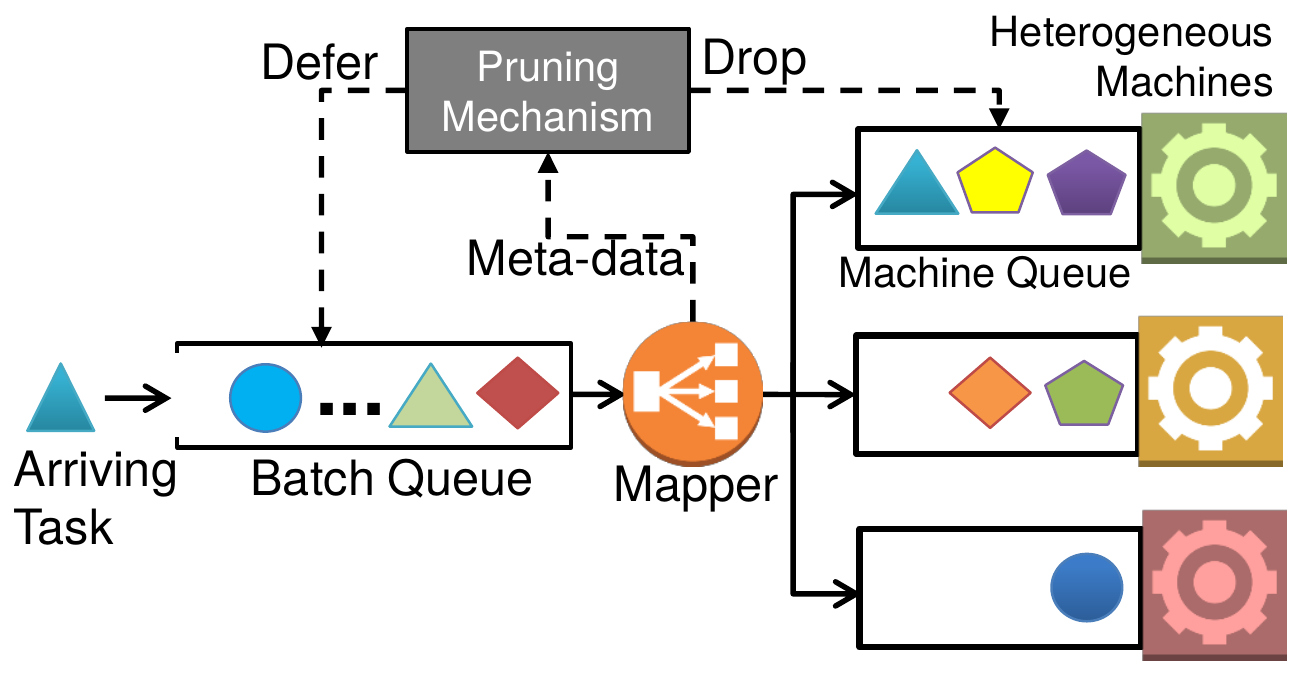} 
    \label{fig:BatchSchedulePrune}
    
    \caption{\small{Batch-mode resource allocation operates on batches of tasks upon task completion (and task arrival when machines queues are not full).  Pruning mechanism can be plugged into existing resource allocation system. Geometries of different shapes, color, and size represent different processing tasks. }} %Batch-mode resource allocation, with task pruning mechanism attached.
\end{figure}

\section{Calculating Task Completion Time in the Presence of Task Dropping}\label{sec:convolution}

Upon dropping a task in a given machine queue, the completion time PMF of those tasks behind the dropped tasks is improved. Intuitively, dropping a task, whose deadline has passed or has a low chance of success, enables the tasks behind it to begin execution sooner, thus, increasing their probability of success and subsequently, overall robustness of the HC system. Each task in queue compounds the uncertainty in the completion time of the tasks behind it in the queue. Dropping a task excludes its PET from the convolution process, reducing the compound uncertainty as well.  

The pruning mechanism we propose in this research should be able to calculate the impact of dropping a task on the probability of success (\ie success chance) of tasks behind the dropped tasks. In this section, we provide the mathematical model to calculate the completion time and probability of meeting deadline of a task located behind a dropped task.

Recall that each entry $(i,j)$ of PET matrix is a PMF represents the \emph{execution time} of task $i$'s task type on a machine type $j$. In fact, $PET(i,j)$ is a set of impulses, denoted $E_{ij}$, where $e_{ij}(t)$ represents execution time probability of a single impulse at time $t$. Similarly, completion time PMF of task $i$ on machine $j$, denoted $PCT(i,j)$, is a set of impulses, denoted $C_{ij}$, where $c_{ij}(t)$ is an impulse representing the probability of completing task $i$ on machine $j$ at time $t$.

Let $i$ be a task with deadline $\delta_i$ arrives at time $\alpha$ and is given a start time on idle machine $j$. In this case, the impulses in $PET(i,j)$ are shifted by $\alpha$ to form its $PCT(i,j)$~\cite{salehi2016stochastic}. %Let $c_{ij}(t)$ an impulse in the completion time PMF of task $i$ on machine $j$ at time $t$. 
Then, the success chance of task $i$ on machine $j$ is the probability of completing $i$ before its deadline, denoted $p_{ij}(\delta_i)$, and is calculated based on Equation~\ref{eq:0}. 

\begin{equation} \label{eq:0}
    p_{ij}(\delta_i) = \sum_{t=\alpha}^{t\leq \delta_i} c_{ij}(t)
\end{equation}

\begin{figure}[htbp]
  \centering
  \includegraphics[width=\textwidth]{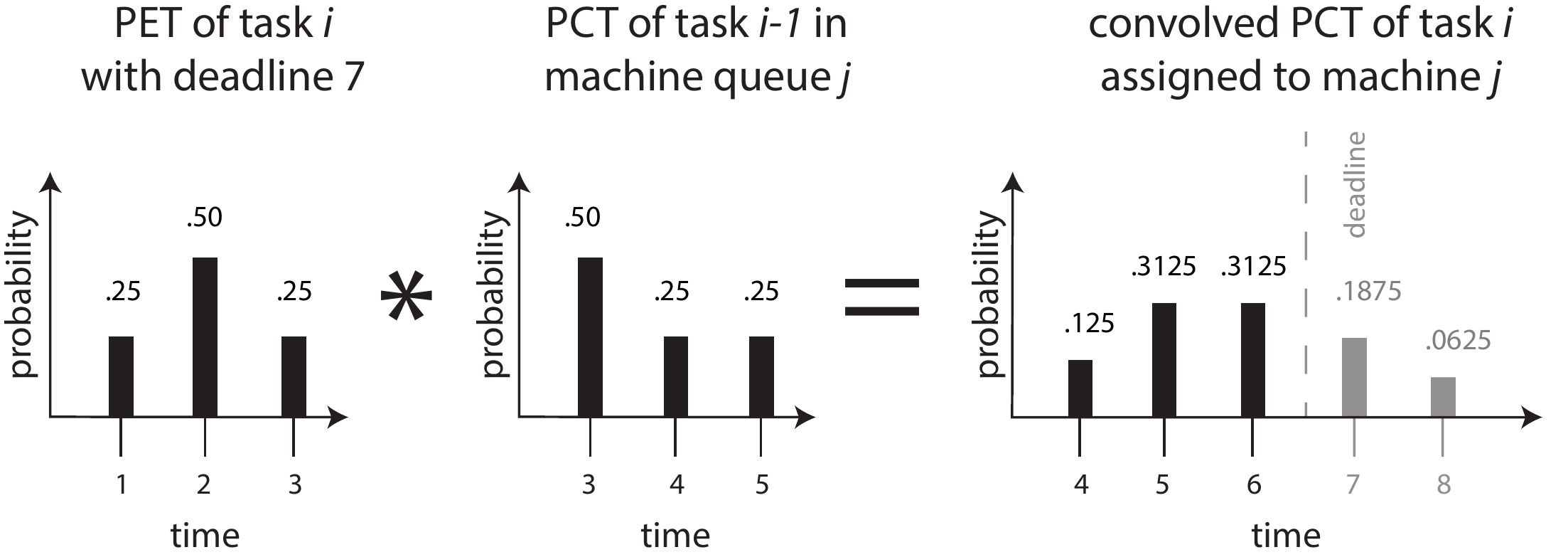}
  \caption{Probabilistic Execution Time (PET) of arriving task $i$ is convolved with the Probabilistic Completion Time (PCT) of the last task on machine $j$ to form $PCT(i,j)$.} \label{fig:conv}

\end{figure}

In case machine $j$ is not idle (\ie it has executing or pending tasks) and task $i$ arrives, the PCT of the last task in machine $j$ (\ie $PCT(i-1,j)$) and $PET(i,j)$ are convolved to form $PCT(i,j)$. This new PMF accounts for execution times of all tasks ahead of task $i$ in the machine queue $j$. For example, in Figure~\ref{fig:conv}, an arriving task $i$ with $\delta_i=7$ is assigned to machine $j$. Then, $PET(i,j)$ is convolved with the PCT of the last task on machine queue $j$ to form $PCT(i,j)$. 

The completion time impulses are generated differently based on the way task dropping is permitted in a system. Three scenarios are possible: \textbf{(A)} Task dropping is not permitted; \textbf{(B)} Only pending tasks can be dropped; and \textbf{(C)} Any task, including the executing one, can be dropped. We note that the initial idea of calculating these completion time PMFs were proposed in~\cite{salehi2016stochastic}. However, in the following, we mathematically model and provide the closed form solution for calculating completion time PMFs. Considering the space limit, interested readers can refer to~\cite{salehi2016stochastic} for further explanations.

\textbf{(A)} Task dropping is not permitted, \ie when all mapped tasks must execute to completion, Equation~\ref{eq:1} is used to calculate the impulses, denoted $c_{ij}^{NoDrop}(t)$, of $C_{ij}$ from the convolution of $PET(i,j)$ and $PCT(i-1,j)$.

\begin{equation} \label{eq:1}
    c_{ij}^{NoDrop}(t)= \sum_{k=1}^{k<t}[e_{ij}(k) \cdotp c_{(i-1)j}^{NoDrop}(t-k)]
\end{equation}

\textbf{(B)} Only pending tasks can be dropped. In this case, the impulses in $PCT(i-1,j)$ that occur after the deadline of task $i$ are not considered in calculating $PCT(i,j)$, as that would indicate task $i$ is dropped due to its deadline passing. Therefore, the formulation changes to reflect the impact of truncated $PCT(i-1,j)$ in the convolution process. Owing to the complexity of calculating $PCT(i,j)$, in this circumstance, we develop a helper function, denoted $f(t,k)$, as shown in Equation~\ref{eq:helper}, that helps Equation~\ref{eq:pending} to discard impulses from  $PCT(i-1,j) \geq \delta_i$. To calculate impulse $c_{ij}(t)$, note that if $t < \delta_i$, then $t-k <\delta_i$. In this case, Equations~\ref{eq:pending} and~\ref{eq:helper} operate the same as Equation~\ref{eq:1}. However, for cases where $t \geq \delta_i$, we use the helper Equation~\ref{eq:helper} to generate an impulse by discarding impulses of  $PCT(i-1,j)\geq \delta_i$. Later, in Equation~\ref{eq:pending}, we add impulses in $i-1$ that occur after $\delta_i$ to account for when task $i-1$ completes at or after $\delta_i$.

\begin{equation} \label{eq:helper}
    f(t,k)= 
\begin{cases}
    0, & \forall (t - k)\geq \delta_i\\
    \\
    e_{ij}(k) \cdotp c_{(i-1)j}^{pend}(t - k),& \forall (t - k)< \delta_i
\end{cases}
\end{equation}
%\vspace{14pt}

\begin{equation} \label{eq:pending}
    c_{ij}^{pend}(t)= 
\begin{cases}
    \sum\limits_{k=1}^{k<t}f(t,k)+c_{(i-1)j}^{pend}(t), & \forall t\geq \delta_i\\
    \\
    \sum\limits_{k=1}^{k<t}f(t,k), & \forall t< \delta_i
\end{cases}
\end{equation}

\textbf{(C)} All tasks (including executing one) can be dropped. In fact, in this case, the completion time impulses are obtained similar to Equation~\ref{eq:pending}. However, the special case happens when $t=\delta_i$ because at this time, if task $i$ has not completed, it is dropped. For the purposes of calculating $PCT(i,j)$ using Equation~\ref{eq:5}, $PCT(i-1,j)$ is guaranteed to be complete by its deadline. Therefore, as Equation~\ref{eq:5} shows, all the impulses after $\delta_i$ are aggregated into the impulse at $t=\delta_i$. We should note that, the discarded impulses, \ie those of task $i-1$ that occur at or after $\delta_i$, must be added to $C_{ij}$, to indicate the probabilities that task $i-1$ completes after task $i$'s deadline.

\begin{equation} \label{eq:5}
    c_{ij}^{evict}(t)= 
\begin{cases}
\sum\limits_{k=t}^{k<\infty}c_{ij}^{pend}(k) + c_{(i-1)j}^{evict}(t), &  t= \delta_i\\
\\
   c_{(i-1)j}^{evict}(t), & \forall t> \delta_i\\
   \\
   \sum\limits_{k=1}^{k<t}f(t,k), & \forall t< \delta_i
\end{cases}
\end{equation}

We note that, calculating completion time and probability of success based on the proposed theory at each mapping event poses a non-negligible overhead to the system. Therefore, in section \ref{sec:approximation}, we propose methods based on approximate computing to mitigate this overhead and making pruning a practical component of a resource allocation system.

%%%%%%%%%%%%%%%%%%%%%%%%%%%%%%%%%%%%%%%%%%%%%

\section{Maximizing Robustness via Pruning Mechanism}\label{sec:solution}

In the beginning of the mapping event, if the system is identified as oversubscribed, the pruning mechanism (aka pruner) examines machine queues. Beginning at the executing task (queue head), for each task in a queue, the success probability (success chance) is calculated. Tasks whose chance of success values are less than or equal to the dropping threshold are removed from the system. Then, the mapping method determines the best mapping for tasks in the batch queue. Prior to assigning the tasks to machines, the tasks with low chance of success are deferred (\ie not assigned to machines) and returned to the batch queue to be considered during the next mapping events. This is in an effort to increase robustness of the system by waiting for a machine with better match to become available for processing the deferred task. To design the pruner for an HC system, three sets of questions regarding deferring and dropping operations are posed that need to be addressed. 

\emph{First}, a set of questions surround the probability thresholds at which tasks are dropped or deferred. How to identify these thresholds are described in Sections~\ref{subsec:dropdefer}--\ref{subsec:defer}. A related question that arises is, should a system-level probability threshold be applied for task dropping? Or, should there be individual considerations based on the characteristics of each task? If so, what characteristics should be considered, and how should they be used in the final determination?

\emph{Second}, there is the matter of when to begin task dropping, and when to cease. That is, how to dynamically determine the system is oversubscribed and transition the pruner to a more aggressive mode to drop unlikely-to-succeed tasks such that the overall system robustness is improved. The answer to this question is provided in Section~\ref{subsec:aggrprun}.

Pruning can potentially lead to unfair scheduling across tasks types---constantly pruning compute-intensive and urgent task types in favor of other tasks to maximize the overall robustness. Hence, the \emph{third} question is how the unfairness across task types can be prevented? Should the system prioritize task types that have been pruned? If so, how much of a concession should be made? We address this question in Section~\ref{sec:mechanism}.

\section{Determining Task Dropping Probability}\label{subsec:dropdefer}
\subsection{Dynamic per-task dropping probability threshold.}~ %can remove word threshold, maybe...
At its simplest, the task dropper can apply uniform dropping threshold for all tasks in a machine queue. However, a deeper analysis tells us that not all tasks have the same effects on the probability of on-time completion for the tasks behind them in queue. This difference can be taken into account to make the best decision about which tasks should stay and which are dropped. 

%As we mentioned earlier, the completion time PMF of a given task $i$ is used for calculating the task robustness. 
In addition to determining task's chance of success, other features of completion time PMF can be valuable in making decisions about probabilistic task dropping. We identify two task-level characteristics that further influence the chance of success of tasks located behind a given task $i$: (A) the position of task $i$ in machine queue, and (B) the shape (\ie skewness) of completion time PMF of task $i$. 

In fact, the closer a task is to execution (\ie to the head of machine queue), the more tasks are affected by its completion time. For instance, with a machine queue size of six, an executing task affects the completion time of five tasks queued behind it, where the execution time of a task at the end of the queue affects no tasks. Therefore, the system should apply a higher dropping threshold for tasks close to queue head. 

Skewness is defined as the measure of asymmetry in a probability distribution and is calculated based on Equation~\ref{eq:skew}, as explained in~\cite{bayes2007bayesian}. In this equation, $N$ is the sample size of a given PMF, $Y_i$ is an observation, $\bar{Y}$ is the mean of observations, and $\sigma$ is the standard deviation of the observations. A negative skewness value means the tail is on the left side of a distribution whereas a positive value means that the tale is on the right side. Generally, $|S| \geq 1$ is considered highly skewed, thus, we define $s$ as bounded skewness and we have $-1 \leq s \leq 1$. 
\vspace{-2pt}
\begin{equation}\label{eq:skew}
S = \frac{\sqrt{N(N-1)}}{N-2} \times \frac{\sum_{i=1}^{n} {(Y_i-\bar{Y})^3}/N}{\sigma^3} 
\end{equation}
%\vspace{1pt}

\begin{figure} [htpb]
  \centering
  \begin{subfigure}[htpb]{0.8\textwidth}
  \centering
    \includegraphics[width=\textwidth]{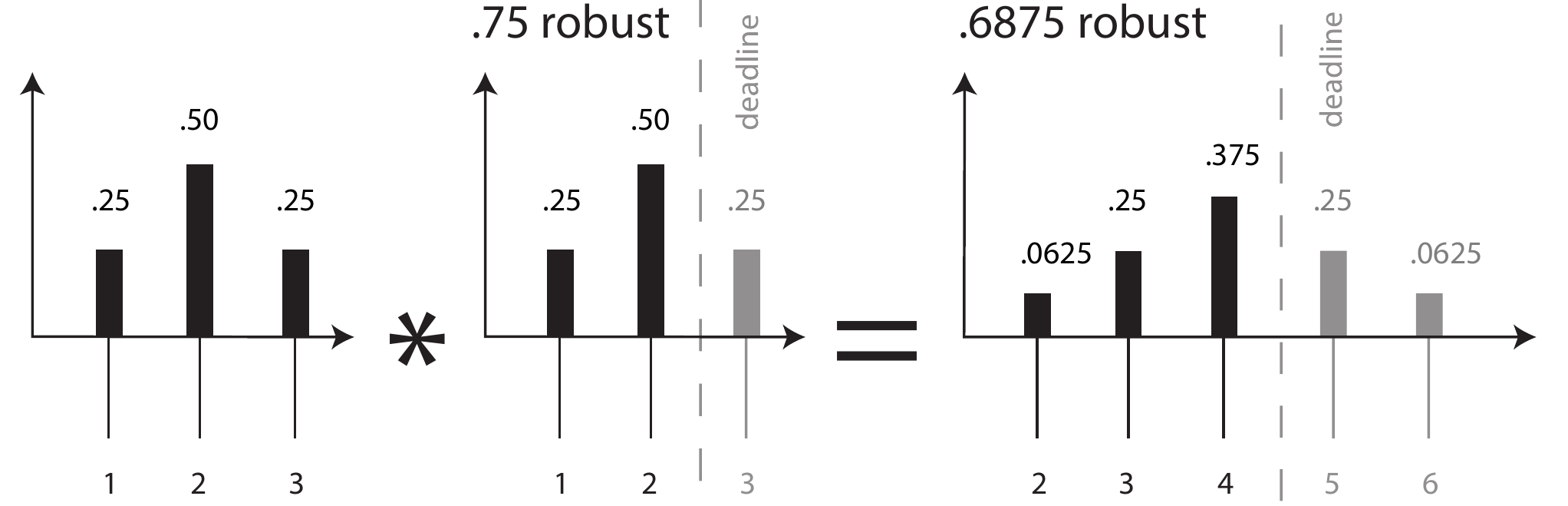}
    \caption{No Skew}
    \label{fig:noskew}
  \end{subfigure}

  \begin{subfigure}[htpb]{0.8\textwidth}
  \centering
    \includegraphics[width=\textwidth]{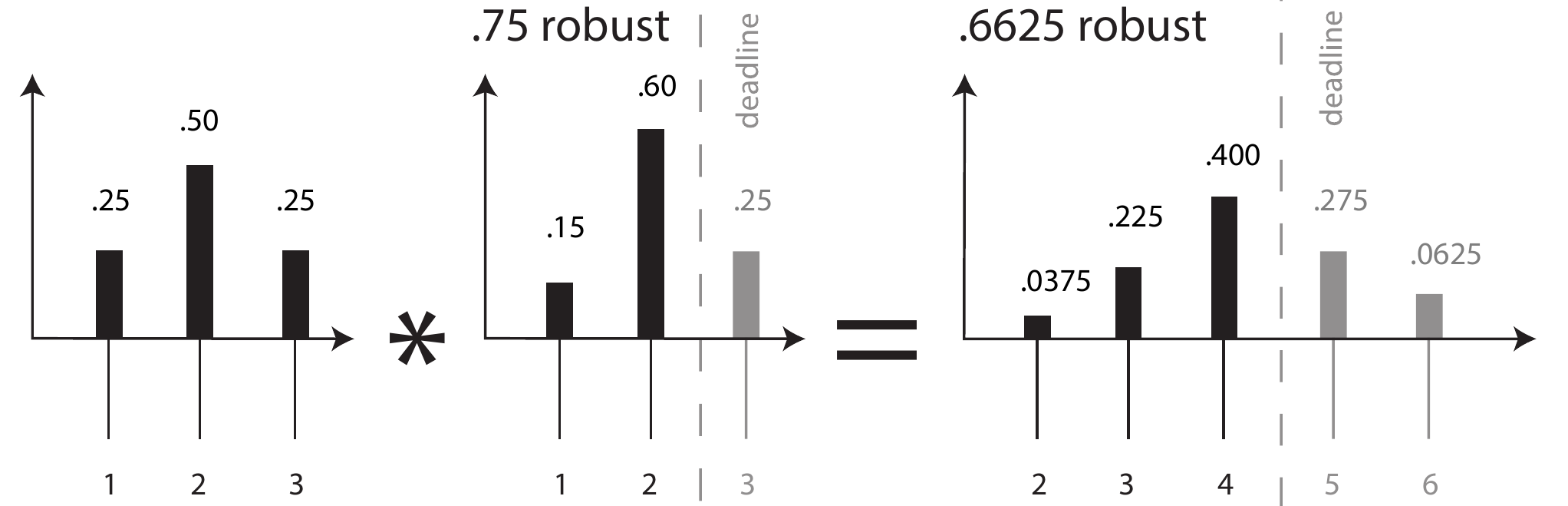}   
    \caption{Left Skew}
    \label{fig:leftskew}
  \end{subfigure}
  
  \begin{subfigure}[htpb]{0.8\textwidth}
  \centering
    \includegraphics[width=\textwidth]{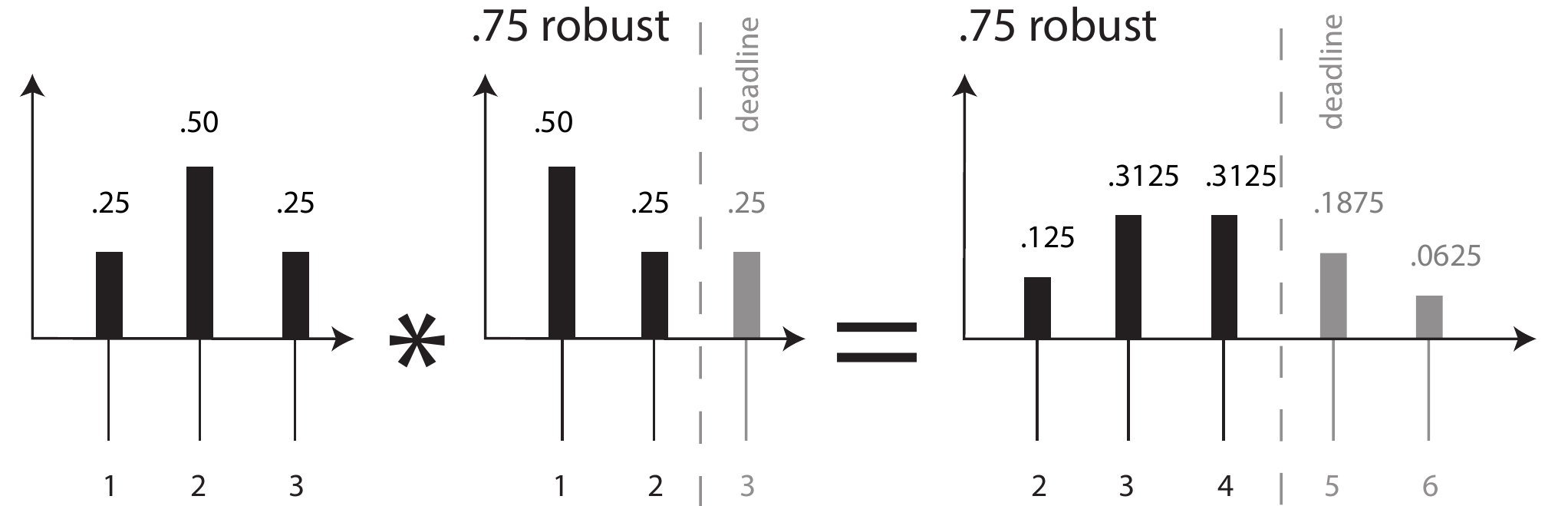}
    \caption{Right Skew}
    \label{fig:rightskew}
  \end{subfigure}
  
  \caption{Demonstration of effect of task $i$'s skewness on completion time PMF of task $i+1$ (right-most PMFs) with a deadline 5 ($\delta_{i+1}=5$). The left-most PMFs show execution time PMF of task $i+1$ and the middle ones show completion time PMF of task $i$ ($\delta_i=3$).} \label{fig:skewsplain}
\end{figure}

%The sign of skewness value denotes the direction of the skew. 
A negatively skewed PMF has the majority of probability occurring on the right side of PMF. Alternatively, because the bulk of a probability is biased to the left side of a PMF, a positive skew implies that a task is completed sooner than later. The middle PMFs in Figure~\ref{fig:skewsplain} each represents a completion time with a success chance of 0.75, however, they show different types of skewness. Using this information, we can see that two tasks with the same success chance can have different impacts on the success chance of tasks behind them in queue. Tasks that are more likely to complete sooner (\ie positive skewness) propagate that positive impact to tasks behind them in queue. The opposite is true for negatively skewed tasks. Reasonably, we can favor tasks with positive skewness in task dropping. Figure~\ref{fig:skewsplain} shows the effects of different types of skews on the completion times of tasks behind them in queue. Subfigure~\ref{fig:leftskew} shows the negative effects of negative skew whereas Subfigure~\ref{fig:rightskew} shows the positive effect of positive skew on the success chance of the next task in the queue. 

Using the skewness and queue position, the system can adjust a base dropping threshold dynamically, for each task in a machine queue. The adjusted dropping threshold for a given task $i$, denoted $\phi_i$, is calculated based on Equation~\ref{eq:location}. To favor tasks with positively skewed completion time PMF, we negate the skewness ($s_i$). To account for position of task $i$ in machine queue, denoted $\kappa_i$, we divide the negated skewness by the position. Addition of 1 is just to avoid division by zero and $\rho$ is a parameter to scale the adjusted dropping threshold. Ideally, this will allow more tasks to complete before their deadline, leading to a higher robustness in an HC system.
%\vspace{-1pt}
\begin{equation}\label{eq:location}
\phi_i=\frac{-s_i \cdotp \rho}{\kappa_i+1}
\end{equation}
%\vspace{1pt}

This dynamic adjustment of the probability is done only in the dropping stage of the pruner. When it comes to deferring tasks, the task position is always the same (\ie the tail of the queue), and it is too early to consider the shape of the tasks PMF, as there are, as yet, no tasks behind it in queue.

\subsection{Determining task deferring probability.}~\label{subsec:defer}
We discussed that any task that has a chance of success lower than its specified dropping threshold has too low of a success chance to warrant risking of allocating it to a resource. The optimal deferring value, however, is applied to unmapped tasks in the batch queue and should vary based on the workload characteristics. In fact, deferring threshold acts as a throttle that controls the flow of incoming tasks to the HC system. In one hand, a too-high deferring threshold ensures that available machine queue slots are reserved only for tasks with a high chance of success, but can lead to resource under-utilization by leaving the computing resource idle. On the other hand, a too-low deferring threshold allows tasks, potentially with low chance of success, to fill the machine queue slots---preventing the mapping of high-chance incoming tasks. To avoid such scenarios, an appropriate deferring threshold should be dynamically determined based on the characteristics of the workload in the system. In the rest of this section, we describe our approach to dynamically adjust the deferring threshold based on the workload characteristics.  

\emph{Selective factor} (denoted $\Delta$) is defined as the ratio of the number of tasks in batch queue (waiting to be mapped) to the number of empty slots in machine queues. A high selective factor indicates that there are many unmapped tasks, but not enough machine queue slots to accommodate them. In this scenario, task mapping should be more selective and task dropping should be more aggressive to free up machine queue slots for a better-suited tasks that are waiting to be mapped.

Let $\upsilon$ be the deferring threshold in an HC system with $b$ unmapped tasks in its batch queue.
A \emph{competent task} is a task whose maximum success chance across all machines is higher than the $\upsilon$. That is, competent tasks are those that are not deferred because they have decent chance of success. Accordingly, \emph{task competency level} (denoted $\Gamma$) in a batch queue is defined as the ratio of the number of competent tasks to the total number of unmapped tasks and is calculated based on Equation~\ref{eq:competent_lvl2}. High task competency level (close to 1) implies a high percentage of tasks in the batch queue are qualified for mapping, but are not mapped due to inadequate slots in the machine queue. In this case, the deferring threshold should be increased, so that only the highly competent tasks are considered for mapping. Conversely, a low task competency level can be an indication that the task deferring threshold is set too high, \ie the system is too selective, such that the majority of the tasks cannot pass the deferring threshold.

\begin{equation}
    \label{eq:competent_lvl2}
    \Gamma = \frac{1}{b}\sum_{i=1}^{b}
\begin{cases}
    0,& max(p_{ij}(\delta_i) ) < \upsilon | j \in \{0 ..  m\}
    \\
   1,& max(p_{ij}(\delta_i) ) \ge \upsilon | j \in \{0 ..  m\}
\end{cases}
\end{equation}
%=pending robustness
\subsection{Instantaneous robustness.}~
Instantaneous robustness at a given time is defined as the average of chance of success for all tasks exist in the system. Let $m$ be the number of machine queues, $q$ the number of queue slots in each machine queue, and $p_{ij}(\delta_i)$ is the chance of meeting the deadline of task $i$ in the machine queue $j$. Then, instantaneous robustness is calculated based on Equation~\ref{eq:mq_robustness}. Our hypothesis is that maintaining a high instantaneous robustness leads to a high level of overall system robustness. As such, instantaneous robustness can act as a performance indicator for task deferral and mapping heuristics. The system should aim to maintain the high instantaneous robustness level and avoid task mappings that reduce the instantaneous robustness. 

\begin{equation}
    \label{eq:mq_robustness}
    \psi =\frac{1}{m \cdot q} \sum_{j=1}^{m}  \sum_{i=1}^{q} p_{ij}(\delta_i)
\end{equation}

\subsection{Deferring probability threshold.}~
% Considering the three performance indicators together, we propose a procedure to automatically adjust the deferring threshold $\upsilon$ to an appropriate value. 
When the system is not heavily oversubscribed and there are more empty slots in the machine queue than tasks in the batch queue (\ie $\Delta < 1$), the new deferring threshold ($\upsilon_n$) can be reduced from its current value ($\upsilon_c$) to allow more task mappings. Alternatively, when there are more tasks to map than the number of available slots (\ie $\Delta >1$), we act based on the competency level ($\Gamma$). If no task is passing its deferring threshold (\ie $\Gamma=0$), it means that the deferring threshold is high and has to be reduced. Otherwise, in the case of oversubscription, we set the deferring threshold to a value near the instantaneous robustness value. % to prevent mapping unlikely tasks.
The vale of $\theta$ is a constant to adjust the deferring probability threshold. %adjustment unit variable which controls the rate of adjustment per time unit.
Equation~\ref{eq:adapt} formally expresses the way deferring probability threshold is dynamically calculated.

\begin{equation} \label{eq:adapt}
    \upsilon_n= 
\begin{cases}
\upsilon_c -\theta, &  \Delta < 1\\
\\
   \psi - \theta, & \Delta \ge 1,\Gamma \neq 0\\
   \\
   \upsilon_c -\theta, & \Delta \ge 1,\Gamma = 0
\end{cases}
\end{equation}

\subsection{Aggressive pruning by dynamically engaging task dropping.}~\label{subsec:aggrprun}
To maximize robustness of the system, the aggression of the pruning mechanism has to be dynamically adjusted in reaction to the level of oversubscription in the HC system. 
The pruning mechanism considers the number of tasks missed that their deadlines since the past mapping event as an indicator of the oversubscription level in the system. We use the identified oversubscription level as a \emph{toggle} that transitions the pruner to task dropping mode. 
However, in this case, the pruner can potentially toggle to dropping mode as a result of an acute spike in task-arrival, and not a sustained oversubscription state. %To avoid this, the Pruner's toggle does not react to temporary spikes in deadline misses, but only to sustained oversubscription by using a weighted average of deadline misses over the past mapping events. Thereby, the Pruner is slower to engage, and slower to disengage.

To judge the oversubscription state in the system, the pruner operates based on a moving weighted average number of tasks that missed their deadlines during the past mapping events. Let $d_\tau$ the oversubscription level of the HC system at mapping event $\tau$; and \(\mu_{\tau}\) the number of tasks missing their deadline since the past mapping event. Parameter $\lambda$ is tunable and is determined based on the relative weight assigned to the past events. The oversubscription level is the calculated based on Equation~\ref{eq:oversub}. In the experiment section, we analyze the impact of $lambda$ and determine an appropriate value for it.

\begin{equation}\label{eq:oversub}
d_\tau=\mu_\tau \cdotp \lambda + d_{\tau-1} \cdotp (1-\lambda)
\end{equation}

Another potential concern is minor fluctuations about the toggle switching the dropping off and then back on. We employ a Schmitt Trigger~\cite{kader2012advancement} to prevent minor fluctuations around dropping toggle. We set separate on and off values for the dropping toggle. Based on our initial experiments, we determined the Schmitt Trigger to have 20\% separation between the on and off values. For instance, if oversubscription level two or higher signals starting dropping, oversubscription value 1.6 or lower signals stopping it.

%%%%%%%%%%%%%%%%%%%%%%%%%%%%%%%%%%%%%%%%
\section{Pruning Mechanism as Module of a Resource Allocation System}
\label{sec:mechanism}
%\subsection{Overview.}~
%Based on theories proposed in Section~\ref{sec:convolution} and ~\ref{sec:solution}, this section we summarize and package the pruning mechanism as a module that can work in conjunction with mapping heuristics to increase the robustness of the HC system. Then, we introduce two probabilistic mapping heuristics to work in conjunction with the pruning mechanism. 
In this section, with the goal of maximizing the system robustness, theories presented in Sections ~\ref{sec:convolution} and~\ref{sec:solution} are leveraged to design a pruning mechanism as a module of resource allocation system that can work with any mapping heuristic. Then, two probabilistic mapping heuristics, called Pruning Aware Mapper (PAM) and Fair Pruning Aware Mapper (PAMF) are proposed to work along with the pruning mechanism.
 \begin{figure} [htbp]
  \centering
  \includegraphics[width=0.99\textwidth]{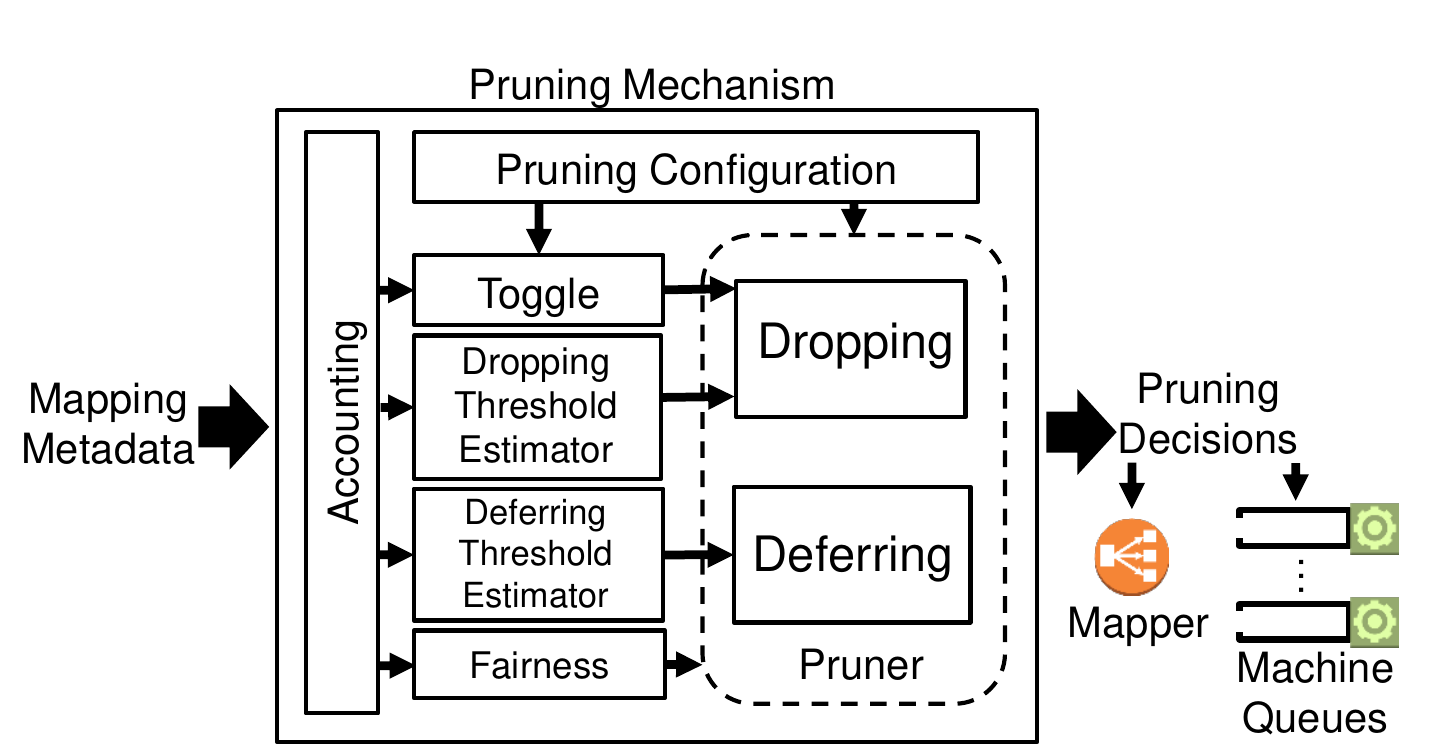}
  \caption{Components of pruning mechanism. Inputs are mapping metadata and outputs are pruning decisions to apply on mapper (task deferring) or machine queues (task dropping). The pruning system is packaged as module in resource allocation systems to function in conjunction with mapping heuristic.} \label{fig:prunearch}
\end{figure}

\subsection{Task pruning mechanism.}~
The overall architecture of the pruning mechanism is shown in Figure~\ref{fig:prunearch}. The \emph{Accounting} module receives meta data (\eg tasks' deadline, PET, and PCT) from the resource allocation system. The meta-data are available for other components to utilize.
The \emph{Toggle} module (in either the default or Schmitt trigger configuration) uses the collected information to measure the oversubscription level of the HC system. It then decides whether or not it is beneficial to engage the ``task dropping''.

%With the goal of maximizing the robustness, the \emph{Pruner} module, first, calculates the chance of success for all tasks, then, enacts the pruning decision on tasks whose chance is lower than the user-defined threshold, specified in \emph{Pruning Configuration}. The \emph{Per-task Threshold} approach modifies the dropping threshold based on the task's skewness and their position in the machine queue.The \emph{Threshold Adaption} module evaluates the oversubscribed situation and adjusts the deferring threshold accordingly.
With the goal of maximizing the robustness, the \emph{Pruner} module enacts the the dropping and deferring sub-modules to prune tasks whose chance of success is lower than the thresholds specified in the  \emph{Pruning Configuration}. Moreover, dropping and deferring thresholds are dynamically adjusted during each mapping event to maximize the system robustness. To this end, the \emph{Dropping Threshold Estimator} modifies the dropping threshold based on each task's skewness and position in the machine queue. Furthermore, The \emph{Deferring Threshold Estimator} module adjusts the deferring threshold based on the oversubscription level.

%The \emph{Fairness} module keeps track of the suffered task types (\ie those are consistently dropped) and adjusts the Pruner to avoid biasness against them. This component is` explained in Section~\ref{subsec:pamf}.

The \emph{Fairness} module is employed to avoid unfair pruning mechanism. This module detects the suffering task types (\ie those that are consistently dropped) and adjusts the pruner to prevent task types being unfairly pruned. The Fairness module is elaborated in Section~\ref{subsec:pamf}. Output of the pruning mechanism is its decision that can be either task dropping (applied on machine queues) or task pruning (applied on unmapped queue). 
 %We successfully added the pruning mechanism to different mapping heuristics. This enables the experiments of existing mapping heuristics with pruning mechanism in Experiment~\ref{subsec:heuristicscmp}.

As mentioned earlier, the pruning mechanism is pluggable, that is, it can be added to any mapping heuristic. In the next part, we explain plugging the pruning mechanism to two new probabilistic-base mapping heuristics, namely PAM and PAMF. 

\subsection{Mapping heuristics with pruning mechanism.}~\label{subsec:heuristics}
%In this part, we develop two mapping heuristics based on the theory explored in this study. Both heuristics are designed to work in conjunction with pruning mechanism. The first heuristic, PAM, leverages chance of success measurement to maximize robustness. However, the second mapping heuristic, in addition to maximizing robustness, aims at achieving fairness across task types.
In this part, two mapping heuristics that work in conjunction with the pruning mechanism are developed. First, a heuristic called Pruning Aware Mapper (PAM) leverages the chance of success to probabilistically maximize the robustness of the system. In this scope, the robustness of the system is defined as the number of tasks completed on time  during the study time. However, only considering the chance of success to maximize the robustness can result in unfair task type completion. As a result, a second mapping heuristics is proposed to achieve maximum robustness balanced with fairness across task types.
%The heuristics are two-phase processes. first phase finds the best machine for each task, by virtue of a per-heuristic objective. In the second phase, from task-machine pairs obtained in the first phase, each heuristic chooses the best task-machine pair for each available machine queue slot. After all slots are filled, virtual mappings are assigned to the machine queues and the mapping method is complete.
The heuristics occur in two phases. In the first phase, for each task, a machine that has the best affinity is determined and a task-machine pair constructed. PAM considers task-machine affinity implicitly via taking the chance of success of a task into consideration. 
Then, the best task-machine pair is selected for mapping and that task is assigned to its paired machine. 
%The pruning mechanism comes into action in two parts:
%First, before making any mapping decision, the pruner analyzes the oversubscription level and performs task dropping on machine queues, if necessary. And between the first and second phases of two phases task mapping procedure, any tasks that do not meet the deferring threshold are deferred. We note that while the pruning mechanism is an optional add-on to any mapping heuristics. PAM and PAMF both are designed specifically to work in conjunction with pruning mechanism.
Note that the pruning mechanism can be plugged in mapping heuristics to maximize the system robustness. The pruning occurs in two steps. First, prior to any mapping decision, the pruner performs task-dropping on machine queues. Next, tasks in the batch queue with chance of success lower than the deferring threshold are deferred, leaving the deferred tasks to remain in the batch queue.

\paragraph*{Pruning aware mapper (PAM):}
%This heuristic uses the PET matrix to calculate task's success chance and then operates in compliance with pruning mechanism such that the mapping decisions are the highly robust (\ie low chance of mapping a task that will eventually get pruned).
In PAM, maximizing the system robustness happens by maximizing each task's chance of success. To this end, the PET matrix is used to determine the chance of success for each task.

%mod for JPDC
%Based on our prior observations and experiments, we conclude that to utilize the chance of success data for tasks mapping effectively, the first phase of PAM should find the machine that offers the highest chance of success. Then the second phase finds the task-machine pair with the lowest completion time and maps it to that machine's virtual queue. 
Based on our prior observations, the machine offering highest chance of success is selected for constructing the task-machine pair during the first phase of the PAM heuristic. Then, in the second phase, the pair with lowest completion time is selected for mapping. 
%Ties are broken by choosing task with the shortest expected execution time. By strategically select a machine based on the maximum chance of success and select task-machine pair by the minimum completion time, the mechanism greedily prefers the mapping pairs that have both a high chance of success and short execution time. This approach maintains the robustness better than mapping heuristics that only utilize either chance of success data or short deadline alone.
In this way, the system prefers to map tasks having both high chance of success and short execution time.

%old
%In the first phase, for each unmapped task, PAM finds the machine offers the highest chance of success. Then, tasks that do not meet the deferring threshold are pruned. The second phase finds the task-machine pair with the lowest completion time and maps it to that machine's virtual queue. Ties are broken by choosing task with the shortest expected execution time. 

%The process repeats until either all tasks in the temporary batch queue are mapped or dropped, or until the virtual machine queues are full. 

\paragraph*{Fair pruning aware mapper (PAMF):}\label{subsec:pamf}
Probabilistic task pruning potentially favors task types with shorter execution times, resulting in unfairness. This is because shorter tasks usually have a higher probability of completion within their deadlines. %In systems where task dropping is practiced, such as live video streaming, dropping tasks of a certain type can disrupt the system.% (such as if all resizing happened, but conversion between encoding formats was deemed too risky to be executed). 
%PAMF heuristic aims at mitigating such unfairness by favoring task types that have suffered from pruning.By adjusting the pruner's dropping and deferring thresholds for tasks of unfairly treated task types, the system can prevent bias against them. 
PAMF is designed to mitigate such unfair task pruning. In this mapping heuristic, thresholds (dropping and deferring) are adjusted for task types unfairly treated. 

We define a \emph{sufferage value} at mapping event $e$ for each task type $f$, denoted $\epsilon_{ef}$, that determines how much to decrease (\ie relax) the base pruning threshold. Note that we define 0 as no sufferage. We define \emph{fairness factor} (denoted $\vartheta$) as a constant value across all task types in a given HC system by which we change sufferage value of task types. This fairness factor denotes how quickly any task's sufferage value changes in response to missing a deadline. A high factor results in large relaxation of probabilistic requirements. Updating the sufferage value occurs upon completion of a task in the system. 

A successful completion of a task of type $f$ in mapping event $e$ results in lowering the sufferage value of task type $f$ by the fairness factor, \ie $\epsilon_{ef}=\epsilon_{(e-1)f}-\vartheta$, whereas for an unsuccessful task we add the fairness factor, \ie $\epsilon_{ef}=\epsilon_{(e-1)f}+\vartheta$.
Note that we limit sufferage values ($\epsilon_{ef}$) to be between 0 to 100\%. The mapping heuristic determines the fair pruning threshold for a given task type $f$ at a mapping event $e$ by subtracting the sufferage value from the base pruning threshold. 

This updated pruning threshold enables PAMF creates a more fair distribution of completed tasks by protecting tasks of unfairly-treated types from pruning. Once we update pruning thresholds for suffered task types, the rest of PAMF functions as PAM. 
%%%%%%%%%%%%%%%%%%%%%5555
%\section{Practicality of Probabilistic Mapping Heuristics}
\section{Practicality of the Pruning Mechanism}
\label{sec:approximation}

One concern when considering the deployment of probabilistic approaches in the mapping of tasks is the extra computational overhead. Repeated convolutions put strains on the machine that handle task mapping, especially when tasks are small and come in large numbers.
To ensure a probabilistic task pruning mechanism and PAM are real-world practical, this section describes some techniques that can be used to mitigate the scheduling overhead. 

\begin{figure} [ht]
  \centering
  \includegraphics[width=0.99\textwidth]{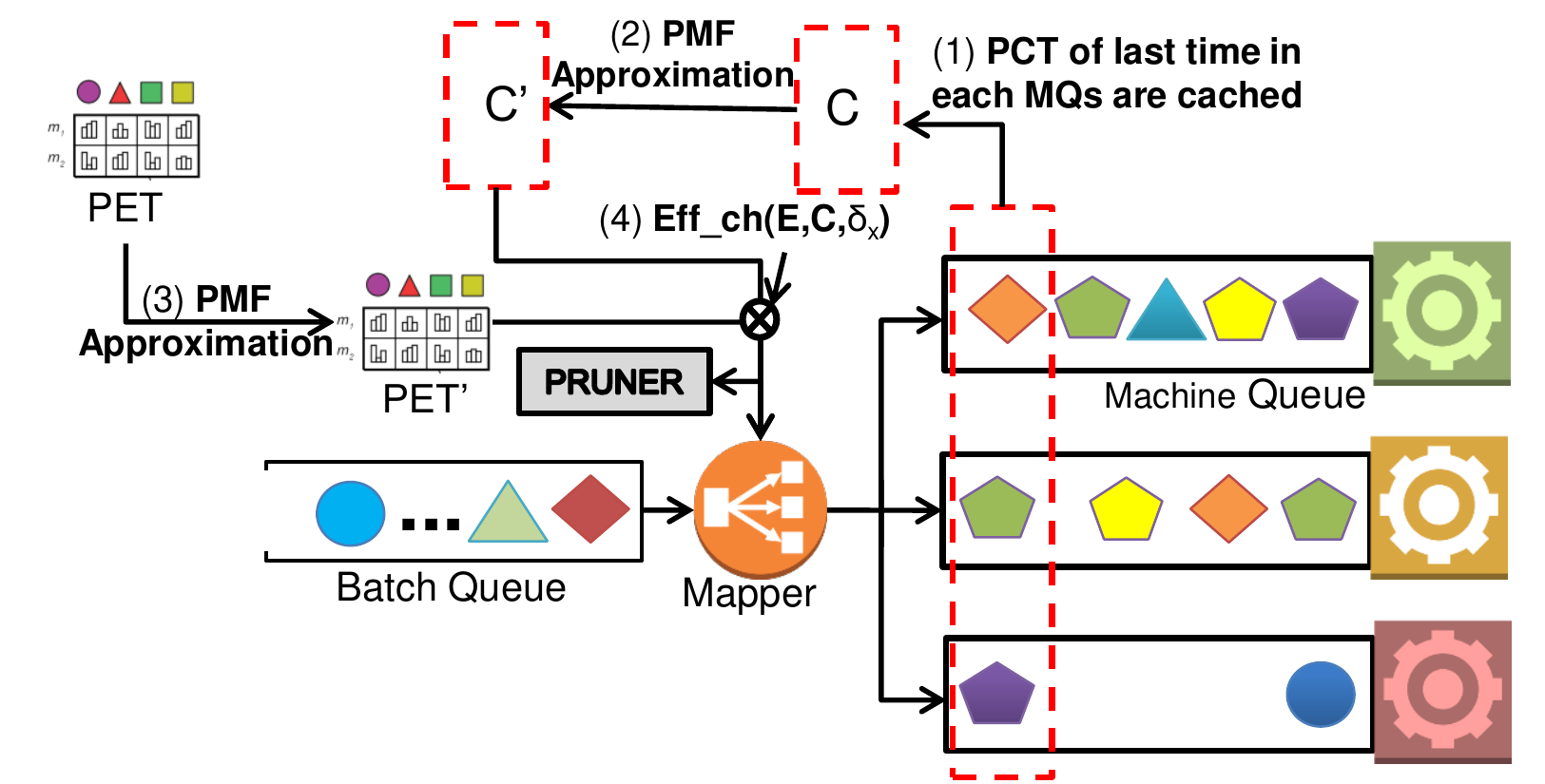}
  \caption{ Overview of optimization strategies, (1) PCT of last task in the machine queue is predetermined before mapping event, (2) and (3) perform PMF compaction (approximation) of last task in machine queue's PCT and arrival task's PET respectively, (4) Chance of success can be calculated by an algorithm with memoization without a complete convolution. } \label{fig:optimize}
\end{figure}

\subsection{Macro-level memoization to reduce redundant calculations.}~
During the first and second phase of PAM (and most probabilistic mapping heuristics), each and every unmapped task's execution time PMF (PET) are convolved with the PCT of the last tasks of each machine queue. Finding PCT is a chain convolution process that starts from the head of the machine queue. Supposing $N$ tasks in each of the $M$ machine queues, to find the PCT of $B$ unmapped tasks, there can be up to $B \cdotp N \cdotp M$ convolutions. Therefore it is recommended to cache the PCT of the last task in each machine queue before the mapping event to remove the repetitive convolutions on the machine queue (this is shown as step (1) in Figure~\ref{fig:optimize}). This reduces the number of convolutions to $B \cdotp M + N \cdotp M $ where $N \cdotp M$ part is cached at the beginning of the mapping event. Note that this caching is only valid for a single mapping event. Once the current time passes, this cache is no longer valid.

Once the PCT of the last task in the machine queue is calculated, it is also possible to perform PMF approximation on the PCT. Which will be explained in the next part.
%with maximum time set as the longest deadline of all tasks in the batch queue to further reduce the scheduling overhead

\subsection{Approximation to reduce convolutions overhead.}~
It is well known that the convolution process can impose a significant computation due to the sheer number of impulses that form the PCT after a chain of compound convolutions. Therefore, to alleviate scheduling overhead, some dynamic programming and approximating techniques are utilized to reduce the time spent in PMF convolution process. In this part, we first introduce a procedure to reduce the number of impulses. We then propose a procedure to replace the last step of the convolution process in a probabilistic mapping heuristic: convolving the PCT of the last task in machine queue and the PET of each unmapped task. Due to the size of PCTs convolved from the machine queue, a large number of unmapped tasks can impose a significant computational overhead, and warrants a customized algorithm.

\begin{figure} [ht]
  \centering
  \includegraphics[width=\textwidth]{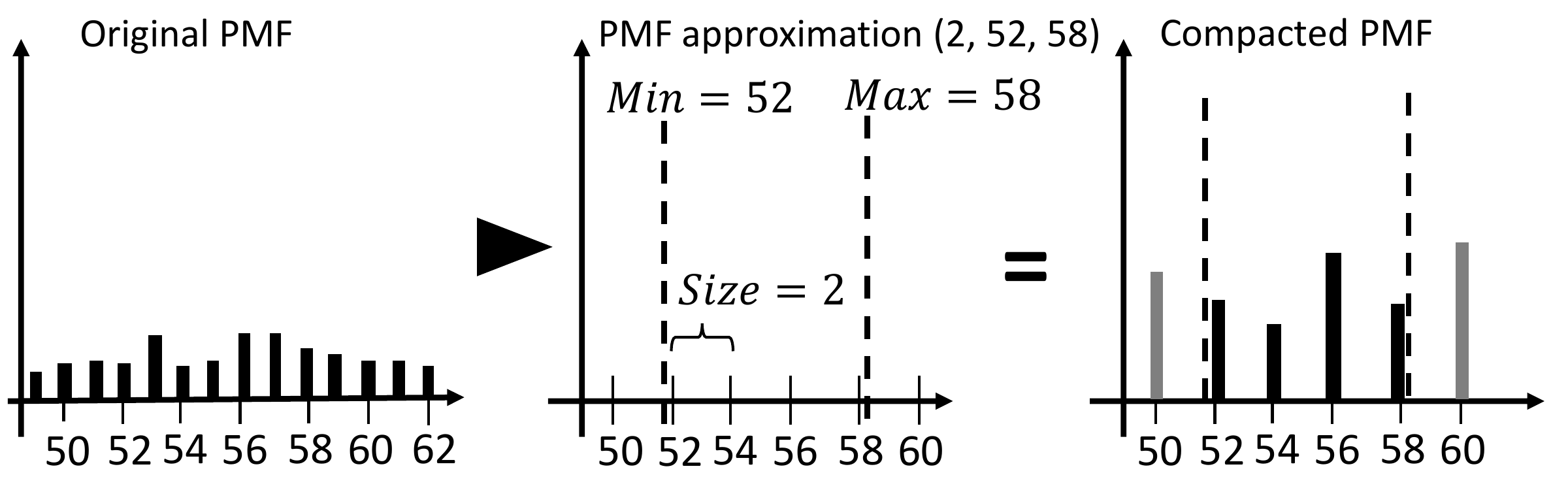}
  \caption{An example of impulse approximation process with bucket size of two and minimum and maximum range set to 52, and 58. Impulses are grouped and combined in 2 time unit interval in the specific range. All impulses that are more than specified max or less than specified min are combined together. } \label{fig:compaction}
\end{figure}

\paragraph*{\textbf{PMF approximation:}}
Convolution process time relates directly to the number of impulses in the PMF. In the case that the PMF is too finely detailed, convolution can be a burden with the calculation of many small impulses. Due to high uncertainty in a heterogeneous computing system, the extra resolution may not yield significantly better decision making. We can therefore, in some calculations, use an approximate PMF which has lower number of impulses than a detailed original PMF. The approximate PMF can be created by combining multiple impulses in a specific range together as shown in Figure~\ref{fig:compaction}. In the approximation process, in the case that we know the range of minimum and maximum time impulses to keep in the distribution, the distribution can also be cropped to the specified range. An example of the case where the relevant range is known is when the impulses of last task in the machine queue's PCTs (step (2) in Figure~\ref{fig:optimize}) are being approximated. The maximum time can be set as the longest deadline of the entire unmapped task without effecting the chance of success measurement in the task mapping process.

\paragraph*{\textbf{Micro-level memoization to reduce convolution overhead:}}

Probabilistic mapping heuristics require calculations of unmapped tasks' chances of success to make mapping decisions. To find these probabilities in a straightforward way, we first compute each unmapped task's PCT (completion time PMF) by convolving the unmapped task's PET against the PCT of the last task of each machine queue (which can be memorized and approximated, as mentioned earlier). Then we measure the resulting PCT against the task's deadline to calculate the chance of success. The PCT of each unmapped task is only calculated to find the potential mapping probability and is not reused. Therefore, to speed up the process, instead of calculating an unmapped task's PCT before measuring its chance of success, we propose a process that directly calculates the success chance from two distributions (PET of the unmapped task and PCT of the last task in a machine queue) without creating a full PCT of the unmapped task first.

Assuming that both distributions' impulses can be iterate through in a sorted order from the earliest time to the latest time (\ie impulses are sorted by their time). The procedure virtually performs a partial convolution on only pairs of impulses that together represent the resulting time less than the specified deadline. And since the impulses are sorted in order, some partial results between the iterations ($memo_c$ in the algorithm) can be memorized.

\begin{algorithm}[H]
\SetKwInput{KwData}{Input}
 \KwData{ $E \leftarrow$ PET of an unmappped task $x$}
 \KwData{ $C \leftarrow$ PCT of the last task in the machine queue $j$}
 \KwData{ $\delta_x \leftarrow$ Deadline of task $x$}
 \KwResult{Chance of convolved distribution finish before deadline }
 $e \leftarrow$ First impulse of $E$\;
 $c \leftarrow$ First impulse of $C$\;
 $k \leftarrow e$\;
 $p_{xj} \leftarrow 0 $   \;
 $memo_c \leftarrow 0 $   \;
 \While{time($k$+1) $ < \delta_x $}{
  $k \leftarrow$ next impulse after $k$ \;
  }
  
  \While{$ k \neq e$}{
    \While{ time($c$)+time($k$) $< \delta_x$}{
  		$memo_c \leftarrow memo_c  $ + chance($c$) \;
  		$ c \leftarrow$ next impulse after $c$\;
  }
  $p_{xj} \leftarrow p_{xj}+ memo_c \cdotp$  chance($k$)\;
  $k \leftarrow$ previous impulse before $k$\;
  }
  
 Return $p_{xj}$\;
 \caption{Efficient Calculation of Probability of Success for Task $x$ on Machine $j$.}
 \label{ALG:combineChance}
\end{algorithm}

 Algorithm~\ref{ALG:combineChance} takes input as the distribution E (PET of an unmapped task $x$), distribution C (cached PCT of the last task in a machine queue $j$), and the deadline $\delta_x$ for measuring the chance of success. $chance(c)$ signifies the probability associated with the impulse $c$ in the distribution $C$, and $time(c)$ signifies the time value associated with the impulse $c$. Line 6-7 finds the last impulse of Distribution E that is less than $\delta_x$. Line 8 and 13 iterate back from the impulse found from line 7 back to the first impulse. Line 9 to 11 combine all chance from C's impulses that when pairing them with a specific impulse of distribution E from line 8 still provides a combined time of less than $\delta_x$. Finally, Line 12 sum the multiplication of impulse k from distribution E (line 8) and the combined chance from lines 9-11. Note that the $memo_c$ and impulse $c$ from line 10 and 11 are not reset on any iteration. The value always carry over from one iteration to the next. The final result is the chance of meeting the deadline $\delta_x$ as if distribution E and distribution C are convolved together.

\begin{figure} [ht]
  \centering
  \includegraphics[width=\textwidth]{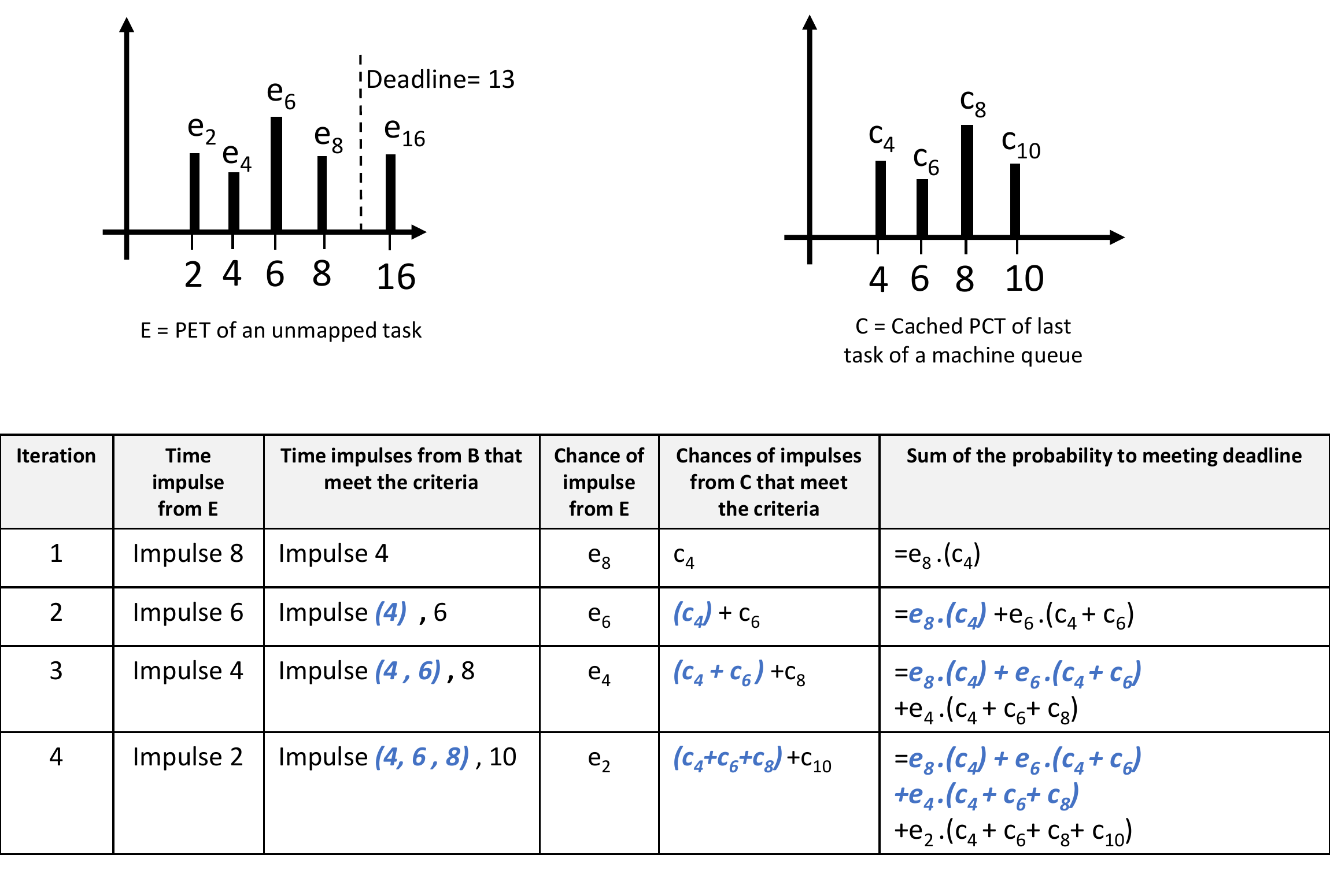}
  \caption{A simplified example of Procedure~\ref{ALG:combineChance}. E is the PET of an unmapped task, C is PCT of the last task in a machine queue. The table goes through each iteration from start to finish. Notions in \textit{\color{blue}{Italic}} are carried from their prior iteration. The carry over notion on the two right most column are stored as a scalar value denoted $memo_c$ and \textbf{$p_{xj}$} in the algorithm, respectively. 
  } \label{fig:graphicExplain}
\end{figure}

 A simplified example of the Algorithm~\ref{ALG:combineChance} is provided in Figure~\ref{fig:graphicExplain}. In this example, task x's deadline is at the time 13, the procedure runs through distribution E and C in 4 iterations where it considers one of E's impulse per iteration. Note that the impulse that has the time 16 is ignored as it is greater than the deadline. During each iteration, it considers C's impulses that can combine with the targeted e impulse and still meeting the deadline. Some partial results are carried over from prior iterations. The right most column is the $p_{xj}$ value after each iteration. And the bottom right most cell contains all the values that constitute the chance of meeting the deadline as if distribution E and distribution C are convolved together.

 Supposing distribution E has $p$ impulses and distribution C has $r$ impulses, the straightforward convolution requires at least $p \cdotp r$ multiplications. Measuring the chance of success also requires another run through combined impulses. Algorithm~\ref{ALG:combineChance} is a combination of the two process together. And it loops through distribution E at most twice and distribution C at most once (rather than $p$ times). The multiplication happens at most $p$ time. Hence we reduce the time complexity from $p \cdotp r$ to $2\cdotp p + r$, significantly speeding up the measurement of the chance of success in probabilistic-based mapping heuristics.

    \section{Experimental of Pruning Mechanism on Commonly Utilized Mapping Heuristics}
%HCW
%\subsection{Overview.}~
%In this part, we first 

%\subsection{Experimental Setup.}~
To evaluate the impact of pruning mechanism on a variety of widely-used mapping heuristics, we conducted a simulation study under various configurations of heterogeneous (in both immediate- and batch-modes) and homogeneous computing systems. For the experiments, Pruning Configurations are set to use Pruning Threshold of 50\% and Fairness factor of 0.05, unless otherwise stated. To accurately analyze the impact of dropping and deferring, we evaluate them both individually and together. 

For each set of experiments, 30 workload trials were performed using different task arrival times built from the same arrival rate and pattern. In each case, the mean and 95\% confidence interval of the results are reported. The experiments were performed using the Louisiana Optical Network Infrastructure (LONI) Queen Bee 2 HPC system~\cite{LONI}.  

%%%%%%%% text below are added after comments to explain overhead
%
While the task completion time estimation involves multiple convolutions which impose calculation overhead, there are multiple implementation techniques that can minimize the overhead of repeated calculation, such as task grouping and memorization of partial results.
Moreover, all the task pruning decisions are made by a dedicated machine which reserved for resource allocation. Therefore, pruning mechanism does not add extra overhead to each HC resources in our experiments.

%As immediate-mode mapping heuristics do not have ability to perform task deferring, they are tested for impact of probabilistic dropping only. 
%When experiment on Dropping toggles, Dropping toggle set to 0 means task dropping procedure is always triggered and Dropping toggle set to 1 means task dropping is triggered if the system missing at least one task since the last mapping event.

\subsection{Workload generation.}~
Twelve SPECint benchmarks were run numerous times on a set of eight machines which were used to generate probabilistic execution time (PET) PMFs~\cite{salehi2016stochastic}. The PMFs were generated by creating a histogram on a sampling of 500 points from a Gamma distribution formed using one of the means, and a shape randomly chosen from the range [1:20]. This was done for each of the twelve benchmarks, on each of the eight machines\footnote{The 8 machines are: Dell Precision 380 3 GHz Pentium Extreme, Apple iMac 2 GHz Intel Core Duo, Apple XServe 2 GHz Intel Core Duo, IBM System X 3455 AMD Opteron 2347, Shuttle SN25P AMD Athlon 64 FX-60, IBM System P 570 4.7 GHz, SunFire 3800, and IBM BladeCenter HS21XM.}, resulting in the eight by twelve machine type to task type PET matrix. The PET matrix remains constant across all of our experiments. 

In each experiment, a determined number of tasks per time unit is fed to the system within a finite time span. For each experiment, the system starts and ends in an idle state. As such, The first and last 100 tasks in each workload trial are removed from the data to focus the results on the portion of the time span where the system is oversubscribed.

 \begin{figure} [ht]
  \centering
  \includegraphics[width=0.7\textwidth]{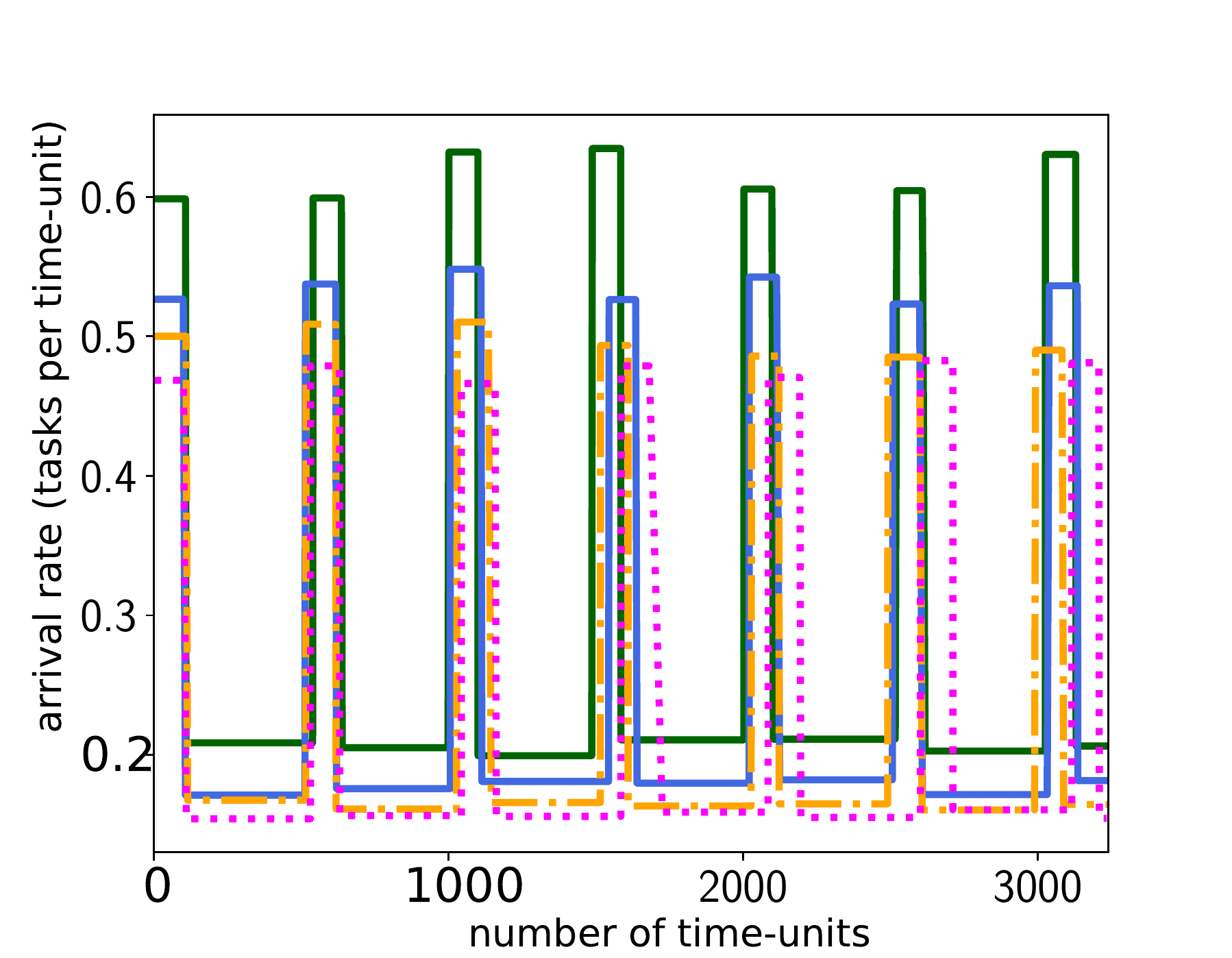}
  \caption{\small{Spiky task arrival pattern used in the experiments. Each color represents one task type. For better presentation, only four task types are shown. The Vertical axis shows the task arrival rate and horizontal axis shows the time span.} \label{fig:arrivalPattern} }
  \vspace{-7px}
\end{figure}

   \begin{figure*}[ht!]
    \centering 
    \subfloat[\small{Immediate-mode mapping heuristics}]{{\includegraphics[width=0.45\textwidth]{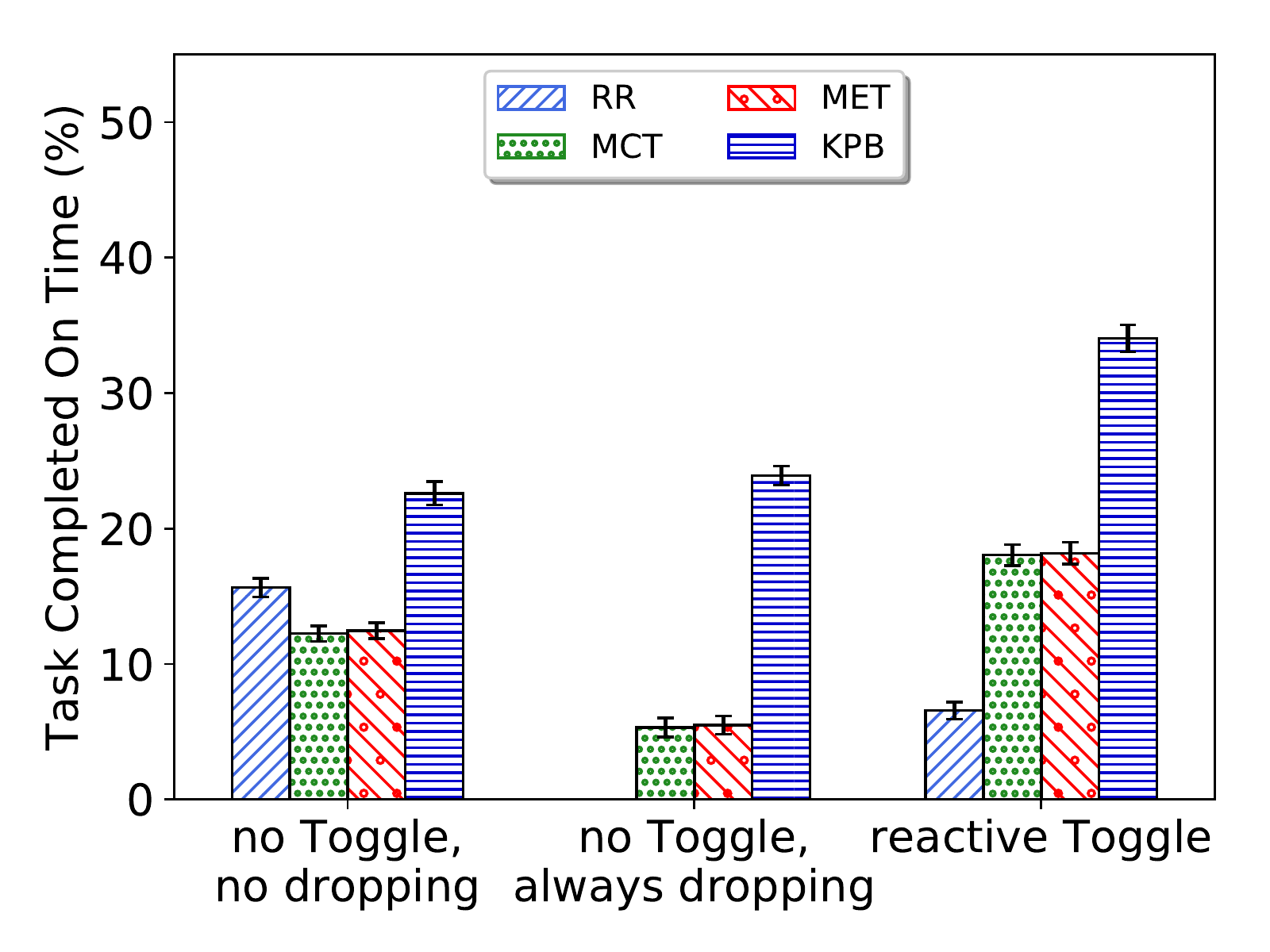} }%\label{fig:BatchSchedule}
    \label{fig:arrivalImmediate}
    }
    \subfloat[\small{Batch-mode mapping heuristics}]{{\includegraphics[width=0.45\textwidth]{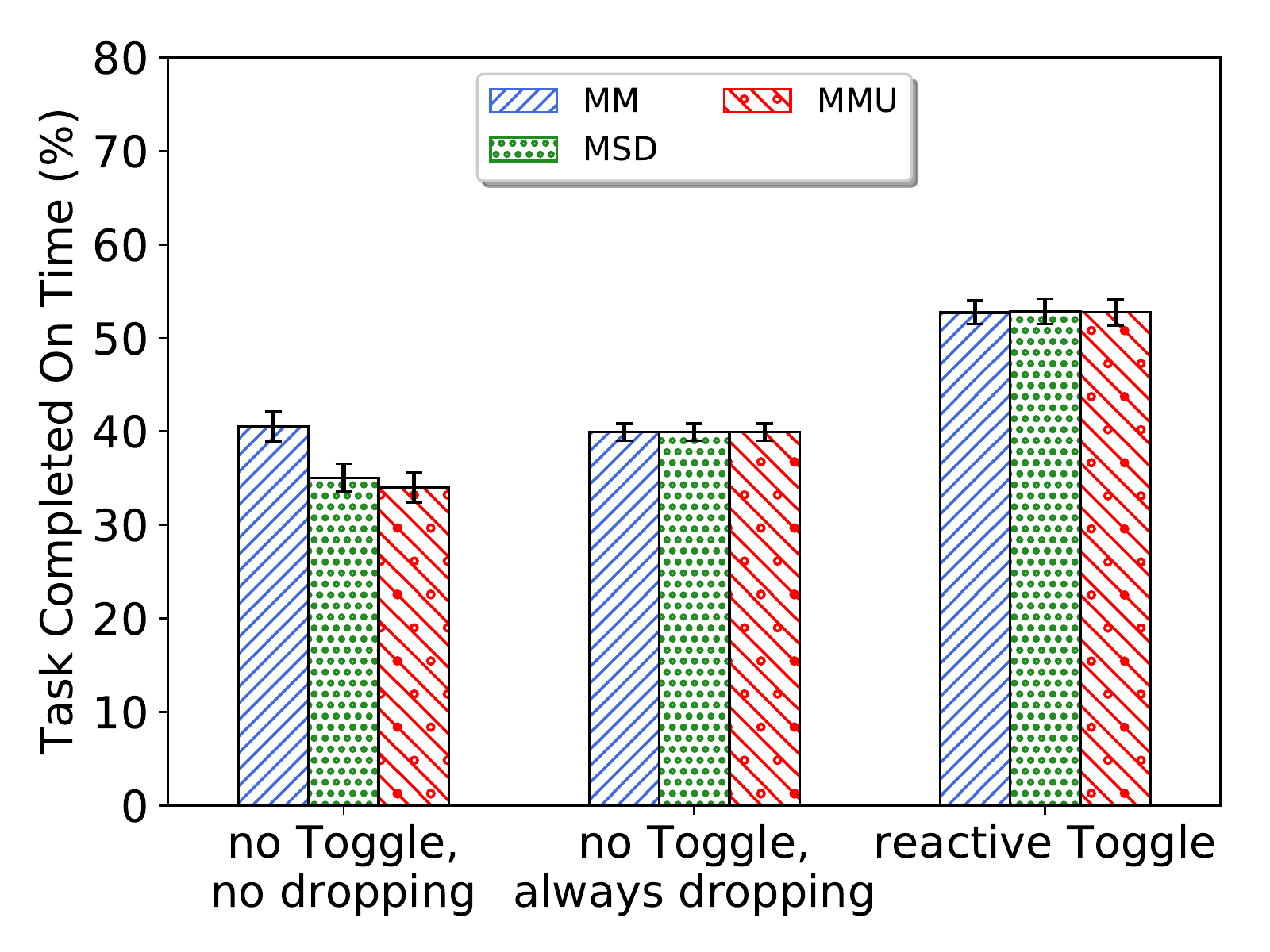} }  
    \label{fig:dropBatch}
	}
    \caption{\small{Impact of employing the Toggle module in a pruning mechanism works with immediate- and batch-mode heuristics. Horizontal axis shows how task dropping is engaged in reaction to oversubscription and vertical axis shows percentage of tasks completed on time. }}  
    \label{fig:toggle}%
    \vspace{-9px}
\end{figure*}

To conduct a comprehensive evaluation, two sets of workload were examined: \textbf{(A)} \emph{Constant rate arrival pattern}: a Gamma distribution is created with a mean arrival rate for all task types. The variance of this distribution is 10\% of the mean. Each task type's mean arrival rate is generated by dividing the number of time units by the estimated number of tasks of that type. A list of tasks with attendant types, arrivals times, and deadlines is generated by sampling from each task type's distribution. \textbf{(B)} \emph{Variable rate (spiky) arrival pattern}: In this case, tasks arrive with variable rates, as shown in Figure~\ref{fig:arrivalPattern}, to mimic arrival patterns observed in HC systems (\eg~\cite{miranda}). The spike times were determined uniformly, from the constant arrival workload, by dividing the workload time span to the number of spikes we want to create. During each spike, task arrival rate rises up to three times more than the base (lull) period. Each spike lasts for one third of the lull period. Since the spiky arrival pattern is frequently observed in real systems, it is our default workload arrival pattern in the experiments. 
For each task, as noted in Equation~\ref{eq:dl}, the deadline is calculated by adding the mean duration for that task type (\(avg_i\)) to the arrival time (\(arr_i\)), and then adding in a slack period based on the mean of all task type's duration multiplied by a tuning parameter (\(\beta \cdotp avg_{all}\)). This slack allows for the tasks to have a chance of completion in an oversubscribed system. In the workload trials, the value of $\beta$ of each task is randomly chosen from the range of $[0.8 , 2.5]$. 

\begin{equation}\label{eq:dl}
    \vspace{-3px}
\delta_i = arr_i + avg_i + (\beta \cdotp avg_{all})
\end{equation}

We carried out experiments under a variety of task arrival rates (oversubscription levels), however, the default rate used for plotting graphs includes 15K tasks that represents a moderately oversubscribed system. All the workload trials are publicly available from \url{git.io/fhSZW} for reproducing purposes.

%The task types each have a mean duration, from which the PET described above is derived. This mean duration is used in generating the deadlines for the tasks that arrive in the system. For a givexcen task, as noted in Equation~\ref{eq:dl}, the deadline is calculated by adding the mean duration for that task type (\(avg_i\)) to the arrival time (\(arr_i\)), and then adding in a slack period based on the mean of all task type's duration multiplied by a tuning parameter (\(\beta \cdotp avg_{all}\)). This slack allows for the tasks to have a chance of completion in an oversubscribed system.

\subsection{Impact of toggle reacting to oversubscription in HC systems.}~
In this experiment, our goal is to evaluate the impact of Toggle module within the pruning mechanism. Recall that the Toggle module is in charge of triggering task dropping operation. As such, we evaluate three scenarios: First, when there is no Toggle module in place and dropping operation is never engaged (referred to as ``no Toggle, no dropping"); Second, when Toggle module is not in place and task dropping is always engaged (referred to as ``no Toggle, always dropping"); Third, when the Toggle module is in place and is aware of (\ie reactive to) oversubscription (referred to as ``reactive Toggle"). In this case, the Toggle module engages task dropping only in observation of at least one task missing its deadline, since the previous mapping event. 

Figure~\ref{fig:arrivalImmediate} shows the results for the immediate-mode mapping heuristics and Figure~\ref{fig:dropBatch} shows them for the batch-mode. In both cases, we can observe that when Toggle functions in reaction to oversubscription, the overall system robustness is improved, regardless of the mapping heuristic deployed. The only exception is RR immediate-mode heuristic. The reason is that RR does not take execution time or completion time into account and it continuously maps tasks with a relatively low chance of success. These mapped tasks are subjected to be removed by task dropping. Without probabilistic task dropping, some of those low-chance tasks can complete on time.
%These mapped tasks with low chance of success also cause frequent miss of deadlines and, therefore, frequent engagement of task dropping.
 We can also observe that in immediate-mode, KPB provides the highest robustness (percentage of tasks completing on time) and also benefits the most from task dropping. This is because it makes more informed mapping decisions after dropping underperforming tasks.
%Another observation is that dropping mechanism remedy poor decision of MMU and MSD, making their robustness catch up with the level of MM.

The experiment testifies that our hypothesis in removing tasks with low chance of success in favor of  other tasks is true and can significantly improves robustness---by up to 12\% in immediate-mode and 19\% in batch-mode. %%%%%

 %%%%%%%%%% defer
\begin{figure}[ht]
  \centering
  \includegraphics[width=0.7\textwidth]{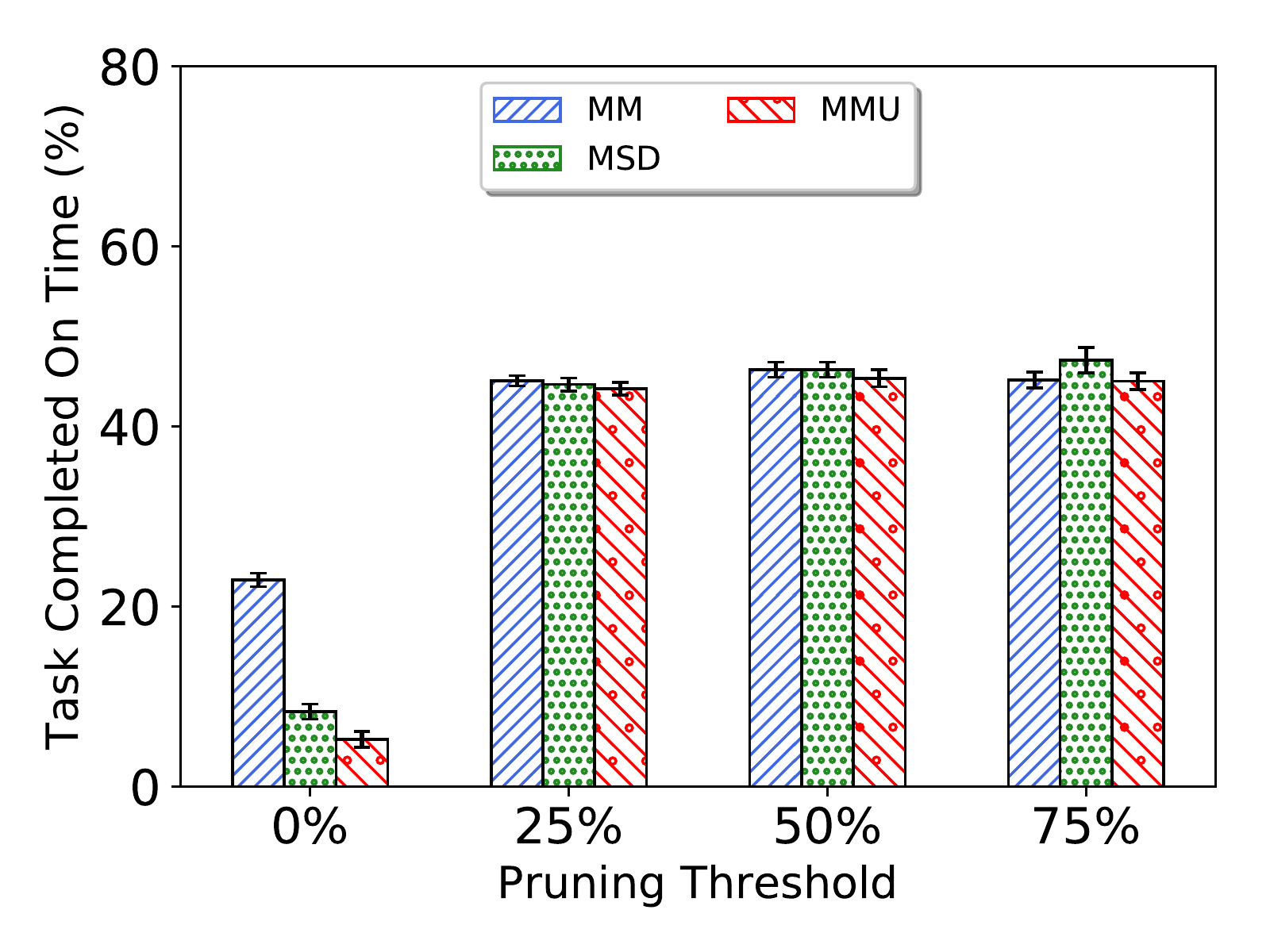}
  \caption{\small{Impact of tasks deferring on batch-mode mapping heuristics in an HC System with workload intensity of 25K. The Vertical axis shows percentage of tasks completed on time. The horizontal axis is the minimum success probability needed for each task to be mapped.   } \label{fig:deferBatch} }
  \vspace{-15px}
\end{figure}
%%%%%%%%%% prune 
   \begin{figure*}[hb!] %
    \subfloat[\small{Constant Arrival Pattern}]{{\includegraphics[width=0.45\textwidth]{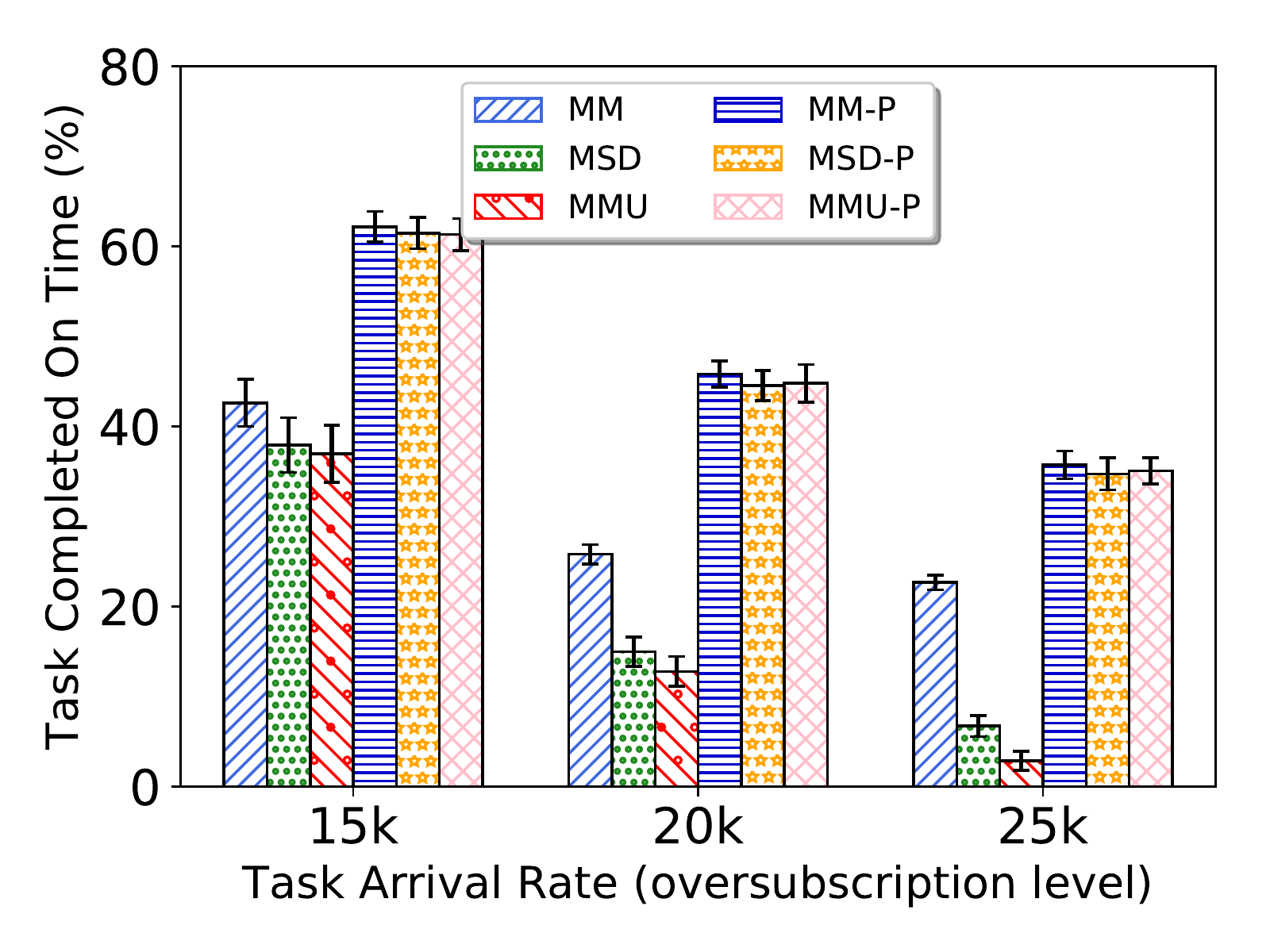} }  
    \label{fig:pruningBatch_c}
	}
	\centering 
    \subfloat[\small{Spiky Arrival Pattern}]{{\includegraphics[width=0.45\textwidth]{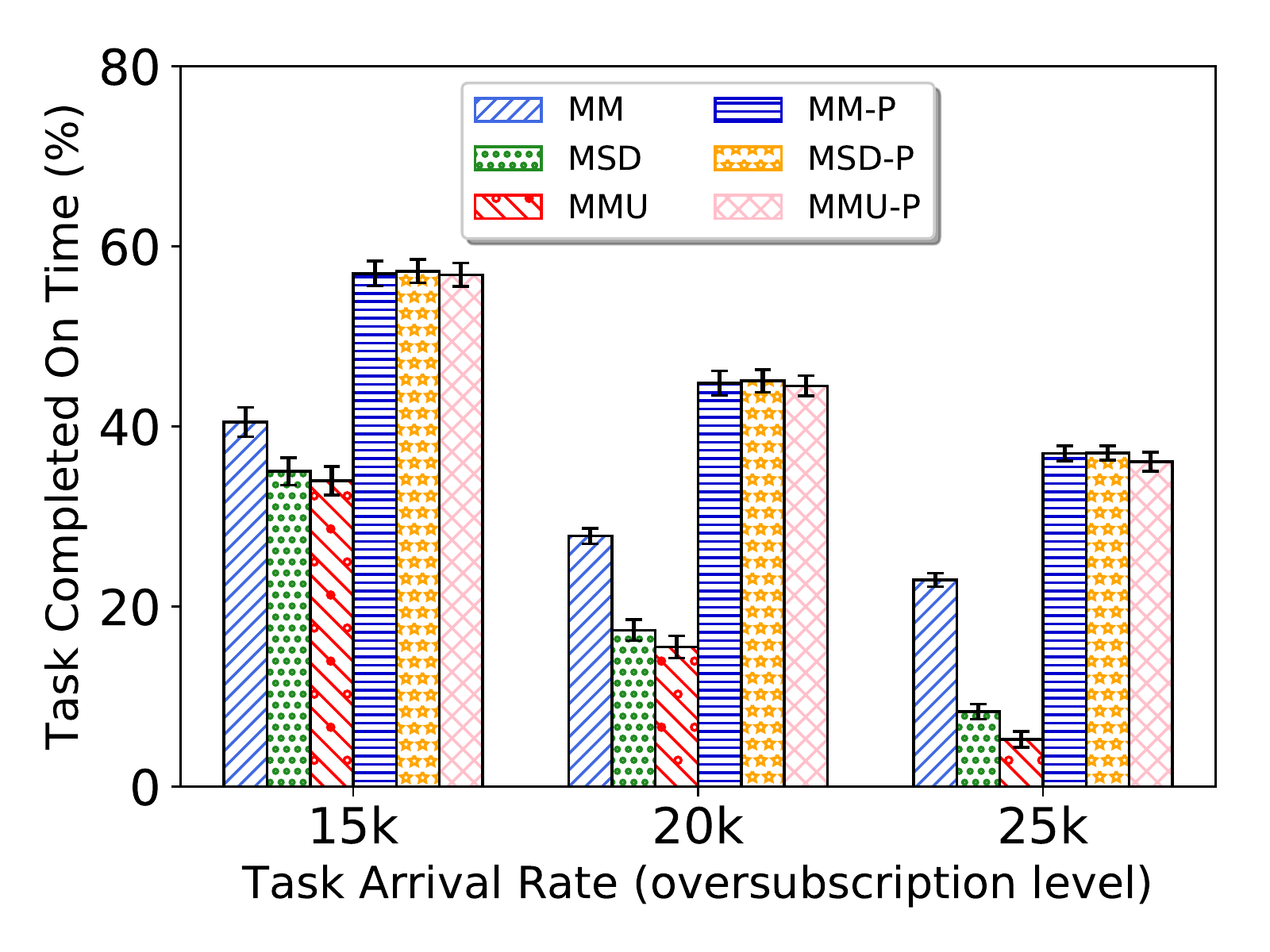} }%\label{fig:BatchSchedule}
    \label{fig:pruningBatch_s}
    }
    \caption{\small{Impact of pruning mechanism on batch-mode heuristics in HC systems. Horizontal axes show the number of tasks arriving within a time unit (\ie oversubscription level). In the legend, ``-P" denotes heuristics use pruning mechanism.
    }}   %of either consistence arrival pattern or spiky arrival pattern  
    \label{fig:pruningBatch}%
\end{figure*}

%%%%%%%%%% prune homo
   \begin{figure*}[ht!]%
    \centering 
    \subfloat[\small{Constant Arrival Pattern}]{{\includegraphics[width=0.45\textwidth]{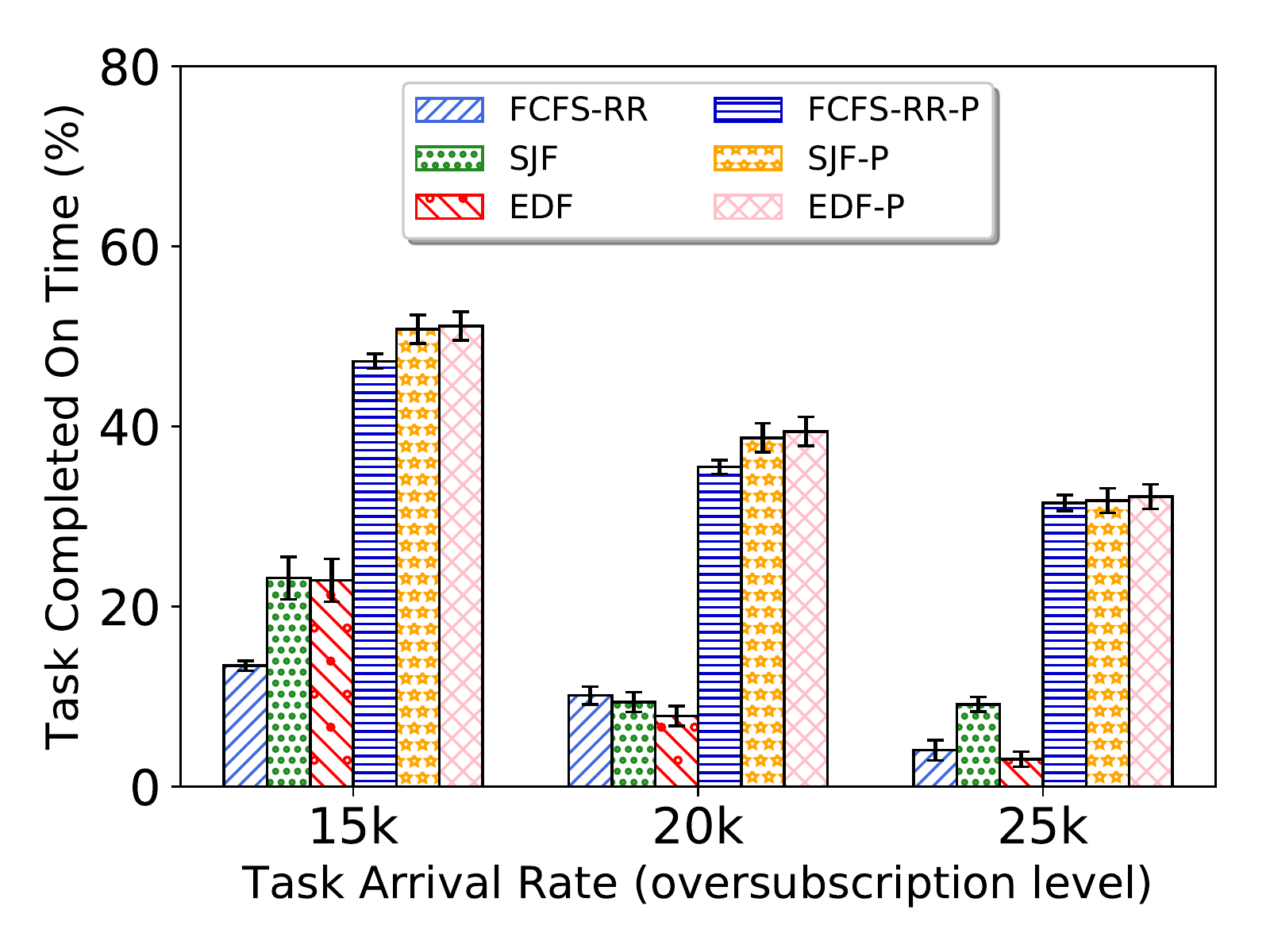} }  
    \label{fig:arrivalHomo_c}
	}
    \subfloat[\small{Spiky Arrival Pattern}]{{\includegraphics[width=0.45\textwidth]{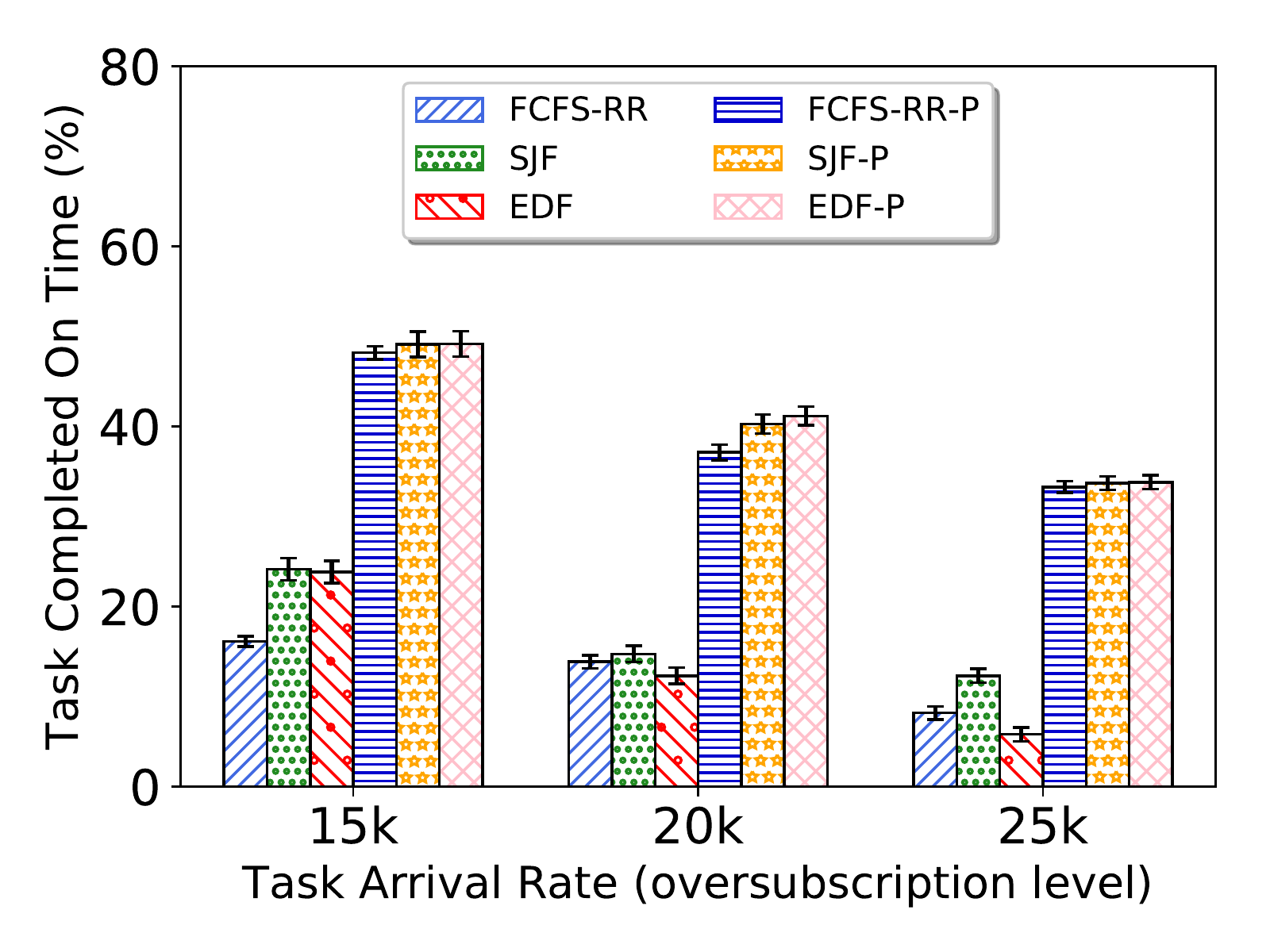} }%\label{fig:BatchSchedule}
    \label{fig:arrivalHomo_s}
    }
    \caption{\small{Impact of pruning mechanism on mapping heuristics of homogeneous systems. Horizontal axes show the number of tasks arriving within a time unit (\ie oversubscription level). In the legend, ``-P" denotes heuristics use pruning mechanism.
    }}    
    \label{fig:arrivalHomo}%
    \vspace{-7px}
\end{figure*}

\subsection{Impact of task deferring on batch-mode heuristics.}~
In this experiment, we evaluate the impact of task deferment within the pruning mechanism. As deferring operation works on the arrival (batch) queue, it can only be enabled for batch-mode heuristics. We conducted the experiment for task pruning threshold set to 0\% (no task pruning), 25\% , 50\%, and 75\%. As the results of this experiment is more prominent under high level of oversubscription, we set the task arrival to 25K tasks in the workload trials.

Figure \ref{fig:deferBatch} shows that, without task deferring (\ie when Pruning Threshold is zero), MM, MSD, and MMU's robustness are the lowest (between 5\% to 23\%). However, as task deferring is employed, all mapping heuristics can attain more than 44\% robustness. This is because pruning mechanism delays mapping of tasks with low chance of success until a more suitable machine becomes available. Hence, machines are utilized only to execute promising tasks, thereby increasing the robustness. 
%removed 'system' in front of robustness to save new line
 Our observation also implies that, for these widely used batch-mode mapping heuristics, by limiting the selection-pool of a mapping-heuristic to likely-to-succeed tasks, task deferring can reduce the performance differences of the heuristics to offer similar robustness, regardless of their algorithmic logic.

%Our observation also implies that, at least for widely used batch-mode mapping heuristics, task deferring can forgive their poor mapping decisions and push them to offer similar robustness, regardless of their algorithmic logic. 

In Figure \ref{fig:deferBatch}, we can see that, in all heuristics, the robustness does not improve for Pruning Thresholds higher than 50\%. In fact, a high Pruning Threshold makes the system conservative in allocating tasks and defers tasks whose completion can improve overall robustness. Therefore, setting Pruning Threshold to 50\% is a proper configuration for the pruning mechanism.

\subsection{Impact of pruning mechanism on batch-mode heuristics.}~
%% can remove 'holistically' or one of 'the' below to save a line...
In this experiment, our goal is to evaluate the impact of the pruning mechanism holistically under various oversubscription levels. We evaluated the system robustness when mapping heuristics are coupled with and without the pruning mechanism. The pruning mechanism is configured with Pruning Threshold of 50\% and Toggle is set to engage task dropping reactively. 

Figure~\ref{fig:pruningBatch} shows that, for all heuristics under both constant and spiky arrival pattern, pruning mechanism improves the robustness. 
Pruning mechanism makes the largest impact for MSD and MMU. These heuristics attempt to map tasks with short deadlines and, thus, low chance of success. By limiting these heuristics to map tasks whose chance is beyond a certain threshold, their overall system robustness is improved.

\subsection{Impact of pruning mechanism on homogeneous systems.}~
In addition to mapping heuristics for heterogeneous system, we also conduct experiments on homogeneous mapping heuristics to evaluate the impact of pruning mechanism. Pruning configurations are set to use reactive Toggle and Pruning Threshold of 50\%.
 
Figure~\ref{fig:arrivalHomo} shows that, in all levels of oversubscription, applying pruning mechanism to homogeneous systems significantly increases system robustness (by up to 28\%) for all mapping heuristics on both constant and spiky arrival pattern. 
Importantly, as the oversubscription level increases, the impact of pruning mechanism is more substantial. With 25K tasks arrival rate, in constant arrival pattern, EDF and SJF can only achieve 4\% and 10\% robustness, respectively. Coupling pruning mechanism into these heuristics raises both the robustness to more than 30\%. The reason is that, similar to heterogeneous systems, pruning mechanism allows the system to avoid mapping unlikely-to-succeed tasks, which appear more often under higher levels of oversubscription. 

Based on the observations of this experiment, we can conclude that, pruning mechanism works equally as well and provides as much benefit to homogeneous systems as to the heterogeneous systems.

%In this study, to reduce simulation execution times, the number of machines comprising the distributed systems is constrained, but, as the calculation of a machine's probability state can be performed at each machine, or by a control node for each subset of machines, the proposed methods are scalable to any number of machines. 
\section{Experimental of Pruning Mechanism with Specifically Designed Pruning-Aware Mapping Heuristics}
\label{sec:evltn}
%\subsection{Overview.}~
To conduct a comprehensive performance evaluation of specifically created Pruning-Aware Mapping heuristics (PAM and PAMF), we simulate a computing system with eight inconsistently heterogeneous machines (\ie $M=8$). %They comprise an inconsistently heterogeneous system where a given machine A can exhibit higher performance for certain task types than machine B, yet machine B may exhibit higher performance on other task types~\cite{Braun01}. 
To generate the probabilistic execution time PMFs (PET), the mean execution time results from twelve SPECint benchmarks on a set of eight bare-metal machines\footnote{The 8 machines are: Dell Precision 380 3 GHz Pentium Extreme, Apple iMac
2 GHz Intel Core Duo, Apple XServe 2 GHz Intel Core Duo, IBM System X 3455 AMD
Opteron 2347, Shuttle SN25P AMD Athlon 64 FX-60, IBM System P 570 4.7 GHz,
SunFire 3800, and IBM BladeCenter HS21XM.} were determined. These mean execution times for each benchmark on each system formed the mean values for our task-machine execution times. The function describing execution time of the tasks on a machine is assumed to be a unimodal distribution; from a gamma distribution using the task-machine mean execution time, and with a shape randomly picked from the range [1:20], 500 execution times were sampled. From these times, a histogram was generated to produce a discrete probability mass function (PMF). This was repeated for each task type on each machine, and the resultant eight machine by twelve task type matrix of PMFs was stored as the PET matrix which remains constant across all of our experiments. We note that other statistical methods can be explored to learn and tweak PMF distributions in an online manner.

\subsection{Workload generation.}~
Our simulation is of a finite span of time units, starting and ending in a state where the system is idle. As the system comes online, and tasks begin to accumulate in the queue, the system is not in the desired state of oversubscription. The same is true of the end of the simulation, when the last tasks are finishing, and no more are arriving to maintain the oversubscribed state. In an effort to minimize the effects of the non-oversubscribed portion of the simulation from the data, the first and last 100 tasks to complete are removed from the results. Only the remaining tasks from the oversubscribed portion of the simulation are used in the analysis.

Based on other workload investigations~\cite{khemka14utility,khemka2015utility}, a gamma distribution is created with a mean arrival rate for all task types that is synthesized by dividing the total number of arriving tasks by the number of task types. The variance of this distribution is 10\% of the mean. Each task type's mean arrival rate is generated by dividing the number of time units by the estimated number of tasks of that type. A list of tasks with attendant types, arrivals times, and deadlines is generated by sampling each task type's distribution.  

%The task types each have a mean execution time, from which the PET described above is derived. This mean execution time is used in generating the deadlines for the tasks that arrive in the system. 
Recall that we consider each task to have an individual hard deadline and it has to be dropped once the deadline is missed.
For a given task $i$, the deadline is calculated as $\delta_i = arr_i + avg_i + (\beta \cdotp avg_{all})$, where \(arr_i\) is the arrival time, \(avg_i\) is the mean execution time for that task type (range from 50 to 200 ms)
, $\beta$ is a slack coefficient, and $avg_{all}$ is the mean of all task type's execution. This slack allows for the tasks to have a chance of completion in an oversubscribed system. 

%In addition to our standard workload, section~\ref{subsec:Adaptive} also includes the customized workloads with tasks' deadline 30\% more urgent and 30\% less urgent than our standard workloads.

%%%%%%%%%%%%%%%%%%%%%%%55
%\section{Performance Evaluation}
%\subsection{Overview.}~
A series of simulations were run using the Louisiana Optical Network Infrastructure (LONI) Queen Bee 2 HPC system~\cite{LONI}. For each set of tests, for each examined parameter, 30 workload trials were performed using different task arrival times built from the same arrival rate and pattern, and the mean and 95\% confidence interval of the results is reported. The arrival rates are listed in terms of number of tasks per time unit.

Each experiment is a set of 30 workload trials, consisting of 1200 tasks per trial. Each of the experiments investigates high levels of oversubscription where few tasks complete successfully using baseline heuristics. Due to frequent task mapping events, each machine in the HC system has a machine-queue size of three, counting the executing task and the dropping toggle as one task. We also evaluated the system with the size of machine-queue equals to six, and the results were consistent with the presented ones. For each of the experiments, unless otherwise noted, the performance metric (and the vertical axis) is the percentage of tasks completed before their deadline (\ie overall robustness).

\begin{figure} [ht]
  \centering
  \includegraphics[width=0.7\textwidth]{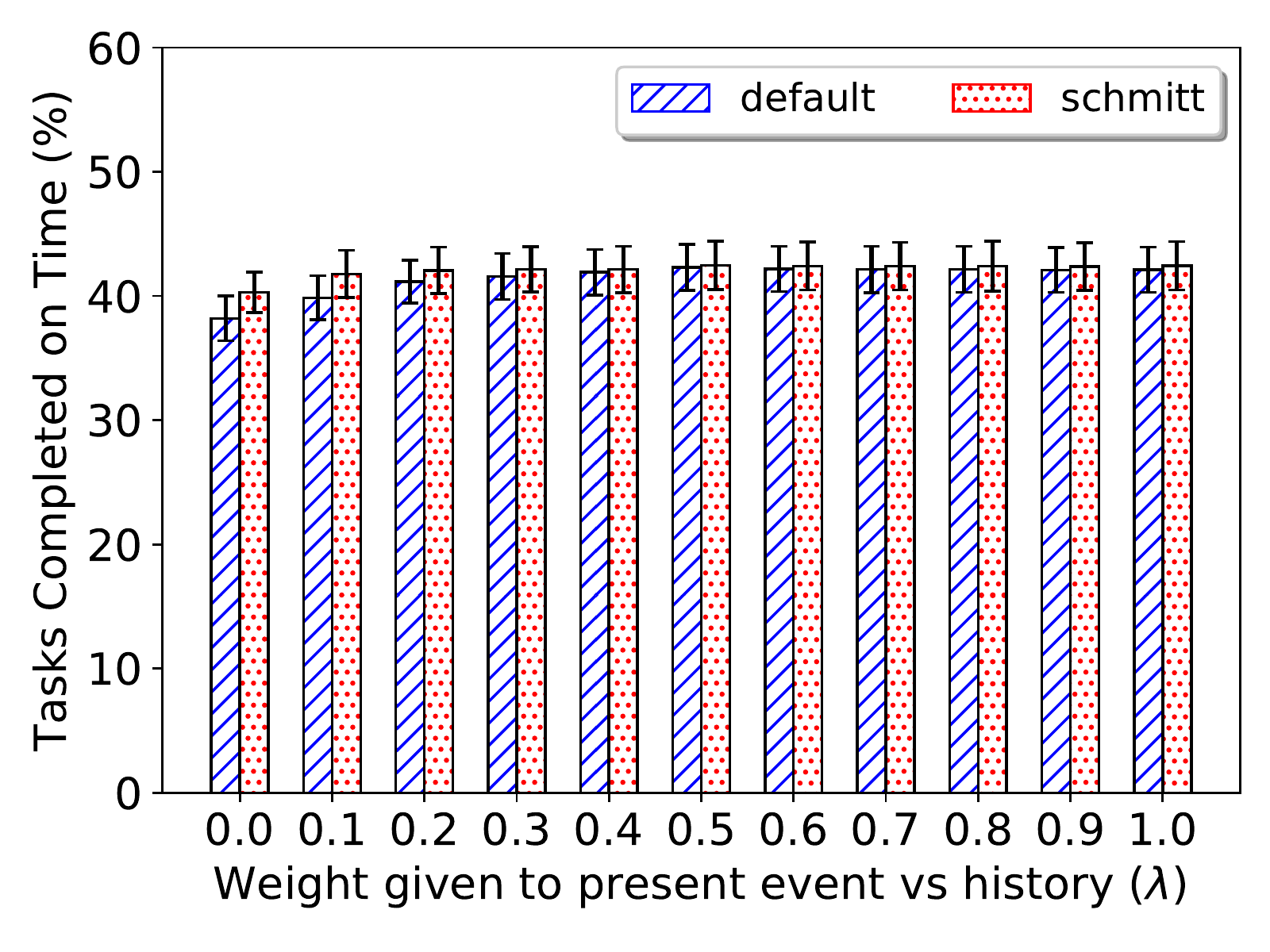}
  \caption{Impact of historical oversubscription observations and Schmitt Trigger on determining oversubscription level of HC system. Horizental axis represents the value of $\lambda$ coefficient in Equation~\ref{eq:oversub}.} \label{fig:schmitt} 
\end{figure}

\subsection{Dynamic engagement of probabilistic task dropping.}~\label{subsec:toggle}
In this experiment, our aim is to appropriately measure the oversubscription level (see Equation~\ref{eq:oversub}) by determining the weight that should be assigned to the number of deadlines missed in the recent mapping event versus the previous values of the oversubscription level. We also evaluate the impact of using Schmitt Trigger as opposed to using a single threshold for dynamic engagement of task dropping. This experiment was conducted under 25k tasks arriving to the system.

Figure~\ref{fig:schmitt} shows that by assigning a higher weight to the number of dropped tasks in the most recent mapping event, the overall robustness of the system is increased from 39.9\% to 42.5\%. This is due in part to the steady nature of task-arrival in our workload trials with only few sudden spikes. While the maximum robustness is reached with $\lambda=0.9$ with Schmitt Trigger, enabling Schmitt Trigger alone make the bigger difference than setting the $\lambda$ to an optimal value. The system robustness of $\lambda=1$ is close enough to the result with $\lambda=0.9$, while ignoring the history tracking altogether which can incur less scheduling overhead. %With Schitt Trigger, the result of little to no $\lambda $ value ($\lambda=0.1 or 0$) are not far off from the best possible case while incurs less scheduling overhead.
%TODO, add lambda=0.0 to the figure then change 0.1 to 0.0 above

We can conclude that under high oversubscription levels, the best results come from taking immediate action when tasks miss their deadlines, and then a steady application of probabilistic task dropping until the situation is decidedly controlled (\ie reaching the lower bound of Schmitt Trigger).

\subsection{Evaluating the mutual impacts of deferring and dropping thresholds.}~\label{subsec:thresh}
%As we discussed in , deferring threshold must be higher than the dropping threshold. 
The goal of this experiment is two-fold: First, it identifies the impact of choosing a proper initial dropping threshold; Second, it evaluates the  impact of deferring threshold on effectiveness of the dropping. 
For that purpose, we disable dynamic deferring threshold and set it statically. Note that, if the workload characteristics is known, it can be helpful to set deferring threshold statically to reduce the pruning overhead. 

\begin{figure} [ht]
  \centering
  \includegraphics[width=0.7\textwidth]{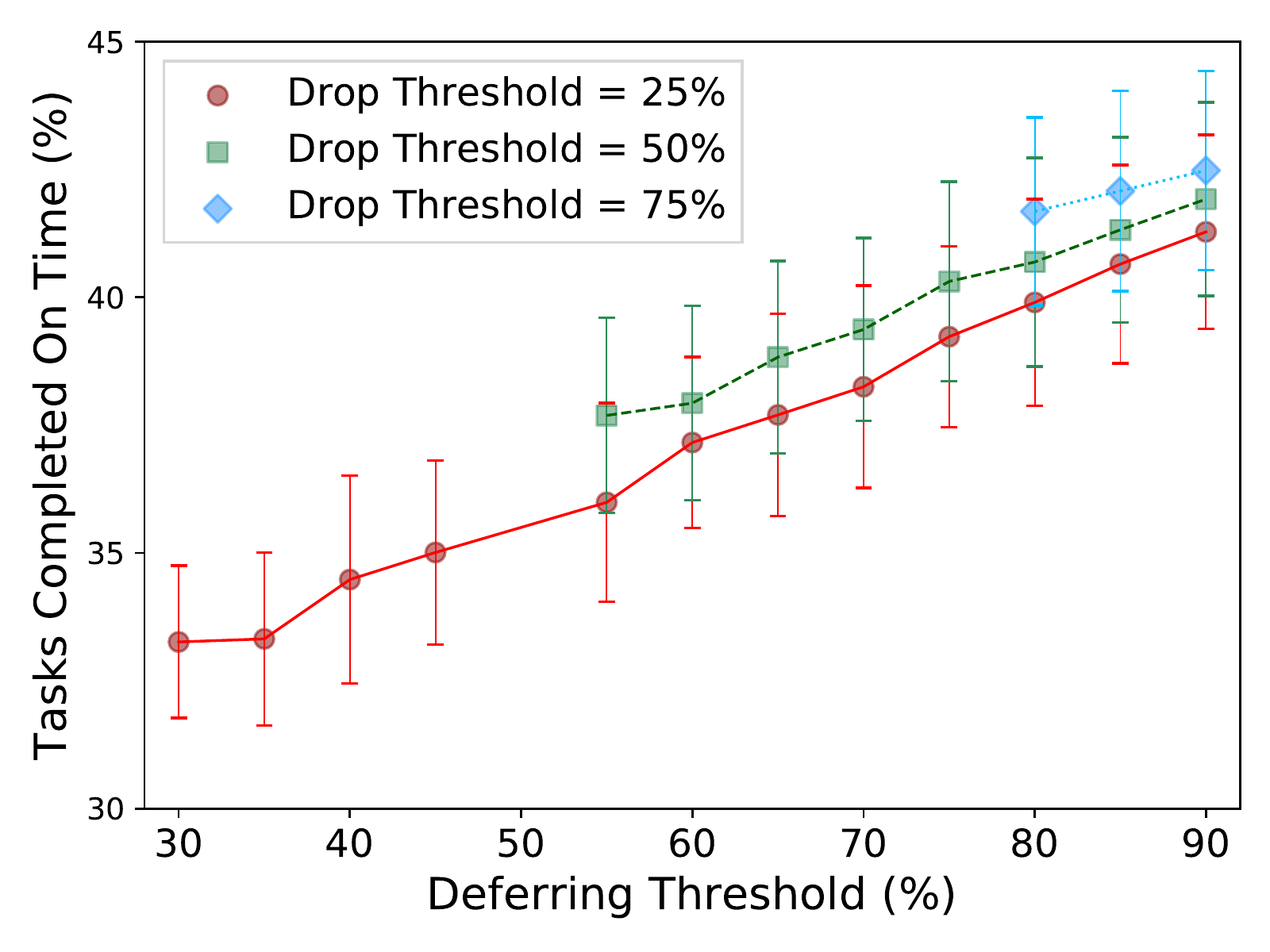}
  \caption{Impact of deferring and dropping thresholds on the system robustness. Dropping threshold is denoted by line type and color. } \label{fig:decouple} 
\end{figure}

A static deferring threshold has to be designated greater than the dropping threshold. Otherwise, a task can be dropped immediately, once it is mapped. Accordingly, to conduct this evaluation, we add a gap value to the initial dropping threshold (\eg a dropping threshold of 50\% would require at least 55\% robustness to map a task to a machine). Three dropping thresholds (25\%, 50\%, and 75\%) are examined and the 5\% gap is increased until the deferring threshold reaches 90\%. The results, shown in Figure~\ref{fig:decouple}, are generated from a workload with 25k tasks. %

Figure~\ref{fig:decouple} validates the experiment assumption, by showing that using a higher deferring threshold leads to higher system robustness. 
In addition, we observe that if the deferring threshold is chosen high enough, deferring operation prevails dropping and diminishes its influence on the system robustness. Specifically, if we choose deferring threshold at 90\%, we obtain a similar system robustness, regardless of the intitial dropping threshold value. It is noteworthy that a higher dropping threshold influences the incurred cost of using an HC system, because they prevent wasting time processing unlikely-to-succeed tasks that have been mapped to the system. Based on the experiment, in the rest of evaluations, initial dropping threshold 50\% is used.

%New for JPDC

    \begin{figure} [ht]
    \label{fig:adaptability} 
  \centering
  \begin{subfigure}[t]{0.49\textwidth}
 \includegraphics[width=\textwidth]{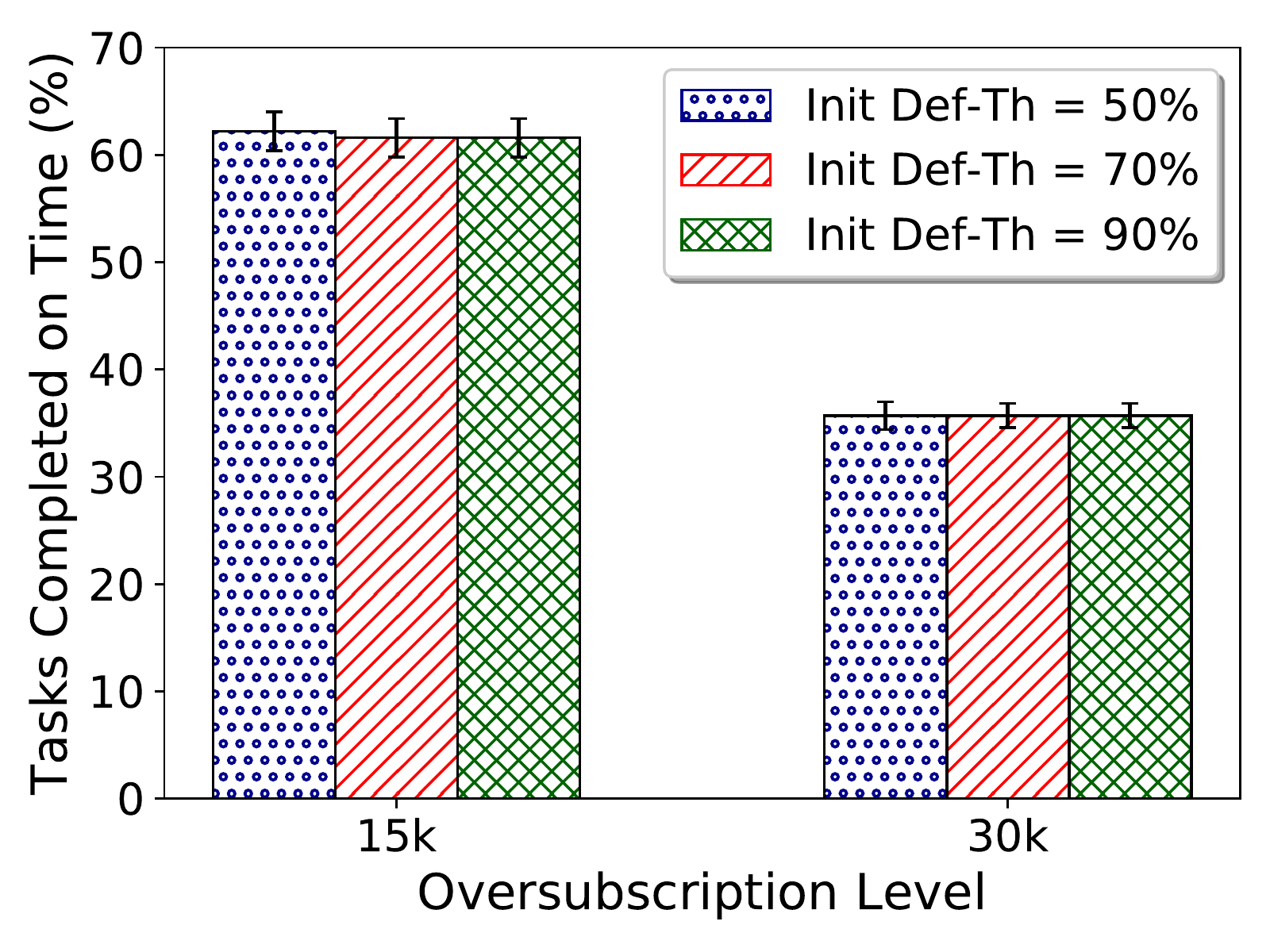}
  \caption{Impact of initial deferring thresholds}
  \label{fig:selfadj} 
  \end{subfigure}
  \begin{subfigure}[t]{0.49\textwidth}
  \includegraphics[width=\textwidth]{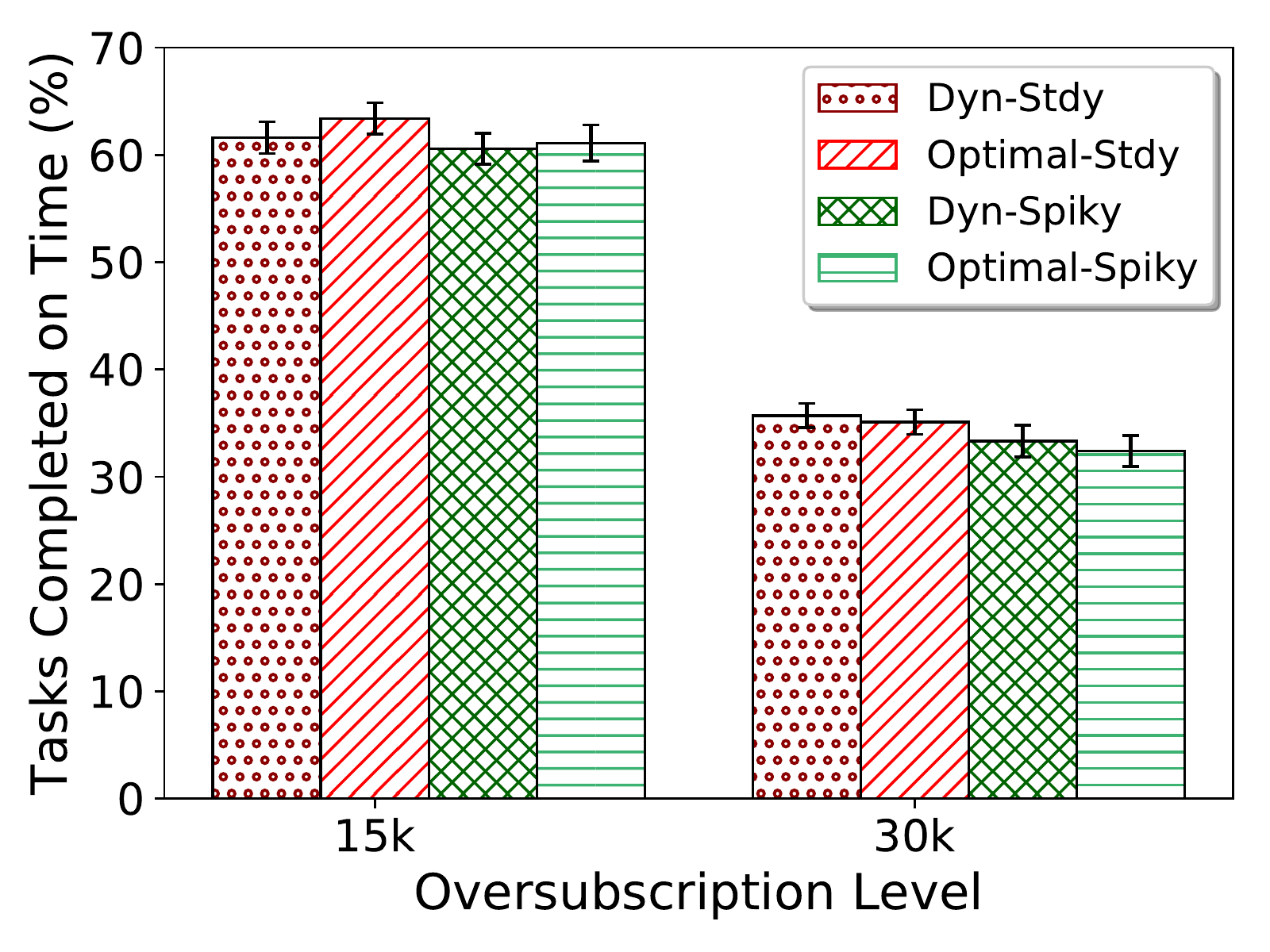}
  %\caption{Robustness of dynamic deferring Threshold (S-ADJ) vs Static optimal threshold (N-ADJ) on steady arrival rate workloads (STD-ARR) and bursty arrival rate workloads (SPK-ARR) }
\caption{Impact of pruning on steady and spiky workloads }          \label{fig:selfadj_workload} 
  \end{subfigure}
  
  \caption{Evaluating the impact of dynamic deferring probability on the system robustness on the system with oversubscription level of 15k and 30k}. (a) The effect of choosing different initial deferring thresholds on the system robustness. (b) Comparing dynamic deferring probability threshold (denoted with \emph{Dyn-} prefix) against the best statically-determined deferring threshold (denoted with \emph{Optimal-} prefix) for both steady (stdy) and spiky workloads.
\end{figure}

\subsection{Evaluating the impact of deferring on various types of workloads.}~
  \label{subsec:Adaptive} 
  In this experiment, our goal is to evaluate effectiveness of the dynamic deferring threshold adjustment in various scenarios. First, we examine if the initial value of dynamic deferring threshold matters for the ultimate system robustness. For that purpose, we vary the initial deferring probability threshold and study the system robustness using PAM heuristic. Specifically, we examined initial deferring thresholds (shown as \emph{Init Def-th}) to 50\%, 70\%, and 90\%. Results of the experiment in Figure~\ref{fig:selfadj} shows that, as the system adjusts the deferring probability threshold dynamically, the initial deferring threshold does not make a difference to the final system robustness values and they are nearly identical, regardless of the initial deferring threshold. 

Second, we compare the performance of PAM when it is geared to a pruning mechanism that uses dynamic deferring threshold against when the pruning mechanism is set to the best experimentally-found deferring threshold value. Note that, in the latter case, the deferring threshold is static and does not change throughout the experiment. Also, note that the deferring threshold is the best for the examined workload and the best value might be different for other workloads. To assure the applicability of the analysis to any workload, we study two types of arriving workloads: (A) Steady task arrival rate (shown as \emph{stdy} in the experiment); and (B) Varying arrival rate (shown as \emph{spiky} in the experiment). The varying arrival rate workload has the same number of total tasks arriving to the system as the steady one, but with burst task arrival periods. That is, within each time interval, the task arrival rate switches between on-peak (\ie high arrival rate) and off-peak (\ie low arrival rate) periods. In summary, by combining static and dynamic deferring threshold and steady or spiky workloads, we evaluate four cases, shown as \emph{dyn-stdy, best-stdy, dyn-spiky,} and \emph{best-spiky}, in Figure~\ref{fig:selfadj_workload}.

Figure~\ref{fig:selfadj_workload} expresses that, in both steady and spiky workloads, the dynamic threshold provides almost the same robustness as to the best-known static deferring threshold. In addition, comparison between steady and spiky workload reveals that the pruning mechanism does not suffer significantly from the uncertainties in task arrival rate. That is, the system shows to be robust against the uncertainties in task arrival rate.

\subsection{Evaluating the impact of fairness factor.}
\begin{figure} [ht]

  \centering
  \includegraphics[width=0.7\textwidth]{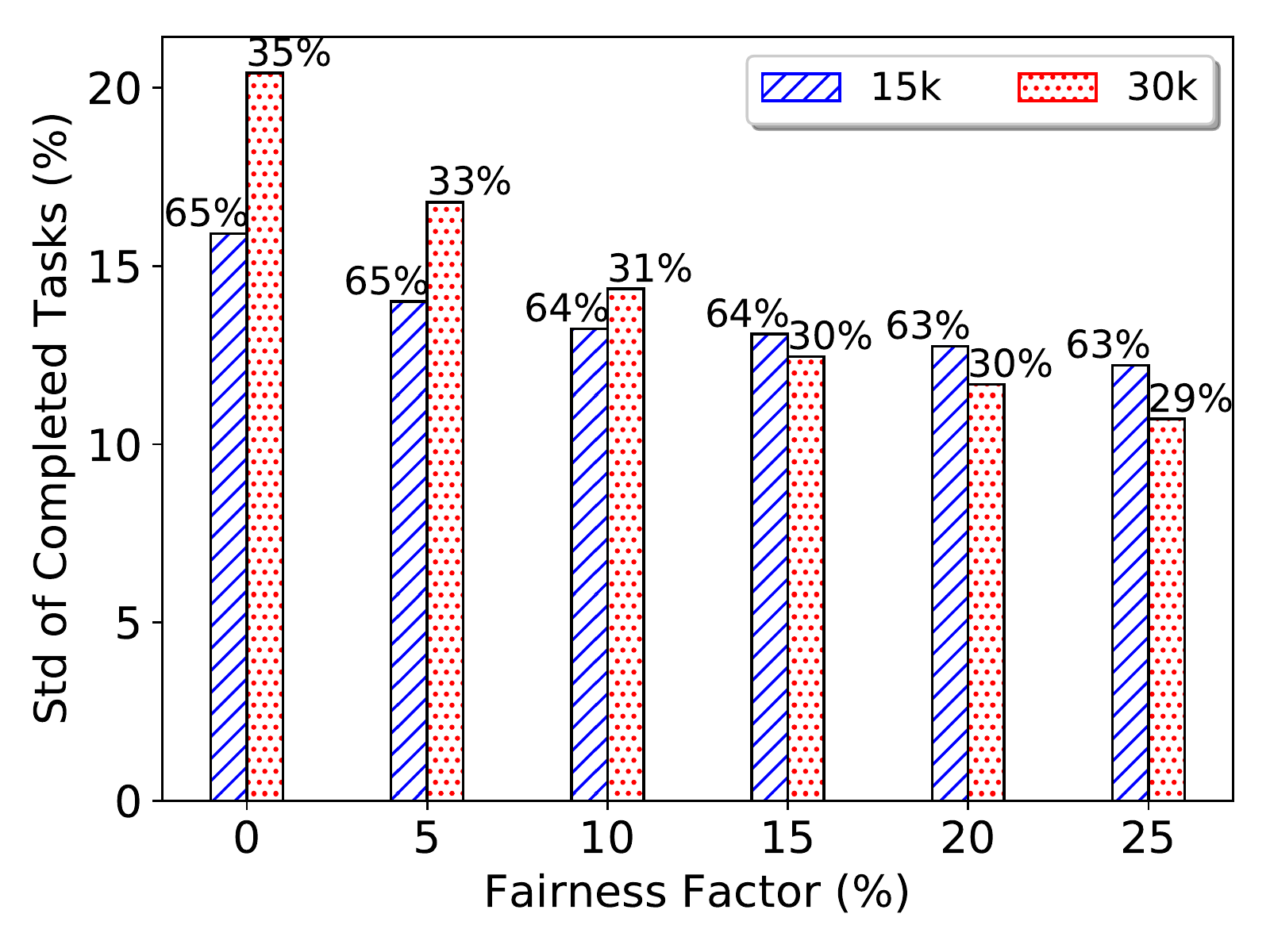}
   \label{fig:fairnessPercent} 

     \caption{Evaluating fairness and robustness on the system with 15k and 30k oversubscription level}. Horizontal axis shows fairness factor modifier to the sufferage value. Vertical axis is the standard deviation of completed task types. Values above bars show robustness.
     \label{fig:fairness} 
\end{figure}

~
Our aim is to study if PAMF heuristic (see Section~\ref{sec:solution}\,~~) alleviates unfairness. We test the system using a fairness factor ranging from 0\% (\ie no fairness adjustment) to 25\%. Recall that this fairness factor is the amount by which we modify the sufferage value for each task type. The sufferage value for a given task type at a given mapping event is subtracted from the required threshold, in an effort to promote fairness in completions amongst task types. For each fairness factor, we report: (A) The variance in percentage of each task type completing on time. The objective is to minimize the variance among these. (B) The overall robustness of the system, to understand the robustness we have to compromise to attain fairness. Robustness value is noted above each bar in Figure~\ref{fig:fairness}. We tested oversubscription level of 25k and 30k tasks.

Figure \ref{fig:fairness} shows that significant improvement in fairness can be attained at the cost of compromising robustness. In the case of 15k oversubscription level, we observe that using 10\% fairness factor results in a remarkable reduction in standard deviation of completed tasks that implies increasing fairness. The standard deviation drops from 16\% to 13.5\%, at the cost of $\simeq$1\% reduction in robustness (from 65\% to 64\%). This compromise in robustness is because deferring fewer tasks in an attempt to improve fairness results in fewer tasks successfully completed overall. 

However, in the case of 30k oversubscription level (and to a lesser extent, 20k and 25k cases that are not shown in the figure), the fairness factor makes more significant differences as the oversubscription increases. This is due to the fact that a higher oversubscription level provides more tasks to select at each mapping event. Therefore there is more possibility to bias the mapping to make the task mapping fairer.

Since high fairness factor value significantly impact on robustness in highly oversubscribed cases, we configure PAMF with 10\% fairness factor in the experiments, which include various oversubscription levels. 

\subsection{Evaluating the impact of pruning mechanism on the system robustness.}~
\label{subsec:heuristicscmp}
In this experiment, we compare the overall robustness offered by PAM and PAMF against baseline heuristics described in Section~\ref{subsec:baseline} and those baseline heuristics retrofitted with probabilistic pruning mechanism. We conducted this evaluation under various oversubscription levels. However, for presentation clarity, we only show oversubscription levels with 15k and 30k tasks. We note that the same pattern is observed with other oversubscription levels evaluated.

\begin{figure} [ht]

\begin{subfigure}[t]{0.49\textwidth}
  \centering
  \includegraphics[width=\textwidth]{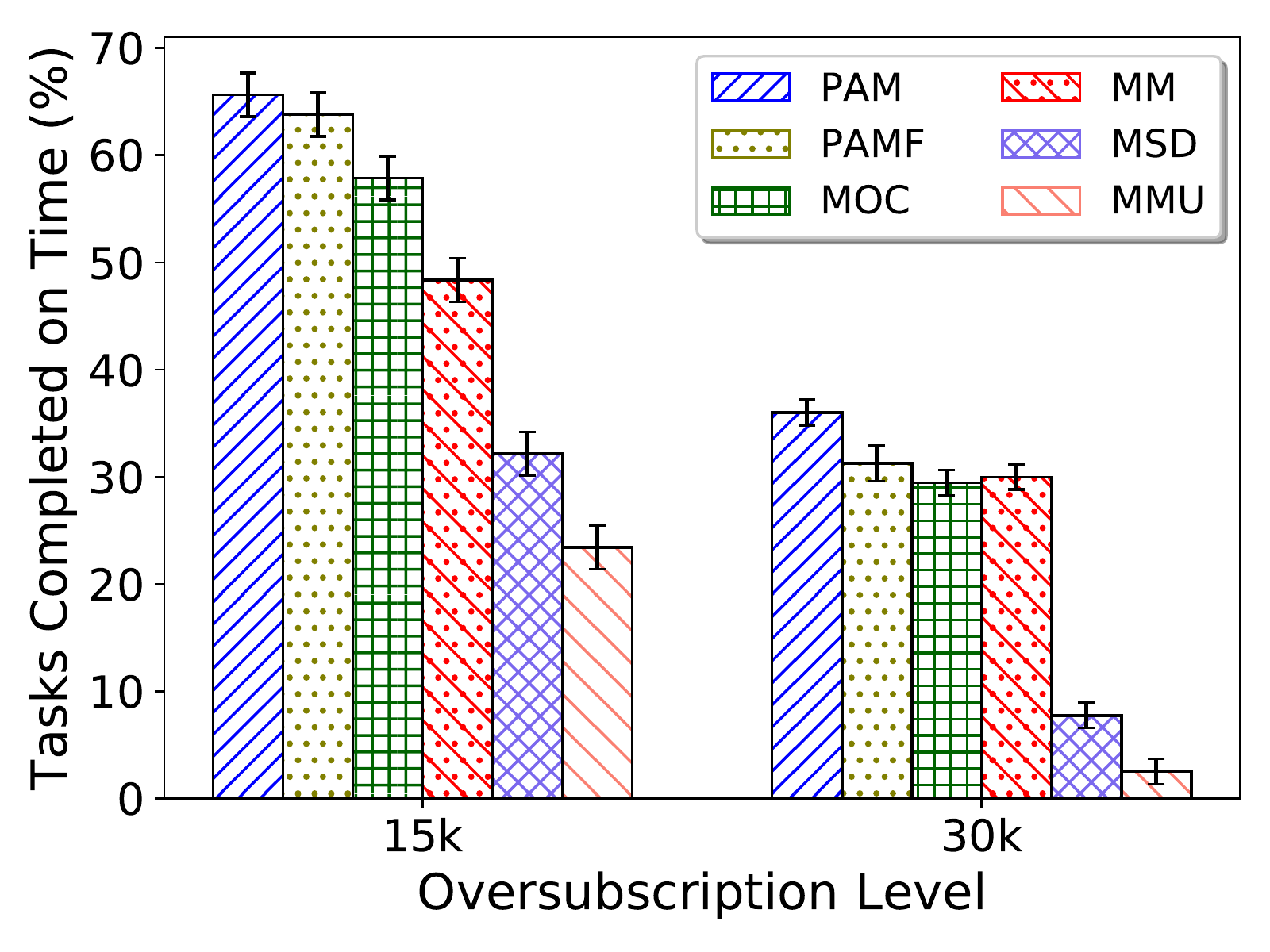}
  \caption{PAM and PAMF vs other heuristics. Horizontal axis shows oversubscription in form of number of tasks.}
   \label{fig:compare} 
  \end{subfigure}
\begin{subfigure}[t]{0.49\textwidth}
  \centering
  \includegraphics[width=\textwidth]{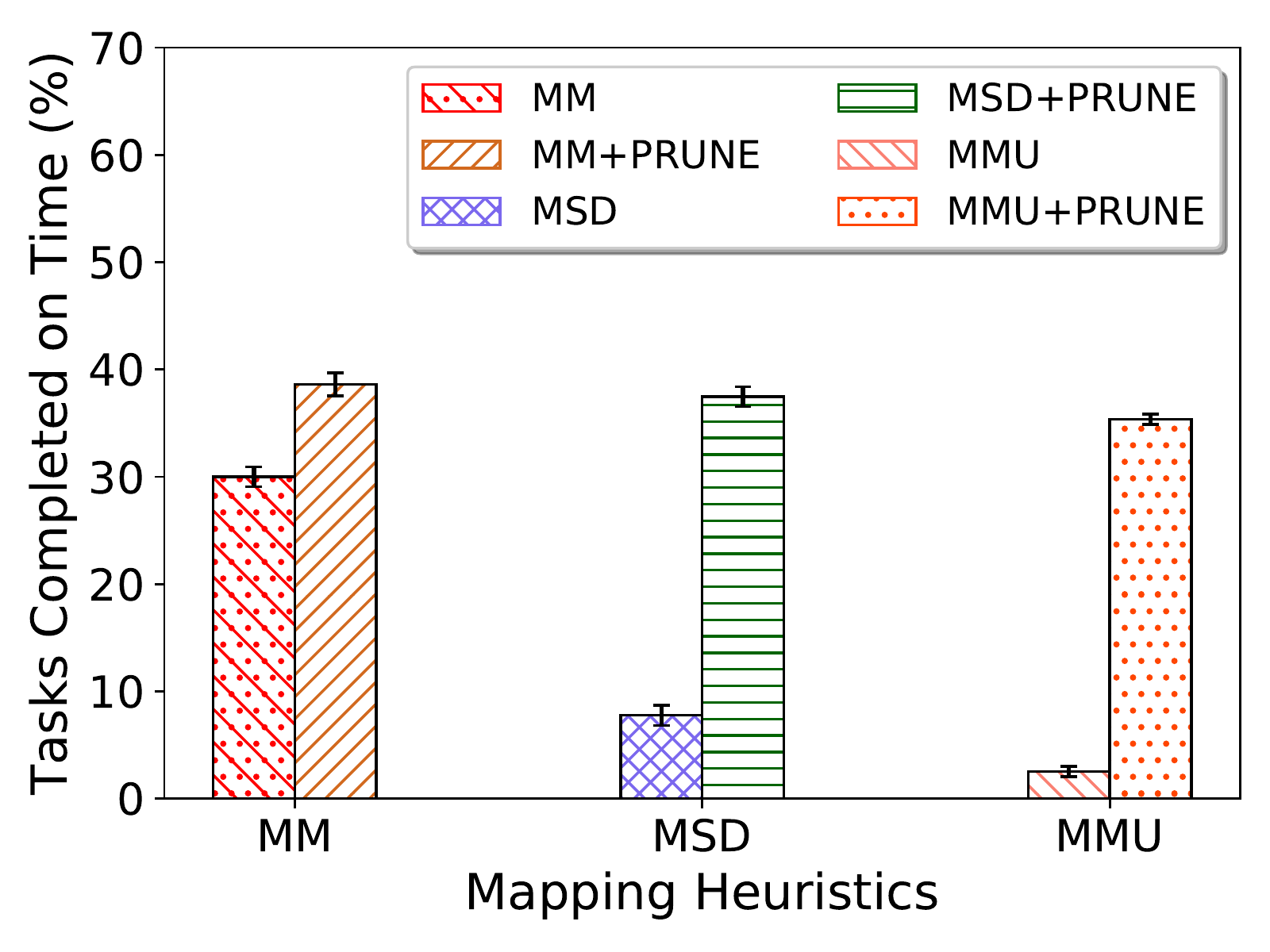}
  \caption{Impact of pruning mechanism when attached to baseline mapping heuristics at 30k oversubscription level.}
   \label{fig:compareConv} 
     \end{subfigure}
     \caption{ Comparison of PAM and PAMF against baseline heuristics with and without pruning mechanism. Vertical axis shows the percentage of task completed on time. }
\end{figure}
%Evaluating the effect of some baseline heuristics, with and without pruning mechanism retrofitted.

In Figure~\ref{fig:compare}, we observe that PAM results in a substantial increase in system robustness in comparison to other heuristics. On oversubscription level of 15k, PAM scores at nearly 67\% robustness and PAMF, trading percentage of tasks completed for types of tasks completed, results in nearly 64\% robustness. MOC, another heuristic that maps tasks based on the  robustness value, is the closest in robustness to PAM, rivaling PAMF, at nearly 58\%. The inability to probabilistically drop tasks leads to wasted processing and delayed task mapping, thereby lowering robustness. With robustness under 50\%, the performance of MinMin lags behind, as it allocates tasks to machines no matter how unlikely they are to succeed. The robustness offered by both MSD and MMU suffers in comparison because these heuristics, instead of maximizing the performance of the most-likely tasks, prioritize tasks whose deadlines or urgency is closest (\ie least likely to succeed tasks). With an oversubscription of 30k tasks, MSD and MMU perform particularly bad because they mostly map tasks that fail to meet their deadlines. When comparing PAM and PAMF against the average of the other four heuristics, PAM and PAMF result in averagely 22\% higher robustness. %, with no ability to recover from an initial mistaken mapping.

%New for JPDC, similar to HCW
In Figure~\ref{fig:compareConv}, we observe that, for all heuristics, adding the pruning mechanism to the existing mapping heuristics improves the robustness. The pruning mechanism makes the largest impact on MSD and MMU. These heuristics occasionally attempt to map tasks with too tight deadlines, thus, resulting in a low chance of success. By limiting these heuristics to map tasks whose chance is beyond a certain threshold, their overall system robustness is significantly improved.

  \begin{figure} [ht]

\begin{subfigure}[t]{0.49\textwidth}
  \centering
  \includegraphics[width=\textwidth]{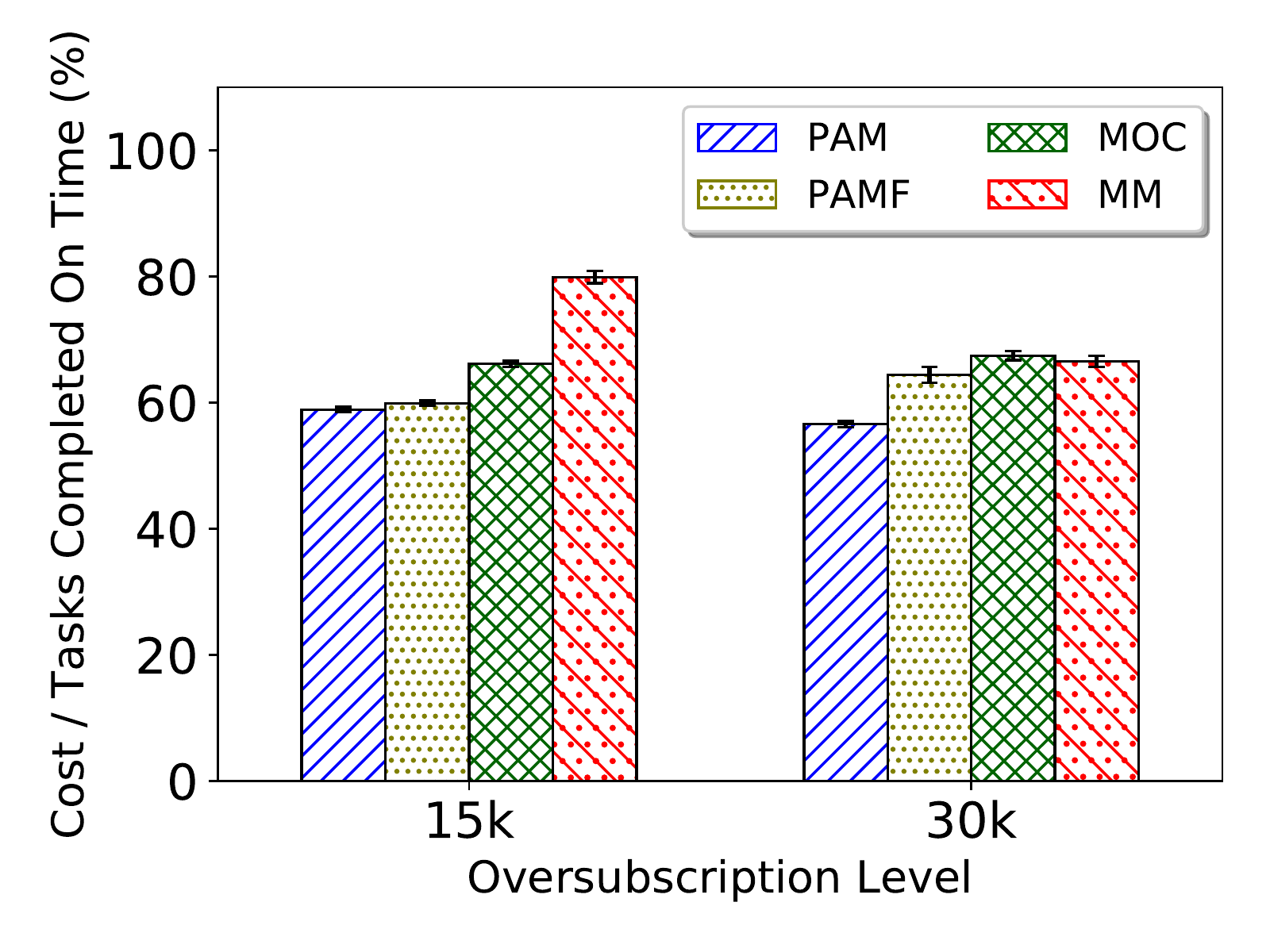}
  \caption{Analysis of the incurred cost}\label{fig:cost} 
\end{subfigure}
\begin{subfigure}[t]{0.49\textwidth}
  \centering
  \includegraphics[width=\textwidth]{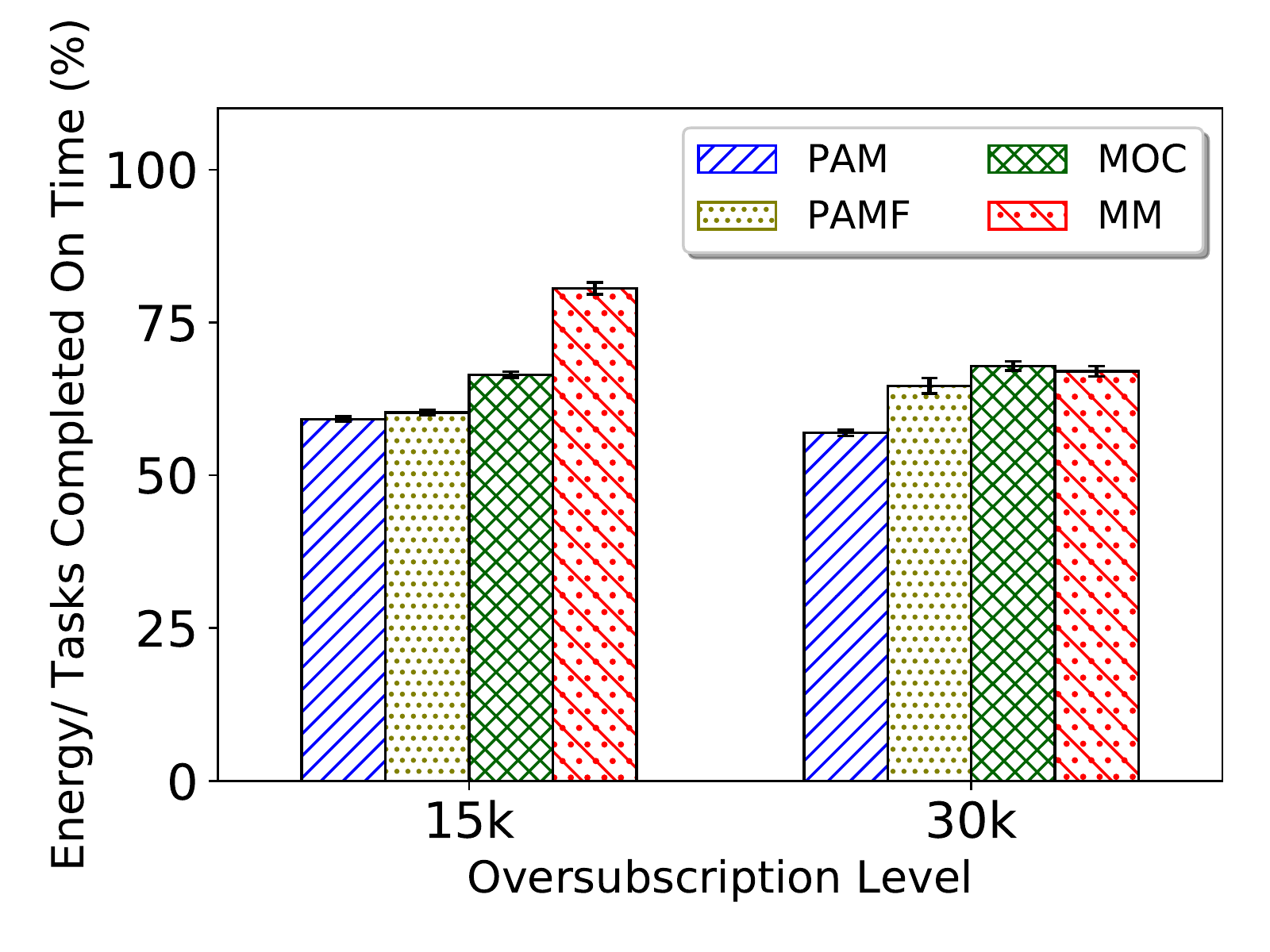}
  \caption{Analysis of the consumed energy}\label{fig:energy} 
\end{subfigure}
\caption{Impact of probabilistic task pruning on the incurred cost and consumed energy of using resources. Horizontal axes show the oversubscription level and the vertical axes, respectively, show the average incurred cost and energy consumed per task completed on time.}
\end{figure}

\subsection{Cost and energy gains of probabilistic task pruning.}~
To investigate the incurred cost of using resources, pricing from Amazon cloud VMs~\cite{aws} has been corresponded to the machines in the simulation. Energy consumption in active and idle states has been roughly estimated based on the machine's specification~\cite{fan2007power}. Specifically, we assume each machine to consume 70\% of their rated power supply when the machine is active and 25\% when it is idle. Each machine's usage time is tracked. The price and energy incurred to process the tasks are divided by the percentage of tasks completed on time to provide a normalized view of the incurred costs and consumed energy.

Figure \ref{fig:cost} and~\ref{fig:energy} suggest that in an oversubscribed system, both PAM and PAMF incur at least 33\% lower cost and energy per completed task than MM. We exclude MMU and MSD from the figure because they are shown to perform poorly in the prior experiment, which makes their cost per task completion ratio unchartable, when compared to other heuristics. %PAM also incurs at least 8\% lower cost than MOC, which is another probabilistic based mapping heuristic. 

While previous tests have shown PAM outperforms other heuristics in terms of robustness in the face of oversubscription, these results demonstrate that the benefits are realized in dollar cost and consumed energy as well, due to not processing tasks needlessly.

%New for JPDC
\subsection{Evaluating the imposed overhead.}~
  \label{subsec:Overhead}
To evaluate task pruning mechanism and PAM's scheduling overhead. We compare PAM that is implemented from the concept introduced in Section~\ref{sec:mechanism} against PAM that utilizes computational-reuse and approximation techniques that we introduced in Section~\ref{sec:approximation} (called approximate PAM and shown as \emph{PAM+APPROX}). First, we compare the task mapping performance in terms of the number of tasks completed on time. Then, we measured and compared the makespan of the simulation, which is directly related to the scheduling overhead. For the sake of accuracy, all the measurements have been carried out on an isolated machine, without any disturbing workload.

  \begin{figure} [ht]
  
  \centering
  
  \begin{subfigure}[t]{0.49\textwidth}
  \centering
 \includegraphics[width=\textwidth]{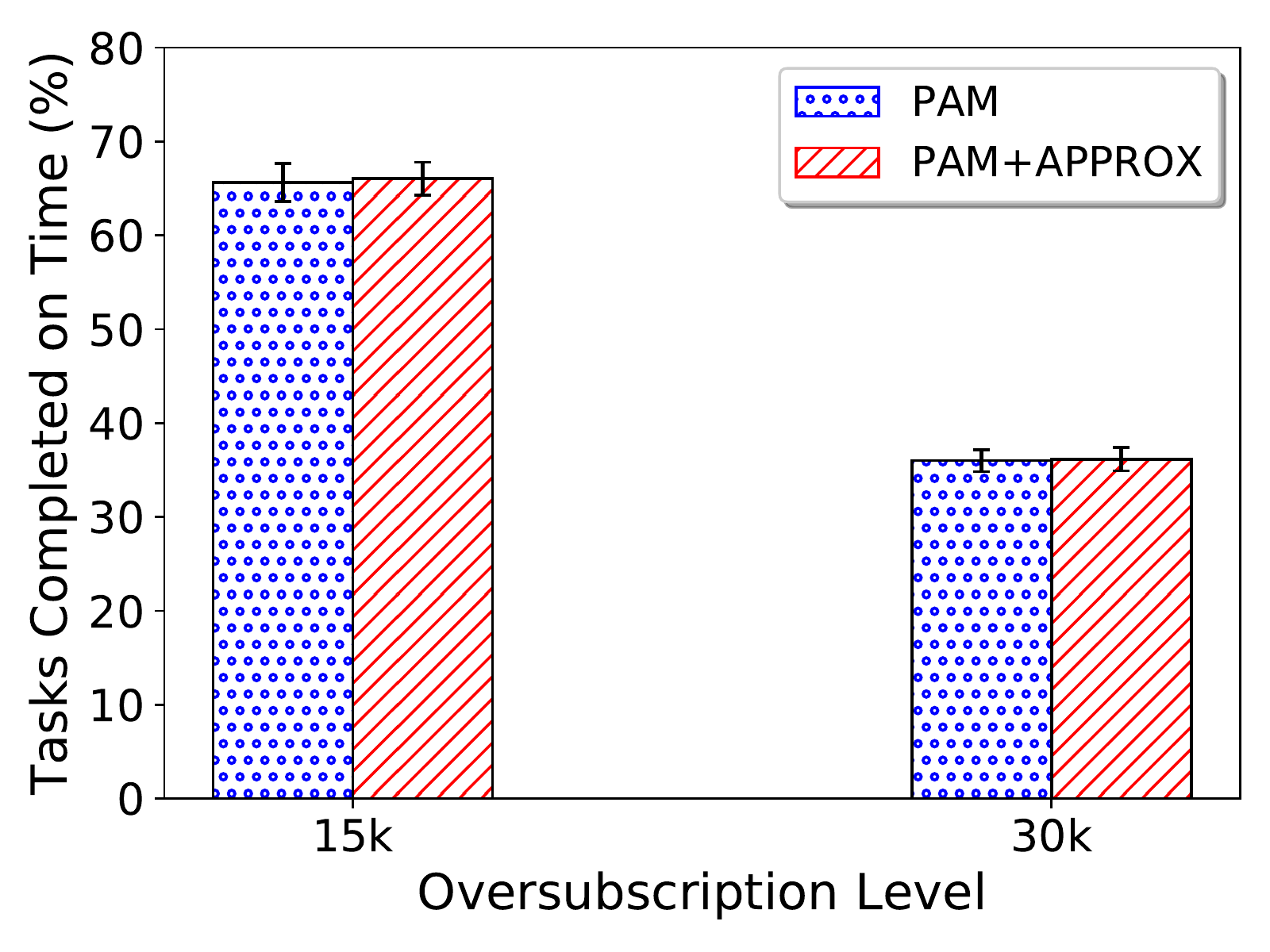}
  \caption{Analysis of obtained robustness}
  \label{fig:approxComple} 
  \end{subfigure}
  \begin{subfigure}[t]{0.49\textwidth}
  \centering
  \includegraphics[width=\textwidth]{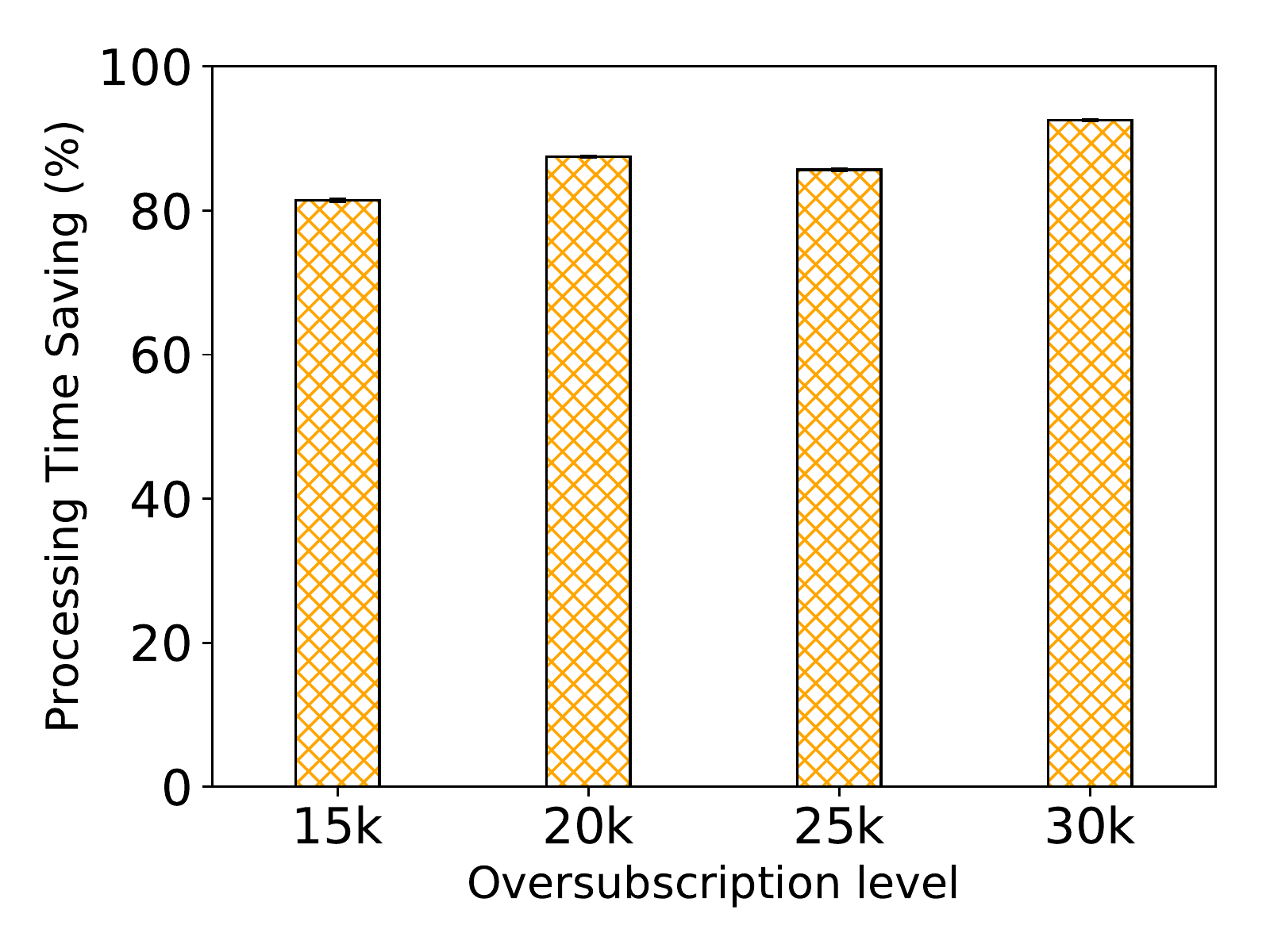}
  \caption{Analysis of imposed overhead}\label{fig:approxTime} 
  \end{subfigure}
  \caption{Impact of reusing and approximation techniques of the pruning mechanism and mapping heuristics on the (a) system robustness and (b) imposed overhead. Horizontal axis in both figures shows the oversubscription level and vertical axis in (a) shows sytem robustness and in (b) shows the percentage of reduction in the imposed overhead.}

\end{figure}

Figure \ref{fig:approxComple} shows that there is no statistically and practically significant difference in the performance of PAM and approximate PAM. This confirms our hypothesis that computational reuse does not change the mapping results, while the approximation introduces only minor rounding errors. However, due to the high uncertainty of the heterogeneous computing system, such approximated rounding induces only minimal changes to mapping decisions.

Figure \ref{fig:approxTime} shows that the approximate PAM performs drastically faster than the PAM implementation. The saving is particularly remarkable on the larger oversubscription cases such as 30k where approximate PAM cuts the processing time out by 93\% when compared to the PAM's implementation (\ie approximate PAM is 13.5 times faster than simple PAM).
Another point we observe is the growth of execution time. PAM's scheduling overhead grows in a more than the linear way in response to the increase in oversubscription level. However, approximate PAM's scheduling overhead is a little lower on 30k than 15k oversubscription level. This is because while more oversubscribed workload puts more tasks in the batch queue at each moment (which make each mapping event slower), the higher oversubscription level also means there are fewer mapping events for the experiment with the same number of tasks arrival. The fewer mapping events and more load per mapping event cancel each other in the case of approximate PAM.

%Approximate PAM's scheduling overhead is not far off from MinMin, which is a widely used mapping heuristics for heterogeneous system. Therefore deploying (well optimized) pruning mechanism in mapping heuristics is not impractical.

    \section{Summary}
The goal of this chapter was to improve robustness of the systems via pruning tasks with low probability of success. We designed a pruning mechanism as part of resource allocation system. For pruning, we determined probability values to either defer or drop a task whose chance of success is low. We enabled the pruning mechanism to determine dropping threshold at the task level and dynamically adjust the deferring threshold based on the characteristics of the arriving workload. We showed that task pruning mechanism improves robustness of the system when work in conjunction with commonly used task mapping heuristics. To gain even higher robustness, we developed a probabilistic mapping heuristic, PAM, that cooperates with the pruning mechanism. We showed that PAM can improve system robustness by on average $\simeq$22\%. We upgraded PAM to accommodate fairness by compromising around four percentage points robustness. We employed approximate computing in calculation of probabilities in the system to reduce the scheduling and pruning overheads (by up to 93\%) and ensure that the mechanism can be used practically. We concluded that: \textbf{(A)} when the system is not oversubscribed, tasks with low chance of success should be deferred (\ie wait for more favorable mapping in the next mapping); \textbf{(B)} When the system is sufficiently oversubscribed, the unlikely-to-succeed tasks must be dropped to alleviate the oversubscription and increase the probability of other tasks succeed; \textbf{(C)} The system benefits from setting higher deferring threshold than dropping threshold. Evaluation results revealed that the pruning mechanism (and PAM) not only improves system robustness but also reduces the cost and energy of using cloud-based HC resources by $\simeq$33\%. The idea of pruning developed in this chapter is generic and can be applied to other HC systems as well. 

%The idea of pruning developed in this chapter is generic and can be plugged to other systems. We plan to extend the probabilistic approach for tasks preemption and its impact on the convolution process. 

In the next chapter we develop the prototype of the Serverless Media Streaming Engine with all the findings of this and all the prior chapters.

%transition to next chapter?
    
\chapter{ Prototype Implementation of Serverless Media Streaming Engine (SMSE)}
\label{section:platform}
%SMSE
%%%%%%%SMSE implementation paper
\section{Overview} %cut from intro, abs
\label{subsection:platform_ov}
In the prior chapters, we studied the mechanisms to perform computational reuse via task merging and approximate processing through task dropping and deferring on a serverless computing platform. In this chapter, we prototype some of those mechanisms in a media stream processing platform using serverless computing resources.

Multimedia streaming is becoming an integral part of many applications, ranging from virtual and augmented reality (VR/AR) \cite{Sharma18,zink19}, 360$^{\circ}$ streaming (\eg Wowza \cite{wowza}), holographic video \cite{holography19}, and gaming (\eg Twitch.tv \cite{twitch}) to e-learning \cite{Delen14}, remote surgery \cite{remotesurgery17}, video conferencing \cite{Fukuda18}, network-based TVs \cite{pierson2015digital}, personal broadcasting (\eg Facebook Live \cite{fblive}), situational awareness via video surveillance \cite{matin_paper,7545004}, and movie industry (\eg Netflix \cite{netflix}). In fact, just video streaming alone is already one of the major services of the Internet and constitutes more than 75\% of the whole Internet traffic~\cite{intro_2,cdnstat}.
%%background link to section C2

Multimedia streaming services grow in popularity and diversity; their demand for hardware and software resources increases. Due to the burden and cost of maintaining such resources, making use of cloud services has become a common practice for stream providers. Currently, stream providers such as Netflix and many others extensively rely on general-purpose cloud services (\eg Amazon cloud \cite{pe_1}) to offer robust and reliable streaming services \cite{liccgrid16,li2018cost,liperformanceanalysis,he16} to the extent that costs of using clouds have become the main source of expenditure for them. Netflix, as an example, is estimated to expend around \$40 Million on Amazon cloud every month \cite{cloudcost}. Another expense is the software development costs that stream providers are incurred due to a lack of high-level programming support from general-purpose cloud providers \cite{cloudcost15}.

CDN and multiple caching techniques have been used with limited effectiveness to reduce cloud resource usage. However, such an approach is not applicable to live media streaming, and it also loses its effectiveness as the number of media customization increases. It is also not the best practice to send out the high-quality master media to perform extensive processing on viewers' thin-clients (\eg smartphones) due to bandwidth, energy, and compute limitations~\cite{lin17}. Therefore, a back-end system to on-demand process the media before sending them to the user must be utilized.

Our goal in this chapter is to develop a special-purpose cloud platform for multi-media streaming to offer flexible services in a robust and cost-efficient manner. The platform enables interactive streaming services through on-demand media stream processing on potentially heterogeneous cloud services in a cost-efficient manner while observing viewers' QoS guarantees.

To this end, we developed an Interactive Serverless-Based Media Streaming Engine (SMSE). SMSE~ facilitate cost-efficient and QoS-aware interactive live or VOD multimedia streaming using serverless cloud service for a different type of subscribers. SMSE~ is designed to be extensible, meaning that the \ssp~will be able to introduce new interactive services on video streams, and the core architecture can accommodate the services while respecting the QoS and cost constraints of the \ssp. To accomodate the large number of available services, each service is stored in \emph{service repository} where stream providers (and end-users) can cherry-pick which of these services they want to have available to them.

%%%%%%%%%%%%%%%%
\section{Architecture} %%contain architecture section, and flexibility section
\begin{figure}[htbp]
 \begin{center}
    \includegraphics[width=\textwidth]{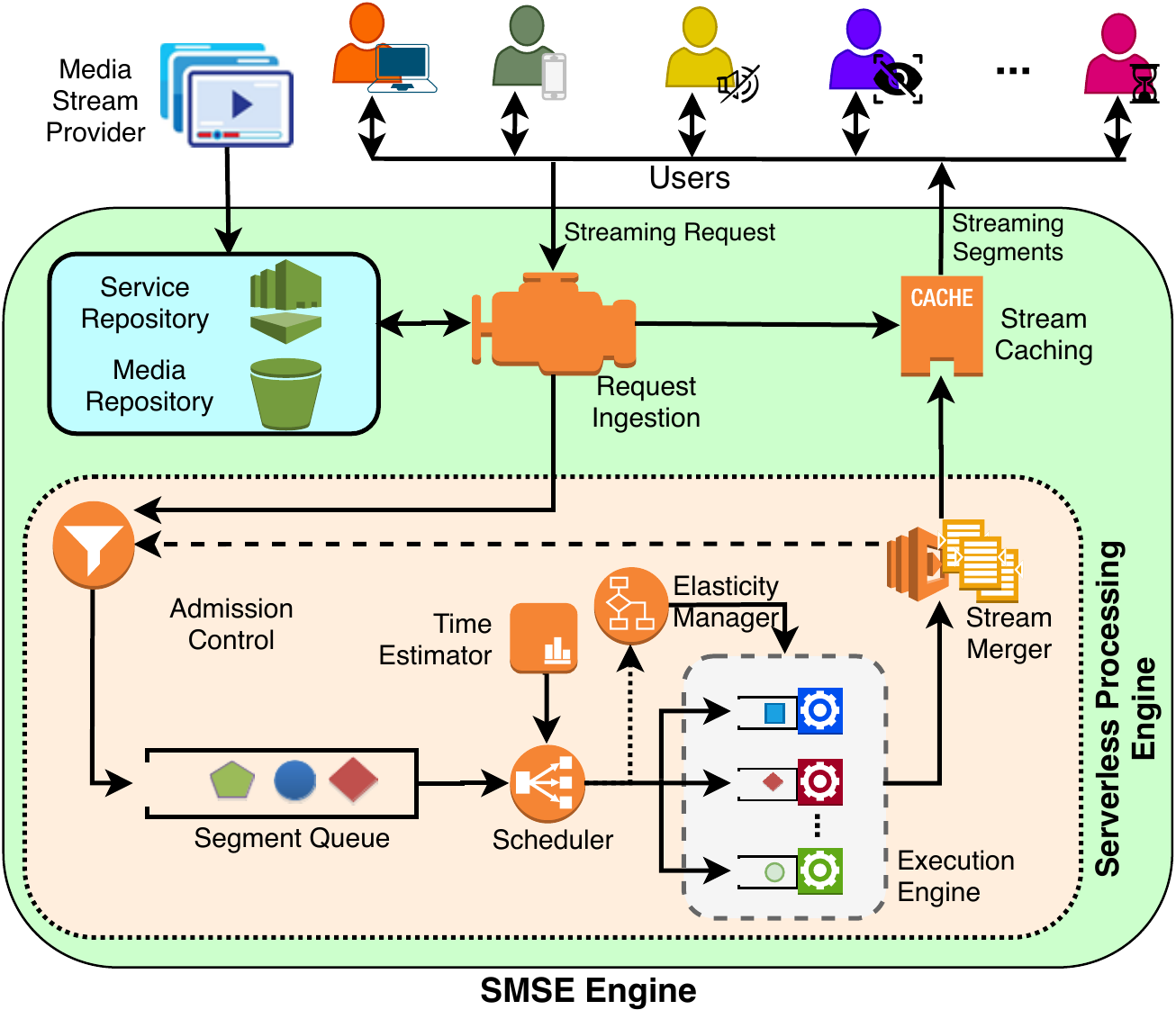}
  \end{center}
  \caption{System components of the Interactive Serverless-Based Media Streaming Engine (SMSE)}
     \label{fig:mid}
\end{figure}

\label{subsection:platform_arch} 
An overview of SMSE is presented in Figure~\ref{fig:mid}. In a simplified use case, the user hits the media playing button on a front-end web page to signify the Request Ingestion component to generate corresponding media transcoding requests. In certain cases where the specified media with exact specifications are already cached in CDN or Video Repository, the content is then sent directly to the user without processing on the cloud. Otherwise, each user request generates multiple \textit{processing requests}. Each of the processing requests is an association between a) each of the pre-splitted media segment required by the media stream. We pre-split the media %in Video Repository 
in such a way that it allows them to be parallel processed independently. b) processing service available in the \textit{Service Repository} and c) processing specification and metadata.  \textit{Admission Control} component assign priority of each processing request. Each processing request waits in the segment queue to be scheduled by \textit{Scheduler} (with execution time information from \textit{Time Estimator} component) to be processed in one of the \textit{Processing Unit}. After the task is scheduled from the segment queue to a processing unit's machine queue, the associated media (such as video segment) is fetched to the Processing unit to get ready for its execution once it reaches the head of the queue. The \textit{Elasticity Manager} monitors the performance of Processing units and adaptively allocates the appropriate amount of resources from the cloud (\eg containers) based on the workload. Finally, once the task is being processed, the resulting media is sent to CDN and the viewer. If the caching component detects that such a processing request is popular, it may also decide to cache the resulting media.
In the event that a segment is delayed (\eg due to failure), the Output Manager asks the Admission Control to resubmit the request with urgent priority. The rest of this section, we go into detail on the details and design priority of each component. 

\subsection{Request ingestion.}~
Request Ingestion handles all processing requests being made by viewers. %In a simulation run, request ingestion can simply perform as a request generator which generate requests at a certain time stamps. 
%In a real working mode, request generator can connect to various web-based video stream player or other application that require media streaming. %The main work of request ingestion in this mode is to check validity of the request.
Each one of the streaming requests usually generates multiple sub-tasks that can be processed in parallel. The Request Ingestion component converts the user request to tasks for the SMSE system to handle with defined deadline. For most media streaming, each request has the individual deadline as the presentation time of the media segment~\cite{li2016vlsc}. %for VOD, soft deadline, live hard deadline

%The processing of the media stream is determined by the characteristics within a viewers' request.

\subsection{Media repository.}~
Media Repository contains a copy of each media available on the platform. Those media segments are usually stored in a high definition format suitable to be reprocessed to other specifications. Once a media stream request is received, the media repository notify the media segment specifications to the request ingestion. Then, once the task is assigned to one of the processing units, a copy of the original media segment is then transfer to the processing unit, waiting to be processed. Media Repository can be augmented with content deduplication and approximate storage features.  %be processed to the characteristics of the media stream request.

\subsection{Service repository.}~
Service Repository manages the types of processing that the SMSE can perform. As shown in Figure~\ref{fig:mid}, new stream processing services can be dynamically defined and extended within SMSE~by third parties (\eg~\ssp). Due to the nature of video processing frameworks, which can have a large framework dependency, SMSE encourages the functions to be offered in the form of long-running containers.

\subsection{Admission control and segment queue.}~
Admission Control is the front gate of the segment queue (queue for tasks waiting to be scheduled by the scheduler) and it is in charge of  assigning a priority level and a certain internal metadata to each video segment based on their level of urgency. Further more, it can be extended to perform other functions such as deduplicating or merging the arriving task to an existing task in the system as extensively studied in Chapter~\ref{section:Reusing}. %explain more about merging in next section.

%Segment queue is a task queue or task pool where tasks are waiting to be scheduled by the scheduler.

%However, while the task scheduler in Chapter~\ref{section:Reusing} focus on homogeneous computing resources, the implementation in this platform can be hetegeneous.

\subsection{Scheduler.}~
Scheduler distributes the arriving segment processing tasks to available processing units by considering the QoS demands of the request against several factors. Those factors include the current workload of each processing unit and the ability of the processing unit to process such task type (called task-machine affinity). %.

The scheduler is the class that provides the most customizability in our platform. While the simple implementation of the scheduler submits tasks to machines in FCFS, round-robin manner, we provide a variety of more sophisticated scheduler implementations. 
%Our default scheduler in SMSE is based-on MMU scheduling policy described in~\ref{subsec:baseline}
%It operates in two-stage where the first stage is to select a task; then, the second stage selects a processing unit to handle the task. In the first stage, we select the task by sorting them by the highest priority first. If there are multiple tasks with the same priority (likely) then selecting the task that is the most urgent first. Urgency~\cite{matin} of task $i$ is calculated as $U_i=1/(\delta_i - E_i)$. 
%Then, on the second stage, we simply select the (available) processing unit that is predicted to have the earliest completion time. 
In addition to scheduling policies explained in Section~\ref{section:bgTaskMap}\,~~, we also implemented the Priority-based scheduling~\cite{li2016high} proposed by Li~\etal with a couple of minor details adjusted. First, not all processing units can process a given task type. Second, even the processing units capable of processing the task type can get too busy to be considered as a candidate (\ie their waiting queue is too long). The scheduler component can be augmented with the probabilistic task dropping and deferring proposed in Section~\ref{section:Reusing}. 
However, not all tasks can be dropped from the system. The SMSE platform supports the processing of both live media streaming and also on-demand media streaming. In the case of live media streaming, the task can be dropped after the deadline as they are no longer useful. In contrast, in the case of on-demand media streaming, the viewers still expect to receive the media segments even after the deadline has passed. Therefore, those tasks still have to be processed even after the deadline.

%%%%%%%%%% explain more about priority-based scheduling

%We encourage SMSE deployment to customize their own scheduler to tailor the scheduling mechanism to their preference.

% \begin{list}
%     \item Highest Priority First
%     \item Most Urgent First
%     \item Shortest Execution Time First
% \end{list}

%Then, the Scheduler module maps the arriving streaming segments to the  available processing units by considering the QoS demands of the request. For instance, in live streaming, segments that miss their deadlines are not valuable and are dropped (ignored) to maintain the liveness, whereas in VOD the segments have to be completed, even after their deadlines. 

\begin{figure}[htbp]
      \vspace{-10pt}%\begin{figure} 
 \begin{center}
    \includegraphics[width=\textwidth]{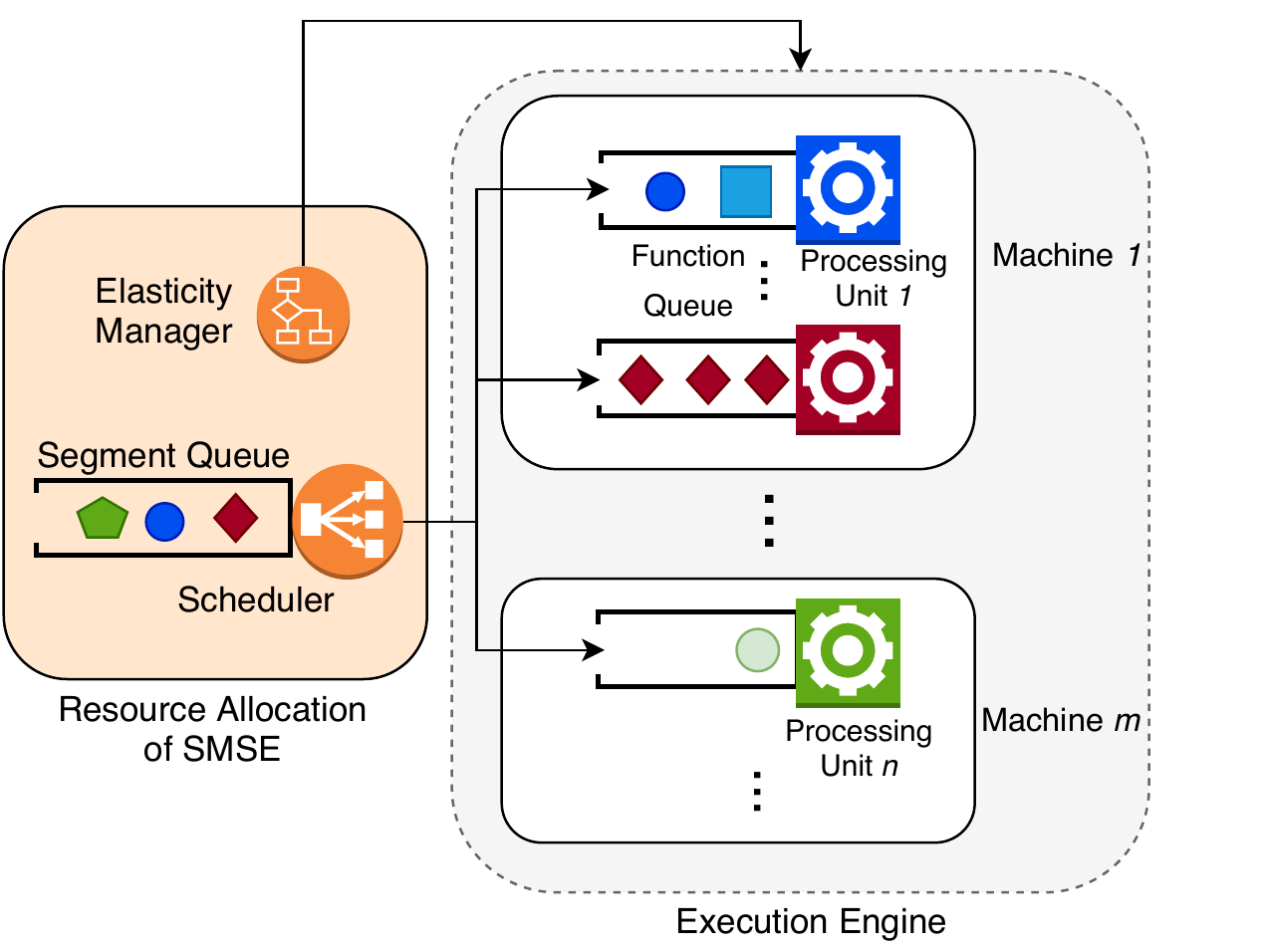}
  \end{center}
  \caption{An example configuration of the elasticity manager controlling several machines. In this simplified example, elasticity manager is suggested to scale up the number of red processing unit (container) for red rectangular shape to keep up with the demand. Geometries of different shapes, color, and size represent different processing tasks. Geometries that stack together signify compounded (merged) tasks.}
      \vspace{-3pt}	\label{fig:resources}
\end{figure}

\subsection{Elasticity manager.}~
%allows the platform to dynamically allocate or deallocate containers based on the current workflow.
The elasticity manager dynamically allocates or deallocates resources based on the current workflow. It can work solely on top of a serverless computing platform, in which case the elasticity manager only adjusts the appropriate budget limit to the serverless platform usage, leaving the actual resource management works to the serverless platform.

Alternatively, the elasticity manager can also manually manage the resource by working on two levels. %: first; scaling the number of compute node, then second; manage the containers on top of each compute node. 
In this way, the elasticity manager scale in and out the number of machines to adjust the computing power. Then, within each machine, the elasticity manager controls the number of processing units of each function individually.  Figure~\ref{fig:resources} shows the elasticity manager working in this configuration. Note that each machine can also be heterogeneous machines that best suit to different type of functions.

%Internally, the Elasticity Manager keeps track of each processing units and provide the scheduler a list of functional interfaces to properly utilize each of the processing unit. A new type of processing unit (such as cloud at remote edge) can be introduced into the system by creating a new \textit{Processing Unit Interface}. Note that each of the processing unit interface also contains the information of available service the processing unit can handle. Such that, the scheduler can avoid schedule unsuitable tasks to the processing units.

\subsection{Processing units.}~
We define a processing unit as the most granular computing resource in the view of the scheduler. A processing unit in the SMSE platform can be an entire machine or a long-running container on top of the (bare-metal or virtual) machine. However, we discourage the use of an entire bare metal machine as a processing unit for two reasons. First, many service functions are not implemented to be highly parallelizable, and therefore, a large part of resources in the bare-metal machine might not be used efficiently. Second, currently, a new function in SMSE is providing as a container image. Using bare machine as a processing unit does not allow the real-time introduction of a new (function) service. %In the future, we envision non-virtualization serverless computing platforms such as those that utilize WebAssembly~\cite{shillaker2020faasm} to become more mature and rival the usability of container-based serverless computing platform. As such, having an entire metal as one processing unit can become a good practice with the help of non-virtualization serverless computing platforms. 

Once started, each processing unit listens to a machine queue (implemented using RabbitMQ) for tasks assigned from the scheduler. The machine queue is vital for efficient usage of processing power by minimizing scheduling and resource waiting time. Upon task arrival to the machine queue, the processing unit fetch the corresponding media (such as video segment) ahead of the task's execution time. All machine queue is managed in a FCFS manner for tasks with normal priority. However, a task with a high priority can skip the line for urgent execution.

%perform the processing of video segments to the appropriate characteristics.
% \subsubsection{Output Manager.}~
% contains cached portions of popular videos in different versions based on their popularity. For parts of a video that are not cached, they are processed on-demand.

\subsection{Time estimator.}~

The time estimator component highlights the modular design of our platform. The time estimator predicts task execution time (and completion time) of each request on each processing unit. For the balance of usability and prediction accuracy, the time estimator model execution time following a normal distribution. Hence, the result of timing predictions has a mean and standard deviation component. However, the class can be overwritten to predict the execution time with Probability Mass Function (PMF) or Probability Density Function (PDF) as necessary. 

We provide two implementations of the Time Estimator to select on run time.
%We produced and recorded these workloads to inform the decision making process within the CVSE.
In an implementation called profile mode, the time estimator simply reads the pre-defined table to find the expected execution time of each media segment, with the specified processing service on the specified processing unit type. This mode is highly deterministic and is suitable for testing several components of the SMSE.

In another implementation called learn-mode, time estimator accumulated historical data from prior task executions to form knowledge for estimating each task type of the later occurrences. Unlike profile mode, The time estimator in this mode does not attempt to discriminate the estimation data between different media segments. In other words, a task that involves the same operation on the same processing unit yields the same time estimation regardless of involved media segments. This mode is useful when a high amount of task type variation is expected, and there is no complete execution time profile available.

\subsection{Output manager and caching system.}~
The output manager resides near the end of the processing pipeline. It keeps track of all requested media segments of all users and making sure all segments are processed properly and in a timely manner. In a case that a certain segment is missing, the output manager component can request the admission control to resend the processing request of the segment in an urgent priority. The output manager also allows the multi-stage (\ie workflow) tasks to be performed in the SMSE. When the output manager found a task that needs to be processed further with a different service to reach the user's specification, it generates the new task accordingly.

In addition, output manager can determine the hotness~\cite{darwich2019cost} (\ie popularity) of each segment. Media segments that are predicted to be requested again in the near future can be cached for reusing in a local caching server or on CDN~\cite{veillon2019f}.

%%%%%%%% Fexibility become just subsubsections
\subsection{Extendability of the system.}
\begin{figure}[htbp]
 \begin{center}
    \includegraphics[width=\textwidth]{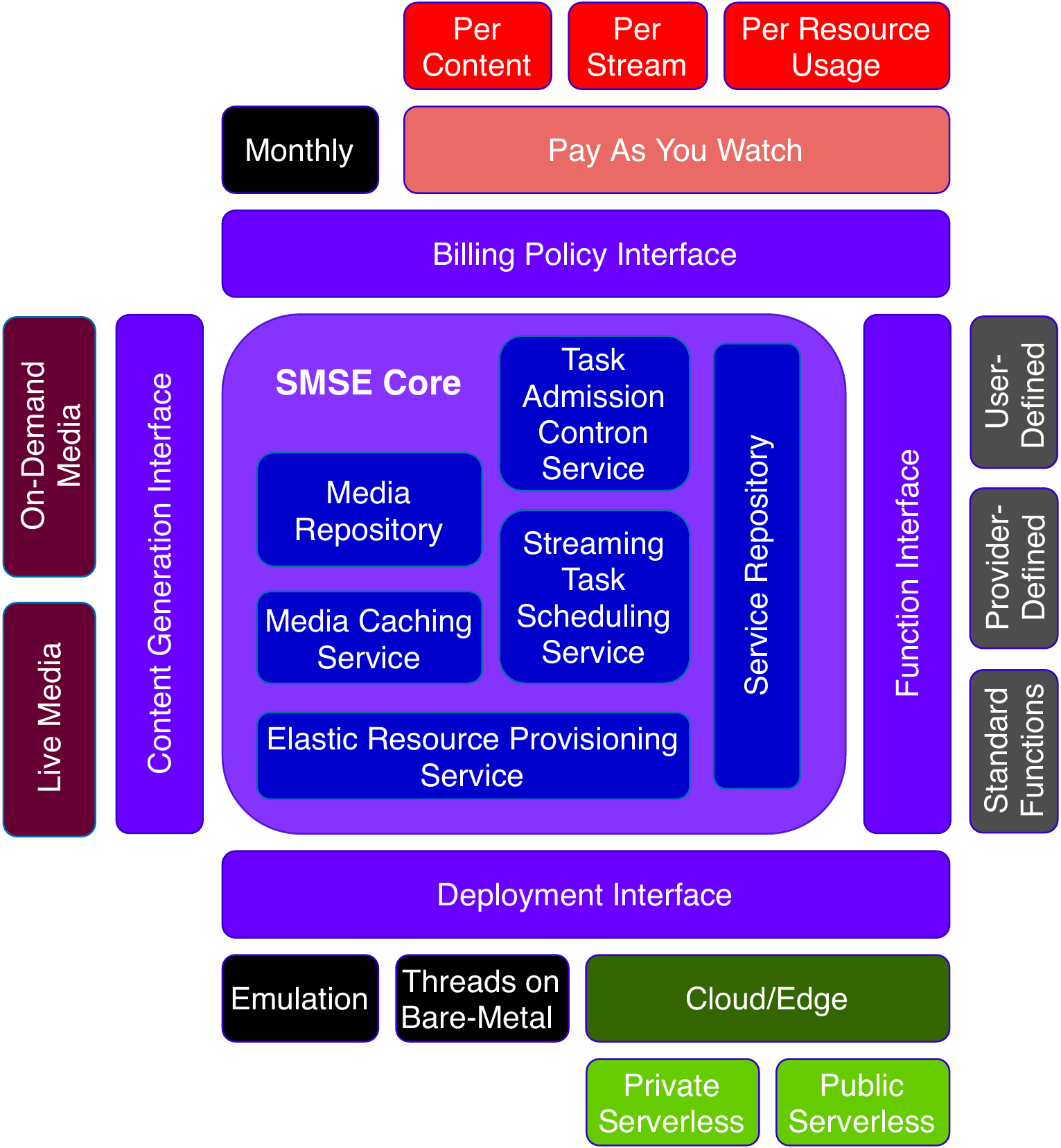}
  \end{center}
  \caption{Expandable Interfaces in Serverless Media Stream Processing Engine (SMSE). Each Box outside of the SMSE Core represents each of the expandable interface to increase the capability and compatibility of the system. }
      \label{fig:class}
\end{figure}

~
While SMSE provides a number of pre-defined media processing services, we expect most of the task types to be created by the service providers.
As such, we structured video processing services within SMSE as an interface that can be extended by third parties as shown in Figure~\ref{fig:class}. For example, if a \ssp~ would like to provide an auto-translation for video subtitles. Then they can define the media processing task and provide SMSE a container image to process such task type. We take the processing service as a long-running container for several reasons. First, many task types for media processing require either a specific version of video processing tools or machine learning frameworks, which is easiest to deliver as a container. Second, regarding the long-running manner, many of the video processing tasks require low latency results. By making service containers standby and ready to execute tasks without the container start-up overhead, the start-up latency is significantly reduced. We evaluate the  overhead of launching one container for one task against reusing the long-running container in section~\ref{subsec:Evalreusecontainer}.

% \begin{figure}[ht]
%       \vspace{-10pt}
%  \begin{center}
%     \includegraphics[width=0.4\textwidth]{figures/C6/conservative_inplace.png}
%   \end{center}
%   \caption{\small{An example of deadline miss-rate during a simulation run.}}
%       \vspace{-3pt}	\label{fig:sim_graph}
% \end{figure}

% \subsection{Emulation Mode.}~
% SMSE currently have two deployment modes. One is the full deployment mode intended to perform as full stack multimedia processing platform which receive users' tasks to execute using cloud computing resources. During the development of the project, to evaluate each component in the system, we also utilize a simulation run to remove the needs of utilizing real cloud computing resources. Later, it evolve into a full simulation mode. In this mode, we switch the time stamp in the system from real time to simulation time. User requests are pre-generated by Ingestion Control and arrive at the certain time stamp. The execution time of each task are sampling in a range estimated by the Time estimator component. The simulation result come as a simulation log as well as a graph showing deadline miss-rate of all tasks during the simulation time. Figure~\ref{fig:sim_graph} shows an example graph plotting the number of task completed on time against task that miss its deadline during a simulation run. As SVSE is modular and easily modifiable. It become our internal simulation tools to evaluate some novel ideas focussing on a certain parts in the system. This include the experimental features like task merging.

\begin{table}[ht]
    \centering
    \begin{tabular}{|c|c|}
     \hline
        \textbf{Component} & \textbf{Tools} \\
         \hline

        Language for Scheduler & Java \\
         \hline
        Language for & \\
        Processing Unit & Python 3\\
         \hline
        Message Queuing & RabbitMQ~\cite{dobbelaere2017kafka}\\
         \hline
        Object Serializer & Google Protobuf~\cite{protobuf} \\
        \hline
        Default Media Processing Tool & FFmpeg~\cite{ffmpeg16} \\
        \hline
        Default Video Streaming Standard & Apple HLS~\cite{hlsDash} \\
         \hline
        Web Player Hosting & Node.js~\cite{tilkov2010node} \\
    \hline
    \end{tabular}
    \caption{Quick reference of technology and tools used in SMSE}
    \label{tab:3rdpartytools}
\end{table}

%%%skip part about merging. it is explained in a whole section 3

%%%%%%%%%%%%%%%%
\section{Performance Evaluation}
\label{section:platform_eval}

\subsection{Experimental setup.}~
To evaluate the prototype of the SMSE platform implemented using tools listed in Table~\ref{tab:3rdpartytools}. %We made SVSE publicly available\footnote{https://github.com/hpcclab/adaptivemerging} for the research community and reproducibility purposes. 
% as the system's only processing units.  
We experiments two parts separately. The first two experiments measure the start up latency of the service containers. The rest of the experiments the SMSE in a full deployment mode (\ie not simulated) using containers performing media processing tasks.

The media repository we used for evaluation includes a set of videos. Videos in the repository set are diverse both in terms of the content types and length. The length of the videos in the repository varies in the range of [10, 220] seconds splitting into 5-110 video segments. The benchmark videos are publicly available for reproducibility purposes\footnote{\url{https://github.com/hpcclab/videostreamingBenchmark}}. More details about the characteristics of the repository videos can be found in Chapter~\ref{section:MergeSaving}. %For each segment of the benchmark videos, we obtained their execution times by executing each micro-service 30 times. 
The processing services in the experiment includes: reducing resolution, adjusting bit rate, and adjusting frame rate. In each case, two conversion parameters are examined. For example, frame rate is changed from 60 fps down to either 30 fps or 24 fps.

To evaluate the system under various workload intensities, user requests are profiled to request for [400, 1,200] media segment processing tasks within a fixed time interval. Other than the first two experiments, all transcoding micro-services are available in the processing units (\ie warm starting micro-services). Processing tasks arrive to the system in a group of up to twenty consecutive segments at a time. To accurately profile common workload arrival pattern observed in the real video steaming systems, each workload repeatedly toggle their arrival rate between base period and high load period where the arrival rate is increased by two folds. Each base period is approximately three times longer than the high load period.

\subsection{Evaluating processing unit configurations.}~
In this part, we compare the start-up overhead of various deployment schemes of a processing unit. In SMSE, the task processing unit can be deployed as a thread, as a virtual machine, and as a container. To cope with the ever-changing demand of each functions, the number of processing units for each function is dynamically adjusted periodically. Therefore, a low start-up overhead of the resource can help to keep up with the surge of demand without resorting to a high amount of hot spare resources.

\begin{figure}[ht]
	\centering
	\includegraphics[width=0.8\textwidth]{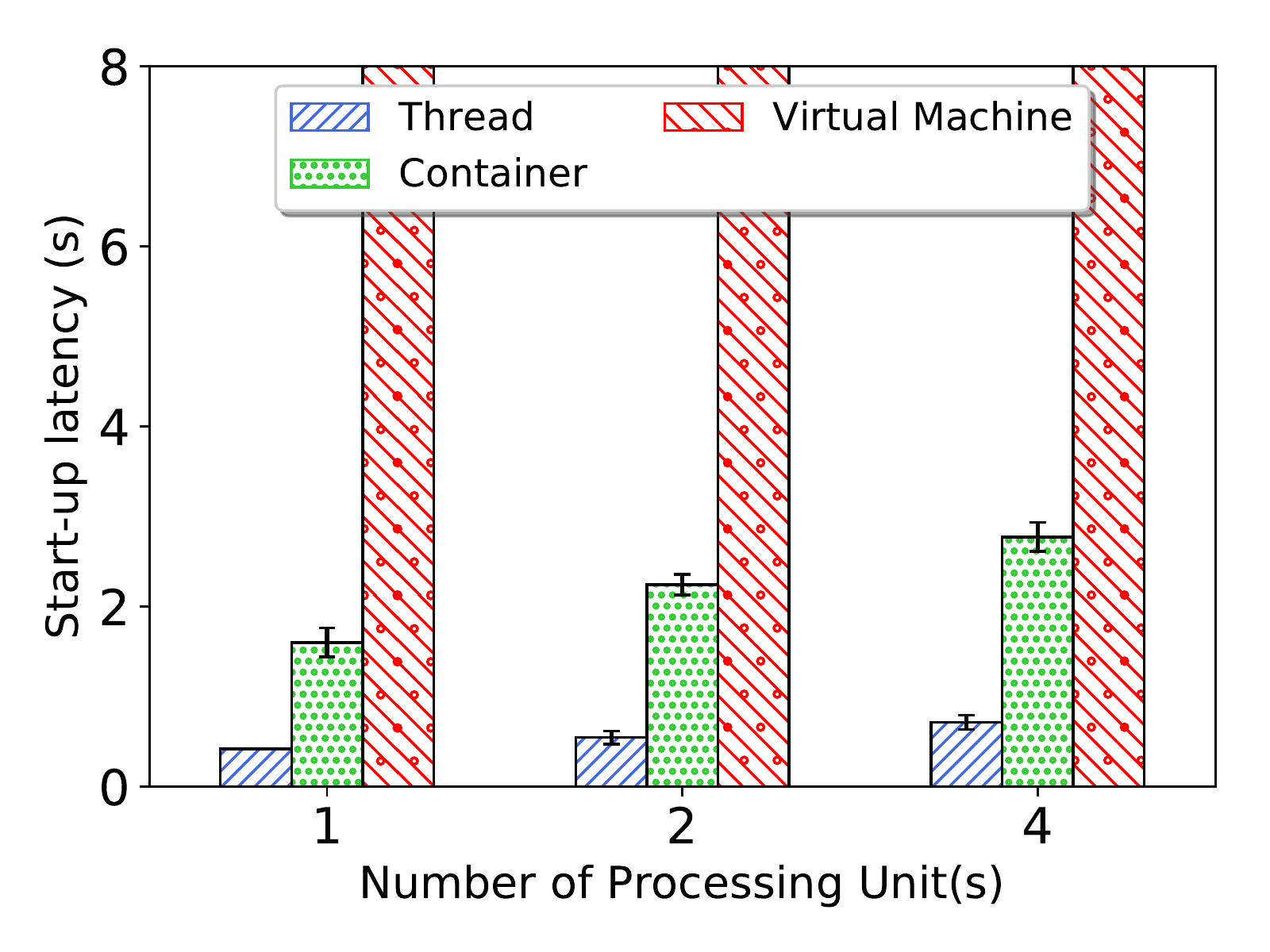} 
	\caption{Comparing the start-up latency of three processing unit deployments: processing unit as a thread, processing unit as a container, processing unit as a virtual machine. X-axis shows the number of processing units to deploy. Y-axis shows the time in seconds to start all the processing units indicated in X-axis.}
	\label{fig:StartUpLatency}%
\end{figure}

\paragraph*{Evaluating startup latency of processing units:}
Figure~\ref{fig:StartUpLatency} shows the start-up latency of processing unit deployments: processing unit as a thread, processing unit as a container, processing unit as a virtual machine. 

The latency of starting virtual machines in our set-up is far beyond other schemes at over 12 seconds for one VM and increases to over 30 seconds for four VMs. Therefore we leave their data out of the chart for the clarity of other configurations. Performance wise, starting a processing unit as a thread has the least start-up overhead. However, such configuration lack isolation and scaling flexibility. Each service processing unit running on a bare-metal machine can interfere with each other, and each of them may require different software libraries that conflict with each other. Deploying the processing unit as a container eliminates such a downside while imposing an acceptable amount overhead.

\paragraph*{Evaluating the effect of container reusing:}
\label{subsec:Evalreusecontainer}
In this part, we evaluate the two approaches to using containers for task processing. The most straight-forward and simple solution is to launch a new container for each task. This imposes a container start-up overhead for every task. The second solution is to reuse a container for multiple tasks by making the processing unit a long-running container. Each of the processing units is assigned its own machine queue, which will be filled by tasks assigned from the scheduler.  In this way, each processing unit (each container) can process multiple tasks with only a single start-up overhead. However, each task still imposes the queueing overhead.

\begin{figure}[ht]
	\centering
	\includegraphics[width=0.8\textwidth]{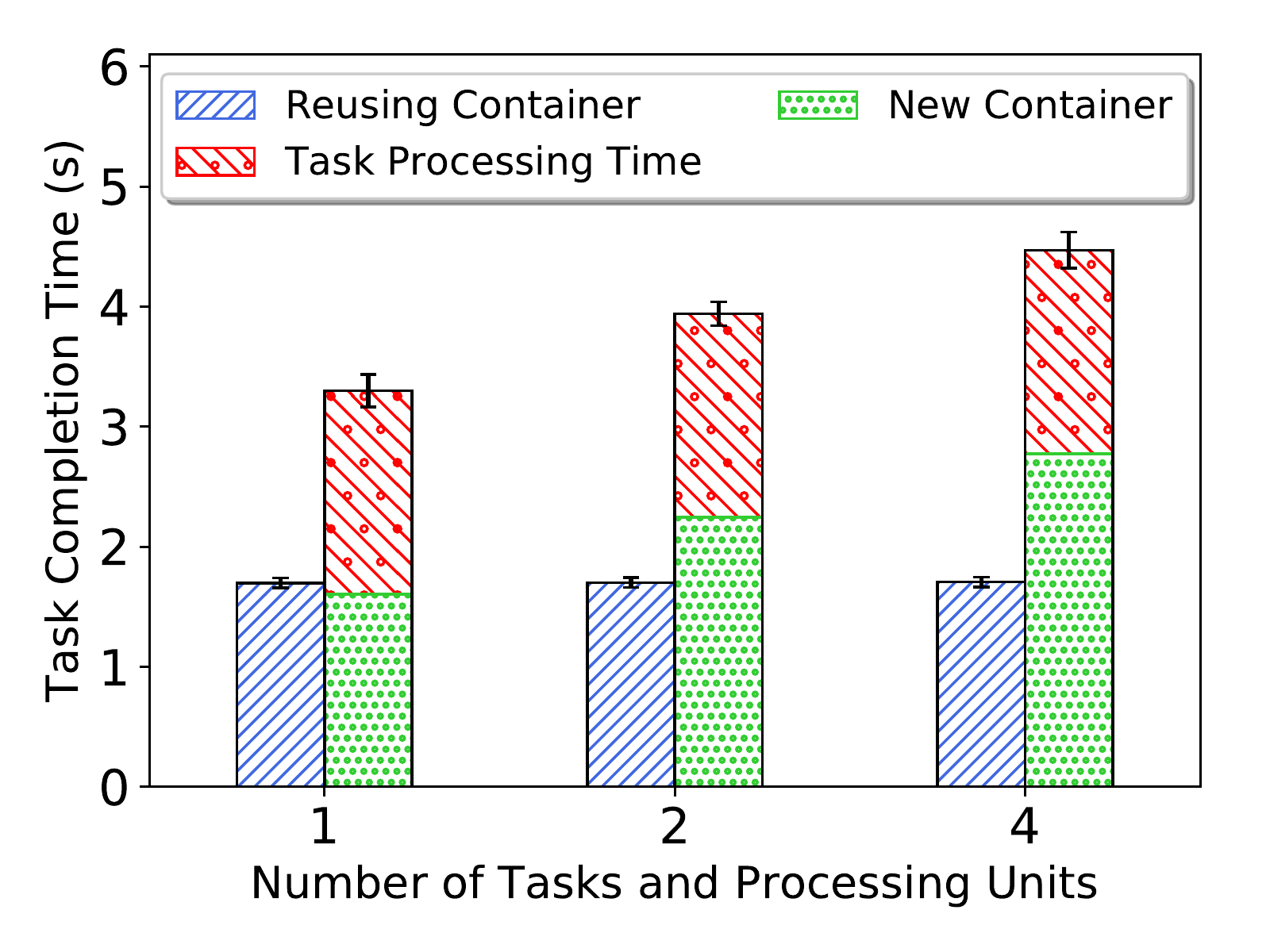} 
	\caption{Comparing the execution time of one to four video processing tasks running on containers in two configurations: reusing existing container and starting a new container for each task.}
	\label{fig:ContainerRoundTrip}%
\end{figure}

Figure~\ref{fig:ContainerRoundTrip} shows the round trip time from assigning one to four tasks to getting the processing result back from the processing units. For the container reusing approach, we found the task processing time to overwhelm the queueing overhead. The round trip messaging latencies are only six to thirty milliseconds and therefore are not plottable when the scale of task processing time is almost two seconds. For the approach that creates a new container for each task, we found the container creation takes a similar or longer time than the time it takes to process the task. As such, creating one container per task execution is highly inefficient. And therefore, the container reusing approach is selected for all further experiments.

% \begin{figure}[ht]
% 	\centering
% 	\includegraphics[width=0.5\textwidth]{figures/C6/ContainerLatency.pdf} 
% 	\caption{\small{Comparing the start-up latency of a task running on container in two configurations: warm-start, and cold-start}}
% 	\label{fig:ContainerLatency}%
% \end{figure}

% \subsubsection{Makespan.}~

% Container vs thread vs VM

% Different services 

% Time to add new service?

\subsection{Evaluating the optimal degree of concurrency on the execution platform.}~
In this part, we evaluate the number of tasks deploying concurrently for optimal performance. Specifically, for a machine with 16 CPU cores, how many concurrent processing units (containers) result in the lowest deadline miss-rate? Note that the function containers in the study can use up to two CPU cores each. 

\begin{figure}[ht]
	\centering
	\includegraphics[width=0.8\textwidth]{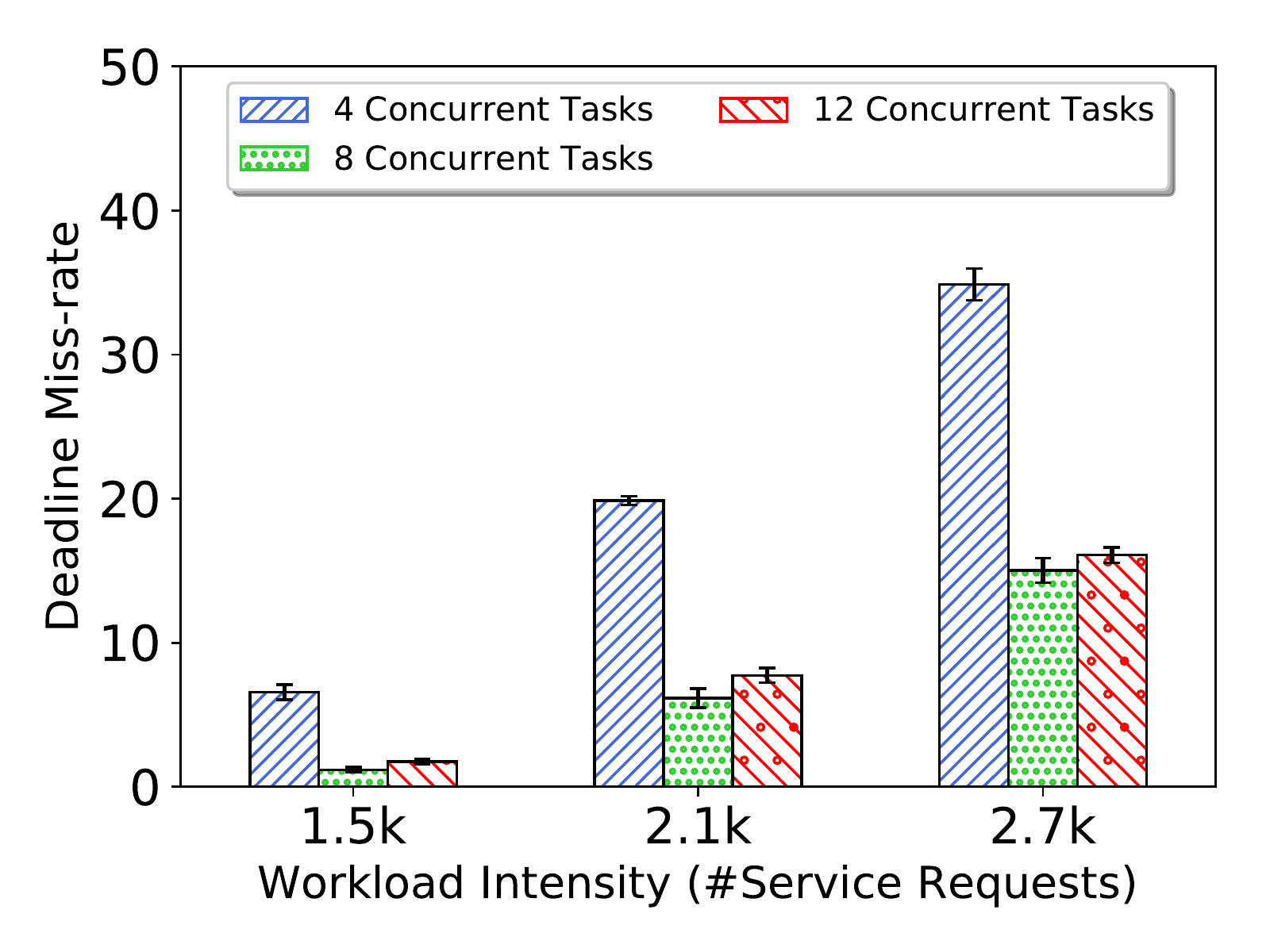} 
	\caption{Comparing the deadline miss-rate of systems with various workload intensity across three configurations. Each configuration specifically runs 4, 8, and 12 concurrent containers on a machine with sixteen CPU cores, respectively. Each container can also use two CPU cores concurrently.}
	\label{fig:DMRScale}%
\end{figure}

Figure~\ref{fig:DMRScale} shows the deadline miss-rate as we increase the number of concurrent containers and the workload intensity. We observe a significant improvement in the deadline miss-rate when the number of concurrent tasks increases from four concurrent containers (which utilize eight CPU cores) to eight concurrent containers (utilize sixteen CPU cores). From there, when increasing the degree of concurrency further to twelve concurrent containers, the deadline miss-rate rises. This is due to two reasons. First, twelve concurrent containers can use up to twenty-four CPU cores while the system only has sixteen. This increases the number of context switches. Second, media processing tasks require a significant amount of memory and storage access that can become more scarce when sharing with more concurrent task executions. %We also observe a near hundred percent memory usage in the case of executing 12 concurrent tasks together.
Based on the results, we believe the optimal performance can be achieved for the media processing execution platform when the number of expected CPU core usage is the same as the number of CPU in the system. The extra concurrency after that point can impact the performance in a negative way.

%scaleability chart
% \begin{figure}[ht]
% 	\centering
% 	\includegraphics[width=0.8\textwidth]{figures/C6/MakespanScale.pdf} 
% 	\caption{\small{Comparing the makespan of completing various workload intensity, with the amount of processing units of 4, 8, and 12 units.}}
% 	\label{fig:MakespanScale}%
% \end{figure}

\subsection{Evaluating scheduling policy.}~
\begin{figure}[ht]
	\centering
	\includegraphics[width=0.8\textwidth]{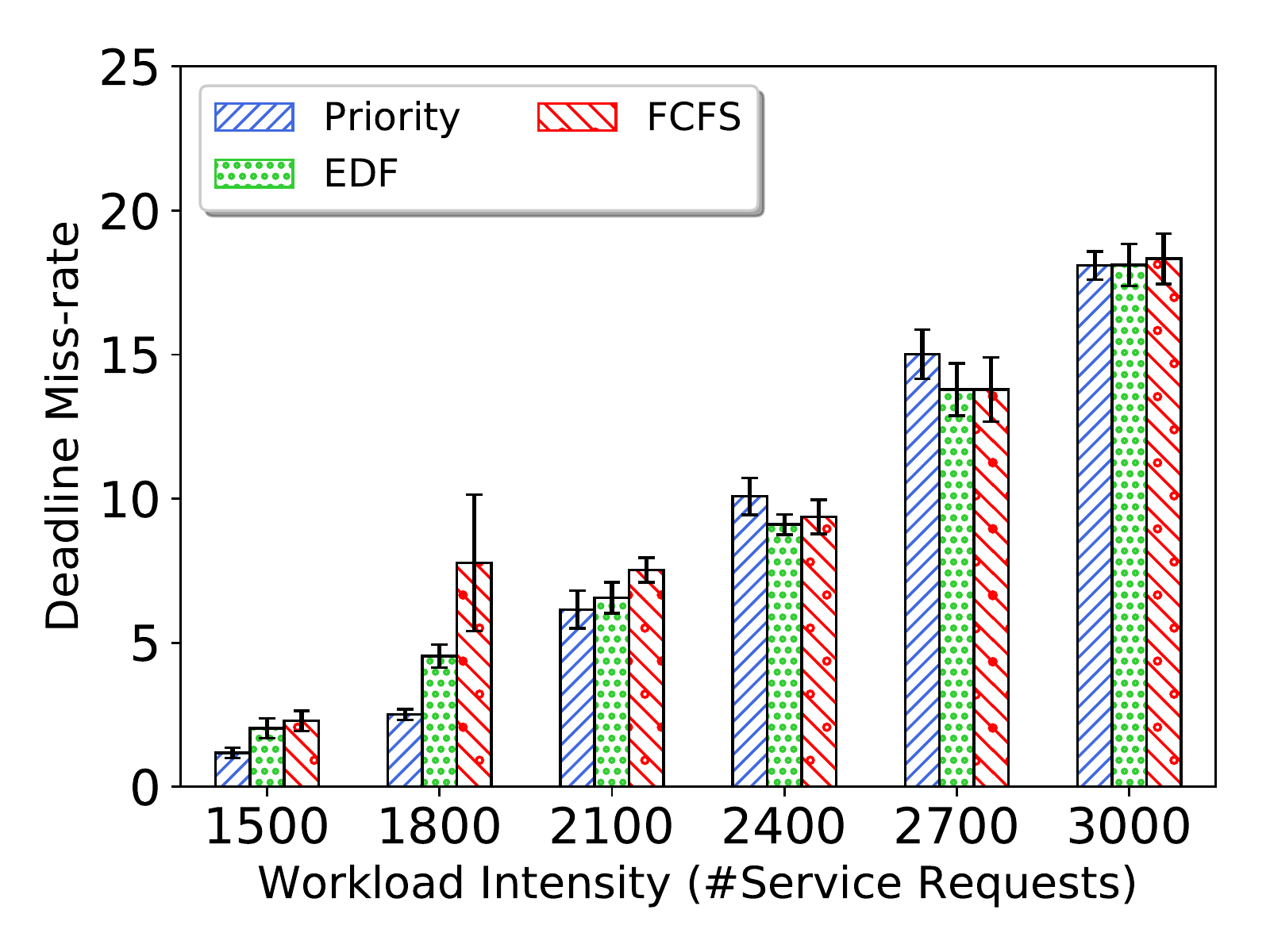} 
	\caption{Comparing the deadline miss-rate of system with various workload intensity, with various scheduling policy.}
	\label{fig:DMRPolicy}%
\end{figure}

%as the System Load Intensity Increases
In this part, we evaluate the perceived QoS of the users in the system with three built-in scheduling policies, namely Priority, First-Come-First-Serve (FCFS), and Earliest-Deadline-First (EDF). Priority-based policy is described in this Section, while other two policies are explained in detail in Section~\ref{sec:CVSEbg}\,. The task executions run on eight concurrent containers using a container reusing execution scheme, as they found to be optimal from the prior experiment. Figure~\ref{fig:DMRPolicy} shows the deadline miss-rate as the workload intensity increases. We found that the deadline miss-rate of three scheduling policies is not significantly different from each other. Therefore, we measure other factors in addition to the deadline miss-rate, namely: start-up delays and fairness.

\paragraph*{\textbf{Evaluating streaming start-up delay:}}
\begin{figure}[ht]
	\centering
	\includegraphics[width=0.8\textwidth]{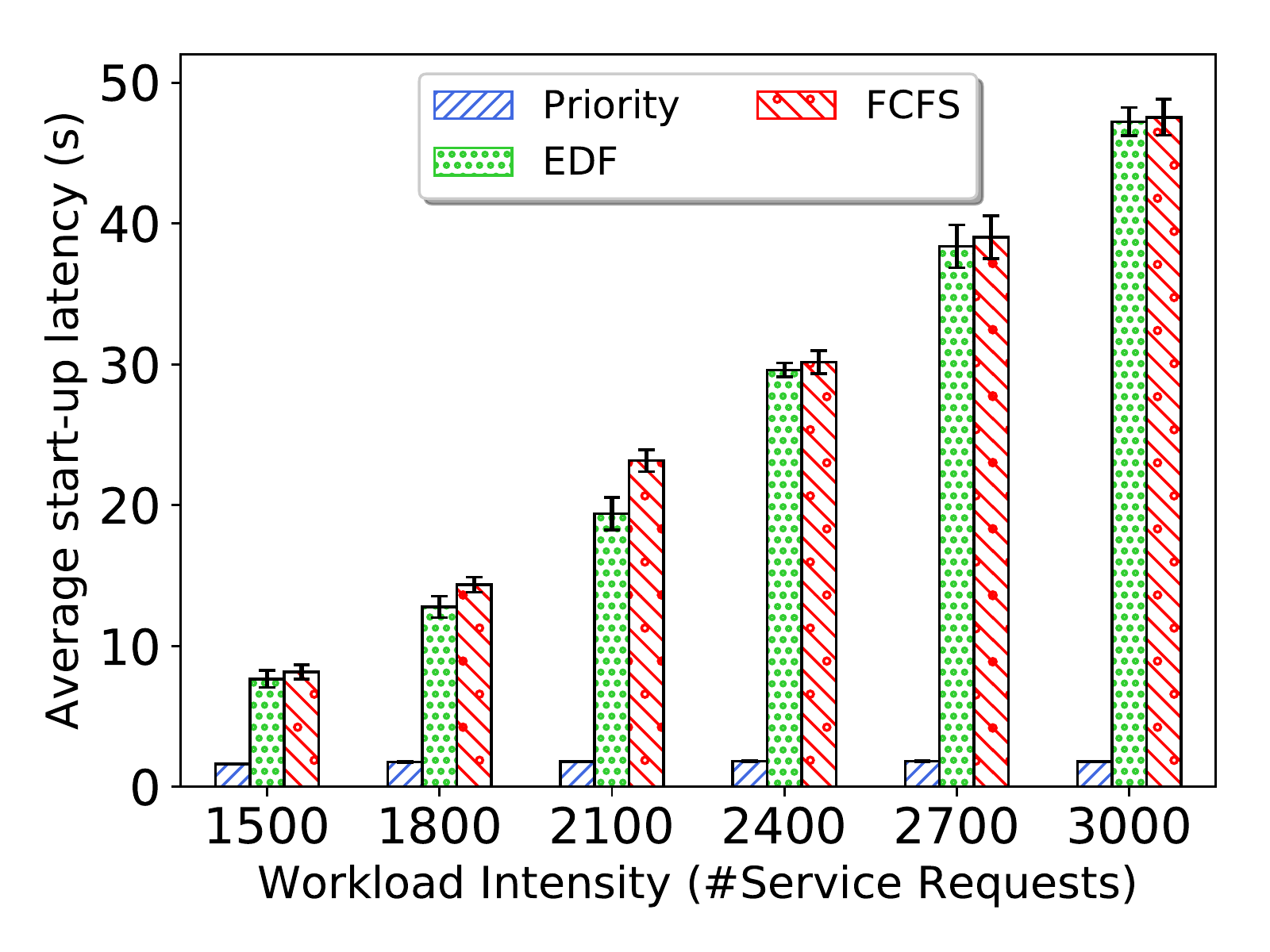} 
	\caption{Comparing the start-up delay of the first video segment of each stream between the systems with different scheduling policy.}
	\label{fig:streamstart}%
\end{figure}

To evaluate the streaming start-up delay, we measure the end-to-end time from when the stream request arrives in the system to the time that the first segment is delivered to the viewer. Figure~\ref{fig:streamstart} shows the start-up delay of systems with different task scheduling policies across various workload intensities. Note that all these workload intensities are causing oversubscription in the system. Specifically, even at 1,500 tasks arrival during the time window, the system already misses a few percentages of task deadlines (see Figure~\ref{fig:DMRPolicy}). By the nature of how each policy operates, the priority-based scheduling policy yields the lowest start-up delay by far. This is because the priority-based scheduling policy prioritizes the first few segments of each video stream over the later segments. Therefore, this policy likely results in the lowest wait time for each streaming.

%\subsubsection{DMR of Multi users.}~
\paragraph*{\textbf{Evaluating fairness across users:}}

\begin{figure}[ht]
	\centering
	\includegraphics[width=0.8\textwidth]{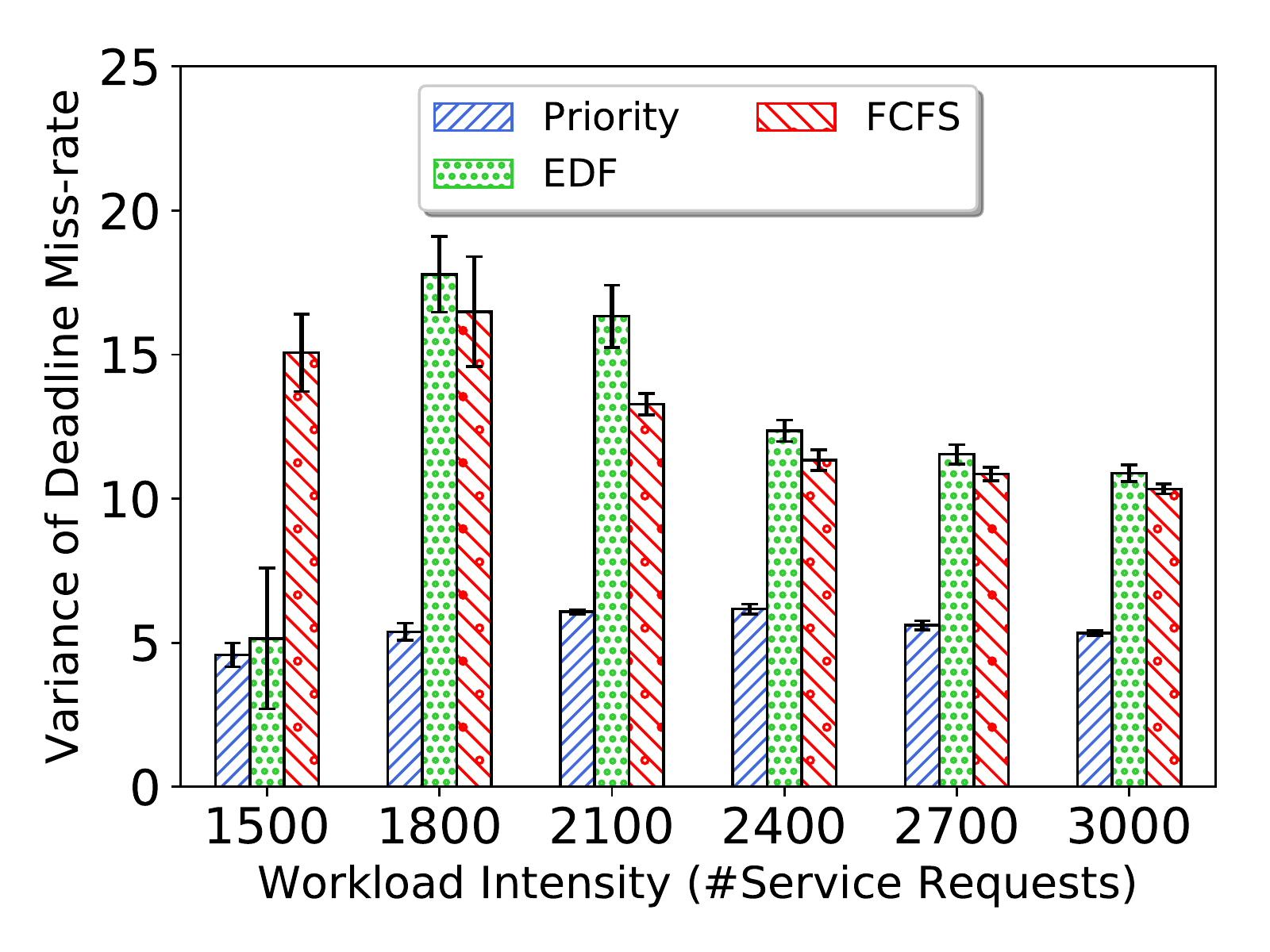} 
	\caption{Comparing the fairness among users as a result of scheduling policies. Y axis shows the suffering variation of each user measured by variance of deadline miss-rate across video streams.}
	\label{fig:C6fairness}%
\end{figure}

In this part, we assume each user to request exactly one media stream. Recalling that deadline miss-rate are similar between each of the scheduling policy, the variance of deadline miss-rate across different streams can indicate unfairness. Specifically, stream with a high deadline miss-rate get unfair treatment compare to another stream that has a lower deadline miss-rate. Based on the result in Figure~\ref{fig:C6fairness}, priority-based scheduling is also the fairest scheduling policy. It makes sure that each user can get their first few media segments in a timely manner. And then, if the system cannot keep up with the workload intensity, all the streams suffer near the end of their streaming.

\section{Summary}
In this chapter, we dealt with the problems of efficiently utilizing serverless computing resources for creating special propose computing platform. 
Specifically, we implements the prototype of the Serverless Media Stream Processing Engine that is highly extensible. The platform enables interactive streaming services through on-demand media stream processing on potentially heterogeneous cloud services in a cost-efficient manner while observing viewers' QoS guarantees. 

The prototype developed in this Chapter is the motivational application which helps evaluating the concepts in other chapters.
    \chapter{Conclusion and Future Research Directions}
\label{section:thesiscon}
\vspace{14pt}
This chapter summarizes the research and findings of this dissertation. Additionally, research topics that have emerged during the course of this research but is have not been addressed are discussed. 

\section{Discussion}
In this dissertation, our main objective goal was to investigate an efficient serverless cloud computing system. This leads to realization of potential inefficiency of processing similar or duplicated task requests and their corresponding effect to oversubscription issue. As such, two approach of dealing with oversubscription issue is proposed. 

In chapter~\ref{section:MergeSaving}, we studied the potential of reusing computation via merging similar tasks to reduce their overall execution-time in the clouds.
Considering video processing context, we built a video benchmarking dataset and evaluated the parameters that influence the merge-saving. We observed that merging similar video processing tasks can save up to 31\% (for merging two tasks) of the execution-time that implies a significant cost saving in the cloud. We also learned that the merge-saving gain becomes negligible, when degree of merging is greater than three. Then, we leveraged the collected observations to train a machine learning method based on Gradient Boosting Decision Trees (GBDT) to predict the merge-saving of unforeseen task merging cases. The fine-tuned prediction model can provide up to 93\% accurate resource saving prediction.

In chapter~\ref{section:Reusing}, we alleviate the oversubscription of the system via merging arriving requests with others (exact or similar) requests in the system. There are two challenges of this approach: First, how to identify identical and similar requests in an efficient manner? Second, how to perform (or not perform) merging to achieve the best QoS in the system? 
To address the first challenge, we identified three main levels of similarity that requests can be merged. Then, we developed a method to detect different levels of request similarity within a constant time complexity. To address the second challenge, we developed a method that determines, based on system oversubscription condition, how to perform the merge operation so that the deadlines of other requests in the system are likely least affected. Experimental results demonstrate that the proposed system can reduce the overall execution time of requests by more than 9\%. Hence, cloud resources can be deployed for a shorter time. This benefit comes with improving QoS of the users as well. We also concluded that when the level of oversubscription in the system is high, merging requests aggressively (\ie without considering the impact on other requests) helps in maintaining QoS and makes the system more robust. Conversely, with lower levels of oversubscription, merging requests should be carried out with consideration of the impact on other requests to not cause unnecessary impact the QoS.

In chapter~\ref{section:DropDefer} we remedy the oversubscription impact on QoS and the cost/energy per on-time task completion by emblace the task pruning mechanism where we prioritize and prune away tasks that have low chance of completing on time from clogging the resources that can be used on other tasks which have higher chance of on time completion. We designed a pruning mechanism as a stand-alone component of the resource allocation system. For pruning, we determined probability values to either defer or drop a task whose chance of success is low. We enabled the pruning mechanism to determine dropping threshold at the task level and dynamically adjust the deferring threshold based on the characteristics of the arriving workload. We developed a probabilistic mapping heuristic, PAM, that cooperates with the pruning mechanism. We showed that PAM can improve system robustness by on average $\simeq$22\%. We upgraded PAM to accommodate fairness by compromising around four percentage points robustness. We employed approximate computing in calculation of probabilities in the system to reduce the scheduling and pruning overheads (by up to 93\%) and ensure that the mechanism can be used practically. We concluded that: \textbf{(A)} when the system is not oversubscribed, tasks with low chance of success should be deferred (\ie wait for more favorable mapping in the next mapping); \textbf{(B)} When the system is sufficiently oversubscribed, the unlikely-to-succeed tasks must be dropped to alleviate the oversubscription and increase the probability of other tasks succeed; \textbf{(C)} The system benefits from setting higher deferring threshold than dropping threshold. Evaluation results revealed that the pruning mechanism (and PAM) not only improves system robustness but also reduces the cost and energy of using cloud-based HC systems by $\simeq$33\%.

Finally, the implementation details of a serverless-based cloud media processing platform is detailed in Chapter~\ref{section:platform}. Although this implementation of serverless computing platform is domain-specific to media processing. Most of the general design features can expand and apply to process other type of computation as well. The design goal of this platform is not only to be efficient but also modular and flexible. A wide variety of task mapping heuristics is supported. Task execution time can be modelled as normal distribution (through mean and standard deviation) or other type of distribution (through Probability Mass Function). Multiple computing resources are supported and the application interface is provided to expand the capability further. The platform is used for evaluating components in-focus of other chapter.

In conclusion, an efficient serverless computing platform that is capable of merging similar task requests and prune tasks to maintain QoS in the event of oversubscription can be particularly useful to serve as a foundation to the future generation of FaaS cloud computing system.

\section{Future Works}
Based on our findings during the development of a serverless computing platform. There are several points where the work could be expanded upon that were not covered in this dissertation.

\subsection{Optimal workflow formation to maximize reusability.}~
% in Task's Workflow
All the tasks in this study are independent tasks with no data dependency between each other. However, in many use cases of Serverless cloud, each FaaS task requests can be a part of the complex workflow. This is especially true when the functions are developed following CI/CD guideline. When the task dependency is considered, it open up a few branches of research topic to be explored.

First, the task pruning mechanism must understand the implication of dropping or delaying a dependency task to its dependent tasks. The task pruning mechanism can no longer consider all tasks to be equally important and can drop them independently.

Second, to reach the final output of each workflow, there can be multiple configurations of the intermediate steps along the workflow. For example to get a censored video in a lower resolution, the video can change its resolution first before applying censorship later or vice versa where we apply censorship first then changing resolution. In each configuration, the resulting execution time and the resusability of the intermediate data varies. Task mapping heuristics should dynamically optimize task's workflow, by scheduling the order of functions in a way that the total execution time is minimized and reusability of the result in the intermediate steps before reaching the final step is enhanced.

\subsection{Smart prediction of the merge-saving for new functions and/or new machine types.}~
In Chapter~\ref{section:MergeSaving}, we found the resource saving of Codec changing operation to scale different to those of VIC operations as the degree of merging increase. On top of that, some of the Codec changing operations also behave different to the others. We manually categorize those operations that behave similarly together as a group. However, a systematic and automatic test and categorization process for an unknown operation should be studied. Such that, the Merge-Saving Predictor can perform a few tests on an unknown operation to categorize the operation in a group which share the same behavior to ease the prediction.

Similar categorization and reusing of similar machine's data can be explored to create the execution profile data of new machine type also.

\subsection{Fairness for domain-specific functions.}~
In Chapter~\ref{section:DropDefer}, we augmented pruning-aware mapping heuristics with a fairness module. Such fairness module tries to balance the number of completion among task types and or users. 
However, just the equalizing the number of task completion or deadline-miss rate still considered unfair. As each domain have various QoS concerns, domain-specific fairness models should be explored.

% \subsection{User-Defined Task Prioritization and Optimization Goal}
% Our study in Chapter~\ref{section:DropDefer} assume all tasks to have the same priority and give the same utility once completed, Chapter~\ref{section:Reusing} study some task request mapping heuristics that work based on simplified priority calculation based on their deadline and execution time. In many systems, both task priority and utility function can vary. For example, a large public serverless cloud computing platform use data throughput and cost margin as their optimization goal and the main factor in task utility calculation of each task. While another smaller private computing platform can emphasize the completion of task with high priority first, then executing tasks with lower priority as the energy budget allows. Task mapping heuristics, pruning mechanism and merging mechanism should be further refined to take user-defined task priority into their consideration. The utility gain of each task and optimization goal should also be customizable.

\subsection{Semantic deduplication in media streaming.}~
Our motivational application is the interactive media streaming system which process the media to the user's specification upon request. This is done to avoid storing multiple redundant versions of the same media in multiple configurations. However, while we can avoid pre-processing and storing multiple pre-transcoded versions of the same media, the users can still upload redundant media of the same content to the system. An example of such case is the video of popular football match. Many users upload the exact same video to their channels creating duplicated content. Alternatively, they can also voice over with their comments on the same video creating a similar video (same pictures, different audio). A storage system that is capable of deduplicating the same media content and approximately store similar content can reduce the redundancy in the system further.

\subsection{Semantic request approximation.}~
Consider a task request that want a video segment to be transcoded from their original 4K resolution to be 1080p in an oversubscribed system. As the system is oversubscribed, the task cannot be completed in a timely manner. Chapter~\ref{section:Reusing} allow the task request to combine with other similar request (\eg same video segment in 720p) to form a compound request which still can also miss its deadline if the saving from request merging is not significant enough. Chapter~\ref{section:DropDefer} can drop the task if it is considered infeasible and then supply the user with a baseline minimal version of the result (\eg same video segment in 240p). Between the two approach there is a third unexplored alternative.
 
Providing that there is a similar task request already scheduled in the system. If the request merging with accurate request specification does not yield a timely completable request. Instead of dropping such request and supply the user with a bare minimal result, the system can explore with the options of marginally compromising requests' specification accuracy (substitute some request parameters with similar value) to enable more computational sharing with other existing requests. Such specification compromising can allow the the user to get a less than ideal result that is still better than the bare minimum version of the result if the task request is dropped.

\subsection{Merge-aware task scheduler for under-loaded systems.}~
The proposed task merging system in for serverless cloud assume the system to be oversubscribed. Thus focussing on resolving QoS issue that caused by oversubscription. However, in a case when the system is not oversubscribed (\ie resources are highly scalable), the system that emphasize on completing as many tasks as soon as possible can miss out on potential cost saving merges. For instance, if an arriving task is scheduled to execute too early, it cannot be merged with a soon to arrive tasks. In contrast, delaying the execution to the latest possible moment to still catch the QoS increase the chance of it to be merge with others. We believe a merge-aware task scheduler can increase the cost efficiency of executing tasks on serverless system.

\subsection{Failure recovery in serverless computing systems.}~
Fault tolerant in most computing system can be achieved through check-pointing and restoring of the faulted computing resource. However, in serverless computing paradigm, the exact restoration of the failed hardware is not required. Also, each tasks is generally run for a short time. And specifically in our case, task does not required to take the exact same step to complete its execution. For example, a compound (merged) task in a failed machine's scheduling queue can be re-issued as separate tasks (or each of them combine with other tasks) on different machines to complete within the deadline.

%In the future, we plan to extend this work by considering the impact of using heterogeneous computing resources in the system. 

\bibliographystyle{amsalpha}
\bibliography{A-main-thesis-file}

%\appendix
%   \include{G-append1}
%   \include{G-append2}

\closingpages
    \setlinespacing{1}
 \begin{description}
    \setlength{\parsep}{-3pt}
    \setlength{\itemsep}{-1.7pt}
    \setlength{\parindent}{0pt}
    \setlength{\labelsep}{-5pt}
    \item Denninnart, Chavit.$\ $ 
                Bachelor of Engineering, Kasetsart University, 2013;
                Master of Science, University of Louisiana at Lafayette,
                2015;
                Doctor of Philosophy, University of Louisiana at Lafayette,
                2020
    \item Major:$\ $                  \Major
    \item Title of Dissertation:$\ $  \Title
    \item Dissertation Director:$\ $  \Supervisor
    \item Pages in Dissertation:$\ $  217;  Words in Abstract:$\ $  333
 \end{description}
\setlinespacing{1}
 \begin{theabstract}
 	\indent Cloud-based serverless computing systems, either public or privately provisioned, aim to provide the illusion of infinite resources and abstract users from details of the allocation decisions.

With the goal of providing a low cost and a high QoS, the serverless computing paradigm offers opportunities that can be harnessed to attain the goals. Specifically, our strategy in this dissertation is to avoid redundant computing, in cases where independent task requests are similar to each other and for tasks that are pointless to process. We explore two main approaches to (A) reuse part of computation needed to process the services and (B) proactively pruning tasks with a low chance of success to improve the overall QoS of the system. 

For the first approach, we propose a mechanism to identify various types of ``mergeable'' tasks, which can benefit from computational reuse if they are executed together as a group. To evaluate the task merging configurations extensively, we quantify the resource-saving magnitude and then leveraging the experimental data to create a resource-saving predictor.   
We investigate multiple tasks merging approaches that suit different workload scenarios to determine when it is appropriate to aggregate tasks and how to allocate them so that the QoS of other tasks is minimally affected.
      
For the second approach, we developed the mechanisms to skip tasks whose chance of completing on time is not worth pursuing by drop or defer them. 
We determined the minimum chance of success thresholds for tasks to pass to get scheduled and executed. We dynamically adjust such thresholds based on multiple characteristics of the arriving workload and the system's conditions. 
We employed approximate computing to reduce the pruning mechanism's computational overheads and ensure that the mechanism can be used practically. The developed task pruning components are modular and are compatible with various scheduling policies.
      
We develop both approaches in the context of a serverless on-demand media processing platform. 
However, the ideas and approaches developed in this dissertation are generic and can be applied to serverless systems in other contexts.

 \end{theabstract}

    % Note: For proper spacing, the {} is necessary after the \Author command.
\begin{biosketch}
   \indent \Author{} received his
   Bachelor of computer engineering in the spring of 2013 from Kasetsart University in Thailand. He immediately began his pursuit of a master's degree in computer science in the fall of 2013 at the University of Louisiana at Lafayette. He completed it in 2015, and continued further to complete his doctorate in computer science in 2020.  
   His research interests are: serverless cloud computing, cloud-based video streaming, task scheduling, approximate computing, and computational reuse. Denninnart was also a member of Ragin Cajun Cycling Team with some podium results.
\end{biosketch}

\end{document}